\documentclass[12pt]{article}
\usepackage{a4wide,epsfig}
\usepackage{latexsym}
\renewcommand{\thefootnote}{\fnsymbol{footnote}}
\renewcommand {\theequation}{\arabic{section}.\arabic{equation}}
\renewcommand {\thefigure}{\arabic{section}.\arabic{figure}}
\renewcommand {\thetable}{\arabic{section}.\arabic{table}}
\newcommand{\li}{\mathop{{\mbox{Li}}_4}\nolimits}

\newcommand{\gsim}{\;\rlap{\lower 3.5 pt \hbox{$\mathchar \sim$}} \raise 1pt
 \hbox {$>$}\;}
\newcommand{\lsim}{\;\rlap{\lower 3.5 pt \hbox{$\mathchar \sim$}} \raise 1pt
 \hbox {$<$}\;}

\newcommand{\qsla}{q\hspace{-.5em}/\hspace{.2em}}
\newcommand{\psla}{p\hspace{-.5em}/\hspace{.2em}}

\newcommand{\lmM}{l_{\mu M}}

\newcommand{\smM}{\mbox{\small{\it M}}}

\newcommand{\logqmms}{l_{qm}}

\newcommand{\logmsms}{l_{ms}}
\newcommand{\lMs}{L_{ms}}

\newcommand{\logqmums}{l_{q\mu}}
\newcommand{\logmum}{l_{\mu m}}
\newcommand{\Lw}{L_\omega}

\begin{document}    

\title{\vskip-3cm{\baselineskip14pt
\centerline{\normalsize\hfill DESY 02--004}
\centerline{\normalsize\hfill hep-ph/0201075}
}
\vskip.7cm
Results and Techniques of Multi-Loop Calculations
\vskip2em
}
\author{
  {Matthias Steinhauser}
  \\[3em]
  { II. Institut f\"ur Theoretische Physik,}\\ 
  { Universit\"at Hamburg, D-22761 Hamburg, Germany}
}
\date{}
\maketitle

\begin{abstract}
\noindent
In this review some recent multi-loop results obtained in the
framework of perturbative Quantum Chromodynamics (QCD) and 
Quantum Electrodynamics (QED) are discussed.
After reviewing the most advanced techniques used for the computation
of renormalization group functions, we consider the decoupling of heavy
quarks. In particular, an effective method for
the evaluation of the decoupling constants is presented and explicit
results are given.
Furthermore the connection to observables involving a scalar Higgs boson is
worked out in detail. An all-order low energy theorem is derived which
establishes a relation between the coefficient functions in the hadronic
Higgs decay and the decoupling constants. We review the radiative
corrections of a Higgs boson into gluons and quarks and present
explicit results up to order $\alpha_s^4$ and $\alpha_s^3$,
respectively.
In this review special emphasis is put on the applications of asymptotic
expansions. A method is described which combines expansion terms of
different kinematical regions with the help of conformal mapping and
Pad\'e approximation.
This method allows us to proceed beyond the present scope of exact
multi-loop calculations.
As far as physical processes are concerned, we review the computation of
three-loop current correlators in QCD taking into account 
the full mass-dependence. In particular, we concentrate on the evaluation
of the total cross section for the production of hadrons in $e^+e^-$
annihilation. The knowledge of the complete mass dependence at order
$\alpha_s^2$ has triggered a bunch of theory-driven analyses of the
hadronic contribution to the electromagnetic coupling evaluated at
high energy scales. The status is summarized in this review.
In a further application four-loop diagrams are considered which
contribute to the order $\alpha^2$ QED corrections to the $\mu$ decay.
Its relevance for the determination of the 
Fermi constant $G_F$ is discussed. Finally the
calculation of the three-loop relation between the 
$\overline{\rm MS}$ and on-shell quark mass definitions is presented
and physical applications are given.
To complete the presentation, some technical details are presented in
the Appendix, where also explicit analytical results are listed.
\end{abstract}

\vspace{2em}

\centerline{(To appear in Physics Reports)}

\thispagestyle{empty}
\newpage

\setcounter{page}{1}

\tableofcontents

\renewcommand{\thefootnote}{\arabic{footnote}}
\setcounter{footnote}{0}


\section{Introduction}
\setcounter{equation}{0} 
\setcounter{figure}{0} 
\setcounter{table}{0} 

Nowadays the Standard Model (SM) of elementary particle physics is
well established. Some parts of it (e.g. the properties of the $Z$
boson) have been tested to an accuracy
far below the percent level --- mostly at the CERN Large-Electron-Positron
collider (LEP, Geneva), at the SLAC Linear Collider (SLC,
Stanford) and at the Fermilab TEVATRON (Chicago). Up to now
no significant 
deviation between theory and experiment has been found.
For other parts of the SM, related to CP violation and quark mixing, the
$B$ factories like BaBar at SLAC, Belle at KEK (Tsukuba) or HERA-B at DESY
(Hamburg) will provide deeper insight, and significant improvements in
the determination of the corresponding parameters will be obtained.
Currently mainly the scalar sector of the SM
eludes from direct experimental observation. This affects 
both the generation of the particle masses and the existence of the
Higgs boson itself.
Probably
Run II of the TEVATRON and certainly the Large Hadron Collider (LHC) at
CERN will provide more sureness --- not only in connection to the
Higgs sector but also to possible extensions of the SM.
Once the Higgs boson would be discovered it immediately would become
subject to precision measurements. In particular at a future $e^+e^-$
linear collider such as DESY TESLA, 
a precise study of its properties would be possible.

In the recent years there has been an enormous development in the
evaluation of radiative corrections. 
It is fair to say that the major part of it was initiated by the
fundamental works of 't~Hooft and Veltman in 
1972~\cite{'tHooft:1972fi,'tHooft:1972ue,'tHooft:1979xw}
where dimensional regularization 
(see also~\cite{BolGia72})
was established as a powerful tool in
the evaluation of multi-loop 
diagrams\footnote{Dimensional regularization
applied to infra-red divergences and mass singularities has first
been considered in~\cite{DRirmass}.}.
Since that time a whole industry has been formed to develop techniques
for the computation of complicated Feynman integrals. 
At one-loop order the procedure of the computation has been
systematically studied quite some time
ago~\cite{Sirlin:1980nh,Passarino:1979jh,Hollik:1990ii,Denner:1993kt}.
Nevertheless also nowadays it is not completely
straightforward to evaluate an
arbitrary one-loop diagram --- in particular if many legs and
complicated momentum configurations are involved.
One can easily imagine that at two and more loops one arrives quite
soon at the limit where the occuring mathematical expressions
can not be solved.
At two-loop order certain classes of diagrams can still be treated
by either 
using a combination of analytical simplifications and fast numerical
routines, like in the case of two-point function with several non-zero
masses~\cite{Weiglein:1994hd},
or applying purely analytical methods,
like in the case of massless digrams with four external
legs~\cite{Gehrmann:2001ih}.
However, at three-loop order it is essentially only 
possible to solve one-scale integrals.
A systematic study at four or more loops is still missing.

QCD, the field theoretical realization of the strong interaction,
constitutes an important part of the SM and also of most of its 
extensions. At low energies the coupling
constant of QCD, $\alpha_s$, is large and perturbative calculations
are not possible. However, due to the phenomenon of asymptotic freedom
the value of $\alpha_s$ gets smaller with raising energy and
perturbation theory is an appropriate tool to evaluate radiative
corrections.

Up to now the vast majority of the multi-loop calculations have been
performed in the framework of QED and QCD.
One reason is certainly that the calculations are simpler as compared
to the full SM since there are less
parameters. Furthermore, there is a strong hierarchy both in the 
quark and lepton masses which also simplifies the calculations.
On the other hand, the higher-order corrections are indeed necessary.
In QED there exist precise experiments which require 
high theoretical precision and although the coupling constant is quite
small sometimes high loop orders are necessary.
For example, in the case of the anomalous
magnetic moment of the electron, four-loop corrections are needed to
match the experimental precision.
In QCD the coupling is roughly a factor of ten bigger. Nevertheless it
is often still small enough to perform a perturbative expansion.
However,
the higher order terms are significant and can not be neglected
in the cases where high precision is required.

In this work some recent developments in the calculation of multi-loop
diagrams are reviewed. 
Thereby we will mention the most important methods which
have been used in the computation of higher order quantum corrections
and explain a few selected ones in greater detail. At the same time, we
discuss the present theoretical status of important physical
quantities. In particular, the renormalization group functions in the
modified minimal subtraction scheme~\cite{tHo73,BarBurDukMut78}
($\overline{\rm MS}$) are provided up to the four-loop order. As
is well-known, four loop running must be accompanied by three-loop
matching at quark thresholds. The corresponding decoupling relations
are presented in Section~\ref{sec:dec}.

The hadronic Higgs decay is closely connected to the decoupling
relations as we will show is Section~\ref{sec:dim4}.
Parts of the quantum corrections can be computed in the
framework of an effective Lagrangian where the coefficients can be
determined from the decoupling relations. The origin of this
miraculous connection lies in the use of the dimension-four operators,
which constitute an important ingredient of the effective
Lagrangian. Another application of the dimension-four operators are the
quartic mass corrections to the cross section
$\sigma(e^+e^-\to\mbox{hadrons})$, which is also discussed in
Section~\ref{sec:dim4}.

The last issue is again picked up in Section~\ref{sec:pade}, where also
QCD corrections to the production of hadrons in $e^+e^-$ annihilation
are computed. Putting together all terms one arrives at a complete
picture up to the quartic mass corrections of order $\alpha_s^3$. The
main purposes of Section~\ref{sec:pade} are practical applications of
asymptotic expansions\footnote{See Appendix~\ref{sub:ae} for details.}. 
Besides the diagonal correlators also the
non-diagonal ones are considered.

The Fermi constant, $G_F$, the mass of the $Z$ boson, $M_Z$, and the
electromagnetic coupling constant, $\alpha$, are the best known 
parameters of the SM. $M_Z$ has been measured at LEP with an accuracy
of a few per mille to be $M_Z=91.1876\pm0.0021$~GeV. In this review we
want to discuss quantum corrections to the other two parameters.
An essential ingredient to the running of the electromagnetic coupling
from $q^2=0$ to $q^2=M_Z^2$ is provided by the cross section
$\sigma(e^+e^-\to\mbox{hadrons})$. The correction terms 
discussed earlier have been used to obtain so-called theory-driven
results. The different approaches are discussed.
As further applications we present the status of the QED corrections
to the muon decay and the relation between the $\overline{\rm MS}$ and
the on-shell quark mass.
Let us in the following discuss the individual issues in more detail.

As far as radiative corrections are concerned,
a crucial role is played by the renormalization group functions. 
In particular the functions $\beta(\alpha_s)$ and $\gamma_m(\alpha_s)$
governing the running of the coupling and the quark masses 
comprise a significant part of the higher quantum corrections.
They are in particular very important to re-sum large
logarithms to all orders in perturbation theory.
Only a few years ago the four-loop terms of order $\alpha_s^4$ have
been evaluated for $\beta(\alpha_s)$~\cite{RitVerLar97_bet} and 
$\gamma_m(\alpha_s)$~\cite{Che97_gam,LarRitVer97_gam}. The latter
has been computed by two groups using completely independent methods.
Both methods are based on the fact that the pole part of a
logarithmically divergent diagram is independent of the masses or
momenta. In~\cite{Che97_gam} this is exploited together with the 
technique of infra-red re-arrangement (IRR) in order to obtain a
factorization of the four-loop integrals into massless three-loop and
massive one-loop ones. In Refs.~\cite{RitVerLar97_bet,LarRitVer97_gam}
all lines were assigned to the same mass, and all external momenta were
set to zero, which leads to a special class of bubble diagrams. From
them only the pole parts have to be computed. 
In Section~\ref{sec:rge} we want to review both methods and 
explicitly demonstrate the way they work.

In this review
special emphasis is put on the construction of effective theories
in the framework of QCD.
In Section~\ref{sec:dec} an effective QCD Lagrangian is constructed for
the case where one of the quarks is much heavier than the others.
The construction is made explicit by specifying the relations between
the parameters in the full and effective theories.
These relations provide at the same time the well-known decoupling
constants which have to be applied in QCD every time a particle
threshold is crossed. The most prominent example for their necessity
is probably the computation of $\alpha_s(M_\tau)$ from  
$\alpha_s(M_Z)$ or vice versa. In the latter case five quarks are
active whereas in the former one only three quarks are
present in the effective QCD Lagrangian.

In Section~\ref{sec:dim4} a slightly different point of view is adopted.
Here the scalar operators of dimension four are considered
in QCD. In a first step they are used to construct an effective Lagrangian
describing the coupling of an intermediate-mass 
Higgs boson to quarks and gluons. 
The top quark is considered as heavy and manifests itself in the
coefficient functions of the effective Lagrangian. Once the latter has
been found, the imaginary part of the Higgs boson correlator in the
effective theory leads to the total decay rate. As a central result of
Section~\ref{sec:dim4} we derive a low-energy theorem which
considerably simplifies the computation of the coefficient functions
as they are related to the decoupling constants of QCD evaluated in
Section~\ref{sec:dec}. 
In Section~\ref{sub:hggbfm} the background field method is
introduced as a convenient tool for the computation of higher-order
corrections. As an example, the coefficient functions describing 
the decay of the Higgs boson into gluons is also computed in this
framework. 

In the second part of Section~\ref{sec:dim4} another important
application of the scalar dimension four operators is discussed,
namely the quartic mass corrections to the cross section
$\sigma(e^+e^-\to\mbox{hadrons})$.
Mass corrections of order $(m^2/s)^0$ and $(m^2/s)^1$ are obtained
relatively easy
as in QCD there are no non-trivial operators of dimension less than
four.
However, the quartic corrections require the
inclusion of the dimension-four operators with all their
renormalization and mixing properties. We will explain the techniques
and present results obtained recently at order $\alpha_s^3$.

The last part of this review, Section~\ref{sec:pade}, is devoted to the
discussion of some results obtained with the help of asymptotic
expansion accompanied with conformal mapping and Pad\'e
approximation.
This method has been developed in the recent years and has been applied
successfully to a number of important processes. 
The underlying idea is the following: 
only in rare cases it is possible to compute three-loop diagrams
involving more than one scale. However, if a certain hierarchy exists
between the scales it is promising to apply an asymptotic expansion.
This effectively reduces the number of scales present in the integrals
which are subsequently significantly simplified.

In particular we will discuss the corrections of order $\alpha_s^2$ 
to the photon polarization function. Its imaginary part is directly
connected to the physical quantity
$R(s)\equiv\sigma(e^+e^-\to\mbox{hadrons})/\sigma(e^+e^-\to\mu^+\mu^-)$. 
The application of our method leads to the full mass dependence.
Combining the results with the quartic corrections given in
Section~\ref{sub:as3m4}, one obtains a prediction for $R(s)$ up to
and including ${\cal O}(\alpha_s^3 m^4/s^2)$.
Only recently also the non-diagonal current correlator formed by a
massive and a massless quark has been computed. In this application
special emphasis lies on the extraction of information about the
threshold behaviour which has some relevance in the framework of
heavy-quark effective QCD.

As an application of the knowledge of $R(s)$ to high perturbative
order, we discuss the evaluation of $\alpha(M_Z^2)$.
The electromagnetic coupling is defined at vanishing momentum
transfer. However, its evolution to high energies constitutes the
dominant part of the radiative corrections to electroweak
observables. 
The accurate determination of $\alpha(M_Z^2)$ is
thus essential for any precise test of the theory. At the same time the
indirect determination of the masses of heavy, hitherto unobserved
particles, e.g.~the Higgs boson or supersymmertic
particles, depends critically on
this parameter. Of particular importance in this context is the hadronic
vacuum polarization. It is nearly as large as the leptonic contribution,
but cannot yet be computed perturbatively. However, it may be 
related through dispersion relations to the cross
section for hadron production in electron-positron annihilation.
The integrand can thus be
obtained from data, phenomenological models and/or perturbative QCD,
whenever applicable.
In Section~\ref{subsub:delal} we will discuss the developments in the
evaluation of $\alpha(M_Z^2)$ which took place in the recent two to
three years due to the knowledge of the complete mass dependence
of $R(s)$ at order $\alpha_s^2$ (cf. Section~\ref{subsub:R}).

$G_F$ is defined through the muon lifetime, and the decay
of the muon, as a purely leptonic process, is rather clean --- both 
experimentally and theoretically. The one-loop corrections of order $\alpha$
were computed more than 40 years ago~\cite{KinSir59Ber58}, whereas only 
recently the two-loop corrections of order $\alpha^2$ have been 
evaluated~\cite{RitStu99,SeiSte99}. 
The large gap in time shows that this calculation 
is highly non-trivial. The inclusion of the two-loop terms greatly
reduced the relative
theoretical error of $1.5\times 10^{-5}$ which was an estimate of the size of 
the missing corrections. The remaining error on $G_F$ now reads  
$0.9\times 10^{-5}$ and is of pure experimental nature. Upcoming experiments 
will further improve the accuracy of the muon lifetime measurement and
therefore 
the  ${\cal O}(\alpha^2)$ corrections to the muon decay are very important and 
constitute a crucial ingredient from the theoretical side. 
In Section~\ref{sub:mudec} we discuss the results obtained with the
help of asymptotic expansion.

In the SM the quark masses have still relatively big uncertainties. 
This is mainly due to the confinement property of QCD which
prevents the production of free quarks.
It is also important to have a convenient definition of the quark
mass in order to perform a comparison between theory and experiment.
Recently there has been quite some activity connected to
the precise determination of the bottom- and top-quark masses.
The bottom-quark mass is determined with the help of 
QCD sum rules where a proper mass definition helps to reduce the error.
In the case of the top quark, studies have been performed for 
an $e^+e^-$ collider with a center-of-mass energy in the 
threshold region of top-quark-pair production. An energy scan which
provides the measurement of the total production cross section
would provide an error of about 100~MeV in the top-quark mass.
Also here a special mass definition has to be employed.
In both cases the three-loop on-shell--$\overline{\rm MS}$ conversion
formula is needed in order to obtain the 
corresponding $\overline{\rm MS}$ quark mass. The latter is 
important for processes not connected to the threshold.
In Section~\ref{sub:msos} these issues are discussed in detail.


\section{\label{sec:rge}Renormalization group functions in QCD}
\setcounter{equation}{0} 
\setcounter{figure}{0} 
\setcounter{table}{0} 

In perturbative QCD 
the renormalization group functions play
a very important role.
In particular the $\beta$ and
$\gamma_m$ functions governing the running of the strong coupling and
the quark masses, respectively, are indispensable when evaluating
physical observables.
For this reason we decided to discuss the techniques used for the
computation of the four-loop contributions 
to $\beta$~\cite{RitVerLar97_bet} and
$\gamma_m$~\cite{Che97_gam,LarRitVer97_gam} in more detail.

In the $\overline{\rm MS}$ scheme the knowledge of the renormalization
group functions is equivalent to the knowledge of the corresponding
renormalization constant. Thus, in order to compute a renormalization
constant at $n$-loop order it is sufficient to evaluate the 
ultra-violet (UV) poles of $n$-loop diagrams.
Nevertheless it is often also quite useful to have also a handle
on the infra-red (IR) poles.
More details on the UV and IR structure of Feynman diagrams are given in 
Section~\ref{sub:UVIR}. Afterwards, in Sections~\ref{sub:IRR}
and~\ref{sub:massint}, two practical methods are described
which have been applied at the four-loop level. 


\subsection{\label{sub:UVIR}Ultra-violet and infra-red counterterms}

Before we want to consider explicit examples in the next two sections
the theoretical background needed for the higher-loop calculation of
renormalization group functions is introduced in this Subsection.
In particular we want to demonstrate the interplay between UV and IR
divergences in dimensional regularization
and show their connection to asymptotic expansions.
We refrain from presenting the material in a mathematical rigorous
framework, for which we refer to the original literature,
but exemplify the important points at explicit diagrams.

As the properties discussed in this section are independent of the
particle type we consider only scalar propagators
of the form
\begin{eqnarray}
  \frac{1}{M^2 - p^2}
  \,,
\end{eqnarray}
where also $M=0$ is allowed.
We furthermore assume that the integration momenta are denoted by
$k_1, k_2, \ldots$ and introduce the abbreviation
$\int_i\equiv\mu^{4-D}\int {\rm d}^D k_i/(2\pi)^D$.

\begin{figure}[t]
  \begin{center}
    \begin{tabular}{ccc}
      \leavevmode
      \epsfxsize=5.cm
      \epsffile[165 300 447 513]{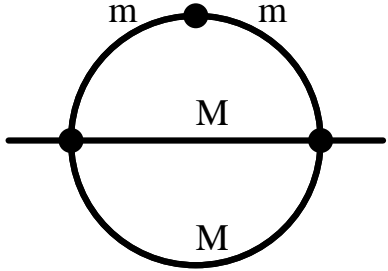}
    \end{tabular}
  \end{center}
  \caption{\label{fig:Rex}
    Scalar two-loop diagram. 
    }
\end{figure}

In order to remove the UV divergences the
so-called $R$ operation has been introduced~\cite{BogPar57,Hep66}.
It is a recursive subtraction scheme where the UV divergences are
removed from the Feynman integrals in a way compatible to adding local
counterterms to the Lagrangian.
Formally, the $R$ operation can be written as a sum where each term is
a product of operators acting on subsets of disjoint
one-particle-irreducible subgraphs. In this way counterterms are
generated. They have to be inserted into the vertices of the diagrams
which remain after shrinking the corresponding subgraphs to a point.
Non-trivial diagrammatic examples can, e.g., be found in the text
books~\cite{Collins,Muta}. 
The $R$ operation applied to a one-loop propagator-type integral 
with external momentum $q$ leads to the equation
\begin{eqnarray}
  R\left[\int_1 \frac{1}{[M^2-k_1^2][M^2-(k_1+q)^2]} \right]
  &=&
  \int_1\frac{1}{[M^2-k_1^2][M^2-(k_1+q)^2]} + Z^{(1)}
  \nonumber\\
  &=& \mbox{finite}
  \,,
  \label{eq:Z1}
\end{eqnarray}
from which the renormalization constant $Z^{(1)}$ is determined.
In the $\overline{\rm MS}$ scheme it reads
\begin{eqnarray}
  Z^{(1)} &=& \frac{1}{16\pi^2} \left( - \frac{1}{\varepsilon} \right)
  \,.
  \label{eq:Z1_2}
\end{eqnarray}
Once $Z^{(1)}$ is known one can turn to the two-loop order. The
application of the $R$ operation to the diagram shown in 
Fig.~\ref{fig:Rex} leads to
\begin{eqnarray}
   \lefteqn{
   R\left[ \int_1\int_2
           \frac{1}{[M^2-(k_1-k_2)^2][M^2-(k_1+q)^2][m^2-k_2^2]^2} \right]
   }
   \nonumber\\&=&\mbox{}
   \int_1\int_2
   \frac{1}{[M^2-(k_1-k_2)^2][M^2-(k_1+q)^2][m^2-k_2^2]^2}
   + Z^{(1)} \int_2\frac{1}{[m^2-k_2^2]^2}
   + Z^{(2)}
  \nonumber\\
  &=& \mbox{finite}
  \,,
  \label{eq:Z2}
\end{eqnarray}
which fixes $Z^{(2)}$ to
\begin{eqnarray}
  Z^{(2)} &=& \left(\frac{1}{16\pi^2}\right)^2
  \left[ \frac{1}{2\varepsilon^2} - \frac{1}{2\varepsilon} \right]
  \,.
  \label{eq:Z2_2}
\end{eqnarray}
A formal definition of the $R$ operation can, e.g., be found
in~\cite{Che91}.

There are two special features of dimensional regularization which
prove to be a powerful tool, especially in the evaluation of
renormalization group functions.
The first one is that dimensional regularization is able 
to regularize simultaneously both UV and IR divergences. 
Furthermore, as was
realized in~\cite{Col75}, in renormalization schemes based on minimal
subtraction all UV counterterms are polynomial in the
momenta\footnote{This has to be the case for each meaningful renormalization
prescription.} 
and masses.
This means that the divergence of 
logarithmically divergent Feynman integrals are polynomial in
$1/\varepsilon$ and there is no dependence on a dimensionful scale.
As a consequence, in the computation of the corresponding coefficients 
one is free to make arbitrary re-arrangements of
masses and external momenta --- provided no IR divergences are introduced.  
This was for the first time observed in Ref.~\cite{Vla80} 
(see also~\cite{CheKatTka80,CasKen82,Che91}).

In the example of Eqs.~(\ref{eq:Z1}) and~(\ref{eq:Z2}) 
this means that the masses $M$ could be set to zero without changing
the results for $Z^{(1)}$ and $Z^{(2)}$.

This procedure is often referred to as infra-red re-arrangement 
(IRR)~\cite{CheKatTka80}. One
has in mind a transformation of the IR structure of a diagram
in such a way that its UV divergent part can be easily computed.
Next to nullifying masses or momenta, also differentiations
with respect to masses are allowed in order to achieve the simplifications.

The requirement that no IR divergences may be introduced is quite
restrictive and makes the application of the IRR very tedious.
Often one has to remain with Feynman integrals which are not
as simple as one would like to have them.
To overcome this disadvantage the $R$ operation has been
generalized and the so-called $R^\star$ operations has been 
developed~\cite{CheTka82,CheSmi84}
in order to deal not only with the UV but also with the IR
divergences. Thus arbitrary re-arrangements of masses and momenta are
allowed as the $R^\star$ operation takes care of all occuring IR 
divergences.

The $R^\star$ operation can be written as 
$R^\star = R \tilde{R} = \tilde{R} R$,
where $R$ is responsible for the UV divergences and $\tilde R$
subtracts the IR ones.
For $\tilde{R}$ there exists a recursive definition which is in
close analogy to the one of the $R$ operation.
The precise definition and the criterion for the subgraphs, which have
to be considered while applying $\tilde{R}$, is quite involved and
requires the introduction of a lot of mathematical
terminology~\cite{CheSmi84}.
The way $\tilde{Z}$ --- the renormalization constant generated by
$\tilde{R}$ --- is computed in practice 
can best be seen by looking at the
examples discussed below.
At this point we only want to mention that a renormalization constant 
$\tilde{Z}$ for the whole diagram only appears for scaleless
integrals. As a consequence, the $\tilde{R}$ operation does not
commute with the limit of masses or external momenta going to zero
(cf. Eqs.~(\ref{eq:ir1}) and~(\ref{eq:tilZ2}) below).

The $R^\star$ operation is essential to prove the following
theorem~\cite{CheSmi84}:
\begin{center}
\begin{minipage}{14cm}
 {\em
 Any UV counterterm of a $(l+1)$-loop Feynman integral can be written
 in terms of the poles and finite parts of appropriately constructed
 $l$-loop massless propagator-type integrals.
 }
\end{minipage}
\end{center}
This theorem is very powerful. It states that --- at least in
principle --- all renormalization group functions at four-loop order
can be computed from the knowledge of massless three-loop two-point
functions. The latter are, e.g., provided by the package {\tt
MINCER}~\cite{mincer}.
In practice there are problems connected to the (in general) large 
number of contributing diagrams as the prescriptions 
for the $R^\star$ operation given
in~\cite{CheTka82,CheSmi84} have to be applied individually to each of
them. Thus, for practical applications further improvements are
necessary.
We will come back to this point later.

Let us consider the one-loop integral of Eq.~(\ref{eq:Z1})
with $M=0$ and $q=0$. By definition the resulting massless tadpole integral 
is set to zero in dimensional regularization. 
On the other hand, the application
of the $R^\star$ operation gives
\begin{eqnarray}
  0 &=& R^\star\left[ \int_1 \frac{1}{[-k_1^2][-k_1^2]} \right]
  \nonumber \\
    &=& \tilde{R} \left[ \int_1 \frac{1}{[-k_1^2][-k_1^2]} + Z^{(1)} \right]
  \nonumber \\
    &=& \int_1 \frac{1}{[-k_1^2][-k_1^2]} + \tilde{Z}^{(1)} + Z^{(1)}
  \nonumber \\
    &=& \tilde{Z}^{(1)} + Z^{(1)}
  \,.
  \label{eq:Z1Z1til}
\end{eqnarray}
In the first step the $R$ operation is applied resulting in the
diagram itself and the counterterm $Z^{(1)}$ rendering the expression
in the square brackets of the second line UV finite.
The subsequent application of $\tilde{R}$ generates the counterterm
$\tilde{Z}^{(1)}$ which corresponds to the IR divergence of
one-loop integral. Once the application of $R^\star$ is resolved
scaleless integrals are set to zero which leads to the last line of
Eq.~(\ref{eq:Z1Z1til}).
Using Eq.~(\ref{eq:Z1Z1til}) in combination with
Eq.~(\ref{eq:Z1_2}), one obtains for the IR
renormalization constant of the one-loop two-point function
\begin{eqnarray}
  \tilde{Z}^{(1)} &=& \frac{1}{16\pi^2} \frac{1}{\varepsilon}
  \,.
  \label{eq:tilZ1_2}
\end{eqnarray}
In analogy, applying the $R^\star$ operation to the diagram of
Fig.~\ref{fig:Rex} with all masses and external momenta set to zero
leads to
\begin{eqnarray}
  0 &=& R^\star\left[\int_1\int_2
        \frac{1}{[-k_1^2][-k_2^2]^2[-(k_1-k_2)^2]}\right]
  \nonumber\\
    &=& \tilde{R}\left[\int_1\int_2
        \frac{1}{[-k_1^2][-k_2^2]^2[-(k_1-k_2)^2]}
        + Z^{(1)}\int_2\frac{1}{[-k_2^2]^2}
        + Z^{(2)}\right]
  \nonumber\\
    &=& \tilde{Z}^{(2)} + Z^{(1)} \tilde{Z}^{(1)} + Z^{(2)}
  \,.
  \label{eq:ir1}
\end{eqnarray}
This equation defines the IR renormalization constant $\tilde{Z}^{(2)}$.

Let us come back to the two-loop diagram of
Fig.~\ref{fig:Rex} with $m=0$. In this limit the diagram contains both
an IR and UV divergent subdiagram.
The application of $R^\star$ generates the following terms
\begin{eqnarray}
   \lefteqn{
   R^\star\left[\int_1\int_2 
   \frac{1}{[M^2-(k_1-k_2)^2][M^2-(k_1+q)^2][-k_2^2]^2} \right]
   }
   \nonumber\\&=&\mbox{}
   \tilde{R}\left[
     \int_1\int_2
     \frac{1}{[M^2-(k_1-k_2)^2][M^2-(k_1+q)^2][-k_2^2]^2}
     + Z^{(1)} \int_2\frac{1}{[-k_2^2]^2}
     + Z^{(2)}
   \right]
   \nonumber\\&=&\mbox{}
   \int_1\int_2
   \frac{1}{[M^2-(k_1-k_2)^2][M^2-(k_1+q)^2][-k_2^2]^2}
   + \tilde{Z}^{(1)} \int_1\frac{1}{[M^2-k_1^2][M^2-(k_1+q)^2]}
   \nonumber\\&&\mbox{}
   + Z^{(1)} \tilde{Z}^{(1)}
   + Z^{(2)}
  \nonumber\\
  &=&
  \left(\frac{1}{16\pi^2}\right)^2
  \Bigg\{
  \left[ -\frac{1}{2\varepsilon^2} 
         -\frac{1}{2\varepsilon}\left(-1+2\ln\frac{\mu^2}{M^2}\right)
  \right]
  +
   \frac{1}{\varepsilon}
   \left(\frac{1}{\varepsilon}+\ln\frac{\mu^2}{M^2}\right) 
  \nonumber\\&&\mbox{}
  +
   \left(-\frac{1}{\varepsilon}\right)
   \frac{1}{\varepsilon}
  +\left[ 
   \frac{1}{2\varepsilon^2} 
   - \frac{1}{2\varepsilon} 
   \right]
  + \mbox{finite terms}
  \Bigg\}
  \nonumber\\
  &=& \mbox{finite}
  \,,
  \label{eq:tilZ2}
\end{eqnarray}
where $\tilde{R}$ after the second equal sign acts on the IR divergent 
integral $\int_2 1/[-k_2^2]^2$ and generates the factors
$\tilde{Z}^{(1)}$. Note that no term $\tilde{Z}^{(2)}$ appears as the
original integral involves the scales $M$ and $q$.
After the third equal sign the results of
Eqs.~(\ref{eq:Z1_2}),~(\ref{eq:Z2_2}),~(\ref{eq:tilZ1_2})
and of the Appendix (\ref{eq:Pab})
have been used in order to explicitly show the finiteness.
In general the logic is the other way around: one chooses the masses
and external momenta in such a way that the very diagram, i.e. the first
term in the second line of
Eq.~(\ref{eq:tilZ2}), can easily be evaluated and uses
Eq.~(\ref{eq:tilZ2}) in order to determine $Z^{(2)}$.

At the end of this Subsection we want to work out the connection of
the $R^\star$ operation to the asymptotic expansion with respect to
large masses.
The hard-mass procedure provides a prescription on how to 
evaluate Feynman integrals where one of the internal masses is much
larger than the others (cf. Appendix~\ref{sub:ae}). 
Thus one can adopt the point-of-view to 
introduce light masses which regularize all IR divergences 
and apply the hard-mass procedure. 
For illustration let us consider the diagram in Fig.~\ref{fig:Rex}
with one of the propagators carrying mass $M$ doubled in order to
avoid UV divergences which keeps the formula more transparent. 
For $m\not=0$ the diagram is also IR finite.
Applying the hard-mass procedure in the limit
$M^2\gg m^2,q^2$ and keeping only the leading terms in $m^2/M^2$ and
$q^2/M^2$ leads to
\begin{eqnarray}
   \lefteqn{
   \int_1\int_2
           \frac{1}{[M^2-(k_1-k_2)^2][M^2-(k_1+q)^2]^2[m^2-k_2^2]^2}
   }
   \nonumber\\&=&\mbox{}
   \int_1\int_2
   \frac{1}{[M^2-(k_1-k_2)^2][M^2-k_1^2]^2[-k_2^2]^2}
   +\int_1\frac{1}{[M^2-k_1^2]^3}
    \int_2\frac{1}{[m^2-k_2^2]^2}
  +\ldots
  \,,
  \nonumber\\
  \label{eq:Z2hmp_1}
\end{eqnarray}
where the ellipses represent terms of order $q^2/M^2$ and $m^2/M^2$.
The general rules and explicit examples for the hard-mass procedure are
discussed in Appendix~\ref{sub:ae}. The application of
Eq.~(\ref{eqasexp}) to the two-loop diagram at hand leads to the two
contributions which are listed in the second line of
Eq.~(\ref{eq:Z2hmp_1}). The first term corresponds to the
whole diagram which according to the rules of the hard-mass procedure
has to be expanded in the small quantities $q$ and $m$. To our
approximation this means to simply nullify both $q$ and $m$. In the
second contribution the hard subgraph consists of the one-loop
subdiagram where all lines carry the heavy mass $M$. The expansion in
the external momenta leads to $\int_1 1/[M^2-k_1^2]^3$ which finally
leads to the second term on the right-hand side of Eq.~(\ref{eq:Z2hmp_1}).
Note that the first integral is IR divergent whereas the second one is
UV divergent. Their connection to the $R$ and $\tilde{R}$ operation
becomes clear after adding and subtracting the term
$Z^{(1)}\int_1 1/[M^2-k_1^2]^3$
\begin{eqnarray}
   \mbox{Eq.~(\ref{eq:Z2hmp_1})}
   &=&
   \int_1\int_2
   \frac{1}{[M^2-(k_1-k_2)^2][M^2-k_1^2]^2[-k_2^2]^2}
   -\int_1\frac{1}{[M^2-k_1^2]^3}Z^{(1)}
  \nonumber\\&&\mbox{}
   +\int_1\frac{1}{[M^2-k_1^2]^3}
    \left(\int_2\frac{1}{[m^2-k_2^2]^2}+Z^{(1)}\right)
  +\ldots
  \nonumber\\  
  &=&
  \tilde{R}\left[   
   \int_1\int_2
   \frac{1}{[M^2-(k_1-k_2)^2][M^2-k_1^2]^2[-k_2^2]^2}
  \right]
  \nonumber\\&&\mbox{}
   +\int_1\frac{1}{[M^2-k_1^2]^3}
   R\left[ \int_2\frac{1}{[m^2-k_2^2]^2} \right]
  +\ldots
  \,,
  \label{eq:Z2hmp}
\end{eqnarray}
where in the last step Eq.~(\ref{eq:Z1Z1til}) has been used.
In Eq.~(\ref{eq:Z2hmp}) the correspondence between the IR and UV
divergences,
which are introduced through the hard-mass procedure, 
can nicely be observed.
Furthermore, it can be seen how they have to be combined in order to
arrive at the final form which contains the application of the $R$ and
$\tilde{R}$ operations.


\subsection{\label{sub:IRR}Global infra-red re-arrangement
  and the quark anomalous dimension}

In the previous Subsection it has been demonstrated that the 
IRR in connection with the $R^\star$ operation provides a very
powerful tool for the computation of renormalization group functions
at higher orders. In this Subsection we want to discuss its
practical application in the case of the
quark anomalous dimension $\gamma_m$.

In the $\overline{\rm MS}$ scheme the running of the quark masses is governed
by the function $\gamma_{m}(\alpha_s)$
\begin{eqnarray}
  \mu^2\,\frac{d}{d\mu^2}m^{(n_f)}(\mu)
  &=&
  m^{(n_f)}(\mu)\,\gamma_m^{(n_f)}\left(\alpha_s^{(n_f)}\right) 
  \,\,=\,\,
  -m^{(n_f)}(\mu)\,\sum_{i\ge0} \gamma_{m,i}^{(n_f)}
  \left(\frac{\alpha_s^{(n_f)}(\mu)}{\pi}\right)^{i+1}
  \,,
  \nonumber\\
  \label{eq:defgamma}
\end{eqnarray}
where the coefficients $\gamma_{m,i}$ are known up to the four-loop
order~\cite{Tar81,Tar82,Larin:massQCD,Che97_gam,LarRitVer97_gam}
\begin{eqnarray}
  \gamma_{m,0}^{(n_f)} &=& 1\,,
  \nonumber\\
  \gamma_{m,1}^{(n_f)} &=& \frac{1}{16}\left[ \frac{202}{3}
    - \frac{20}{9} n_f \right]\,,       
  \nonumber \\
  \gamma_{m,2}^{(n_f)} &=& \frac{1}{64} \left[1249+\left( - \frac{2216}{27} 
      - \frac{160}{3}\zeta_3 \right)n_f 
    - \frac{140}{81} n_f^2 \right]\,,
  \nonumber \\
  \gamma_{m,3}^{(n_f)} &=& \frac{1}{256} \left[ 
    \frac{4603055}{162} + \frac{135680}{27}\zeta_3 - 8800\zeta_5
    +\left(- \frac{91723}{27} - \frac{34192}{9}\zeta_3 
      + 880\zeta_4 
    \right.\right.
  \nonumber \\
  &&{}+ \left.\left.
      \frac{18400}{9}\zeta_5 \right) n_f
    +\left( \frac{5242}{243} + \frac{800}{9}\zeta_3 
      - \frac{160}{3}\zeta_4 \right) n_f^2
  \right.
  \nonumber \\&& \left.\mbox{}
    +\left(- \frac{332}{243} + \frac{64}{27}\zeta_3 \right) n_f^3 \right]
  \,,
\end{eqnarray}
with $\zeta_3\approx1.202\,057$,
$\zeta_4=\pi^4/90$ and $\zeta_5\approx1.036\,928$.
The superscript $n_f$ indicates the dependence on the number of quarks.

For the computation of $\gamma_m$ one has to know the quark mass
renormalization constant in the $\overline{\rm MS}$ scheme,
$Z_m$, which relates the bare mass, $m^0$, to
the renormalized one through
\begin{eqnarray}
  m^0 &=& Z_m m
  \,.
  \label{eq:Zmdef}
\end{eqnarray}
$Z_m$ can be obtained form the vector and scalar parts of the quark
propagator. Its inverse reads in bare form
\begin{eqnarray}
  \left( S_F^0(q) \right)^{-1} &=& 
  i \left[ m^0 - \qsla - \Sigma^0(q) \right]
  \nonumber\\
  &=&
  i \left[ m^0 \left( 1 - \Sigma_S^0 \right) 
  - \qsla \left( 1 + \Sigma_V^0 \right)\right]
\label{eq:sfinv0}
\,,
\end{eqnarray}
where the functions $\Sigma_S^0$ and $\Sigma_V^0$
depend on the external momentum $q$, the bare mass $m^0$ and on the bare
strong coupling constant $\alpha_s^0$.
From the requirement that the renormalized quark propagator is finite
one gets in the $\overline{\rm MS}$ scheme
the following two equations 
\begin{eqnarray}
  Z_2 &=& 1 - K_\varepsilon\left[ Z_2 \Sigma_V^0 \right]
  \,,
  \nonumber\\
  Z_2Z_m &=& 1 + K_\varepsilon\left[ Z_2 Z_m \Sigma_S^0 \right]
  \,,
  \label{eq:Z2Zm}
\end{eqnarray} 
where $Z_2$ is the wave function renormalization of the quark.
The operator $K_\varepsilon$ extracts the poles in $1/\varepsilon$.
The Eqs.~(\ref{eq:Z2Zm}) are solved recursively for $Z_2$ and $Z_m$,
and $\gamma_m$ is computed from Eq.~(\ref{eq:defgamma}) using
(\ref{eq:Zmdef}) and the fact that $\mu^2 {\rm d} m_0/{\rm d}\mu^2 = 0$.
This leads to
\begin{eqnarray}
  \gamma_m &=& - \mu^2 \frac{{\rm d}}{{\rm d}\mu^2} \ln Z_m
  \nonumber\\
           &=& -\beta^\varepsilon(\alpha_s)
        \pi\frac{\partial}{\partial\alpha_s} \ln Z_m
  \,,
\end{eqnarray}
where $\beta^\varepsilon(\alpha_s)=-\varepsilon+\beta(\alpha_s)$ 
is the $D$-dimensional $\beta$ function and $\beta(\alpha_s)$ is defined
below in Eq.~(\ref{eq:defbeta}).

\begin{figure}[t]
  \begin{center}
    \begin{tabular}{c}
      \leavevmode
      \epsfxsize=14.cm
      \epsffile[57 425 530 740]{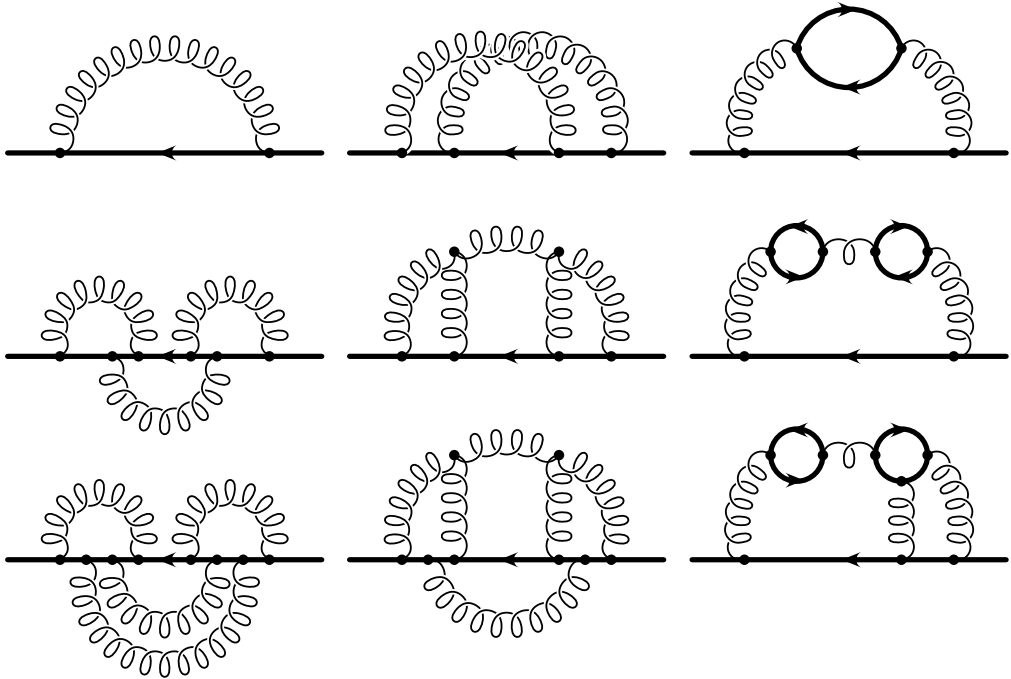}
    \end{tabular}
  \end{center}
  \caption{\label{fig:fpdiags}
    Sample diagrams contributing to the fermion propagator.
    }
\end{figure}

For the computation in the $\overline{\rm MS}$ scheme
one has to evaluate the pole part
of the fermion propagator. Some sample diagrams 
contributing at one, two, three and four loops
are shown in Fig.~\ref{fig:fpdiags}.
As the diagrams contributing to
$\Sigma_S^0$ and $\Sigma_V^0$ are logarithmically divergent
the computation can be performed by setting all internal
masses to zero and keeping the external momentum finite.
This leads to massless $l$-loop propagator-type integrals for the
evaluation of $\gamma_{m,l-1}$. All occuring integrals 
are free from IR divergences.
At one-, two- and three-loop order the
package {\tt MINCER}~\cite{mincer} can be used in order to perform the
computation. However, this method fails to be practical 
for the computation of
$\gamma_{m,3}$ as currently massless four-loop integrals are not
available. 
Alternatively one could set the external momentum to zero but keep the
quark mass finite. Again the technology is available to perform the 
three-loop calculation~\cite{matad}, however, the corresponding
four-loop vacuum diagrams are currently again out of range.

A tempting approach for the computation of $\gamma_{m,3}$, 
which actually was considered in~\cite{Che97_gam} (see also~\cite{Che96}),
is the following: due to the properties of the IRR~\cite{Vla80}
one can set to zero the external momentum and choose 
an arbitrary subset of the 
internal lines to have a non-zero mass.
A clever choice is to allow only the quark propagator which is
attached to the left vertex a non-zero mass $M$.
This has the advantage that the $l$-loop integrals
can be solved by iterating a massive one-loop vacuum 
and a $(l-1)$-loop massless propagator-type integral where the
external momentum of the latter exactly corresponds to the 
loop momentum of the former. Thus, even at four-loop order
at most massless three-loop two-point integrals have to be
evaluated. It is sufficient to compute their finite parts,
as only the $1/\varepsilon$ poles of the $l$-loop integral are needed.
However, there are two subtleties connected to this choice of 
IR structure. The first one is connected to the asymmetry one 
introduces due to the choice of the massive line.
As a consequence the ``left'' vertex has to be renormalized
differently from the ``right'' one. This we will explicitly see 
in the formulae we derive below.

The second disadvantage is the occurrence of IR divergences.
At this point the idea is to use the $R^\star$ operation
in order to subtract them. However, an effective evaluation of the
roughly 6000 diagrams is only possible if the computation of the IR
renormalization constant can be performed in a global way.
This was achieved in Ref.~\cite{Che97_gam} and will be described in the
following. 

As the explicit formulae for $Z_2$ and $Z_m$ are not yet available in the
literature and as they will be published elsewhere~\cite{Che:priv}
we present the derivation of a formula for the renormalization constant
of the vector current correlator, $Z^{\rm em}$, which is quite similar
to the one of $Z_2$ and $Z_m$.
$Z^{\rm em}$ evaluated at four-loop order
immediately leads to corrections of order $\alpha_s^3$ to
the cross section $\sigma(e^+e^-\to \mbox{hadrons})$~\cite{Che97_R}.

Our starting point for the computation of $Z^{\rm em}$
is the renormalized vector current polarization 
function\footnote{For a precise definition see Eq.~(\ref{eqpivadef}).}
which can be cast in the form
\begin{eqnarray}
  \Pi^{\mu\nu}(q) &=& \left(\frac{Z_V}{Z_2}\right)^2 \Pi^{0,\mu\nu}(q) 
                   + Z^{\rm em}\left(-q^2 g^{\mu\nu}+ q^\mu q^\nu\right)
  \,.
  \label{eq:pimunu}
\end{eqnarray}
$\Pi^{0,\mu\nu}(q)$ is the bare correlator as indicated by the
index ``0''. 
$Z_2$ corresponds to the wave function renormalization constant
and $Z_V$ is the renormalization constant for the vector current 
$j^v=\bar{\psi}\gamma^\mu\psi$ defined through
\begin{eqnarray}
  \left(j^{v}\right)\Bigg|_{\rm ren.} &=& 
  \frac{Z_V}{Z_2} \left( j^{v} \right)\Bigg|_{\rm bare}
  \,.
\end{eqnarray}
Note that $Z^{\rm em}$ as defined in
Eq.~(\ref{eq:pimunu}) receives contributions starting from one-loop order.

\begin{figure}[ht]
  \begin{center}
    \begin{tabular}{cc}
      \leavevmode
      \epsfxsize=7.cm
      \epsffile[70 118 530 450]{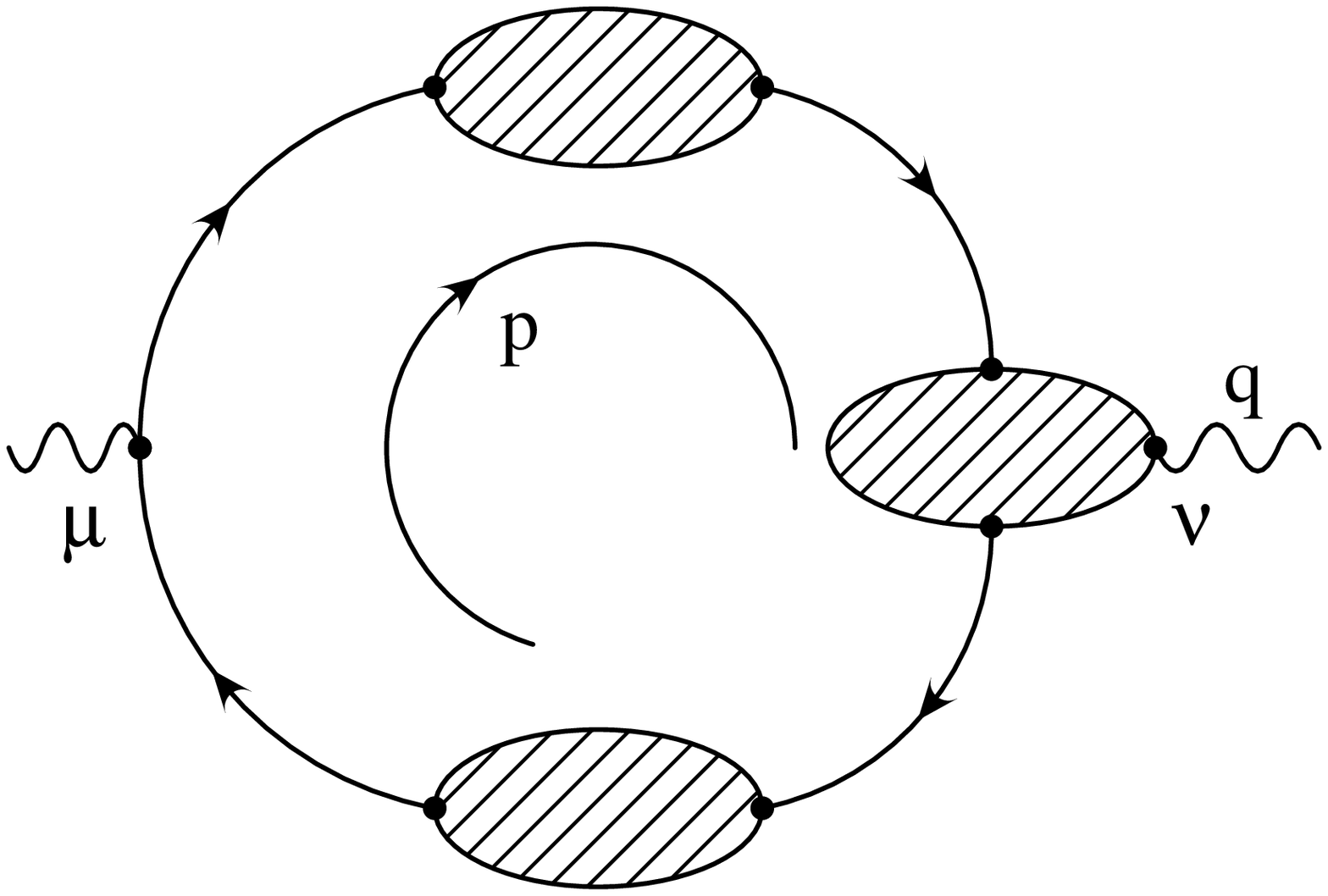}
      &
      \leavevmode
      \epsfxsize=7.0cm
      \epsffile[72 180 540 436]{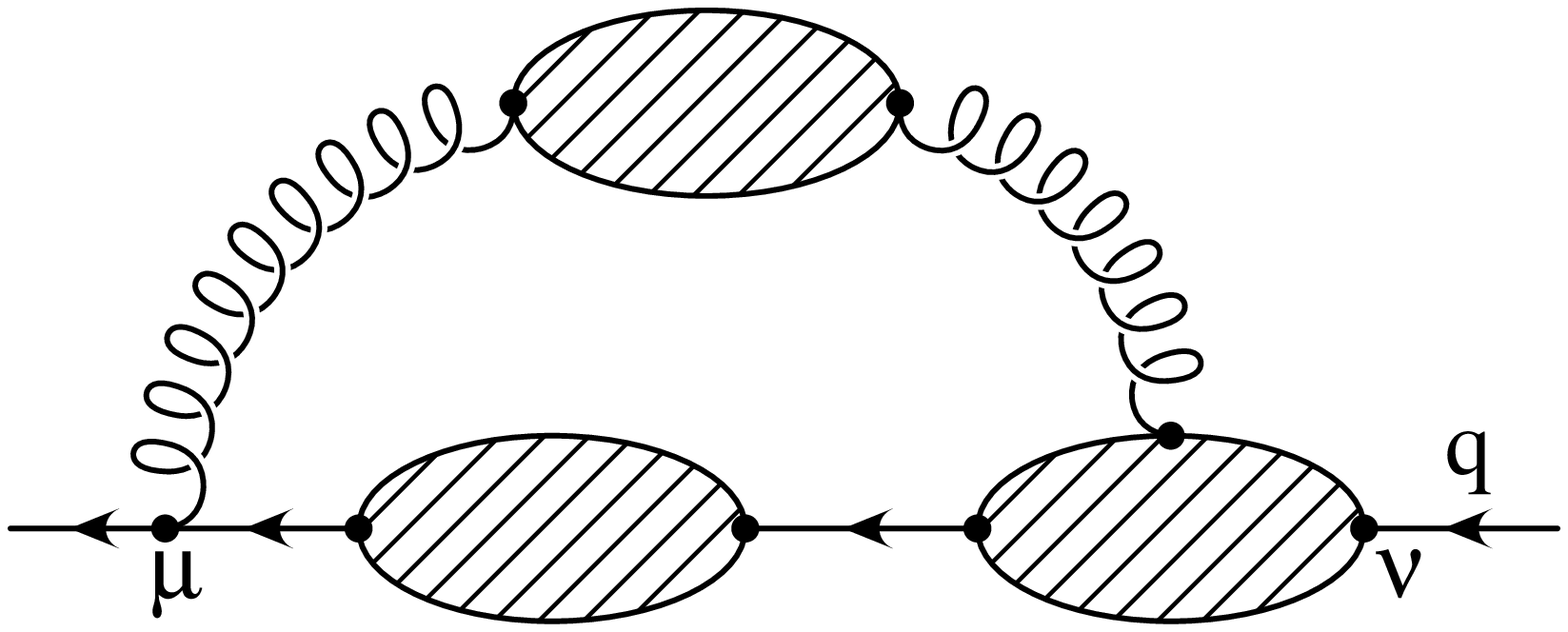}
      \\
      (a) & (b)
    \end{tabular}
  \caption{\label{fig:fpropZ2}(a) Graphical representation of
  Eq.~(\ref{eq:Zem_a}). In the case of the fermion propagator the
  analogous diagram looks like the one in (b).
  In both cases an artificial mass is introduced in the propagator of
  the fermion line attached to the left vertex.
  }
  \end{center}
\end{figure}

In order to derive a formula for the computation of $Z^{\rm em}$
we use for $\Pi^{0,\mu\nu}(q)$ 
a Dyson-Schwinger-type representation containing 
the full fermion propagator, $G^0$,
and the proper photon-quark vertex function, $\Gamma^{0,\mu}$,
leaving out one integration over the final loop momentum which we call
$p$.
The resulting diagrammatic representation 
is visualized in Fig.~\ref{fig:fpropZ2}(a).
After contraction with $g_{\mu\nu}$ one obtains
\begin{eqnarray}
  \Pi^{\mu}_{\mu}(q) &=& 
            - \left(\frac{Z_V}{Z_2}\right)^2
              \mbox{Tr}\left[
                  \int\frac{{\rm d}^D p}{(2\pi)^D}
                  i \gamma_\mu
                  G^0(p,\alpha_s^0)
                  \Gamma^{0,\mu}(p,q,\alpha_s^0)
                  G^0(p+q,\alpha_s^0)
            \right]
  \nonumber\\&&\mbox{}
                   + Z^{\rm em} q^2 \left(1-D\right)
\,,  
\label{eq:Zem_a}
\end{eqnarray}
where the minus sign in front of the first term on the right hand side
accounts for the closed fermion loop corresponding to the $p$ integration.
In Eq.~(\ref{eq:Zem_a}) all colour indices have been suppressed and
the dependence of $G^0$ and $\Gamma^{0,\mu}$ on the bare coupling
$\alpha_s^0$ is made explicit.

Next we exploit the finiteness of $\Pi^\mu_\mu$ in Eq.~(\ref{eq:pimunu})
and the fact that $Z^{\rm em}$ only contains poles in
$1/\varepsilon$. 
This leads to an explicit formula for $Z^{\rm em}$ which reads
\begin{eqnarray}
  Z^{\rm em}\left(D-1\right) 
  &=& - K_\varepsilon\Bigg\{
            \left(\frac{Z_V}{Z_2}\right)^2
            \frac{\Box_q}{2D}\mbox{Tr}\Bigg[
                  \int\frac{{\rm d}^D p}{(2\pi)^D}
                  i \gamma_\mu
\nonumber\\&&\qquad\qquad\qquad\mbox{}
                  G^0(p,\alpha_s^0)
                  \Gamma^{0,\mu}(p,q,\alpha_s^0)
                  G^0(p+q,\alpha_s^0)
            \Bigg]\Bigg|_{q\to0}
            \Bigg\}
  \label{eq:Z2_1}
  \,.
\end{eqnarray}
In this equation we consider the limit $q\to0$ on the right-hand
side. This is possible as in the $\overline{\rm MS}$ scheme
$Z^{\rm em}$ does not depend on any dimensional scale.
In order to evaluate the four-loop contribution to the polarization
function $G^0$ and $\Gamma^{0,\mu}$ have to be inserted at
tree-level and at one-, two- and three-loop order.

Let us next have a closer look to the renormalization constant $Z_V$.
For the ``right'' vertex we have $Z_V=Z_2$. However,
due to the artificial mass which we want to introduce in the fermion
line connected to the ``left'' vertex
the equation $Z_V=Z_2$ can not be applied immediately.
Instead we write for the ``left'' vertex $Z_V=1+\delta Z_V$ and thus
obtain from Eq.~(\ref{eq:Z2_1})
\begin{eqnarray}
  Z^{\rm em}\left(D-1\right) 
  &=& - K_\varepsilon\Bigg\{
            \frac{1}{Z_2}
            \frac{\Box_q}{2D}\mbox{Tr}\Bigg[
                  \int\frac{{\rm d}^D p}{(2\pi)^D}
                  i \gamma_\mu
                  \frac{p^2}{p^2-M^2}
\nonumber\\&&\qquad\qquad\qquad\mbox{}
                  G^0(p,\alpha_s^0)
                  \Gamma^{0,\mu}(p,q,\alpha_s^0)
                  G^0(p+q,\alpha_s^0)
            \Bigg]\Bigg|_{q\to0}
\nonumber\\&&\mbox{}
            + \frac{\delta Z_V}{Z_2}
            \frac{\Box_q}{2D}\mbox{Tr}\Bigg[
                  \int\frac{{\rm d}^D p}{(2\pi)^D}
                  i \gamma_\mu
                  G^0(p,\alpha_s^0)
                  \Gamma^{0,\mu}(p,q,\alpha_s^0)
                  G^0(p+q,\alpha_s^0)
            \Bigg]\Bigg|_{q\to0}
            \Bigg\}
  \,.
  \nonumber\\
  \label{eq:Z2_2a}
\end{eqnarray}
In the first line the presence of the artificial mass is
made explicit in the factor $p^2/(p^2-M^2)$ which effectively replaces
the massless propagator by a massive one. On the other hand, in the
second equation $M$ can be set to zero, as $\delta Z_V$ already
contains the effect of $M$.
In a next step we want to perform the limit $q\to0$.
In particular, we use $M^2\gg q^2$ and apply the hard-mass procedure
to the first term in Eq.~(\ref{eq:Z2_2a}) which transforms it to
\begin{eqnarray}
  Z^{\rm em}\left(D-1\right) 
  &=& - K_\varepsilon\Bigg\{
            \frac{1}{Z_2}
            \frac{\Box_q}{2D}\mbox{Tr}\Bigg[
                  \int\frac{{\rm d}^D p}{(2\pi)^D}
                  i \gamma_\mu
                  \frac{p^2}{p^2-M^2}
\nonumber\\&&\qquad\qquad\qquad\mbox{}
                  G^0(p,\alpha_s^0)
                  \Gamma^{0,\mu}(p,q,\alpha_s^0)
                  G^0(p+q,\alpha_s^0)
            \Bigg]\Bigg|_{q=0}
\nonumber\\&&\mbox{}
          +
            \frac{1}{Z_2}\tilde{R}\Bigg[\Gamma_{f\bar{f}\gamma}(0,0)
            \frac{\Box_q}{2D}\mbox{Tr}\Bigg[
                  \int\frac{{\rm d}^D p}{(2\pi)^D}
                  i \gamma_\mu
\nonumber\\&&\qquad\qquad\qquad\mbox{}
                  G^0(p,\alpha_s^0)
                  \Gamma^{0,\mu}(p,q,\alpha_s^0)
                  G^0(p+q,\alpha_s^0)
            \Bigg]\Bigg|_{q=0}\Bigg]
\nonumber\\&&\mbox{}
            + \frac{\delta Z_V}{Z_2}\tilde{R}\Bigg[
            \frac{\Box_q}{2D}\mbox{Tr}\Bigg[
                  \int\frac{{\rm d}^D p}{(2\pi)^D}
                  i \gamma_\mu
\nonumber\\&&\qquad\qquad\qquad\mbox{}
                  G^0(p,\alpha_s^0)
                  \Gamma^{0,\mu}(p,q,\alpha_s^0)
                  G^0(p+q,\alpha_s^0)
            \Bigg]\Bigg|_{q=0}\Bigg]
            \Bigg\}
  \,.
  \nonumber\\
   \label{eq:Z2_3}
\end{eqnarray}
where $\gamma^\mu \delta\Gamma_{f\bar{f}\gamma}(0,0)$ 
represents the sum of all one-particle-irreducible (1PI) 
hard subgraphs containing the mass $M$.
It gets contributions starting from one-loop order.
In Eq.~(\ref{eq:Z2_3}) we introduced the $\tilde{R}$ operation in
order to treat the IR divergences of those terms which lead to
massless tadpoles for $q\to0$. Note that in general also the first
term in (\ref{eq:Z2_3}) contains IR divergences. However, they
originate from the hard-mass procedure and are correlated to the UV
poles of the second term in Eq.~(\ref{eq:Z2_3})
(cf. the discussion around Eq.~(\ref{eq:Z2hmp})).

The IR divergences which occur in the last two terms of
Eq.~(\ref{eq:Z2_3}) for $q=0$ 
are taken care by introducing appropriate
renormalization constants. They can be determined in a global manner
by setting $q=0$
in Eq.~(\ref{eq:Z2_1}) and applying the $\tilde{R}$ operation.
As the left-hand side is unaffected we have

\begin{eqnarray}
  Z^{\rm em}\left(D-1\right) 
  &=& - K_\varepsilon\Bigg\{
            \tilde{R}\Bigg[
            \frac{\Box_q}{2D}\mbox{Tr}\Bigg[
                  \int\frac{{\rm d}^D p}{(2\pi)^D}
                  i \gamma_\mu
\nonumber\\&&\qquad\qquad\qquad\mbox{}
                  G^0(p,\alpha_s^0)
                  \Gamma^{0,\mu}(p,q,\alpha_s^0)
                  G^0(p+q,\alpha_s^0)
            \Bigg]\Bigg|_{q=0}
            \Bigg]
            \Bigg\}
  \,.
  \label{eq:Z2_4}
\end{eqnarray}
The application of $\tilde{R}$ generates only one term, namely the
IR counterterm, as we set massless tadpoles to zero.
This counterterm is a global one as it treats the IR divergences of
the sum of all diagrams. We want to mention that this convenient
aspect is new as compared to older calculations where the
$\tilde{R}$ operator has been used. In
Ref.~\cite{GorKatLar91SurSam91}, e.g., the $\tilde{R}$ operator has
been applied to each diagram individually which in practice is quite
tedious. 

Inserting Eq.~(\ref{eq:Z2_4}) into~(\ref{eq:Z2_3}) finally leads to
\begin{eqnarray}
  Z^{\rm em}\left(D-1\right) 
  &=& - K_\varepsilon\Bigg\{
            \frac{1}{Z_2}
            \frac{\Box_q}{2D}\mbox{Tr}\Bigg[
                  \int\frac{{\rm d}^D p}{(2\pi)^D}
                  i \gamma_\mu
                  \frac{p^2}{p^2-M^2}
\nonumber\\&&\qquad\qquad\qquad\mbox{}
                  G^0(p,\alpha_s^0)
                  \Gamma^{0,\mu}(p,q,\alpha_s^0)
                  G^0(p+q,\alpha_s^0)
            \Bigg]\Bigg|_{q=0}
\nonumber\\&&\mbox{}
            +\frac{(1-D)Z^{\rm em}}{Z_2}
               \left(\delta\Gamma_{f\bar{f}\gamma}(0,0) + \delta Z_V \right)
            \Bigg\}
  \,.
  \label{eq:Z2_5}
\end{eqnarray}
Note that at this point the relation $\delta Z_V=Z_2-1$ can be
used.

\begin{figure}[ht]
\leavevmode
\begin{center}
  \begin{tabular}{cccc}
   \epsfxsize=3cm
   \epsffile[140 270 470 540]{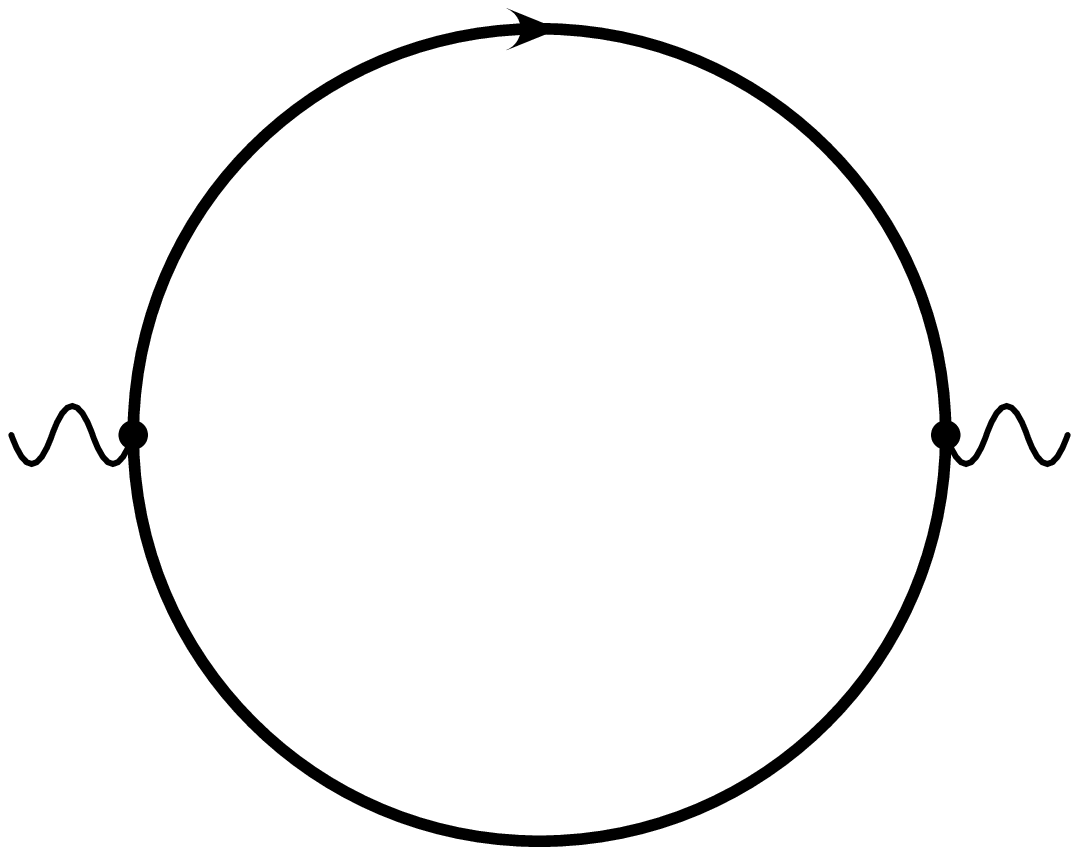}
   &
   \epsfxsize=3cm
   \epsffile[140 270 470 540]{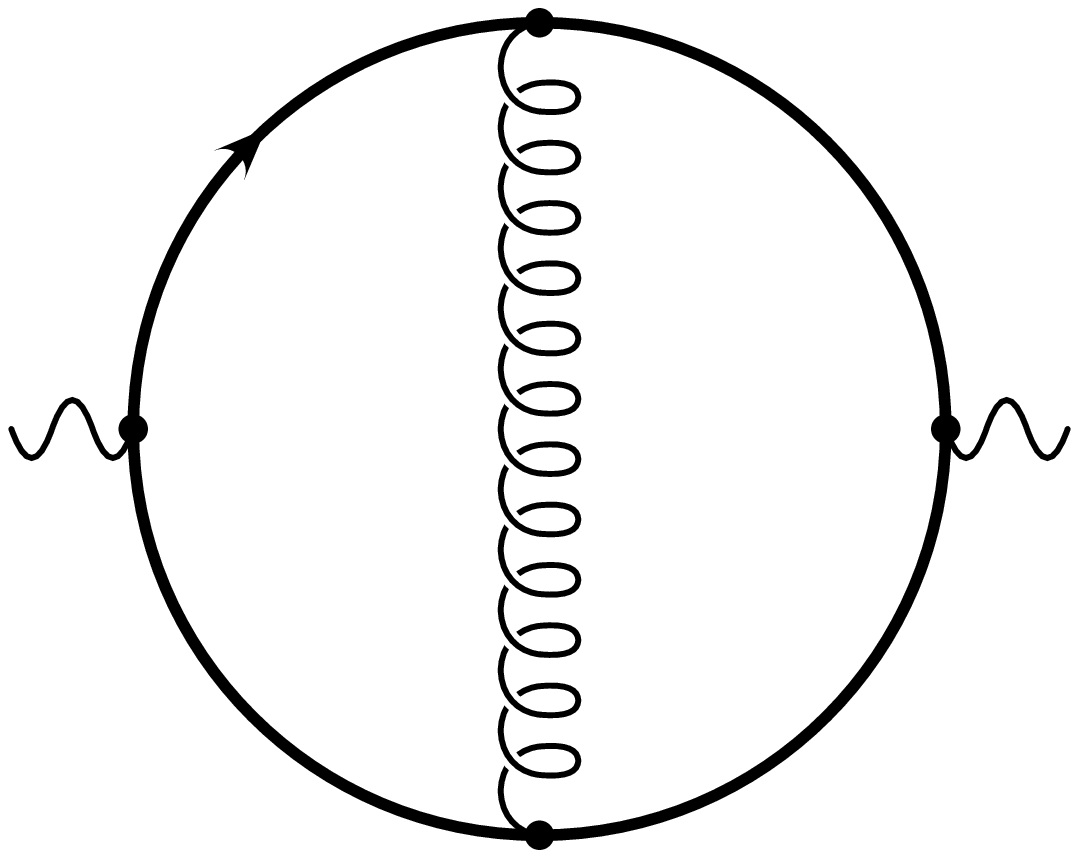}
   &
   \epsfxsize=3cm
   \epsffile[140 270 470 540]{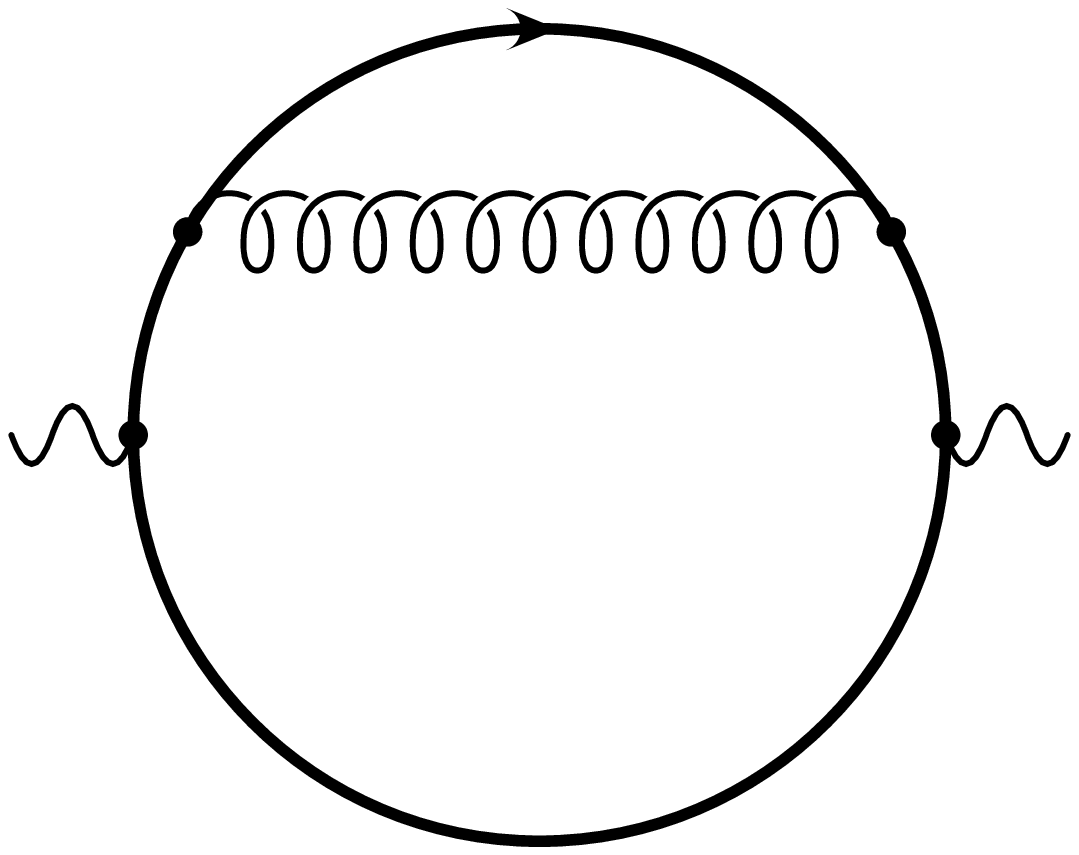}
   &
   \epsfxsize=3cm
   \epsffile[140 270 470 540]{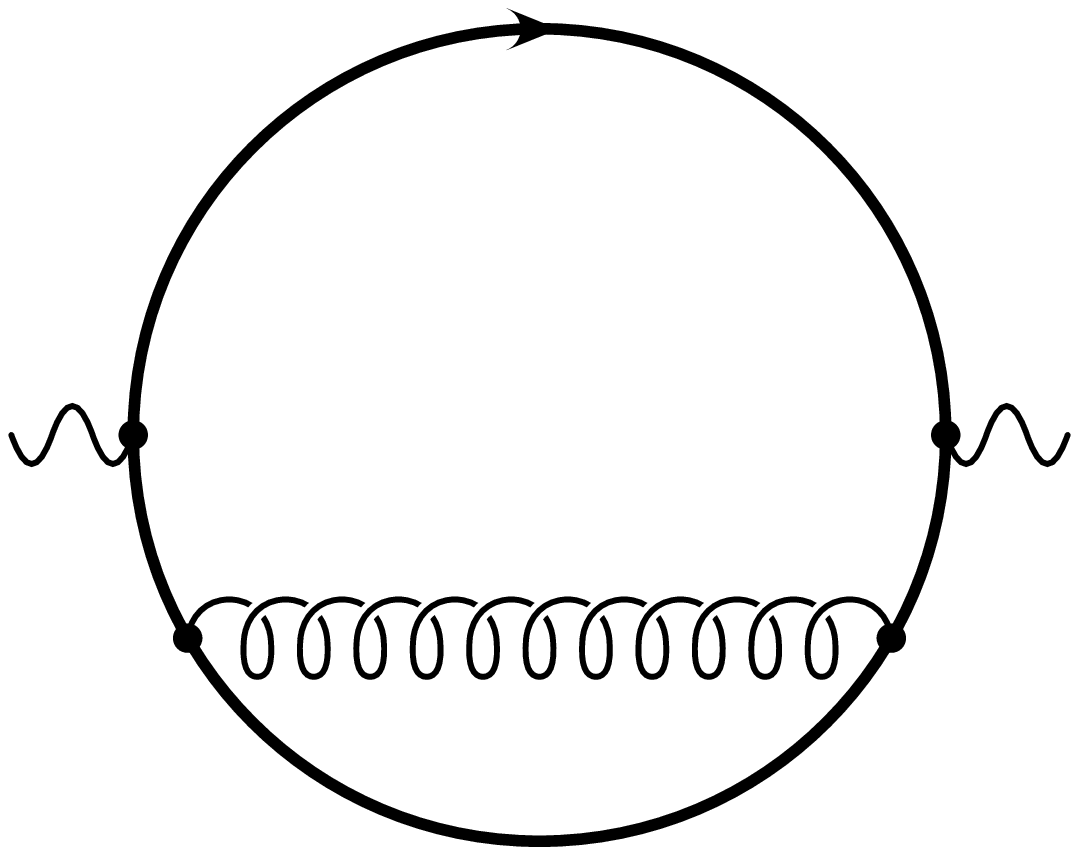}
  \end{tabular}
  \caption{\label{fig:photonprop}One- and two-loop diagrams contributing 
  the vector current correlator.}
\end{center}
\end{figure}

At one-loop order only the first line of Eq.~(\ref{eq:Z2_5})
contributes where all renormalization constants can be set to one.
Furthermore the functions $G^0$ and $\Gamma^{0,\mu}$ take their
tree-level values and only the one-loop diagram shown in 
Fig.~\ref{fig:photonprop} has to be evaluated for external momentum zero
and finite quark mass in one of the fermion lines. 
In this case no IR divergences occur.

A non-trivial contribution from the terms in the second line of
Eq.~(\ref{eq:Z2_5}) occurs for the first time at two-loop order.
In particular, for $\delta\Gamma_{f\bar{f}\gamma}$ the one-loop
photon-fermion vertex corrections
have to be evaluated where one of the quarks carries mass 
$M$ and the other one is massless. All external momenta equal to zero.
In the three two-loop diagrams (see Fig.~\ref{fig:photonprop})
contributing to the first line 
we consequently assign a mass $M$ to the fermion
(not the anti-fermion)
line attached to the ``left''
vertex and set the masses in all other propagators to zero.
For vanishing external momentum in each diagram a massless one-loop
two-point function can be identified which can be solved with the help
of Eq.~(\ref{eq:Pab}) in Appendix~\ref{sub:single}.
The external momentum coincides with the loop momentum of the remaining
one-loop vacuum integral which is also expressible in terms of
$\Gamma$ functions (cf. Eq.~(\ref{eq:Va})).

In general,
for a $l$-loop calculation the renormalization constants occuring in
Eq.~(\ref{eq:Z2_5}) are only needed at loop order $(l-1)$.
The loop integrals which are necessary to compute the first
expression in Eq.~(\ref{eq:Z2_5}) are
massless $(l-1)$-loop two-point functions and one-loop vacuum integrals.
For the evaluation of $\delta\Gamma_{f\bar{f}\gamma}$, $(l-1)$-loop 
vacuum integral are needed. In particular,
for $l=4$ the occuring integrals are very well studied
(cf. Refs.~\cite{mincer,matad}).

In the above equations the bare coupling constant is used 
as a parameter. It has to be expressed through the renormalized version
using the $(l-1)$-loop formula for the renormalization constant 
$Z_g$ (cf. Eq.~(\ref{eq:renconst})).
We want to note that in our case the mass $M$ 
needs not to be renormalized as in the diagrams no subdivergence 
is present which could induce a corresponding counterterm.

Besides the application to 
the vector current correlator, which leads to corrections of order
$\alpha_s^3$ to $R(s)$~\cite{Che97_R}, the  
method described in this Subsection has successfully been applied to
the four-loop fermion propagator and
the correlator of scalar
currents in order to evaluate the 
the four-loop contribution to $\gamma_m$~\cite{Che97_gam}
and corrections of order $\alpha_s^3$ to
the decay of a scalar Higgs boson~\cite{Che97_Higgs}, respectively.

Due to the large number of genuine four-loop diagrams an automation of
the computation is mandatory. 
The four-loop calculations of~\cite{Che97_R,Che97_Higgs,Che97_gam}
have been performed with the help of the package {\tt
GEFICOM}~\cite{geficom}. Within this framework the computation of
$\delta\Gamma_{f\bar{f}\gamma}$ 
is straightforward as one- two- and three-loop
vacuum integrals are directly accessible.
For the four-loop contributions, like in the first line of
Eq.~(\ref{eq:Z2_5}), some tricks are necessary. 
In particular completely massless four-loop diagrams are generated.
In a next step a {\tt Mathematica}~\cite{math} program is used to
identify the ``left'' vertex and to introduce the mass $M$. At this
point the topology of the massless three-loop diagram is fixed and the
mapping of the momenta according to the notation of {\tt
MINCER}~\cite{mincer} can be performed.
One has to exploit that the whole diagram is logarithmically divergent
and the mass dimension is given by the artificial mass $M$.
Then the massless integration and at the end also the one-loop massive
one can be performed.


\subsection{\label{sub:massint}Massive 
  vacuum integrals and the $\beta$ function}

A different approach has been employed
in~\cite{RitVerLar97_bet,LarRitVer97_gam}
in order to compute the four-loop contribution
to the $\beta$ and $\gamma_m$ functions.
Also here the basic property of the IRR has been exploited and the
integrals have been simplified by modifying the IR behaviour of the
Feynman diagrams.
This time, however, a different attitude has been adopted than in the
previous section:
in all denominators, also the ones of the gluons and ghosts, 
a common mass parameter, $M$, has been introduced (see
also~\cite{Chetyrkin:1998fm}).
This avoids completely the IR divergences.
On the other side, however, both gauge invariance and useful Ward
identities are broken by this method. 
Moreover, multiplicative renormalization, which is very convenient in
practical renormalization is lost. Nevertheless the $R$ operation
applied to the individual diagrams still works and in principle it can
be used to compute the overall renormalization constant.
However, this is not at all practical, especially as in the case of
the $\beta$ function roughly $50\,000$ diagrams have to be
considered~\cite{RitVerLar97_bet} and an automatic treatment is
absolutely mandatory. This is achieved by introducing effective
vertices and propagators which incorporate all lower-order
renormalization constants~\cite{RitVerLar97_bet}. Furthermore
counterterms have to be introduced which correspond to a
renormalization of the gluon and ghost mass and an overall
renormalization constant for the gluon propagator. 

In the following we want to illustrate this method by considering
two-loop QCD corrections to the photon propagator and evaluate the
corrections to the wave function renormalization defined through
\begin{eqnarray}
  A^{0,\mu} &=& \sqrt{Z_3^\gamma} A^{\mu}
  \,,
\end{eqnarray}
where $A^\mu$ is the photon field.
For convenience we also introduce the renormalization constants
$Z_1^\gamma$ and $Z_2^\gamma$ via
\begin{eqnarray}
  \psi^0 &=& \sqrt{Z_2^\gamma} \psi
  \nonumber\\
  e^0 \bar{\psi^0}\gamma^\mu\psi^0 A_\mu^0 &=&
  Z_1^\gamma e \bar{\psi}\gamma^\mu\psi A_\mu
  \,,
\end{eqnarray}
where $e$ is the electromagnetic charge and $\psi$ is the fermion field.
The corrections up to order $\alpha\alpha_s$ to
$Z_3^\gamma$ are obtained from the diagrams of
Fig.~\ref{fig:photonprop} which contribute to the photon polarization
function.
The latter can be written in the form
\begin{eqnarray}
  \Pi^{\mu\nu}(q) &=& \left(-g^{\mu\nu}+\frac{q^\mu q^\nu}{q^2}\right)
                      \Sigma_T(q^2)
                      + \frac{q^\mu q^\nu}{q^2} \Sigma_L(q^2)
  \,.
\end{eqnarray}

The canonical way to compute $Z_3^\gamma$ would be to consider all
particles as massless and evaluate the occuring massless two-point
function in dimensional regularization. 
In this case $\Sigma_T(q^2)$ is proportional to $q^2$ and 
$\Sigma_L(q^2)\equiv 0$ which is due to gauge invariance.
One observes that the sum of the bare two-loop diagrams only contains
a simple pole in $\varepsilon$ and that it is independent
of the QCD gauge parameter, $\xi$. Furthermore, due to the Ward identity
connecting the photon-quark vertex to the quark self energy no
renormalization is necessary and the bare diagrams of
Fig.~\ref{fig:photonprop} already lead to the final answer
where the formula
\begin{eqnarray}
  Z_3^\gamma &=& 
  1 - K_\varepsilon\left(\frac{\Sigma_T(q^2)}{q^2}Z_3^\gamma\right)
  \,,
\end{eqnarray}
can be used.

In the method proposed
in~\cite{RitVerLar97_bet,LarRitVer97_gam,Chetyrkin:1998fm} 
a common mass $M$ is introduced in all lines
for the computation of the renormalization group functions.
As a consequence gauge invariance is broken and 
$\Sigma_L(q^2)$ is not zero any more. Furthermore, both 
$\Sigma_T(q^2)$ and $\Sigma_L(q^2)$ contain terms proportional to $M^2$.
One also observes that the sum of the bare
diagrams contains a pole of the form $\xi/\varepsilon$.
Clearly the $R$ operation applied to the individual diagrams would
still lead to the correct answer. However, we want to ``immitate''
multiplicative renormalization as close as possible and introduce
effective vertices and propagators. 
In the case of the photon-quark vertex we write
\begin{eqnarray}
  \gamma^\mu \longrightarrow \gamma^\mu Z_1^\gamma 
             \,\,=\,\,       \gamma^\mu (1+\delta Z_1^\gamma)
  \,.
  \label{eq:effver}
\end{eqnarray}
which actually corresponds to multiplicative renormalization.
The quark propagator is modified to
\begin{eqnarray}
  \frac{\psla}{p^2-M^2} \longrightarrow
  \frac{\psla}{p^2-M^2} \left(1-\psla \delta Z_2^\gamma\frac{
                                \psla}{p^2-M^2}\right) 
  \,.
  \label{eq:effprop}
\end{eqnarray}
Note that for $M=0$ the terms proportional to 
$\delta Z_1^\gamma$ and $\delta Z_2^\gamma$ are proportional to the
Born diagram and a cancellation takes place due to the Ward identity
$Z_1^\gamma=Z_2^\gamma$.
For $M\not=0$, $Z_1^\gamma$ and $Z_2^\gamma$ have to be determined form
the vertex correction and quark self-energy, respectively.
To our approximation they take the same values as for $M=0$, namely
\begin{eqnarray}
  Z_1^\gamma &=& Z_2^\gamma \,\,=\,\, 
  1 + \frac{\alpha_s}{\pi} \frac{C_F}{\varepsilon}
      \left(\frac{1}{4}\xi-\frac{1}{4}\right)
  \,,
\end{eqnarray}
where $\xi$ is the gauge parameter defined in Eq.~(\ref{eq:gluprop}).

The use of Eqs.~(\ref{eq:effver}) and~(\ref{eq:effprop}) for the
computation of the one-loop diagram of Fig.~\ref{fig:photonprop}
induces terms of order $\alpha\alpha_s$ which render the pole part
independent of $\xi$.
However, $\Pi^{\mu\nu}$ still contains terms proportional to
$g^{\mu\nu} M^2$. They are removed by introducing a local
mass counterterm for the photon
of the form $M^2 A_\mu A^\mu$. 
Also for the gluon field a similar counterterm has to be introduced.
It is contained in the effective gluon propagator as it 
is needed for the cancellation of subdivergences.

Finally one arrives at
\begin{eqnarray}
  Z_3^\gamma &=& 1 - \frac{\alpha}{\pi}\left(
                         \frac{1}{3\varepsilon}
                        + \frac{\alpha_s}{\pi} C_F \frac{1}{8\varepsilon}
                     \right)
  \,.
\end{eqnarray}

In Ref.~\cite{RitVerLar97_bet} this method has been applied to obtain
the four-loop contribution of the QCD $\beta$ function.
It has been applied to the ghost-gluon vertex, the gluon propagator
and the ghost propagator in order to obtain the corresponding
renormalization constants $\tilde{Z}_1$, $Z_3$ and $\tilde{Z}_3$
(cf. Eq.~(\ref{eq:renconst}))
and finally the one for $\alpha_s$ via
\begin{eqnarray}
  Z_g &=& \frac{\tilde{Z}_1}{\tilde{Z}_3\sqrt{Z_3}}
  \,.
\end{eqnarray}
The $\beta$ function is defined through
\begin{eqnarray}
  \mu^2\frac{{\rm d}}{{\rm d}\mu^2}
  \frac{\alpha_s^{(n_f)}(\mu)}{\pi}
  &=&
  \beta^{(n_f)}\left(\alpha_s^{(n_f)}\right)
  \,\,=\,\,
  - \sum_{i\ge0}
  \beta_i^{(n_f)}\left(\frac{\alpha_s^{(n_f)}(\mu)}{\pi}\right)^{i+2} 
  \,,
  \label{eq:defbeta}
\end{eqnarray}
where $n_f$ is the number of active flavours.
For completeness we want to list the results for the
coefficients which are given by~\cite{gro,jon,tar,RitVerLar97_bet}
\begin{eqnarray}
  \beta_0^{(n_f)} &=&\frac{1}{4}\left[ 11 - \frac{2}{3} n_f\right]
  \,,
  \nonumber\\
  \beta_1^{(n_f)} &=&\frac{1}{16}\left[ 102 - \frac{38}{3} n_f\right]
  \,,
  \nonumber \\
  \beta_2^{(n_f)} &=&\frac{1}{64}\left[\frac{2857}{2} - \frac{5033}{18} n_f
    + \frac{325}{54} n_f^2\right]
  \,,
  \nonumber \\
  \beta_3^{(n_f)} &=&\frac{1}{256}\left[  \frac{149753}{6} + 3564 \zeta_3 
    + \left(- \frac{1078361}{162} - \frac{6508}{27} \zeta_3 \right) n_f
  \right.\nonumber\\&&\mbox{}
  + \left( \frac{50065}{162} + \frac{6472}{81} \zeta_3 \right) n_f^2
  +\left.  \frac{1093}{729}  n_f^3\right]
  \,.
  \label{eq:betafct}
\end{eqnarray}
$\zeta$ is Riemann's zeta function, with values $\zeta_2=\pi^2/6$
and $\zeta_3\approx1.202\,057$.

One of the main new achivements of Ref.~\cite{RitVerLar97_bet} is the
treatment of the four-loop vacuum diagrams. The task is simplified due
to the fact that only the divergent parts in $\varepsilon$ are needed.
The method of integration-by-parts~\cite{CheTka81} 
has been used to derive recurrence relations which reduce a general
four-loop integral to a linear combination of simple integrals and 
two (difficult) master integrals.


\section{\label{sec:dec}Decoupling of heavy particles}
\setcounter{equation}{0} 
\setcounter{figure}{0} 
\setcounter{table}{0} 

Quantum corrections to processes involving only light degrees of
freedom contain in general the whole particle spectrum. In particular
also heavy particles with masses much larger than the
energy scale of the considered process contribute.
It is highly desirable
that in the limit where the heavy mass, $M$, goes to infinity
its contribution to the light-particle Green function must tend to
zero like $\mu/M$ where $\mu$ is a typical scale of the process.
This is exactly the content of the so-called decoupling theorem
which is proven in~\cite{AppCar75}.

\begin{figure}[ht]
  \begin{center}
  \epsfxsize=10.0cm
  \epsffile[25 208 587 584]{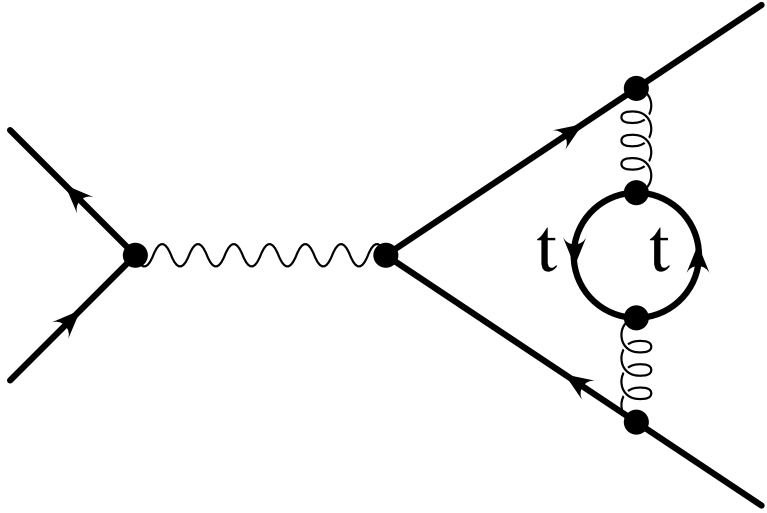}
  \caption{\label{fig:eeqqheavy}Feynman diagram contribution to the
  process $e^+e^-\to b\bar{b}$ which gives rise to top-quark-dependent
  terms.
  }
  \end{center}
\end{figure}

To be more precise let us consider an example, namely the
production of heavy quarks in $e^+e^-$ annihilation.
Due to the hierarchy in the quark masses there is a
clear separation into light and heavy. 
For center-of-mass energies, $\sqrt{s}$,
of about 40~GeV we are well below the production threshold of top quarks and
thus we expect their influence to be suppressed by $\sqrt{s}/M_t$.
Analogously the contribution of bottom quarks to the 
production of charm quarks close to the
threshold must be proportional to
$\sqrt{s}/M_b$.
In our example the heavy quarks enter the first time at order
$\alpha_s^2$ via the diagram pictured in Fig.~\ref{fig:eeqqheavy}.
If the contribution of this diagram is computed in a
momentum-subtraction scheme one indeed observes this 
behaviour\footnote{Actually the proof of the decoupling theorem
  in~\cite{AppCar75} is performed for a momentum-subtraction scheme.}.
However, in this example, and also in most other QCD processes it is
much more convenient to use a mass-independent renormalization prescription,
like the MS~\cite{tHo73} or its popular modification, the 
$\overline{\rm MS}$~\cite{BarBurDukMut78} scheme.
These schemes are characterized through the fact that their
renormalization group functions are mass-independent which makes 
renormalization group improvements much more transparent.
It also has the advantage that the computation of the renormalization group
functions themselves is significantly simplified 
(cf. Section~\ref{sec:rge}).
On the other hand there is the big drawback that the decoupling
theorem of Appelquist and Carazzone~\cite{AppCar75}
does not hold in mass-independent schemes.
This is due to the mass-independence of the 
renormalization group functions which implies 
that, e.g. in case of QCD,
the top quark and the down quark have identical contributions.

Note that in the broad classes of momentum subtraction schemes
the decoupling theorem is valid. However, the calculations are much
more complicated and in general a coupled system of differential
equations involving also the quark masses and the gauge parameter
has to be solved in order to obtain the running of the couplings.

Coming back to our example this means that the cross section 
$\sigma(e^+ e^-\to b\bar{b}+{\rm gluons})$ 
evaluated in the $\overline{\rm MS}$ scheme 
with $\sqrt{s}=40$~GeV does not
behave like $\sqrt{s}/M_t$ but still contains logarithms of the form
$\ln(s/M_t^2)$, which arise at order $\alpha_s^2$ from 
the diagram in Fig.~\ref{fig:eeqqheavy}.

Clearly, both from the theoretical and practical point of view this is
not acceptable. The way out is the explicit construction of an
effective Lagrangian where the heavy particles are integrated out.
This means that the dynamical degree of freedom of
the heavy quark is removed,
which manifestly leads to power-suppressed contributions of the latter.
This will be performed in Section~\ref{sub:effL} for the case of QCD.
The effective Lagrangian only depends on light degrees of
freedom where the couplings\footnote{Here we mean ``normal''
coupling constants, but also masses, gauge parameters, etc..} 
are multiplicatively re-scaled by the so-called decoupling
constants. These constants are universal and have been computed in the
case of QCD up to order $\alpha_s^3$~\cite{CheKniSte98}.

Concerning the practical consequences we again want to consider the
example of $e^+e^-$ annihilation.
For a calculation at order $\alpha_s^2$ one computes all relevant
diagrams including the one of Fig.~\ref{fig:eeqqheavy}.
As mentioned above the result diverges proportional to the logarithm of
the heavy quark mass. 
Now one has to remember that the coupling\footnote{Note, that,
  in the case where the light quark masses are neglected, only 
  $\alpha_s$ remains as a parameter.} 
has to be changed according to the known rules 
which describe the transition to the effective Lagrangian.
Thus one arrives at a physical observable expressed in terms of
parameters of the effective Lagrangian and it can be explicitly checked
that the dependence on the heavy quark mass, $M_t$, is
power-suppressed --- in the case at hand it goes like $s/M_t^2$.

Pioneering work in the computation of the decoupling constants has
been done in~\cite{Wei80}. In Ref.~\cite{BerWet82Ber83} the decoupling
constant for $\alpha_s$ has been computed at the two-loop order. 
The crucial idea of the method is based on the fact that the
decoupling theorem~\cite{AppCar75} works in momentum subtraction
schemes. Thus after relating the corresponding coupling constant to 
$\alpha_s$ defined in the $\overline{\rm MS}$ scheme both in the full and
the effective theory, it is possible to derive differential equations
for the decoupling constant. They can easily be solved.
For the corresponding integration constant a two-loop calculation is needed.

Thirteen years later the authors of~\cite{LarvRiVer95} evaluated the
corrections of order $\alpha_s^3$ 
to the total decay rate of the $Z$ boson
induced by a heavy top quark.
In order to make the decoupling explicit the two-loop result
of~\cite{BerWet82Ber83} is needed. 
However, it turned out that after expressing the decay rate in terms
of effective parameters the top quark did not decouple.
Thus in~\cite{LarvRiVer95} a second evaluation of 
the decoupling constant for $\alpha_s$ has been performed
with a different result as in~\cite{BerWet82Ber83}. The result
of~\cite{LarvRiVer95} was confirmed in~\cite{CheKniSte98}.
In the meantime, the authors of Ref.~\cite{BerWet82Ber83} 
have revised \cite{Ber97}
their original analysis and have found agreement with Ref.~\cite{LarvRiVer95}.

The method of Ref.~\cite{LarvRiVer95} for the computation of the
decoupling relations is based on the evaluation of a top quark
contribution to a physical quantity. In particular the authors
of~\cite{LarvRiVer95} considered corrections to the massless quark
propagator with an additional zero momentum operator insertion.
In order to obtain the decoupling relations to order $\alpha_s^2$ a
three-loop calculation corresponding to the order $\alpha_s^3$
corrections is necessary.
Correspondingly, the decoupling relation to order $\alpha_s^3$ would
require a four-loop calculation. In our method, which is described
below, only the computation of
three-loop vacuum diagrams are necessary.


\subsection{\label{sub:effL}Construction of an effective Lagrangian in QCD}

The main idea of effective theories is that the dynamics at low
energies does not depend on the details of the dynamics of high
energies.
Thus the low-energy physics can be described using an effective
Lagrangian which does not depend on the additional degrees of freedom
present at high energies. The only effect of the high-energy
parameters are modified couplings of the effective Lagrangian
with respect to the full one.

In the following we want to describe the construction of the 
effective Lagrangian in QCD with one heavy quark of mass $m_h$
and  $n_l$ light quarks.

Our starting point is the full QCD Lagrangian which reads
\begin{eqnarray}
  {\cal L}^{\rm QCD} &=&
  - \frac{1}{4} G^{a,\mu\nu} G^a_{\mu\nu}
  + \sum_{f=1}^{n_f} 
    \bar{\psi}_f \left( i\,\,/\hspace{-.7em}D - m_f \right) \psi_f 
  \nonumber\\&&\mbox{}
  - \frac{1}{2(1-\xi)} \left(\partial^\mu G_\mu^a\right)^2
  + \partial^\mu \bar{c}^a
      \left(\partial_\mu c^a-g_s f^{abc} c^b G_\mu^c\right) 
  \,,
  \label{eq:L_QCD}
\end{eqnarray}
where the field strength tensor is defined through
$G^{a,\mu\nu}=\partial^\mu G^{a,\nu} - \partial^\nu G^{a,\mu}
+ g_s f^{abc} G^{b,\mu}G^{c,\nu}$.
$f^{abc}$ are the structure constants of the QCD gauge group,
$g_s=\sqrt{4\pi\alpha_s}$ is the QCD gauge coupling,
$\psi_f$ is a quark field with mass $m_f$,
$G^{a,\mu}$ is the gluon field,
$c^a$ is the Faddeev-Popov-ghost field, and
$n_f=n_l+1$ is the total number of quark flavours.
$D_\mu=\partial_\mu-ig_s (\lambda^a/2) G_\mu^a$ is the covariant
derivative in the fundamental representation
and $\lambda^a$ are the Gell-Mann matrices.
For convenience we list the gluon propagator resulting
from~(\ref{eq:L_QCD}) 
\begin{eqnarray}
  D_g(q) &=&
  \frac{i}{q^2+i\epsilon}\left(-g^{\mu\nu}+\xi\frac{q^\mu q^\nu}{q^2}\right)
  \label{eq:gluprop}
  \,.
\end{eqnarray}
In this convention $\xi=0$ corresponds to Feynman gauge and $\xi=1$ to
Landau gauge.

For later use we define the renormalization constants connecting the
bare and renormalized quantities in
Eq.~(\ref{eq:L_QCD}):
\begin{eqnarray}
g_s^0\,\,=\,\,\mu^{\varepsilon}Z_gg_s\,,\qquad
&
m_q^0\,\,=\,\,Z_mm_q\,,\qquad
&
\xi^0-1\,\,=\,\,Z_3(\xi-1)\,,
\nonumber\\
\psi_q^0\,\,=\,\,\sqrt{Z_2}\psi_q\,,\qquad
&
G_\mu^{0,a}\,\,=\,\, \sqrt{Z_3} G_\mu^a\,,\qquad
&
c^{0,a}\,\,=\,\,\sqrt{\tilde{Z}_3}c^a\,.
\label{eq:renconst}
\end{eqnarray}
In addition one introduces the renormalization constants 
of the quark-gluon, three-gluon, four-gluon and gluon-ghost vertex
which are denoted by
$Z_{1F}$, $Z_1$, $Z_4$ and $\tilde{Z}_1$.
The Slavnov-Taylor identities connecting the different 
renormalization constants can, e.g., be found in~\cite{Muta}.

It is clear that in the effective Lagrangian ${\cal L}^{\rm QCD}_{\rm eff}$
all explicit trace to the heavy quark must have disappeared. However,
the mathematical structure must be identical to the one of 
${\cal L}^{\rm QCD}$ in Eq.~(\ref{eq:L_QCD}).
This is because ${\cal L}^{\rm QCD}$ represents
the most general Lagrangian
describing the interaction of quarks and gluons and respecting the
symmetry properties imposed by the Becchi-Rouet-Stora-Tyutin
invariance~\cite{BRST}.
Of course, this also has to be respected by a $n_l$-flavour theory
described by ${\cal L}^{\rm QCD}_{\rm eff}$.
Nevertheless the parameters and fields of ${\cal L}^{\rm QCD}_{\rm
eff}$ are different from the ones of the full theory. It is convenient
to define the corresponding relations in analogy to the
renormalization constants of Eq.~(\ref{eq:renconst}) and introduce
multiplicative factors --- the so-called decoupling constants $\zeta_i^0$.
Thereby it is advantageous to consider the decoupling relations
in the bare theory
which is indicated by the superscript zero.
The renormalization is performed afterwards.
Thus we define 
\begin{eqnarray}
  \begin{array}{lll}
  g_s^{0,\prime} = \zeta_g^0 g_s^0\,,\qquad
  &
  m_q^{0,\prime} = \zeta_m^0m_q^0\,,\qquad
  &
  \xi^{0,\prime}-1 = \zeta_3^0(\xi^0-1)\,,
  \\ \\
  \psi_q^{0,\prime}  = \sqrt{\zeta_2^0}\psi_q^0\,,\qquad
  &
  G_\mu^{0,\prime,a} = \sqrt{\zeta_3^0}G_\mu^{0,a}\,,\qquad
  &
  c^{0,\prime,a}     = \sqrt{\tilde\zeta_3^0}c^{0,a}
  \,,
  \end{array}
  \label{eq:decconst}
\end{eqnarray}  
where the primes mark the quantities of the effective $n_l$-flavour theory.

Taking into account these considerations we can
write down a defining equation for the bare
effective Lagrangian in terms of the full Lagrangian with re-scaled
parameters: 
\begin{eqnarray}
  {\cal L}^{\rm QCD}_{\rm eff}(g_s^0,m_q^0,\xi^0;
                               \psi^0_q,G^{0,a}_\mu,c^{0,a};
                               \zeta_i^0)
  &=& {\cal L}^{\rm QCD}(g_s^{0,\prime},m_q^{0,\prime},\xi^{0,\prime};
                         \psi^{0,\prime}_q,G^{0,\prime,a}_\mu,c^{0,\prime,a})
  \,,
  \label{eq:L_QCD_eff}
\end{eqnarray}
where 
${\cal L}^{\rm QCD}$ is given in Eq.~(\ref{eq:L_QCD}),
$q$ represents the $n_l$ light-quark flavours and $\zeta_i^0$
collectively denotes all bare decoupling constants of
Eq.~(\ref{eq:decconst}).
Once they are explicitly computed the effective Lagrangian is
completely determined.
Green functions of light fields obtained from ${\cal L}^{\rm QCD}$
agree with the ones of ${\cal L}^{\rm QCD}_{\rm eff}$ up to terms
suppressed by inverse powers of the heavy quark mass.

In the language of effective theories the computation of the
decoupling constants is referred to as matching
calculation. It can be performed in a more or less complicated way.
As it is even nowadays highly non-trivial to apply the methods
of~\cite{BerWet82Ber83} and~\cite{LarvRiVer95} at order $\alpha_s^3$,
we developed a procedure which relates the $n$-loop decoupling
constants of Eq.~(\ref{eq:decconst}) to $n$-loop massive one-scale
integrals. It will be described in the next Subsection.


\subsection{Computation of the decoupling constants}

In order to compute the decoupling constants of Eq.~(\ref{eq:decconst})
we have to find convenient Green functions which have to be considered
both in the effective and full theory. 
For this reason we define the bare two-point functions for quarks,
gluons and ghosts as follows:
\begin{eqnarray}
  \frac{1}{m-\psla-\Sigma^0(p)}
  &=&
  i\int {\rm d}^4x\, e^{ipx} \langle T\psi_q^0(x)\bar{\psi}_q^0(0)
  \rangle
  \,,
  \nonumber\\
  \frac{\delta^{ab}\left(-g_{\mu\nu}+\frac{p^\mu p^\nu}{p^2}\right)}
  {-p^2\left(1+\Pi^0_G(p^2)\right)}
  + \ldots
  &=&
  i\int {\rm d}^4x\, e^{ipx} \langle T G^{0,a}_\mu(x) G^{0,b}_\nu(0)\rangle
  \,,
  \nonumber\\
  \frac{\delta^{ab}}{-p^2\left(1+\Pi^0_c(p^2)\right)}
  &=&
  i\int {\rm d}^4x\, e^{ipx} \langle T c^{0,a}(x)\bar{c}^{0,b}(0)\rangle
  \label{eq:2pfunc}
  \,.
\end{eqnarray}
The ellipses
in the case of the gluon propagator 
indicate the longitudinal part which we are not interested in.

From the two-point functions of Eqs.~(\ref{eq:2pfunc}) we will be able
to obtain $\zeta_2^0$, $\zeta_m^0$, $\zeta_3^0$ and
$\tilde{\zeta}_3^0$. In order to get a relation involving $\zeta_g^0$
one has at least to consider three-point functions where the coupling $g_s$
already appears at Born level.
As the vertex between the gluon and ghost is the least complex one we
will take it for the computation. 
In amputated form it is defined through
\begin{eqnarray}
  p^\mu g_s^0\left\{-if^{abc}\left[1+\Gamma_{G\bar cc}^0(p,k)\right]
  \right\}
  + \ldots
  &=&i^2\int {\rm d}x{\rm d}y\,{\rm e}^{i(p\cdot x+k\cdot y)}
  \left\langle Tc^{0,a}(x)\bar c^{0,b}(0)G^{0,c,\mu}(y)\right\rangle^{\rm
  1PI}
  \,,
  \nonumber\\
  \label{eq:Gccdef}
\end{eqnarray}
where $p$ and $k$ are the outgoing four-momenta of $c$ and $G$, 
respectively.
The ellipses indicate other colour structures we are not interested
in. Note that we pull out a factor $g_s^0$ on the left-hand side of 
Eq.~(\ref{eq:Gccdef}) as it is already present at Born level.

\begin{figure}[ht]
\begin{center}
\epsfxsize=15cm
\epsffile[136 634 470 722]{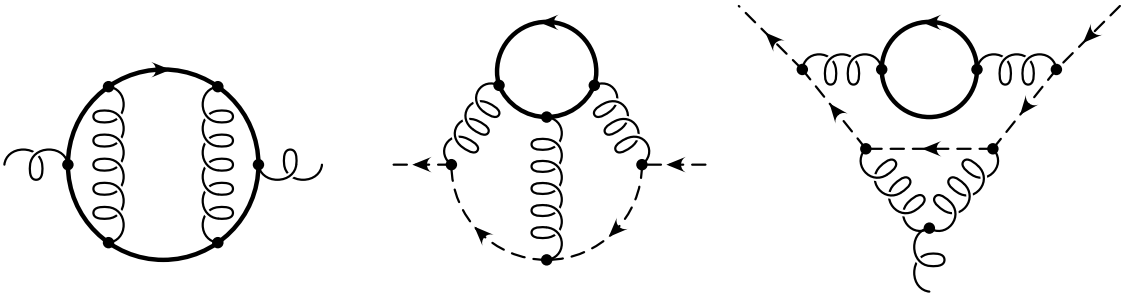}
\caption{\label{fig:zetag}
  Typical three-loop diagrams pertinent to $\Pi_G^{0,h}(0)$,
  $\Pi_c^{0,h}(0)$, and $\Gamma_{G\bar cc}^{0,h}(0,0)$.
  Bold-faced, loopy, and dashed lines represent heavy quarks $h$,
  gluons $G$,  
  and Faddeev-Popov ghosts $c$, respectively.}
\end{center}
\end{figure}

Let us start with the decoupling constant for the gluon field. It is
obvious that the gluon propagator constitutes a good candidate to
compute $\zeta_3^0$. Up to terms of order $1/m_h$ we have the
following chain of equations
\begin{eqnarray}
  \frac{\delta^{ab}\left(-g_{\mu\nu}+\frac{p^\mu p^\nu}{p^2}\right)}
  {-p^2\left(1+\Pi_G^0(p^2)\right)}
  &=&
  i\int {\rm d}^4x e^{ipx} \langle T G^{0,a}_\mu(x) G^{0,b}_\nu(0)\rangle
  \nonumber\\
  &=&
  \frac{1}{\zeta_3^0}
  i\int {\rm d}^4x e^{ipx} 
             \langle T G^{0,\prime,a}_\mu(x) G^{0,\prime,b}_\nu(0)\rangle
  \nonumber\\
  &=&
  \frac{1}{\zeta_3^0}\,\,
  \frac{\delta^{ab}\left(-g_{\mu\nu}+\frac{p^\mu p^\nu}{p^2}\right)}
  {-p^2\left(1+\Pi_G^{0\prime}(p^2)\right)}
  \,.
  \label{eq:decglu}
\end{eqnarray}
where in the second step Eqs.~(\ref{eq:decconst}) has been used. Note
that $\Pi_G^{0\prime}(p^2)$ only contains light degrees of freedom
whereas $\Pi_G^{0}(p^2)$  also receives virtual contributions from the heavy
quark $h$.
Eq.~(\ref{eq:decglu}) provides a formulae for $\zeta_3^0$
\begin{eqnarray}
  \zeta_3^0 &=& \frac{1+\Pi_G^0(p^2)}{1+\Pi_G^{0\prime}(p^2)}
  \label{eq:decglu2}
  \,.
\end{eqnarray}
From the construction of the effective theory it is clear that the
decoupling constants do not depend on the momentum transfer. Thus also
the right-hand side 
of~(\ref{eq:decglu2}) has to be independent of the external
momentum $p$. This means that we can choose any convenient momentum
for the computation. In particular it is possible to choose $p=0$.
This has the advantage that only vacuum diagrams have to be
considered. Since we work in the framework of dimensional
regularization all scaleless integrals can be set to zero. As a
consequence we have $\Pi_G^{0\prime}(0)=0$
and the contribution to $\Pi_G^{0}$
is given by the diagrams containing at least
one heavy quark line. In the following we attach to
the corresponding contributions
an additional index ``h'' and refer to it as the ``hard part''. 
Finally we arrive at the compact formula
\begin{eqnarray}
  \zeta_3^0 &=& 1 + \Pi_G^{0,h}(0)
  \label{eq:zeta30}
  \,.
\end{eqnarray}
The $n$-loop contribution to $\zeta_3^0$ is related to the
$n$-loop vacuum diagrams where the scale is given by the mass of
the heavy quark.
At one-loop order only one diagram contributes to $\zeta_3^0$. At two
loops there are already three diagrams and at three-loop order
altogether 189 diagrams have to be taken into account. A typical
example is pictured in Fig.~\ref{fig:zetag}.
We want to mention that the bare decoupling constants  
may contain non-local terms like
$\ln(\mu^2/m_h^2)/\varepsilon$, which is
in contrast to the renormalization constants.

After the computation of the bare diagrams the parameters (in our case
$\alpha_s$, $\xi$ and the heavy quark mass, $m_h$) have to be expressed in
terms of their renormalized counterparts. The finite decoupling
constant is obtained from Eq.~(\ref{eq:decconst}) after expressing the bare
fields in terms of the renormalized ones via Eq.~(\ref{eq:renconst})
\begin{eqnarray}
  G_\mu^\prime &=& \sqrt{\frac{Z_3\zeta_3^0}{Z_3^\prime}} G_\mu
              \,\,=\,\, \sqrt{\zeta_3} G_\mu
  \label{eq:zeta3}
  \,.
\end{eqnarray}
Note that $Z_3^\prime$ depends on $\alpha_s^\prime$ and $\xi^\prime$.
They have to be transformed to $\alpha_s$ and $\xi$ with the help of
$\zeta_g$ and $\zeta_3$, respectively, which are
needed up to $(l-1)$-loop accuracy 
if Eq.~(\ref{eq:zeta3}) is considered at $l$-loop order.

The ghost propagator can be treated in complete analogy and one
arrives at the following formula for the bare decoupling constant for the
ghost field
\begin{eqnarray}
  \tilde{\zeta}_3^0 &=& 1 + \Pi_c^{0,h}(0)
  \label{eq:tilzeta30}
  \,.
\end{eqnarray}
There is no diagram which contributes at one-loop order and one at
order $\alpha_s^2$. Also at three-loop order the number of diagrams is
moderate and amounts to 25.

The renormalized version of Eq.~(\ref{eq:tilzeta30}) reads
\begin{eqnarray}
  \tilde{\zeta}_3 &=& \frac{\tilde{Z}_3\tilde{\zeta}_3^0}{\tilde{Z}_3^\prime}
  \label{eq:tilzeta3}
  \,.
\end{eqnarray}

\begin{figure}[ht]
  \begin{center}
  \epsfxsize=15cm
  \epsffile[74 624 569 725]{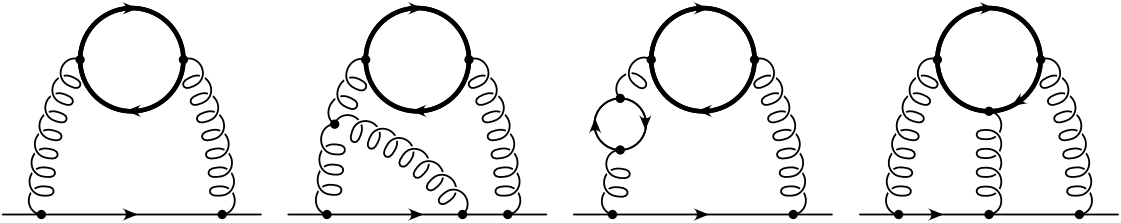}
  \caption{\label{fig:sigh}Typical three-loop diagrams 
  pertinent to $\Sigma_V^{0,h}(0)$ and
  $\Sigma_S^{0,h}(0)$.
  Solid, bold-faced, and loopy lines represent massless quarks $q$,
  heavy quarks $h$, and gluons $G$, respectively.}
  \end{center}
\end{figure}

In order to obtain expressions for $\zeta_2$ and $\zeta_m$ one
considers the light-quark 
propagator which leads to the following chain of
relations:
\begin{eqnarray}
  \frac{1}{m-\,\,/\hspace{-.5em}p-\Sigma(p)} 
  &=&
  \frac{1}
  {m\left(1-\Sigma_S^0(p^2)\right)
  -\,\,/\hspace{-.5em}p\left(1+\Sigma_V^0(p^2)\right)}
  \nonumber\\
  &=&
  \frac{1}
  {\zeta_2^0\zeta_m^0m\left(1-\Sigma_S^{0\prime}(p^2)\right)
  -\zeta_2^0\,\,/\hspace{-.5em}p\left(1+\Sigma_V^{0\prime}(p^2)\right)} 
  \,.
\end{eqnarray}
Nullifying the external momentum $p$ in the self energies leads to
\begin{eqnarray}
  \zeta_2^0 &=& 1+\Sigma_V^{0,h}(0)
  \label{eq:zeta20}
  \,,
  \\
  \zeta_m^0 &=& \frac{1-\Sigma_S^{0,h}(0)}{1+\Sigma_V^{0,h}(0)}
  \label{eq:zetam0}
  \,,
\end{eqnarray}
and the finite expressions are obtained 
from\footnote{Note that the same symbol is also used for Riemann's
zeta function $\zeta_2=\pi^2/6$. However, as they appear in
a completely different context confusion is not possible.}
\begin{eqnarray}
  \zeta_2 &=& \frac{Z_2\zeta_2^0}{Z_2^\prime}
  \label{eq:zeta2}
  \,,
  \\
  \zeta_m &=& \frac{Z_m\zeta_m^0}{Z_m^\prime}
  \label{eq:zetam}
  \,.
\end{eqnarray}
Similarly to $\Pi_c^{0,h}(0)$
there are no one-loop diagrams contributing to $\Sigma_V^{0,h}(0)$ and
$\Sigma_S^{0,h}(0)$ and at two- and three-loop order again
one and 25 diagrams, respectively, have to be considered.
Typical specimen are depicted in Fig.~\ref{fig:sigh}.
Actually, through three loops, Eq.~(\ref{eq:zetam}) simplifies to
$\zeta_m^0=1-\Sigma_V^{0,h}(0)-\Sigma_S^{0,h}(0)$.
It should be noted that the vector and scalar parts separately still depend on
the QCD gauge parameter $\xi$, but $\xi$ drops out in their sum, which is a
useful check for our calculation.

The derivation of a formula for $\zeta_g^0$ is slightly more
involved. As a starting point we choose the full, i.e. non-amputated
gluon-ghost Green function and get (using Eqs.~(\ref{eq:decconst})) 
the following equations 
\begin{eqnarray}
  g_s^0 G^{\mu,abc}_{G\bar{c}c}(p,k) 
  &=& i^2\int {\rm d}x{\rm d}y\,{\rm e}^{i(p\cdot x+k\cdot y)}
  \left\langle Tc^{0,a}(x)\bar c^{0,b}(0)G^{0,c,\mu}(y)\right\rangle
  \nonumber\\
  &=& \frac{1}{\tilde{\zeta}_3^0\sqrt{\zeta_3^0}}
      \,\,
      i^2\int {\rm d}x{\rm d}y\,{\rm e}^{i(p\cdot x+k\cdot y)}
  \left\langle Tc^{0,a}(x)\bar c^{0,b}(0)G^{0,c,\mu}(y)\right\rangle^\prime
  \nonumber\\
  &=& \frac{1}{\tilde{\zeta}_3^0\sqrt{\zeta_3^0}}
      g_s^{0,\prime} G^{\mu,abc,\prime}_{G\bar{c}c}(p,k) 
  \nonumber\\
  &=& \frac{\zeta_g^0}{\tilde{\zeta}_3^0\sqrt{\zeta_3^0}}
      g_s^{0} G^{\mu,abc,\prime}_{G\bar{c}c}(p,k) 
  \label{eq:Gcc}
\end{eqnarray}
where $p$ and $k$ are the outgoing four-momenta of $c$ and $G$,
respectively.
At this point we should amputate the Green functions by multiplying
with the inverse propagators of the external gluon and ghost
fields in the full theory. 
From the decoupling relations derived above we get
an additional factor $(\tilde\zeta_3^0)^2\zeta_3^0$ on
the right-hand side of Eq.~(\ref{eq:Gcc}) which leads to
\begin{eqnarray}
  \left[1+\Gamma_{G\bar cc}^0(p,k)\right]
  &=& 
  \zeta_g^0\tilde\zeta_3^0\sqrt{\zeta_3^0}
  \left[1+\Gamma_{G\bar cc}^{0\prime}(p,k)\right]
  \,.
\end{eqnarray}
Here, $\Gamma_{G\bar cc}^0(p,k)$ is defined through the
1PI part of the amputated 
gluon-ghost Green function as introduced in Eq.~(\ref{eq:Gccdef}).
Setting to zero the external momenta we obtain
\begin{eqnarray}
  \zeta_g^0&=&\frac{\tilde\zeta_1^0}{\tilde\zeta_3^0\sqrt{\zeta_3^0}}
  \,,
  \label{eq:zetag0}
\end{eqnarray}
where
\begin{eqnarray}
  \tilde\zeta_1^0&=&1+\Gamma_{G\bar cc}^{0,h}(0,0)
  \,,
\end{eqnarray}
is the decoupling constant belonging to the gluon-ghost vertex.
Again, the renormalized version of the decoupling constant is
obtained with the help of the renormalization constants in the full
and effective theory as
\begin{eqnarray}
  \zeta_g &=& \frac{Z_g\zeta_g^0}{Z_g^\prime}
  \,.
\end{eqnarray}
Thus, in order to compute $\zeta_g$ one has to evaluate the decoupling
constant for the gluon propagator, the ghost propagator and the
gluon-ghost vertex. We could have chosen also another vertex
involving the strong coupling, e.g. the quark-gluon or the three-gluon
vertex, and would have arrived at a similar expression as in 
Eq.~(\ref{eq:zetag0}).
This is in complete analogy to the renormalization constants
where due to Slavnov-Taylor identities
the various renormalization constants are related to each other
(see, e.g., Ref.~\cite{Muta}).

We should mention that
$\Gamma_{G\bar cc}^{0,h}(0,0)$ has no one-loop contribution and 
there are five diagrams at two loops, which, however, add up to zero.
A non-zero contribution is obtained at order $\alpha_s^3$ where 228
diagrams contribute. A typical representative is shown 
in Fig.~\ref{fig:zetag}.

At order $\alpha_s^2$
the three contributions to $\zeta_g^0$ are still separately
independent of the gauge parameter $\xi$, so that the $\xi$
independence of their combination 
does not provide a meaningful check for our calculation.
The situation changes at ${\cal O}(\alpha_s^3)$, where all three parts
separately depend on $\xi$ and only their proper combination is $\xi$
independent as is required for a physical quantity.
In the calculation this has been used as a check.

In Eqs.~(\ref{eq:zeta30}),~(\ref{eq:tilzeta30}),~(\ref{eq:zeta20}),
(\ref{eq:zetam0}) and~(\ref{eq:zetag0}) the bare decoupling constants 
$\zeta_3^0$, $\tilde{\zeta}_3^0$, $\zeta_2^0$, $\zeta_m^0$ and
$\zeta_g^0$ are expressed in terms of vacuum diagrams. Thus, if the
former are to be computed at order $\alpha_s^3$ the latter need to be
known at three-loop order. In the recent years three-loop vacuum
diagrams have been studied extensively and a {\tt FORM}~\cite{form}
package, {\tt MATAD}~\cite{matad},
has been written which allows for an automated computation. 

At this point we would like to make a comment on the different kind of
poles which appear in the calculation. If we choose zero external
momentum there are in general both UV and IR 
poles in the individual diagrams. However, the renormalization
constants in Eq.~(\ref{eq:renconst}) only contain UV divergences. 
Thus the two
different kind of poles have to be identified in order to arrive at a
finite expression for the decoupling constants.
This is a special feature of dimensional regularization which 
allows to treat simultaneously UV and IR divergences and
to set to zero scaleless integrals.
More details to this context can be found in Ref.~\cite{GorLar87}
and references cited therein.


\subsection{Results}

In the following we list the analytical results for the renormalized
decoupling constants $\zeta_g$ and $\zeta_m$,
which relate the physical parameters 
in the full theory to their counterparts
in the effective theory.
In Appendix~\ref{app:decconst} we also provide the (gauge parameter
dependent) results for $\zeta_2$ and $\zeta_3$.

As already mentioned above, the computation can be reduced to one-scale
vacuum integrals (see also Appendix~\ref{sub:single}). However, the
large number of diagrams requires the automation of the
computation. Actually the computation of $\zeta_g$ was one of the
first application of the package {\tt GEFICOM}~\cite{geficom}
(cf. Appendix~\ref{sub:aut}), which combines several 
stand-alone program packages in order to automate the computation from the
generation of the diagrams to the summation of the individual results.
For the evaluation of the decoupling constants
all diagrams have been generated automatically with the {\tt Fortran}
program {\tt QGRAF}~\cite{qgraf} and the integrations have been
performed using the {\tt FORM}~\cite{form} package {\tt MATAD}~\cite{matad}.

Note, that in
contrast to the renormalization constants $Z_i$ in Eq.~(\ref{eq:renconst}),
the decoupling constants $\zeta_i^0$ also receive contributions from the
finite parts of the loop integrals.
Thus, at ${\cal O}(\alpha_s^3)$, we are led to evaluate three-loop tadpole
integrals also retaining their finite parts.

For illustration of the formalism derived in the previous Subsection
we want to compute the lowest-order contribution to $\zeta_2$ which is
of ${\cal O}(\alpha_s^2)$ and comes from the first diagram in
Fig.~\ref{fig:sigh}.
According to Eq.~(\ref{eq:zeta20}) the vector part has to be evaluated
for zero external momentum. This is conveniently done with the help of
\begin{eqnarray}
  \Sigma_V^{0,h}(0) &=& 
  \frac{1}{4p^2}\mbox{Tr}\left[\psla\Sigma^{0,h}(p)\right]\Big|_{p=0}
  \,,
\end{eqnarray}
where a simple Taylor expansion up to linear order has to be performed
for $\Sigma^{0,h}(p)$.
As only vacuum integrals are involved in the computation one can
easily perform the tensor reduction in the numerator
and then use Eq.~(\ref{eq:Vabc})
in order to arrive at the result for the bare decoupling constant
\begin{eqnarray}
  \zeta_2^0 &=& 1 + \left(\frac{\alpha_s}{\pi}\right)^2 C_F T \left[
    -\frac{1}{16\varepsilon} +\frac{5}{96} 
      - \frac{1}{8}\ln\frac{\mu^2}{m_h^2}
    + {\cal O}\left(\varepsilon\right)
  \right]
  \,.
  \label{eq:zeta202l}
\end{eqnarray}
Now Eq.~(\ref{eq:zeta2}) can be used to get a finite result.
Therefore the ratio $Z_2/Z_2^\prime$ is needed to order $\alpha_s^2$
where the parameters $\alpha_s^\prime$ and $\xi^\prime$ in
$Z_2^\prime$ have to be expressed in terms of $\alpha_s$ and $\xi$.
with the help of Eqs.~(\ref{eq:decconst}). Actually, to our order
the contributions from $\zeta_g$ and $\zeta_3$ exactly cancel and
we are left with
\begin{eqnarray}
  \frac{Z_2}{Z_2^\prime} &=& 1 + 
  \left(\frac{\alpha_s}{\pi}\right)^2 \frac{C_F T}{16\varepsilon}
  \,.
  \label{eq:Z2Z2p}
\end{eqnarray}
Inserting Eqs.~(\ref{eq:zeta202l}) and~(\ref{eq:Z2Z2p}) 
in~(\ref{eq:zeta2}) finally leads to
\begin{eqnarray}
  \zeta_2^{\rm MS} &=& 1 + \left(\frac{\alpha_s}{\pi}\right)^2 C_F T \left[
      \frac{5}{96} - \frac{1}{8}\ln\frac{\mu^2}{m_h^2}
  \right]
  \,,
\end{eqnarray}
which agrees with Eq.~(\ref{eq:zeta2res}).

In the same way also the scalar part of the quark self energy can be
treated in order to obain with the help of
Eq.~(\ref{eq:zetam0}) a result for $\zeta_m$. Taking also the
three-loop diagrams into account we obtain
\begin{eqnarray}
\zeta_m^{\rm OS}&=&1
+\left(\frac{\alpha_s^{(n_f)}(\mu)}{\pi}\right)^2
\left(\frac{89}{432} 
     -\frac{5}{36}\ln\frac{\mu^2}{M_h^2}
     +\frac{1}{12}\ln^2\frac{\mu^2}{M_h^2}
\right)
+\left(\frac{\alpha_s^{(n_f)}(\mu)}{\pi}\right)^3
\left[
\frac{1871}{2916} 
\right.
\nonumber\\&&\left.\mbox{}
- \frac{407}{864}\zeta_3
+\frac{5}{4}\zeta_4
- \frac{1}{36}B_4
+\left(\frac{121}{2592}
- \frac{5}{6}\zeta_3\right)\ln\frac{\mu^2}{M_h^2}
+ \frac{319}{432}\ln^2\frac{\mu^2}{M_h^2}
\right.
\nonumber\\&&\left.\mbox{}
+ \frac{29}{216}\ln^3\frac{\mu^2}{M_h^2}
+ n_l\left(
\frac{1327}{11664}
- \frac{2}{27}\zeta_3
- \frac{53}{432}\ln\frac{\mu^2}{M_h^2}
- \frac{1}{108}\ln^3\frac{\mu^2}{M_h^2}
\right)
\right]
\nonumber\\
&\approx&1
+0.2060\left(\frac{\alpha_s^{(n_f)}(M_h)}{\pi}\right)^2
+\left(1.4773+0.0247\,n_l\right)
\left(\frac{\alpha_s^{(n_f)}(M_h)}{\pi}\right)^3.
\label{eq:zetamOS}
\end{eqnarray}
where $\zeta_3\approx1.202\,057$ and
\begin{eqnarray}
B_4&=&16\li\left({1\over2}\right)-{13\over2}\zeta_4-4\zeta_2\ln^22
+{2\over3}\ln^42
\nonumber\\
&\approx&-1.762\,800
\,,
\end{eqnarray}
with $\li{}$ being the quadrilogarithm, is a constant typical for three-loop
vacuum diagrams~\cite{Bro92}. 
$n_l=n_f-1$ is the number of light-quark flavours, and
$M_h$ is the on-shell mass of the heavy quark $h$.
For the numerical evaluation in the last line of Eq.~(\ref{eq:zetamOS}),
we have chosen $\mu=M_h$.
The ${\cal O}(\alpha_s^2)$ term of Eq.~(\ref{eq:zetamOS}) is computed in
Ref.~\cite{BerWet82Ber83} and the ${\cal O}(\alpha_s^3)$ term 
can be found in~\cite{CheKniSte98}.

The proper combination of the 
three ingredients entering the calculation of $\zeta_g^0$, namely
the hard parts of the gluon and ghost propagators and the 
gluon-ghost vertex correction, lead to
\begin{eqnarray}
\left(\zeta_g^{\rm OS}\right)^2&=&1
+\frac{\alpha_s^{(n_f)}(\mu)}{\pi}
\left(
-\frac{1}{6}\ln\frac{\mu^2}{M_h^2}
\right)
+\left(\frac{\alpha_s^{(n_f)}(\mu)}{\pi}\right)^2
\left(
-\frac{7}{24} 
-\frac{19}{24}\ln\frac{\mu^2}{M_h^2}
+\frac{1}{36}\ln^2\frac{\mu^2}{M_h^2}
\right)
\nonumber\\
&&\mbox{}+\left(\frac{\alpha_s^{(n_f)}(\mu)}{\pi}\right)^3
\left[
-\frac{58933}{124416}
-\frac{2}{3}\zeta_2\left(1+\frac{1}{3}\ln2\right)
-\frac{80507}{27648}\zeta_3
-\frac{8521}{1728}\ln\frac{\mu^2}{M_h^2}
\right.\nonumber\\
&&\left.\mbox{}-
\frac{131}{576}\ln^2\frac{\mu^2}{M_h^2}
-\frac{1}{216}\ln^3\frac{\mu^2}{M_h^2} 
+n_l\left(
\frac{2479}{31104}
+\frac{\zeta_2}{9}
+\frac{409}{1728}\ln\frac{\mu^2}{M_h^2} 
\right)
\right]
\nonumber\\
&\approx&1
-0.2917\left(\frac{\alpha_s^{(n_f)}(M_h)}{\pi}\right)^2
+\left(-5.3239+0.2625\,n_l\right)
\left(\frac{\alpha_s^{(n_f)}(M_h)}{\pi}\right)^3
\,.
\label{eq:zetagOS}
\end{eqnarray}
The ${\cal O}(\alpha_s^3)$ term in Eq.~(\ref{eq:zetagOS}) is
published~\cite{CheKniSte98}.
Leaving aside this term, the results in
Eq.~(\ref{eq:zetagOS}) can be found in 
Refs.~\cite{BerWet82Ber83,LarvRiVer95,Ber97}.

Notice that the ${\cal O}(\alpha_s^3)$ terms of $\zeta_m$ and $\zeta_g$ depend
on the number $n_l$ of light (massless) quark flavours.
However, this dependence is feeble.

The generalization of Eqs.~(\ref{eq:zetamOS}) and~(\ref{eq:zetagOS}), 
appropriate for the gauge group SU($N_c$), is listed in 
Appendix~\ref{app:decconst}.
There, also the results expressed in terms of the 
$\overline{\rm MS}$ quark mass can be found.
In Ref.~\cite{Ste98_higgs} the leading Yukawa corrections proportional
to $G_F M_t^2$ have been computed for $\zeta_m$ and $\zeta_g$.
In Appendix~\ref{app:decconst} also these results are listed.


\subsection{Applications}

\subsubsection*{Cross section \boldmath{$\sigma(e^+e^-\to b\bar{b})$}}

At this point we would like to pick up the example mentioned at
the beginning of this section, namely the ${\cal O}(\alpha_s^2)$
corrections to the cross section $e^+e^-\to b\bar{b}$. If we
consider center-of-mass energies where the first five quarks can be
neglected one obtains\footnote{Note that this result can immediately
  be obtained from the example considered in Appendix~\ref{sub:ae},
  Eq.~(\ref{eq:dbhmpres}).}
\begin{eqnarray}
  R(s) &=& 3 Q_b^2\, \Bigg\{1+\frac{\alpha_s^{(6)}(\mu)}{\pi}
       + \left(\frac{\alpha_s^{(6)}(\mu)}{\pi}\right)^2\left[
         c_2 -\frac{1}{6} \ln\frac{\mu^2}{M_t^2} 
         + \frac{s}{4M_t^2}\left( 
             \frac{176}{675}-\frac{8}{135}\ln\frac{s}{M_t^2} 
                           \right) 
  \right.\nonumber\\&&\left.\mbox{}
         + {\cal O}\left(\frac{s^2}{M_t^4}\right)
         \right]
         \Bigg\}
       + {\cal O}(\alpha_s^3)
       \,,
  \label{eq:Rsas2}
\end{eqnarray}
where $N_c=3$ has been chosen and
at order $\alpha_s^2$ only the contribution from
the diagram in Fig.~\ref{fig:eeqqheavy} is displayed explicitly. All other
diagrams are summed in the constant 
$c_2$~\cite{CheKatTka79DinSap79CelGon80}.
In Eq.~(\ref{eq:Rsas2}) the definition of the coupling constant 
still includes the top quark as indicated by the superscript ``(6)''.
Otherwise the computation of the diagram in Fig.~\ref{fig:eeqqheavy}
would hardly be possible. One recognizes that for $M_t\to\infty$
the contribution from the top quark raises logarithmically. 
Note that the choice $\mu=M_t$ does not help as it introduces 
$\ln M_t$ terms in $c_2$. 
At this point Eqs.~(\ref{eq:decconst}) and~(\ref{eq:zetagOS}) 
can be used to replace
$\alpha_s^{(6)}$ in favour of $\alpha_s^{(5)}$ which leads to
\begin{eqnarray}
  R(s) &=& 3 Q_b^2\, \Bigg\{1+\frac{\alpha_s^{(5)}(\mu)}{\pi}
       + \left(\frac{\alpha_s^{(5)}(\mu)}{\pi}\right)^2\left[
         c_2
         + \frac{s}{4M_t^2}\left( 
             \frac{176}{675}-\frac{8}{135}\ln\frac{s}{M_t^2} 
                           \right) 
  \right.\nonumber\\&&\left.\mbox{}
         + {\cal O}\left(\frac{s^2}{M_t^4}\right)
         \right]
         \Bigg\}
       + {\cal O}(\alpha_s^3)
       \,.
  \label{eq:Rsas2dec}
\end{eqnarray}
Now the top quark is decoupled, i.e. its contribution goes to zero for
$M_t\to \infty$
as $R(s)$ is expressed in terms of $\alpha_s^{(5)}$ which is 
the parameter of the effective theory.

\subsubsection*{Determination of $\alpha_s^{(5)}(M_Z)$ and
  $m_q^{(5)}(M_Z)$ from measurements at the $\tau$ mass scale}

In the previous example the decoupling relation was only needed to
one-loop order. However, the three-loop terms 
will be indispensable in order to relate the QCD predictions for
different observables at next-to-next-to-next-to-leading order.
Meaningful estimates of such corrections already
exist~\cite{Sam95,CheKniSir97}.

Once the corrections of order $\alpha_s^4$ for $R(s)$
are known they can be used to
determine $\alpha_s^{(5)}(M_Z)$ from the knowledge of
$\alpha_s^{(4)}(M_\tau)$ and compare it with other measurements.
One would use $R(s)$ to extract\footnote{The described procedure can
  also be applied to $\alpha_s^{(3)}(M_\tau)$. Only for simplicity and
  transparency we have chosen $\alpha_s^{(4)}(M_\tau)$.}
$\alpha_s^{(4)}(M_\tau)$ with an
accuracy of order $\alpha_s^4$ from the data. Then one would use the 
four-loop $\beta$ function~\cite{RitVerLar97_bet}
in order to perform the running to the bottom-quark threshold.
There the three-loop matching relations would be necessary for a
consistent decoupling. Using again four-loop running finally leads to 
$\alpha_s^{(5)}(M_Z)$.
In the following we will illustrate 
this procedure. However, instead of determining $\alpha_s^{(4)}(M_\tau)$
via $R(s)$ we directly assume a value for 
$\alpha_s^{(4)}(M_\tau)$
and evaluate $\alpha_s^{(5)}(M_Z)$ for different choices of the
matching scale $\mu^{(5)}$.
For the three-loop evolution in
connection with two-loop matching this has been done in Ref.~\cite{Rod93}.
We are in a position to explore the situation at 
the next order. It is instructive to include in the analysis also the 
tree-level and one-loop matching.

\begin{figure}[t]
\begin{center}
\epsfxsize=\textwidth
\epsffile[90 275 463 579]{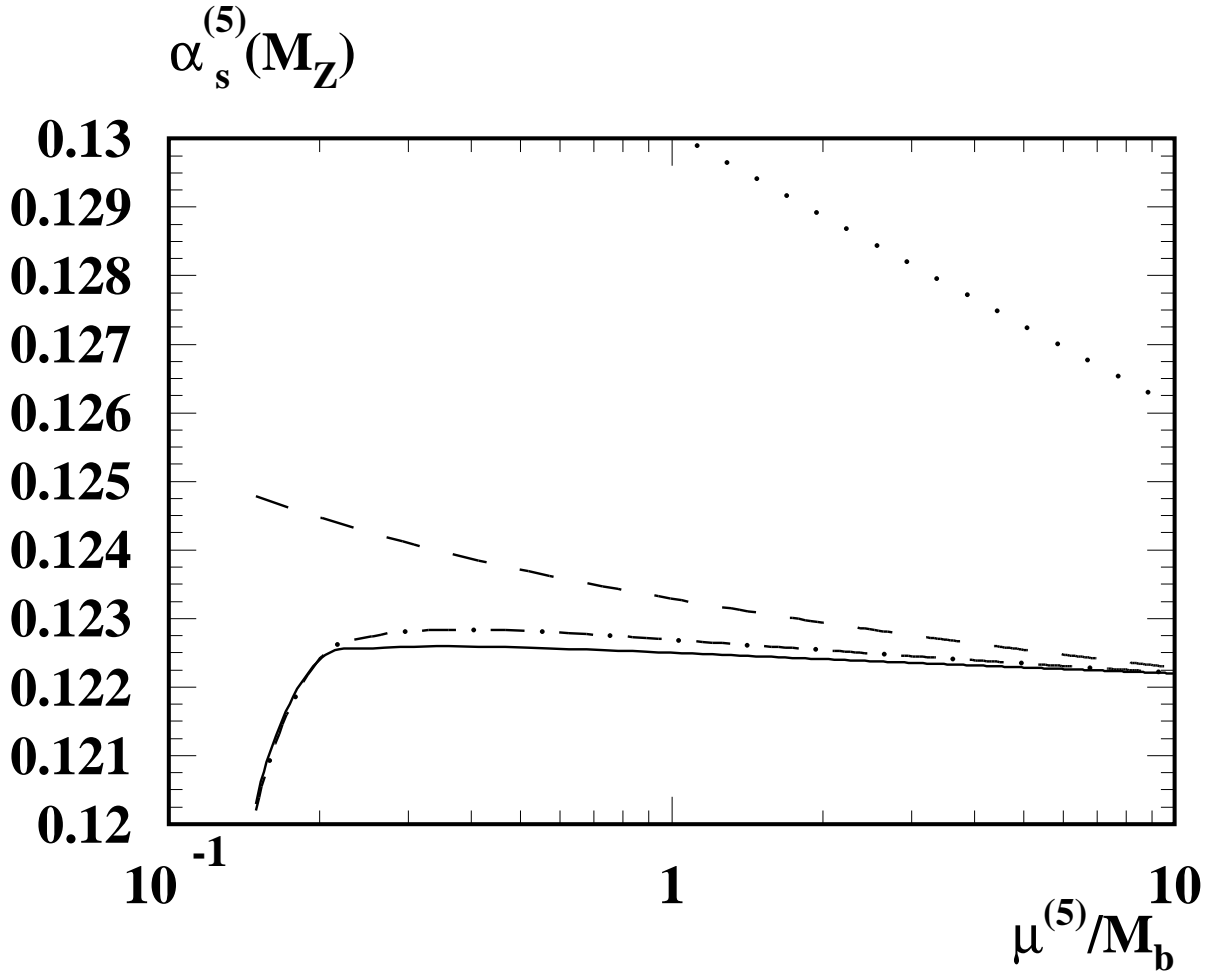}
\caption{\label{fig:alsmz}
  $\mu^{(5)}$ dependence of $\alpha_s^{(5)}(M_Z)$ calculated from
  $\alpha_s^{(4)}(M_\tau)=0.36$ and $M_b=4.7$~GeV using 
  Eq.~(\ref{eq:defbeta}) at
  one (dotted), two (dashed), three (dot-dashed), and four (solid) loops in
  connection with Eqs.~(\ref{eq:decconst}) and 
  (\ref{eq:zetagOS}) at the respective
  orders.}
\end{center}
\end{figure}

Going to higher orders, one expects, on general grounds, that the relation
between $\alpha_s^{(n_f-1)}(\mu^\prime)$ and $\alpha_s^{(n_f)}(\mu)$, where
$\mu^\prime\ll\mu^{(n_f)}\ll\mu$, becomes insensitive to the choice of the
matching scale, $\mu^{(n_f)}$, as long as $\mu^{(n_f)}={\cal O}(m_h)$.
In the above-mentioned situation
we consider the crossing of the bottom-quark threshold.
In particular, we study how the $\mu^{(5)}$ dependence of the relation
between $\alpha_s^{(4)}(M_\tau)$ and $\alpha_s^{(5)}(M_Z)$ is reduced as we
implement four-loop evolution with three-loop matching.
Our procedure is as follows.
We first calculate $\alpha_s^{(4)}(\mu^{(5)})$ by exactly integrating
Eq.~(\ref{eq:defbeta}) with the initial condition
$\alpha_s^{(4)}(M_\tau)=0.36$,  
then obtain $\alpha_s^{(5)}(\mu^{(5)})$ from Eqs.~(\ref{eq:decconst})
and (\ref{eq:zetagOS}) with $M_b=4.7$~GeV, and finally compute
$\alpha_s^{(5)}(M_Z)$ with Eq.~(\ref{eq:defbeta}).
For consistency, $N$-loop evolution must be accompanied by $(N-1)$-loop 
matching, i.e.\ if we omit terms of ${\cal O}(\alpha_s^{N+2})$ on the 
right-hand side of Eq.~(\ref{eq:defbeta}), we need to discard those of
${\cal O}(\alpha_s^N)$ in Eq.~(\ref{eq:zetagOS}) at the same time.
In Fig.~\ref{fig:alsmz}, 
the variation of $\alpha_s^{(5)}(M_Z)$ with $\mu^{(5)}/M_b$
is displayed for the various levels of accuracy, ranging from one-loop to
four-loop evolution.
For illustration, $\mu^{(5)}$ is varied rather extremely, by almost two orders
of magnitude.
While the leading-order result exhibits a strong logarithmic behaviour, the
analysis is gradually getting more stable as we go to higher orders.
The four-loop curve is almost flat for $\mu^{(5)}\gsim1$~GeV.
Besides the $\mu^{(5)}$ dependence of $\alpha_s^{(5)}(M_Z)$, also its absolute 
normalization is significantly affected by the higher orders.
At the central matching scale $\mu^{(5)}=M_b$, we encounter a rapid, monotonic
convergence behaviour.

\begin{figure}[t]
\begin{center}
\epsfxsize=\textwidth
\epsffile[99 275 463 562]{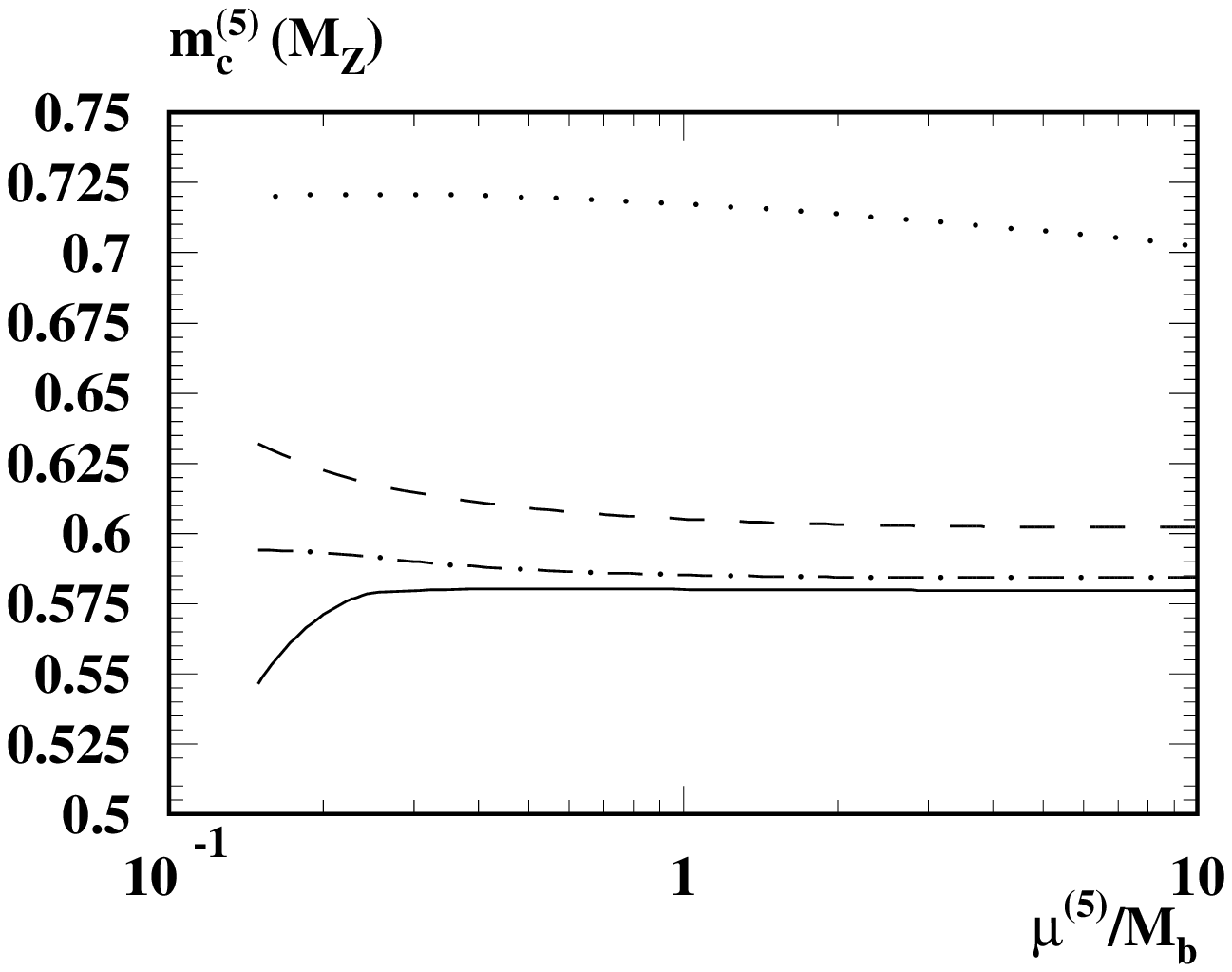}
\caption{\label{fig:mcmz}
  $\mu^{(5)}$ dependence of $m_c^{(5)}(M_Z)$ calculated from
  $\mu_c=m_c^{(4)}(\mu_c)=1.2$~GeV, $M_b=4.7$~GeV, and
  $\alpha_s^{(5)}(M_Z)=0.118$ using Eq.~(\ref{eq:defgamma}) 
  at one (dotted), two
  (dashed), three (dot-dashed), and four (solid) loops in connection with
  Eqs.~(\ref{eq:decconst}) and (\ref{eq:zetamOS}) at the respective orders.}
\end{center}
\end{figure}

Similar analyses may be performed for the light-quark masses as well.
For illustration, let us investigate how the $\mu^{(5)}$ dependence of the
relation between $\mu_c=m_c^{(4)}(\mu_c)$ and $m_c^{(5)}(M_Z)$ changes under
the inclusion of higher orders in evolution and matching.
As typical input parameters, we choose $\mu_c=1.2$~GeV, $M_b=4.7$~GeV, and
$\alpha_s^{(5)}(M_Z)=0.118$.
We first evolve $m_c^{(4)}(\mu)$ from $\mu=\mu_c$ to $\mu=\mu^{(5)}$ 
with the help of
Eq.~(\ref{eq:defgamma}), then obtain $m_c^{(5)}(\mu^{(5)})$ 
from Eqs.~(\ref{eq:decconst})
and (\ref{eq:zetamOS}), and finally evolve $m_c^{(5)}(\mu)$ from
$\mu=\mu^{(5)}$ to $\mu=M_Z$ using Eq.~(\ref{eq:defgamma}).
In all steps, $\alpha_s^{(n_f)}(\mu)$ is evaluated with the same values of
$n_f$ and $\mu$ as $m_c^{(n_f)}(\mu)$.
In Fig.~\ref{fig:mcmz}, we show the resulting values of $m_c^{(5)}(M_Z)$ 
corresponding to $N$-loop evolution with $(N-1)$-loop matching for 
$N=1,\ldots,4$.
Similarly to Fig.~\ref{fig:alsmz}, we observe a rapid, monotonic convergence
behaviour at the central matching scale $\mu^{(5)}=M_b$.
Again, the prediction for $N=4$ is remarkably stable under the variation of
$\mu^{(5)}$ as long as $\mu^{(5)}\gsim1$~GeV.

The various formulae describing 
the running and the decoupling of $\alpha_s$ and the quark masses
are implemented in the program package {\tt RunDec}~\cite{rundec}. 
It is realized in {\tt Mathematica} and
provides a convenient possibility to check, e.g., the figures of this
Subsection and eventually update the numerical input values.
In particular,
Fig.~\ref{fig:alsmz} can easily
be reproduced with the help of the procedure {\tt AlL2AlH[]}.
After loading {\tt RunDec} the command 
\verb|AlL2AlH[0.36,1.777,{{5,4.7,mu5}},91.187,l]|
provides the result for $\alpha_s^{(5)}(M_Z)$ where the matching has
been performed at the scale \verb|mu5|.
$l=1,2,3,4$ corresponds to the number of loops 
used for the evolution and 
the values $\alpha_s^{(4)}(M_\tau)=0.36$, $M_\tau=1.777$~GeV,
$M_b=4.7$~GeV and $M_Z=91.187$~GeV have been adopted.


\section{\label{sec:dim4}Scalar dimension-four operators in QCD}
\setcounter{equation}{0} 
\setcounter{figure}{0} 
\setcounter{table}{0} 

In the last section we constructed an effective QCD Lagrangian which
results from integrating out a heavy quark. From the knowledge of
its structure we determined the coefficient functions (i.e. the
decoupling constants) by computing bare Green functions in the full
and effective theory.

A different point of view is adopted for the
construction of the so-called non-relativistic QCD
(NRQCD)~\cite{Bodwin:1995jh}.
It has been developed in the context of heavy quarkonium physics in order to
separate the relativistic scales associated with the mass of the heavy
quark, $M$, from the non-relativistic ones which are of the order 
$Mv$, where $v$ is the velocity of the quark.
As a result one obtains an effective Lagrangian
which is ordered in inverse powers of $M$. More recently a further
step has
been undertaken and potential NRQCD (pNRQCD) has been introduced in
order to account for the separation of the scales $Mv$ and $Mv^2$. For
recent overviews we refer to~\cite{Brambilla:2000cs,Pen01}.

In this section a different approach is considered.
It is based
on Wilson's operator product expansion (OPE)~\cite{Wil69}
which is a powerful method for the construction of effective theories. 
The basic idea is to introduce local operators ${\cal O}_i$ of
appropriate dimension. They are formed by the light degrees of
freedom and parameterize the long-distance behaviour.
The operators are accompanied by coefficient functions which contain
the remnant of the large parameters of the theory.

As a simple example of the OPE one can consider the decay amplitude of
the muon in the SM. Due to the fact that the momentum transfer is
much smaller than the mass of the $W$ boson, $M_W$, the former can be
neglected with respect to the latter. This leads to the famous 
four-fermion interaction which is generated from dimension-six
operators. Thus, for dimensional reasons the coefficient functions 
contain a factor $1/M_W^2$.

In QCD the scalar operators of dimension four have been studied in great 
detail~\cite{Klu75,Nie75,Spi84}. 
A comprehensive survey concerning their renormalization
properties and mixings is performed in~\cite{Spi84}.
In particular, all renormalization constants of the operators have
been expressed in terms of renormalization constants which already
appear in the QCD Lagrangian (cf. Eq.~(\ref{eq:renconst})).

In this Section we want to discuss two applications of the 
scalar dimension-four operators. In the first one we consider
the decay of an intermediate-mass Higgs boson 
with $M_W\lsim M_H\lsim 2M_W$
into quarks or gluons. Here the top
quark is considered to be heavy and thus it will only
contribute to the
coefficient functions. On the other hand, the scale in the operators is
given by the mass of the Higgs boson.
A somehow complementary situation is considered in the second example:
the quartic quark mass corrections to the vector current correlators.
Here the mass of the quark sets the scale in the operators. It is 
considered to be small as compared to the external momentum which
manifests itself in the coefficient functions.

For definiteness we want to list the operators of dimension four 
in this introductory part and briefly discuss their renormalization.
In~\cite{Spi84} the operators are classified into gauge-invariant and
non-gauge-invariant ones. Furthermore a distinction is made whether the
operators vanish or not by virtue of the equation of motion.
The gauge-invariant operators 
read\footnote{For consistency, the operators should also have a
  prime as a superscript as they are built by quantities of the
  effective theory. However, we refrain from introducing this
  additional index.}
\begin{eqnarray}
  {\cal O}^{0}_1&=&\left(G^{0,\prime,a}_{\mu\nu}\right)^2 \,,
  \nonumber\\
  {\cal O}^{0}_2 &=& 
  m_q^{0,\prime}\bar\psi_{q}^{0,\prime}\psi_{q}^{0,\prime} \,,
  \nonumber\\
  {\cal O}^{0}_3&=&
  \bar\psi_{q}^{0,\prime}
  \left(i\not\!\!D^{0,\prime}-m_{q}^{0,\prime}\right)
  \psi_{q}^{0,\prime} 
  \,,
  \nonumber\\
  {\cal O}_6^{0} &=& \left(m_q^{0,\prime}\right)^4 
  \,,
  \label{eq:op1}
\end{eqnarray}
where $G^{0,a}_{\mu\nu}$ and $D_\mu$ are defined after Eq.~(\ref{eq:L_QCD}).
Note that ${\cal O}^{0}_3$ vanishes after the application of 
the equation of motion.

In order to obtain a closed system two more
operators have to be considered
\begin{eqnarray}
  {\cal O}^{0}_4&=&G_\nu^{0,\prime,a}
  \left(\nabla^{ab}_\mu G^{0,\prime,b\mu\nu}
  +g_s^{0,\prime}\sum_{i=1}^{n_l}\bar\psi_{q_i}^{0,\prime}
  \frac{\lambda^a}{2}\gamma^\nu\psi_{q_i}^{0,\prime}\right)
  -\partial_\mu \bar{c}^{0,\prime,a}\partial^\mu c^{0,\prime,a}
  \,,
  \nonumber\\
  {\cal O}^{0}_5&=&
  (\nabla^{ab}_\mu \partial^\mu \bar{c}^{0,\prime,b}) c^{0,\prime,a}
  \,,
  \label{eq:op3}
\end{eqnarray}
where 
$\nabla_\mu^{ab}=\delta^{ab}\partial_\mu-g_sf^{abc}G_\mu^c$ 
is the covariant derivative acting on the gluon and ghost fields.
The operators in Eq.~(\ref{eq:op3}) 
are not gauge-invariant and thus do not contribute to physical
observables.

From the practical point of view the operators
of Eqs.~(\ref{eq:op1}) and~(\ref{eq:op3}) 
define new Feynman rules which can be read off from
Eqs.~(\ref{eq:op1}) and~(\ref{eq:op3}). E.g., in
Section~\ref{sub:hadrHiggs} we have to compute the correlator 
$\langle {\cal O}_1 {\cal O}_1 \rangle$ up to three-loop order. 
This makes it necessary to extract the Feynman rules for the
coupling of ${\cal O}_1$ to one-, two-, three- and four gluons
from Eq.~(\ref{eq:op1}).
We refrain from listing them explicitly but consider as illustrative
examples the coupling of ${\cal O}_1^0$ and ${\cal O}_4^0$ to two
gluons. This will be useful further below in
Section~\ref{sub:higgs} in order to demonstrate the evaluation of
the corresponding coefficient functions.

\begin{figure}[ht]
\leavevmode
\begin{center}
\epsfxsize=5cm
\epsffile[145 345 455 435]{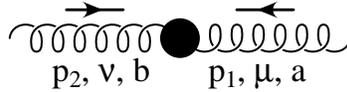}
\caption{\label{fig:O1O4FR}Coupling to two gluons. The solid circle
either represents ${\cal O}_1^0$ or ${\cal O}_4^0$ and
$a$ and $b$ are the colour indices of the gluons. The momenta $p_1$
and $p_2$ are incoming.
  }
\end{center}
\end{figure}

The coupling to two gluons is obtained from those terms of
Eqs.~(\ref{eq:op1}) and~(\ref{eq:op3}) containing two
gluon fields. In the case of ${\cal O}_1^0$ and ${\cal O}_4^0$
they read
\begin{eqnarray}
  {\cal O}_1^0\Big|_{\mbox{\tiny two gluons}}
  &=& 2\left(\partial_\mu  G_\nu^{0,\prime,a} \partial^\mu G^{\nu,0,\prime,b}
            -\partial_\mu  G_\nu^{0,\prime,a} \partial^\nu G^{\mu,0,\prime,b}
       \right)
  \,,
  \nonumber\\
  {\cal O}_4^0\Big|_{\mbox{\tiny two gluons}}
  &=& G_\nu^{0,\prime,a} \Box G^{\nu,0,\prime,b}
    - G_\nu^{0,\prime,a} \partial^\nu \partial_\mu G^{\mu,0,\prime,b}
  \,.
\end{eqnarray}
If we adopt the momenta flow as shown in Fig.~\ref{fig:O1O4FR} this
results in the following Feynman rules for the vertices
\begin{eqnarray}
  V_{gg{\cal O}_1}^{\mu\nu,ab}(p_1,p_2) &=& 
  i \delta^{ab} 4 \left[ - g^{\mu\nu} p_1\cdot p_2 + p_1^\nu p_2^\mu \right]
  \,,
  \nonumber\\
  V_{gg{\cal O}_4}^{\mu\nu,ab}(p_1,p_2) &=& 
  i \delta^{ab} 
     \left[ -g^{\mu\nu} \left( p_1\cdot p_1 + p_2\cdot p_2 \right) 
           + p_2^\mu p_2^\nu + p_1^\mu p_1^\nu \right]
  \,.
  \label{eq:O1O4FR}
\end{eqnarray}

In the applications we discuss below the 
operators ${\cal O}_3$, ${\cal O}_4$ and ${\cal O}_5$ and
the corresponding coefficient functions only appear in bare
form.
Thus we will not specify their renormalization, which can be
found in~\cite{Spi84}, and concentrate in the following
on ${\cal O}_1$, ${\cal O}_2$ and ${\cal O}_6$. The relation between
the bare and the renormalized operators reads
\begin{eqnarray}
  {\cal O}_n &=& \sum_m Z_{nm} {\cal O}_m^0
  \,,
\end{eqnarray}
where the indices $n$ and $m$ adopt the values 1, 2 and 6.
Due to the equality
\begin{eqnarray}
  \sum_n C_n^0 {\cal O}_n^0 &=& \sum_n C_n {\cal O}_n
  \,,
\end{eqnarray}
we obtain the renormalization prescription for the coefficient
functions as
\begin{eqnarray}
  C_n &=& \sum_m \left(Z^{-1}\right)_{mn} C_m^0 
  \,.
\end{eqnarray}
The anomalous dimension matrix pertaining to $Z_{nm}$ is defined
through
\begin{eqnarray}
  \mu^2 \frac{{\rm d}}{{\rm d}\mu^2} {\cal O}_n
  &=& \sum_m \gamma_{mn} {\cal O}_m
  \,.
\end{eqnarray}
It is connected to the renormalization matrix through
\begin{eqnarray}
  \gamma_{nm} &=& 
  \sum_k \left(\mu^2 \frac{{\rm d}}{{\rm d}\mu^2}Z_{mk}\right)
  \left(Z^{-1}\right)_{kn}
  \,,
\end{eqnarray}
and reads in explicit form~\cite{Klu75,Nie75,Spi84}
\begin{eqnarray}
    \gamma &=&
    \left(
      \begin{array}{ccc}
        -\alpha_s\pi\frac{\partial}{\partial\alpha_s}\frac{\beta}{\alpha_s}
        &
        4\alpha_s\frac{\partial}{\partial\alpha_s}\gamma_m
        &
        4\alpha_s\frac{\partial}{\partial\alpha_s}\gamma_0
        \\
        0 & 0 & - 4 \gamma_0
        \\
        0 & 0 & 4 \gamma_m 
      \end{array}
    \right)
    \,.
    \label{eq:gam_mn}
\end{eqnarray}
Thus $\gamma_{nm}$ can be expressed in terms of the functions
$\beta(\alpha_s)$ (see Eq.~(\ref{eq:defbeta})), 
$\gamma_m(\alpha_s)$ (see Eq.~(\ref{eq:defgamma}))
and the anomalous dimension of the vacuum energy $\gamma_0$
which is given by~\cite{CheSpi87,CheKue94,Che:priv}
\begin{eqnarray}
  \gamma_0 &=& -\frac{3}{16\pi^2}\Bigg\{1 
               + \frac{4}{3}\frac{\alpha_s}{\pi}
               + \left(\frac{\alpha_s}{\pi}\right)^2
                 \left[ \frac{313}{72}
                       -\frac{2}{3}\zeta_3
                       -\frac{5}{12}n_f
                 \right]
  \nonumber\\&&\mbox{}
               + \left(\frac{\alpha_s}{\pi}\right)^3
      \left[ \frac{14251}{1296}
            -\frac{77}{2}\zeta_3
            +\frac{19}{6}\zeta_4
            +\frac{1975}{54}\zeta_5
  \right.\nonumber\\&&\left.\mbox{}
            +n_f\left(
                  -\frac{4109}{1944}
                  -\frac{35}{54}\zeta_3
                  -\frac{16}{9}\zeta_4
                \right)
            +n_f^2\left(
                  -\frac{341}{1458}
                  +\frac{2}{9}\zeta_3
                \right)
     \right]
  \,,
  \label{eq:gamma0}
\end{eqnarray}
where $n_f$ is the number of active quark flavours.

In the formalism presented above it is assumed that there are only
currents coupling to quarks of the same flavour which is sufficient
for the purpose discussed below. The more general case involving also
non-diagonal terms can be found in Ref.~\cite{Har:diss}.


\subsection{\label{sub:higgs}Hadronic Higgs decay}

The coupling of a scalar CP-even Higgs boson to quarks has a very simple
structure as no $\gamma$ matrix is involved. 
It essentially consists of a factor containing the mass of
the quarks and the coupling constant.
This fact makes it very simple to construct an effective
Lagrangian and to derive powerful low-energy theorems as we will see in
this Subsection.

We want to consider QCD corrections to the hadronic decay of a
Higgs boson in the so-called intermediate mass range,
that is we compute the partial decay widths into quarks and gluons
assuming that the top quark is much heavier than all other scales
involved in the process. 
Although we have in mind the top quark as the heavy particle we will
in the following consider a generic heavy quark with mass $m_h$.

To lowest order in the inverse heavy quark mass 
the effective Lagrangian is constructed by the
operators of dimension four which are discussed above.
As we consider light quark mass effects at most in leading order
(which corresponds to an overall quadratic factor in the case $H\to
q\bar{q}$ and to quark mass zero in the gluonic case)
there is no contribution from the operator ${\cal O}_6$ and
thus the anomalous dimension matrix becomes two dimensional.

We consider a theory where we have in addition to the QCD Lagrangian 
of Eq.~(\ref{eq:L_QCD}) a
scalar particle, $H$, which couples to fermions via
\begin{eqnarray}
  {\cal L}_Y &=& -\frac{H^0}{v^0}
  \sum_{q}m_{q}^0\bar\psi_{q}^0\psi_{q}^0\,,
  \label{eq:yuk}
\end{eqnarray}
where the sum runs over all quark flavours. In the limit $m_h\to\infty$
Eq.~(\ref{eq:yuk}) can be written as a sum over the operators given
in Eqs.~(\ref{eq:op1}) and~(\ref{eq:op3})
accompanied by coefficient functions containing the residual dependence
on the top quark:
\begin{eqnarray}
  {\cal L}_{Y,\rm eff} &=& -\frac{H^0}{v^0}\sum_{i=1}^5C_i^0{\cal O}_i^0
  \,.
  \label{eq:eff}
\end{eqnarray}
The relation of the bare coefficient functions and the operators to
their renormalized counterparts can be extracted form the anomalous
dimension matrix given in Eq.~(\ref{eq:gam_mn}). One obtains
for the renormalized operators
\begin{eqnarray}
  {\cal O}_1 &=&
  \left[1+2\left(\frac{\alpha_s^\prime\partial}{\partial\alpha_s^\prime}
  \ln Z_g^\prime\right)\right]{\cal O}_1^0
  -4\left(\frac{\alpha_s^\prime\partial}{\partial\alpha_s^\prime}
  \ln Z_m^\prime\right){\cal O}_2^0
  \,,
  \nonumber\\
  {\cal O}_2 &=& {\cal O}_2^0
  \,,
\end{eqnarray}
and accordingly for the coefficient functions
\begin{eqnarray}
  C_1&=&\frac{1}{1+2(\alpha_s^\prime\partial/\partial\alpha_s^\prime)
  \ln Z_g^\prime}C_1^0
  \,,
  \nonumber\\
  C_2&=&\frac{4(\alpha_s^\prime\partial/\partial\alpha_s^\prime)
  \ln Z_m^\prime}
  {1+2(\alpha_s^\prime\partial/\partial\alpha_s^\prime)\ln Z_g^\prime}C_1^0
  +C_2^0
  \,.
  \label{eq:c1ren}
\end{eqnarray}     
Consequently, the physical part of ${\cal L}_{Y,\rm eff}$
takes the form
\begin{eqnarray}
  {\cal L}_{Y,\rm eff}^{\rm phys} &=& - \frac{H^0}{v^0}
  \left(C_1 {\cal O}_1 + C_2 {\cal O}_2\right)
  \,.
  \label{eq:leff}
\end{eqnarray}
$\alpha_s^\prime$, $Z_g^\prime$ and $Z_m^\prime$ are defined in the
effective theory which is indicated by the prime.
$C_i$ and ${\cal O}_i$ $(i=1,2)$ are individually finite,
but, with the exception of $\left[{\cal O}_2^\prime\right]$, they are not
separately renormalization-group (RG) invariant.
In Ref.~\cite{CheKniSte97hbb}, 
a RG-improved version of Eq.~(\ref{eq:leff}) has 
been constructed by exploiting the RG-invariance of the trace of the 
energy-momentum tensor.
The ratio $H^0/v^0$ receives a finite renormalization factor, which is of
${\cal O}(G_FM_t^2)$.
Its two- and three-loop QCD corrections have been found in Refs.~\cite{Kni94}
and \cite{KniSte95}, respectively.

A closer look to the operators in Eqs.~(\ref{eq:op1})
and~(\ref{eq:op3}) and the effective 
Lagrangian in Eq.~(\ref{eq:eff}) suggests that there should be a
connection between the coefficient functions on one side and the
decoupling constants as introduced in Eq.~(\ref{eq:decconst})
on the other side. Actually it turns out that $C_1$ and $C_2$ are
obtained from derivatives of $\ln\zeta_g$ and $\ln\zeta_m$,
respectively. This nice feature allows for the computation of $C_1$
and $C_2$ to order $\alpha_s^n$ from the knowledge of 
$\zeta_g$ and $\zeta_m$ to order $\alpha_s^{n-1}$~\cite{CheKniSte98}
as we will see below.

The coefficient functions contain the remnants of the heavy
quark. It is thus quite plausible that their computation can be
reduced to Feynman diagrams where the only scale is given by $m_h$.
Actually, the philosophy for 
the derivation of formulae for $C_1^0,\ldots, C_5^0$
is very similar to the case of the decoupling relations.
One again considers Green functions in the full and effective theories
and exploits the fact 
that the coefficient functions do not depend on the
momentum configuration. 
We will see that for a certain Green function in general several
coefficient functions are involved. Thus we will obtain five equations
which can be solved for $C_1^0,\ldots, C_5^0$.

From
Eqs.~(\ref{eq:op1}) and~(\ref{eq:op3}) 
one learns that the Green function
involving one gluon, two ghosts and a zero-momentum insertion of the
operator ${\cal O}_h=m_h^0\bar{h}^0h^0$, which mediates the
coupling to the Higgs boson, only gets contributions 
from ${\cal O}_5^0$. We define the 
bare 1PI Green function in analogy to
Eq.~(\ref{eq:Gccdef})
\begin{eqnarray}
  \lefteqn{p^\mu g_s^0\left\{-if^{abc}
  \left[\Gamma_{G\bar cc{\cal O}_h}^0(p,k,0)\right]
  + \ldots \right\} }
  \nonumber\\
  &=&
  i^2\int {\rm d}x{\rm d}y\,e^{i(p\cdot x+k\cdot y)}
  \left\langle Tc^{0,a}(x)\bar
  c^{0,b}(0)G^{0,c\mu}(y) {\cal O}_h(0) \right\rangle^{\rm 1PI}
  \label{eq:ccGO}
  \,,
\end{eqnarray}
where
the ellipses again represent other colour structures and $p$ and $k$
are the outgoing momenta of the $c$ and $G$, respectively. The third
argument of $\Gamma_{G\bar cc{\cal O}_h}^0$ indicates the zero
momentum of the operator ${\cal O}_h$.

In a next step we express ${\cal O}_h$ in terms of the operators given
in Eqs.~(\ref{eq:op1}) and~(\ref{eq:op3}).
Only ${\cal O}_5^0$ has to be taken into account
as only this operator contains the coupling of a gluon 
to two ghost fields. In
the transitition to the effective theory also Eq.~(\ref{eq:decconst})
has to be considered which leads to a factor 
$\zeta_g^0/(\tilde{\zeta_3}^0\sqrt{\zeta_3^0})$.
Note that in Eq.~(\ref{eq:ccGO}) we are dealing with amputated Green
functions. Thus from the external propagators a factor 
$(\tilde{\zeta}_3^0)^2\zeta_3^0$ arise. Finally we arrive at
\begin{eqnarray}
  p^\mu g_s^0 (-if^{abc}) \Gamma_{G\bar{c}c{\cal O}_h}(p,k,0) +
  \ldots 
  &=&
  p^\mu g_s^0 (-if^{abc}) 
  \zeta_g^0\tilde{\zeta}_3^0\sqrt{\zeta_3^0} C_5^0 + \ldots
  \nonumber\\
  &=&
  p^\mu g_s^0 (-if^{abc}) \tilde{\zeta}_1^0 C_5^0 + \ldots
  \,,
  \label{eq:ccGO_2}
\end{eqnarray}
where the ellipses represent other colour structures
and terms suppressed by the inverse heavy quark mass.
On the right-hand side of Eq.~(\ref{eq:ccGO_2})
we also avoided to write down explicitly 
the contributions beyond tree-level as they vanish for $p,k\to0$
within dimensional regularization.
In this limit only the diagrams involving the heavy quark survive on the
left-hand side and we are left with the formula
\begin{eqnarray}
  \tilde{\zeta}_1^0 C_5^0 &=& \Gamma_{G\bar{c}c{\cal O}_h}^{0,h}(0,0,0)
  \,.
\end{eqnarray}

In order to reduce the number of contributing diagrams and also to
simplify their complexity we exploit that the coupling of the Higgs
boson is proportional to $m_h$. Thus it can simply be generated by
differentiation with respect to $m_h$. With the definition
\begin{eqnarray}
  \partial^0_h &=& \frac{(m_h^0)^2\partial}{\partial(m_h^0)^2}
  \,,
  \label{eq:partial}
\end{eqnarray}
we finally obtain\footnote{We want to mention that in Ref.~\cite{CheKniSte98}
  there are misprints in the corresponding formulae: erroneously
  they contain a factor ``$1/2$'' instead of ``2''. However, the
  initial equations and the final results are correct
  in~\cite{CheKniSte98}.} 
\begin{eqnarray}
  \tilde{\zeta}_1^0 C_5^0 &=& 2 \partial_h^0
  \Gamma^{0,h}_{G\bar{c}c}(0,0) 
  \,.
  \label{eq:cfs1}
\end{eqnarray}

As another example, 
let us consider the derivation of a formula involving $C_1^0$
and $C_4^0$.
The starting point is the 1PI Green function of two gluons which contains a
zero-momentum insertion of the composite operator
${\cal O}_h$.
In momentum space, it reads in bare form
\begin{eqnarray}
  \delta^{ab}\Gamma_{GG{\cal O}_h}^{0,\mu\nu}(p,-p)
  &=&i^2\int {\rm d} x {\rm d} y\,e^{ip\cdot(x-y)}
  \left\langle TG^{0,a,\mu}(x)G^{0,b,\nu}(y){\cal O}_h(0)
  \right\rangle^{\rm1PI}
  \nonumber\\
  &=&\delta^{ab}\left[-g^{\mu\nu}p^2\Gamma_{GG{\cal O}_h}^0(p^2)
  +\mbox{terms proportional to $p^\mu p^\nu$}\right]
  \,,
  \label{eq:ggo1}
\end{eqnarray}
where $p$ denotes the four-momentum flowing along the gluon line.
In the limit $m_h\to\infty$, ${\cal O}_h$ may be written as a linear
combination of the effective operators given in 
Eqs.~(\ref{eq:op1}) and~(\ref{eq:op3}),
so that
\begin{eqnarray}
  \Gamma_{GG{\cal O}_h}^{0,\mu\nu}(p,-p)
  &=&-g^{\mu\nu}p^2    \Gamma_{GG{\cal O}_h}^{0}(p,-p) + \ldots
  \nonumber\\
  &=&\frac{i^2}{8}\int {\rm d} x {\rm d} y\,e^{ip\cdot(x-y)}
  \left\langle TG^{0,a,\mu}(x)G^{0,a,\nu}(y)
  \left(C_1^0{\cal O}^0_1 + C_4^0{\cal O}^0_4\right)
  \right\rangle^{\rm1PI}+\ldots
  \nonumber\\
  &=&
  \frac{i^2}{8}\zeta_3^0
  \int {\rm d} x {\rm d} y\,e^{ip\cdot(x-y)}
  \left\langle TG^{0\prime,a,\mu}(x)G^{0\prime,a,\nu}(y) 
  \left(C_1^0{\cal O}_1^0 + C_4^0{\cal O}_4^0\right)
  \right\rangle^{\rm1PI}+\ldots
  \nonumber\\
  &=&
  -g^{\mu\nu}p^2\zeta_3^0(-4 C_1^0 + 2 C_4^0)
  \left(1+\mbox{higher orders}\right)+\ldots
  \,,
  \label{eq:ggoh}
\end{eqnarray}
where the ellipses indicate terms of ${\cal O}(1/m_h)$ and terms proportional
to $p^\mu p^\nu$.
The factor $1/8$ results for the summation over the colour indices.
In the second step, we have used Eq.~(\ref{eq:decconst}) 
together with the fact that
$\Gamma_{GG{\cal O}_h}^{0,\mu\nu}(p,-p)$ represents an amputated Green
function. 
If we consider the coefficients of the transversal part in the limit $p\to0$,
we observe that the contributions due to the higher-order QCD
corrections on the right-hand side 
of Eq.~(\ref{eq:ggoh}) vanish, as massless tadpoles are set to zero
in dimensional regularization.
In principle also the other operators contribute via loop diagrams. 
However, also these contributions lead to massless tadpoles
and are thus zero.
The relative weight between $C_1^0$ and $C_4^0$ and the prefactor in
the last line of Eq.~(\ref{eq:ggoh}) follow immediately from the
Feynman rules given in Eq.~(\ref{eq:O1O4FR}).
On the left-hand side, only those diagrams survive which contain at least one
heavy-quark line.
Consequently, the hard part of the amputated Green function is given by
\begin{eqnarray}
  \Gamma_{GG{\cal O}_h}^{0,h}(0,0)&=&\zeta_3^0\left(-4C_1^0+2C_4^0\right)
  \,.
  \label{eq:amput}
\end{eqnarray}
Using Eq.~(\ref{eq:partial}) we finally arrive at
\begin{eqnarray}
  \zeta_3^0(-4C_1^0+2C_4^0) &=& -2\partial_h^0 \Pi_G^{0,h}(0)
  \,.
  \label{eq:cfs0}
\end{eqnarray}

In a similar way, we obtain three more relationships, namely
\begin{eqnarray}
  \zeta_2^0C_3^0&=& -2\partial_h^0\Sigma_V^{0,h}(0)\,,
  \nonumber\\
  \zeta_m^0\zeta_2^0(C_2^0-C_3^0)&=&
  1-\Sigma_S^{0,h}(0)-2\partial_h^0\Sigma_S^{0,h}(0)\,,
  \nonumber\\
  \tilde\zeta_3^0(C_4^0+C_5^0)&=&2\partial_h^0\Pi_c^{0,h}(0)
  \,.
  \label{eq:cfs}
\end{eqnarray}
The Eqs.~(\ref{eq:cfs1}),~(\ref{eq:cfs0}) and~(\ref{eq:cfs})
may now be solved for the coefficient functions $C_i^0$ $(i=1,\ldots,5)$.
They are expressed in terms of vacuum integrals with only one mass
scale, namely the heavy quark mass. This is also the case for the 
decoupling constants occuring in Eqs.~(\ref{eq:cfs1}),~(\ref{eq:cfs0})
and~(\ref{eq:cfs}) as discussed in Section~\ref{sec:dec}.

As a simple example let us consider the computation of $C_1$ at
lowest order. Here only one diagram contributes to $\Pi_G^{0,h}(0)$, 
namely the one where the gluon splits into two virtual heavy quarks.
It can be evaluated with the help of Eq.~(\ref{eq:Va}). Expanded up to
finite order in $\varepsilon$ it reads
\begin{eqnarray}
  \Pi_G^{0,h}(0) &=& \frac{\alpha_s^0}{\pi} T \left(
     \frac{1}{3\varepsilon}
   + \frac{1}{3} \ln\frac{\mu^2}{(m_h^0)^2}
   \right)
  \,.
\end{eqnarray}
Differentiating with respect to $m_h^0$ leads to
\begin{eqnarray}
  C_1 &=& -\frac{\alpha_s}{\pi} \frac{T}{6}
  \,.
\end{eqnarray}
Note, that at this order no renormalization is needed and thus the bare
and renormalized quantities coincide. Furthermore, there is no
contribution from $\zeta_3^0$ in Eq.~(\ref{eq:cfs0}) and $C_4^0$ 
contributes for the first time at order $\alpha_s^2$ as can be seen
from Eqs.~(\ref{eq:cfs}).

In the next three Subsections we will describe different methods to
compute the coefficient functions. Finally in
Section~\ref{sub:hadrHiggs} we will review the state-of-the-art
corrections to the hadronic decay width of the Higgs boson.


\subsubsection{\label{sub:dir}Direct 
  calculation of the coefficient functions}

There is the possibility to avoid the occurrence of the
non-physical operators and their coefficient functions 
in the computation of $C_1^0$ and $C_2^0$.
For demonstration let us consider the case of $C_1^0$.

\begin{figure}[ht]
\leavevmode
\begin{center}
\epsfxsize=\textwidth
\epsffile[76 635 552 724]{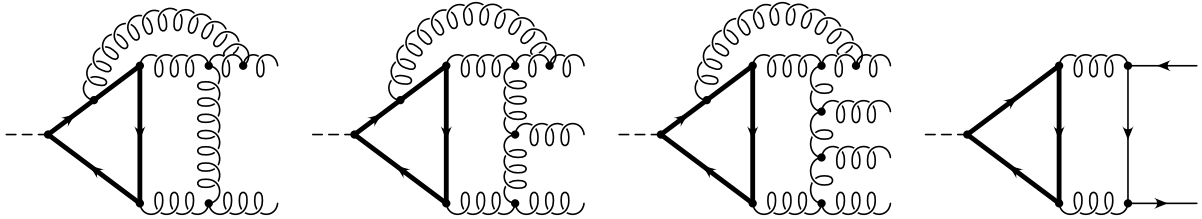}
\caption{Typical Feynman diagrams contributing to the coefficients
$C_1^0$ and $C_2^0$.
Looped, bold-faced, and dashed lines represent gluons, 
heavy quarks, and Higgs 
bosons, respectively.}
\label{fig:aggfig}
\end{center}
\end{figure}

From the definition of the operators in 
Eqs.~(\ref{eq:op1}) and~(\ref{eq:op3})
and the Feynman rules given in~(\ref{eq:O1O4FR})
one can read off that the 1PI Green function 
$\Gamma^{0,\mu\nu}_{GG{\cal O}_h}(p_1,p_2)$,
where $p_1$ and $p_2$ denote the incoming momenta of the gluons,
only receives contributions from $C_1^0{\cal O}_1^0$
if $p_1\not=-p_2$.
In analogy to Eq.(\ref{eq:ggo1}) we can write 
\begin{eqnarray}
  \Gamma^{0,\mu\nu}_{GG{\cal O}_h}(p_1,p_2)
  &=&
  \left(g^{\mu\nu} p_1\cdot p_2 - p_1^\nu p_2^\mu\right)
  \Gamma^0_{GG{\cal O}_h}(p_1,p_2)
  \nonumber\\
  &=&
  \left(g^{\mu\nu} p_1\cdot p_2 - p_1^\nu p_2^\mu\right)
  \zeta_3^0\left(-4C_1^0\right)\left(1 + \ldots\right)
  \,,
\end{eqnarray}
where the ellipses denote terms of order $1/m_h$, possible
other Lorentz structures and higher order loop corrections. The latter 
vanish in the limit $p_1,p_2\to0$ and we are left with
\begin{eqnarray}
  \zeta_3^0 C_1^0 &=& -\frac{1}{4} \Gamma^{0,h}_{GG{\cal O}_h}(0,0)
  \nonumber\\
  &=& -\frac{1}{4} \left(
                   \frac{g_{\mu\nu}p_1\cdot p_2 -p_{1,\nu}p_{2,\mu}
                         -p_{1,\mu}p_{2,\nu}}{(D-2)(p_1\cdot p_2)^2}
                   \Gamma^{0,h,\mu\nu}_{GG{\cal O}_h}(p_1,p_2)
                   \right)\Bigg|_{p_1=p_2=0}
  \,,
  \label{eq:C1dir}
\end{eqnarray}
where the on-shell conditions $p_1^2=p_2^2=0$ have been used.
This equation relates the coefficient function $C_1^0$
directly to a physical amplitude and no Green functions involving
ghost fields have to be considered.
The price one has to pay is that the momenta $p_1$ and $p_2$ 
can only be set to zero after the projection in Eq.~(\ref{eq:C1dir})
has been applied. This complicates the expressions of the
individual diagrams. Furthermore it is not possible to use derivatives
with respect to $m_h^0$ in order to generate the
coupling to the Higgs boson as initially the momentum of the Higgs
boson is not zero. Thus altogether 657 vertex diagrams like the first 
one in Fig.~\ref{fig:aggfig} have to be considered. 

In analogy to the gluon-gluon-Higgs boson
three-point function one could also choose the part of the operator
${\cal O}_1$ involving three or even four gluons. This would lead to
7362 four- and 95004 five-point functions at three-loop level,
where one and no gluon momentum, respectively,  
has to be kept different from zero until the projection is finished.
Sample diagrams are pictured in Fig.~\ref{fig:aggfig}.
In Ref.~\cite{Chetyrkin:1998mw}, where the decay of a pseudo-scalar
Higgs boson has been considered, the 7362 three-loop four-point functions
have been evaluated in order to check the calculation of the
three-point function.
For completeness we want to mention that 
the last diagram in Fig.~\ref{fig:aggfig} is the lowest-order graph
contributing to $C_2$.


\subsubsection{\label{sub:let}Low-energy theorem}

It is tempting to re-express the Green functions on the right-hand side 
of Eqs.~(\ref{eq:cfs1}),~(\ref{eq:cfs0}) and~(\ref{eq:cfs})
in terms of the decoupling constants which were considered in
Section~\ref{sec:dec}. This leads to a straight relation 
between the coefficient functions and the decoupling constants.

For the equations involving $C_1^0$, $C_4^0$ and $C_5^0$ one 
successively obtains
\begin{eqnarray}
  C_5^0 &=& 2\frac{\partial_h^0\tilde{\zeta}_1^0}{\tilde{\zeta}_1^0}
  \,,
  \nonumber\\
  C_4^0 &=& 2\left(
             \frac{\partial_h^0\tilde{\zeta}_3^0}{\tilde{\zeta}_3^0}
           - \frac{\partial_h^0\tilde{\zeta}_1^0}{\tilde{\zeta}_1^0}
             \right)
  \,,
  \nonumber\\
  C_1^0 &=&  \frac{1}{2}
             \frac{\partial_h^0\zeta_3^0}{\zeta_3^0}
           + \frac{\partial_h^0\tilde{\zeta}_3^0}{\tilde{\zeta}_3^0}
           - \frac{\partial_h^0\tilde{\zeta}_1^0}{\tilde{\zeta}_1^0}
  \nonumber\\
        &=& -\partial_h^0 
            \ln\frac{\tilde{\zeta}_1^0}{\tilde{\zeta}_3^0\sqrt{\zeta_3^0}}
  \nonumber\\
        &=& -\partial_h^0 \ln\zeta_g^0
  \,,
  \label{eq:C1C4C5}
\end{eqnarray}
Next, we express $\zeta_g^0$ through renormalized quantities.
Using $\partial_h^0=\partial_h$, we find
\begin{eqnarray}
  -2C_1^0&=&\partial_h\ln(\zeta_g^0)^2
  \nonumber\\
  &=&\partial_h\ln\frac{\alpha_s^{0\prime}}{\alpha_s^0}
  \nonumber\\
  &=&\partial_h\ln(Z_g^\prime)^2+\partial_h\ln\alpha_s^\prime 
  \nonumber\\
  &=&\left[1+\frac{\alpha_s^\prime\partial}{\partial\alpha_s^\prime}
  \ln\left(Z_g^\prime\right)^2\right]\partial_h\ln\alpha_s^\prime\,.
\end{eqnarray}
Identifying the renormalization factor of Eq.~(\ref{eq:c1ren}), we obtain the
amazingly simple relation
\begin{eqnarray}
  -2C_1&=&\partial_h\ln\alpha_s^\prime
  \nonumber\\
  &=&\partial_h\ln\zeta_g^2
  \,.
  \label{eq:c1let1}
\end{eqnarray}
This relation opens the possibility to compute $C_1$ through
${\cal O}(\alpha_s^4)$, since one only needs to know the logarithmic 
contributions of
$\zeta_g$ in this order, which may be reconstructed from its lower-order terms
in combination with the four-loop 
$\beta$~\cite{RitVerLar97_bet} and 
$\gamma_m$~\cite{Che97_gam,LarRitVer97_gam} 
functions.

It is furthermore 
possible to directly relate $C_1$ to the $\beta$ and $\gamma_m$
functions of the full and effective theories.
Exploiting the relation
\begin{eqnarray}
  \beta^\prime(\alpha_s^\prime) 
  &=&\frac{\mu^2d}{d\mu^2}\,\frac{\alpha_s^\prime}{\pi}
  \nonumber\\
  &=&\left[\frac{\mu^2\partial}{\partial\mu^2}
  +\beta(\alpha_s)\frac{\partial}{\partial\alpha_s}
  +\gamma_m(\alpha_s)\frac{m_h\partial}{\partial m_h}
  \right]\frac{\alpha_s^\prime}{\pi}\,,
\end{eqnarray}
where $\alpha_s^\prime=\alpha_s^\prime(\mu,\alpha_s,m_h)$, we find
\begin{eqnarray}
  C_1&=&\frac{\pi}{2\alpha_s^\prime\left[1-2\gamma_m(\alpha_s)\right]}
  \left[\beta^\prime(\alpha_s^\prime)
  -\beta(\alpha_s)\frac{\partial\alpha_s^\prime}{\partial\alpha_s}\right]
  \,.
  \label{eq:c1let}
\end{eqnarray}
In the case of $C_2$, we may proceed along the same lines to obtain
\begin{eqnarray}
  C_2&=&1+2\partial_h\ln\zeta_m
  \nonumber\\
  &=&1-\frac{2}{1-2\gamma_m(\alpha_s)}
  \left[\gamma_m^\prime(\alpha_s^\prime)-\gamma_m(\alpha_s)
  -\beta(\alpha_s)\frac{1}{m_q^\prime}\,
  \frac{\partial m_q^\prime}{\partial\alpha_s}\right]
  \,,
  \label{eq:c2let}
\end{eqnarray}
where $m_q^\prime=m_q^\prime(\mu,\alpha_s,m_h)$.
It should be stressed that Eqs.~(\ref{eq:c1let}) and (\ref{eq:c2let}) 
are valid to all orders in $\alpha_s$.

Fully exploiting present knowledge of the $\beta$ \cite{RitVerLar97_bet} and
$\gamma_m$ \cite{Che97_gam,LarRitVer97_gam} 
functions, we may evaluate $C_1$ and $C_2$
through ${\cal O}(\alpha_s^4)$ via Eqs.~(\ref{eq:c1let}) and (\ref{eq:c2let}).
In the pure $\overline{\mbox{MS}}$ scheme, 
we find~\cite{CheKniSte97,CheKniSte98}
\begin{eqnarray}
C_1^{\rm MS}&=&
-\frac{1}{12}\,\frac{\alpha_s^{(n_f)}(\mu)}{\pi}
\Bigg\{
  1 
+ \frac{\alpha_s^{(n_f)}(\mu)}{\pi}
\Bigg(
\frac{11}{4} 
- \frac{1}{6} \ln\frac{\mu^2}{m_h^2}
\Bigg)
\nonumber\\
&&{}+ \left(\frac{\alpha_s^{(n_f)}(\mu)}{\pi}\right)^2
\Bigg[
\frac{2821}{288} 
- \frac{3}{16} \ln\frac{\mu^2}{m_h^2}
+ \frac{1}{36} \ln^2\frac{\mu^2}{m_h^2}
+ n_l\left(
-\frac{67}{96} 
+ \frac{1}{3} \ln\frac{\mu^2}{m_h^2}
\right)
\Bigg]
\nonumber\\
&&{}+ \left(\frac{\alpha_s^{(n_f)}(\mu)}{\pi}\right)^3
\Bigg[
-\frac{4004351}{62208} 
+ \frac{1305893}{13824}\zeta_3
- \frac{859}{288} \ln\frac{\mu^2}{m_h^2}
+ \frac{431}{144} \ln^2\frac{\mu^2}{m_h^2}
\nonumber\\
&&{}
- \frac{1}{216} \ln^3\frac{\mu^2}{m_h^2}
+  n_l \left(
  \frac{115607}{62208} 
- \frac{110779}{13824}\zeta_3
+ \frac{641}{432} \ln\frac{\mu^2}{m_h^2}
+ \frac{151}{288} \ln^2\frac{\mu^2}{m_h^2}
\right) 
\nonumber\\
&&{}+ n_l^2 \left(
- \frac{6865}{31104} 
+ \frac{77}{1728} \ln\frac{\mu^2}{m_h^2} 
- \frac{1}{18} \ln^2\frac{\mu^2}{m_h^2}
\right)
\Bigg]
\Bigg\}
\nonumber\\
&\approx&
-\frac{1}{12}\,\frac{\alpha_s^{(n_f)}(\mu_h)}{\pi}
\Bigg[1
+ 2.7500
\frac{\alpha_s^{(n_f)}(\mu_h)}{\pi}
+ \left(9.7951 - 0.6979\,n_l\right) 
\left(\frac{\alpha_s^{(n_f)}(\mu_h)}{\pi}\right)^2
\nonumber\\
&&
+ \left(49.1827 - 7.7743\,n_l - 0.2207\,n_l^2 \right)
\left(\frac{\alpha_s^{(n_f)}(\mu_h)}{\pi}\right)^3\Bigg]
\,,
\label{eq:c1}\\
C_2^{\rm MS} &=&
1 
+ \left(\frac{\alpha_s^{(n_f)}(\mu)}{\pi}\right)^2 \Bigg(
  \frac{5}{18} 
- \frac{1}{3} \ln\frac{\mu^2}{m_h^2}
\Bigg)
\nonumber\\
&&{}+ \left(\frac{\alpha_s^{(n_f)}(\mu)}{\pi}\right)^3 \Bigg[
  \frac{311}{1296} 
+ \frac{5}{3}\zeta_3
- \frac{175}{108} \ln\frac{\mu^2}{m_h^2}
- \frac{29}{36} \ln^2\frac{\mu^2}{m_h^2}
\nonumber\\&&\mbox{}
+ n_l\left(
  \frac{53}{216} 
+ \frac{1}{18} \ln^2\frac{\mu^2}{m_h^2}
\right)
\Bigg]
\nonumber\\
&&{}+ \left(\frac{\alpha_s^{(n_f)}(\mu)}{\pi}\right)^4 \Bigg[
  \frac{2800175}{186624} 
+ \frac{373261}{13824}\zeta_3 
- \frac{155}{6}\zeta_4
- \frac{575}{36}\zeta_5
+ \frac{31}{72}B_4
\nonumber\\
&&{}+ \left( 
  -\frac{50885}{2592} 
+ \frac{155}{12}\zeta_3 
\right) \ln\frac{\mu^2}{m_h^2}
- \frac{1219}{216} \ln^2\frac{\mu^2}{m_h^2}
- \frac{301}{144} \ln^3\frac{\mu^2}{m_h^2}
\nonumber\\
&&{}+ n_l \left(
- \frac{16669}{15552} 
- \frac{221}{288}\zeta_3
+ \frac{25}{12}\zeta_4
- \frac{1}{36} B_4
+ \frac{7825}{2592} \ln\frac{\mu^2}{m_h^2} 
+ \frac{23}{48} \ln^2\frac{\mu^2}{m_h^2}
\right.
\nonumber\\
&&
\left.\mbox{}
+ \frac{5}{18} \ln^3\frac{\mu^2}{m_h^2}
\right) 
{}+  n_l^2 \left(
  \frac{3401}{23328} 
- \frac{7}{54}\zeta_3
- \frac{31}{324} \ln\frac{\mu^2}{m_h^2}
- \frac{1}{108} \ln^3\frac{\mu^2}{m_h^2}
\right)
\Bigg]
\nonumber\\
&\approx&
1
+ 0.2778
\left(\frac{\alpha_s^{(n_f)}(\mu_h)}{\pi}\right)^2
+ \left(2.2434 + 0.2454\,n_l\right) 
\left(\frac{\alpha_s^{(n_f)}(\mu_h)}{\pi}\right)^3
\nonumber\\
&&
+ \left(2.1800 + 0.3096\,n_l - 0.0100\,n_l^2 \right)
\left(\frac{\alpha_s^{(n_f)}(\mu_h)}{\pi}\right)^4
\,,
\label{eq:c2}
\end{eqnarray}
where, for simplicity, we have chosen $\mu=\mu_h\equiv m_h(\mu_h)$ 
in the approximate
expressions.
The corresponding results expressed in terms of 
the pole mass $M_h$ read
\begin{eqnarray}
C_1^{\rm OS}&=&
-\frac{1}{12}\,\frac{\alpha_s^{(n_f)}(\mu)}{\pi}
\Bigg\{1 
+ \frac{\alpha_s^{(n_f)}(\mu)}{\pi}
\Bigg(
\frac{11}{4} 
- \frac{1}{6} \ln\frac{\mu^2}{M_h^2}
\Bigg)
\nonumber\\
&&{}+ \left(\frac{\alpha_s^{(n_f)}(\mu)}{\pi}\right)^2
\Bigg[
\frac{2693}{288} 
- \frac{25}{48} \ln\frac{\mu^2}{M_h^2}
+ \frac{1}{36} \ln^2\frac{\mu^2}{M_h^2}
+ n_l\left(
-\frac{67}{96} 
+ \frac{1}{3} \ln\frac{\mu^2}{M_h^2}
\right)
\Bigg]
\nonumber\\
&&{}+ \left(\frac{\alpha_s^{(n_f)}(\mu)}{\pi}\right)^3
\Bigg[
-\frac{4271255}{62208} 
-\frac{2}{3}\zeta_2\left(1+\frac{\ln2}{3}\right)
+ \frac{1306661}{13824}\zeta_3
\nonumber\\
&&{}- \frac{4937}{864} \ln\frac{\mu^2}{M_h^2}
+ \frac{385}{144} \ln^2\frac{\mu^2}{M_h^2}
- \frac{1}{216} \ln^3\frac{\mu^2}{M_h^2}
\nonumber\\
&&{}+  n_l \left(
  \frac{181127}{62208}
+ \frac{1}{9}\zeta_2 
- \frac{110779}{13824}\zeta_3
+ \frac{109}{48} \ln\frac{\mu^2}{M_h^2}
+ \frac{53}{96} \ln^2\frac{\mu^2}{M_h^2}
\right) 
\nonumber\\
&&{}+ n_l^2 \left(
- \frac{6865}{31104} 
+ \frac{77}{1728} \ln\frac{\mu^2}{M_h^2} 
- \frac{1}{18} \ln^2\frac{\mu^2}{M_h^2}
\right)
\Bigg]
\Bigg\}
\nonumber\\
&\approx&
-\frac{1}{12}\,\frac{\alpha_s^{(n_f)}(M_h)}{\pi}
\Bigg[1
+ 2.7500
\frac{\alpha_s^{(n_f)}(M_h)}{\pi}
+ \left(9.3507 - 0.6979\,n_l\right) 
\left(\frac{\alpha_s^{(n_f)}(M_h)}{\pi}\right)^2
\nonumber\\
&&
+ \left(43.6090 - 6.5383\,n_l - 0.2207\,n_l^2 \right)
\left(\frac{\alpha_s^{(n_f)}(M_h)}{\pi}\right)^3\Bigg]
\,,
\label{eq:c1os}\\
C_2^{\rm OS}&=&1 
+ \left(\frac{\alpha_s^{(n_f)}(\mu)}{\pi}\right)^2 \Bigg(
  \frac{5}{18} 
- \frac{1}{3} \ln\frac{\mu^2}{M_h^2}
\Bigg)
\nonumber\\
&&{}
+\left(\frac{\alpha_s^{(n_f)}(\mu)}{\pi}\right)^3 \Bigg[
- \frac{841}{1296} 
+ \frac{5}{3}\zeta_3
- \frac{247}{108} \ln\frac{\mu^2}{M_h^2}
- \frac{29}{36} \ln^2\frac{\mu^2}{M_h^2}
\nonumber\\
&&{}
+ n_l\left(
  \frac{53}{216} 
+ \frac{1}{18} \ln^2\frac{\mu^2}{M_h^2}
\right)
\Bigg]
\nonumber\\
&&{}+ \left(\frac{\alpha_s^{(n_f)}(\mu)}{\pi}\right)^4 \Bigg[
  \frac{578975}{186624} 
- \frac{4}{3}\zeta_2\left(1+\frac{\ln2}{3}\right)
+ \frac{374797}{13824}\zeta_3 
- \frac{155}{6}\zeta_4
\nonumber\\
&&{}
- \frac{575}{36}\zeta_5
+ \frac{31}{72} B_4
+ \left(
  -\frac{83405}{2592}
+ \frac{155}{12} \zeta_3 
\right) \ln\frac{\mu^2}{M_h^2}
- \frac{2101}{216} \ln^2\frac{\mu^2}{M_h^2}
\nonumber\\
&&{}
- \frac{301}{144} \ln^3\frac{\mu^2}{M_h^2}
+ n_l \left(
- \frac{11557}{15552} 
+ \frac{2}{9}\zeta_2
- \frac{221}{288}\zeta_3
+ \frac{25}{12}\zeta_4
- \frac{1}{36} B_4
\right.\nonumber\\
&&{}+\left.
 \frac{9217}{2592} \ln\frac{\mu^2}{M_h^2}
+ \frac{109}{144} \ln^2\frac{\mu^2}{M_h^2}
+ \frac{5}{18} \ln^3\frac{\mu^2}{M_h^2}
\right)
\nonumber\\
&&{}+ n_l^2 \left(
  \frac{3401}{23328} 
- \frac{7}{54}\zeta_3
- \frac{31}{324} \ln\frac{\mu^2}{M_h^2}
- \frac{1}{108} \ln^3\frac{\mu^2}{M_h^2}
\right)
\Bigg]
\nonumber\\
&\approx&1
+ 0.2778
\left(\frac{\alpha_s^{(n_f)}(M_h)}{\pi}\right)^2
+ \left(1.3545 + 0.2454\,n_l \right)
\left(\frac{\alpha_s^{(n_f)}(M_h)}{\pi}\right)^3
\nonumber\\
&&
+ \left(-12.2884 + 1.0038\,n_l - 0.0100\,n_l^2 \right)
\left(\frac{\alpha_s^{(n_f)}(M_h)}{\pi}\right)^4
\,,
\label{eq:c2os}
\end{eqnarray}
where we have put $\mu=M_h$ in the numerical evaluations.
In~\cite{Ste98_higgs} the leading Yukawa corrections of
${\cal O}(\alpha_s^n G_F m_t^2)$ $(n=0,1,2)$ to the coefficient
functions have been evaluated in the SM. The analytical results are
listed in Appendix~\ref{app:decconst}.

With the knowledge of $C_1$ and $C_2$ the construction of the
effective Lagrangian is completed. In Section~\ref{sub:hadrHiggs}
it will be used in order to compute the hadronic decay rate of the
Higgs boson.

Recently the effective Lagrangian has been used in order to consider
the Higgs boson production process via gluon fusion.
In Refs.~\cite{Harlander:2000mg,Harlander:2001is,Catani:2001ic} 
a first step to the
next-to-next-to-leading order QCD corrections has been done.
At this accuracy it is necessary to compute 
two-loop virtual corrections~\cite{Harlander:2000mg} to the process $gg\to H$
using the effective $ggH$ vertex of Eq.~(\ref{eq:leff}).
Thus the coefficient function $C_1$ enters as a multiplicative constant.

For completeness we want to mention that the
low-energy theorem derived in this section has also been specified to
the $\gamma\gamma H$ coupling. The analytical expressions can be found 
in Ref.~\cite{CheKniSte98}.


\subsubsection{\label{sub:hggbfm}$H\to gg$ in the background field method}

An interesting alternative to the considerations 
of the Sections~\ref{sub:dir} and~\ref{sub:let}
is based on the Background Field Method
(BFM)~\cite{BFM,Klu75,Abb81,DenWeiDit94,Gra99}.
In this framework the gluon field is decomposed into 
a quantum and a background part where the former only 
appears as a virtual particle inside the loops.
On the contrary, the background field serves as an external field
in the Green functions.
In the BFM the gauge invariance is maintained while quantizing the theory.
This was one of the main motivations for its development.

A comprehensive discussion for the case of QCD and the computation of
the two-loop $\beta$ function as a practical
application of the BFM can be found in~\cite{Abb81}.
In particular, it is shown, that the relation between the charge
renormalization constant, $Z_g$, (cf. Eq.~(\ref{eq:renconst})) and the 
wave function renormalization constant of the background
gluon, $Z_3^B$, reads
\begin{eqnarray}
  Z_3^B &=& \frac{1}{Z_g^2}
  \,.
  \label{eq:Z3B}
\end{eqnarray}
Thus, only the background gluon self energy has to be computed in order
to obtain the $\beta$ function whereas in the conventional approach
also vertex functions have to be considered. E.g., next to the gluon
propagator also the ghost two-point function and the gluon-ghost vertex 
have to be computed.
The price one has to pay for these kind of simplifications
are more complicated Feynman rules for those vertices involving the
background gluon~\cite{Abb81}. In particular, there are three- and four-gluon 
vertices which contain additional 
terms proportional to $1/(1-\xi)$ where $\xi$ is the
gauge parameter defined in Eq.~(\ref{eq:gluprop}).
As a consequence, it is not possible to choose Landau gauge, which corresponds
to $\xi=1$ and which would avoid the renormalization of $\xi$,
from the very beginning of the calculation. Rather one has to perform the
calculation using a general gauge parameter and either adopt Landau gauge at
the end of the calculation or renormalize the gauge parameter.

\begin{figure}[ht]
\leavevmode
\begin{center}
\epsfxsize=\textwidth
\epsffile[90 580 530 730]{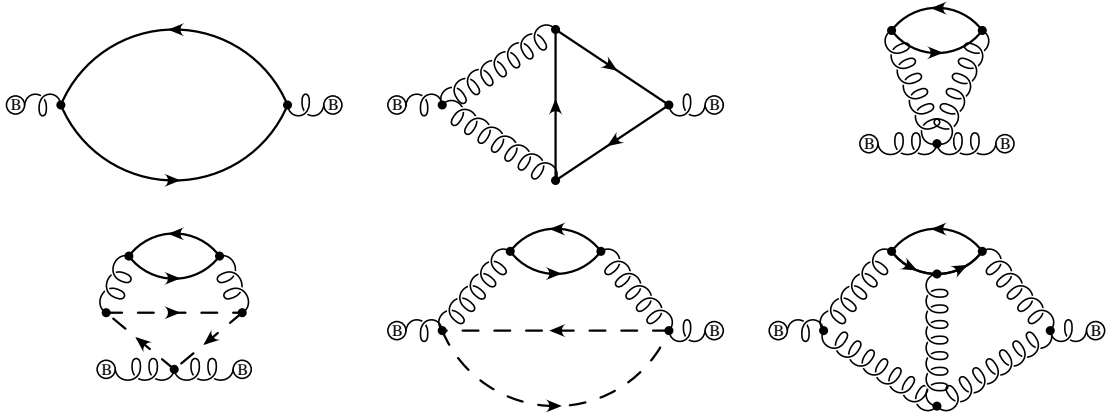}
  \caption{\label{fig:gBgB}Typical Feynman diagrams contributing 
  to the two-point function with external background gluons.
  Looped, solid, and dashed lines represent gluons, heavy quarks, and 
  ghosts, respectively.}
\end{center}
\end{figure}

Concerning the decoupling properties of the background field we can
essentially take over the discussion of Section~\ref{sec:dec}.
In analogy to Eq.~(\ref{eq:zeta30}) one obtains for the decoupling constant of
the background gluon field
\begin{eqnarray}
  \zeta_3^{B0} &=& 1 + \Pi_{G^B}^{0,h}(0)
  \,,
\end{eqnarray}
where $\Pi_{G^B}^{0,h}$ is the hard contribution
of the transversal part of the bare background gluon polarization function.
$\zeta_3^{B0}$ coincides with $\zeta_3^{0}$ at one-loop order
as only the diagram with a heavy quark loop contributes and
the coupling of fermions to background gluons is identical 
to the one of quantum gluons. 
Starting from two loops, however, virtual quantum gluons
appear inside the Feynman diagrams and the analytical 
expressions are different.
In Fig.~\ref{fig:gBgB} some typical one-, two- and three-loop diagrams 
contributing to $\Pi_{G^B}^{0,h}(0)$ are shown.
One arrives at a finite expression for the decoupling constant 
with the help of
\begin{eqnarray}
  \zeta_3^B &=& \frac{Z_3^B \zeta_3^{B0}}{Z_3^{B\prime}}
  \,.
\end{eqnarray}

It is not surprising that one has the
following connection between $\zeta_3^B$ and $\zeta_g$
\begin{eqnarray}
  \zeta_3^B &=& \frac{1}{\zeta_g^2}
  \,,
  \label{eq:zeta3B}
\end{eqnarray}
which is in analogy to Eq.~(\ref{eq:Z3B}).
We explicitly checked this relation and computed $\Pi_{G^B}^{0,h}(0)$ at
three-loop order using a general gauge parameter. After renormalization one
obtains
\begin{eqnarray}
  \zeta_3^{B,{\rm OS}} &=&
  1 +
  \frac{\alpha_s^{(n_l)}(\mu)}{\pi}
  \left(
    \frac{1}{6}\ln\frac{\mu^2}{M_h^2}
  \right)
  +\left(\frac{\alpha_s^{(n_l)}(\mu)}{\pi}\right)^2
  \left(
     \frac{7}{24} 
    +\frac{19}{24}\ln\frac{\mu^2}{M_h^2}
    +\frac{1}{36}\ln^2\frac{\mu^2}{M_h^2}
  \right)
  \nonumber\\
  &&{}+\left(\frac{\alpha_s^{(n_l)}(\mu)}{\pi}\right)^3
  \left[
    \frac{58933}{124416}
    +\frac{2}{3}\zeta_2\left(1+\frac{1}{3}\ln2\right)
    +\frac{80507}{27648}\zeta_3
    +\frac{8941}{1728}\ln\frac{\mu^2}{M_h^2}
  \right.\nonumber\\
  &&{}+\left.
    \frac{511}{576}\ln^2\frac{\mu^2}{M_h^2}
    +\frac{1}{216}\ln^3\frac{\mu^2}{M_h^2} 
    +n_l\left(
      -\frac{2479}{31104}
      -\frac{\zeta_2}{9}
      -\frac{409}{1728}\ln\frac{\mu^2}{M_h^2} 
    \right)
  \right]
  \,,
  \label{eq:invzetagOS}
\end{eqnarray}
which is in agreement with Eq.~(\ref{eq:zetagOS}).
The $219$ diagrams contributing to $\Pi_{G^B}^{0,h}(0)$
have to be compared with the $189+25+228=442$ diagrams 
which are necessary in order to obtain the result of Eq.~(\ref{eq:zetagOS}).

Eq.~(\ref{eq:zeta3B}) also significantly simplifies the computation of the
coefficient function $C_1$. Due to Eq.~(\ref{eq:c1let1}) it is simply obtained
via
\begin{eqnarray}
  C_1 &=& 2 \partial_h\ln\zeta_3^B
  \,.
  \label{eq:c1let1_bfm}
\end{eqnarray}
This coincides with the naive expectation that the effective coupling of the
Higgs boson is generated by taking derivatives of the gluon polarization
function with respect to the heavy quark mass.
Furthermore Eq.~(\ref{eq:c1let1_bfm}) agrees with the 
low-energy theorem for the photon-Higgs interaction as derived
in~\cite{CheKniSte98}.
In this sense we could claim that 
the background gluon is ``more physical'' than the gluon in
the conventional approach.

In this context we want to refer to~\cite{Kniehl:1995tn}
where low-energy theorems in Higgs physics have been considered at
one- and two-loop order. In particular it was realized that the
decay rate of a scalar Higgs boson into two photons can be obtained
by naive differentiation of the photon self energy if the latter is computed in
the framework of the pinch technique.

As a further check on the result for $C_1$ we also perform the direct
calculation by considering the quantity
$\Gamma^{\mu\nu}_{G^BG^B{\cal O}_h}(p_1,p_2)$
which is defined in analogy to Eq.~(\ref{eq:ggo1}):
\begin{eqnarray}
  \delta^{ab}\Gamma_{G^BG^B{\cal O}_h}^{0,\mu\nu}(p_1,p_2)
  &=&i^2\int {\rm d} x {\rm d} y\,e^{i(p_1\cdot x + p_2\cdot y)}
  \left\langle TG^{B,0,a\mu}(x)G^{B,0,b\nu}(y){\cal O}_h(0)
  \right\rangle^{\rm1PI}
  \!\!\!.
  \label{eq:ggo1B}
\end{eqnarray}
The only difference is the presence of external background (instead of
quantum) fields. 
We evaluated the 732 vertex diagrams in the spirit of Eq.~(\ref{eq:C1dir})
and could confirm 
the order $\alpha_s^3$ terms in Eqs.~(\ref{eq:c1}) and~(\ref{eq:c1os}).

Due to the fact that there are no diagrams involving external ghost
fields contributing to Eq.~(\ref{eq:c1let1_bfm})
there is no admixture from ${\cal O}_4^0$ while matching 
$\Gamma^{\mu\nu}_{G^BG^B{\cal O}_h}(p_1,p_2)$ with the effective
theory even for the choice $p_1=-p_2$.
Thus we can immediately set $p_1=-p_2=p$ in Eq.~(\ref{eq:ggo1B}) 
and consider the limit $p\to0$.
No complicated projector like in Eq.~(\ref{eq:C1dir}) is necessary
which simplifies the evaluation of the diagrams with external
background gluons.


\subsubsection{\label{sub:hadrHiggs}Hadronic decay rate of the SM
Higgs boson} 

In the following we present the current status of the total decay rate
of the SM Higgs boson\footnote{For review articles we
 refer to~\cite{Kniehl:1994ay,Spira:1998dg}.}
in the intermediate mass range. This is done on the 
basis of the effective Lagrangian Eq.~(\ref{eq:leff}), where
the top quark takes over the role of the heavy quark $h$.

The total decay rate into hadrons can be cast in the 
form\footnote{The additional index ``q'' for the coefficient function
  $C_2$ indicates that there might be an explicit dependence on the
  flavour through elektroweak corrections.}
\begin{eqnarray}
\Gamma\left(H\to \mbox{hadrons}\right)
&=&
\left(1+\delta_u\right)^2
\bigg\{
  \sum_q
  A_{q\bar q}
  \left[
    \left(1+\Delta^q_{22} \right)\left(C_{2q}\right)^2
    +\Delta^q_{12}\,C_1C_{2q}
    \right]
  \nonumber\\&&\mbox{}
  +A_{gg}
  \,\Delta_{11}\,\left(C_1\right)^2
  +A_{gg}
  \,\Delta^{\rm hdo}_g
  +\sum_q
  A_{q\bar q}
  \left[
      \Delta^{\rm hdo}_q
    + \Delta^{\rm weak}_q\Big|_{x_t=0}
  \right]
\bigg\}
\,,
\nonumber\\
\label{eq:gamhad}
\end{eqnarray}
with
$A_{q\bar q}=3G_FM_Hm_q^2/(4\pi\sqrt2)$
and 
$A_{gg}=4G_FM_H^3/(\pi\sqrt{2})$.
The terms in the first line and the first term 
in the second line proportional to $A_{gg}$ have their origin in 
Eq.~(\ref{eq:leff}). In particular,
the universal corrections $\delta_u$ arise from the renormalization
of the factor $H^0/v^0$ and are known to order 
$\alpha_s^2 G_F m_t^2$~\cite{KniSte95}. 
The factors $\Delta_{ij}$ contain the QED and QCD
corrections from the light degrees of freedom only, while
the terms 
$\Delta^{\rm hdo}$ summarize the corrections coming from
higher dimensional operators. 
They are at least suppressed by a factor
$\alpha_s^2 M_H^2/M_t^2$. 
In Eq.~(\ref{eq:gamhad}) we separately display the weak contribution
where the leading term of order $G_Fm_t^2$ is stripped off. It is denoted
by\footnote{For the definition of $x_t$ see 
  Eq.~(\ref{eq:xt})} $\Delta^{\rm weak}_q\Big|_{x_t=0}$.

\begin{figure}[ht]
 \begin{center}
   \leavevmode
   \epsfxsize=14.0cm
   \epsffile[125 640 500 730]{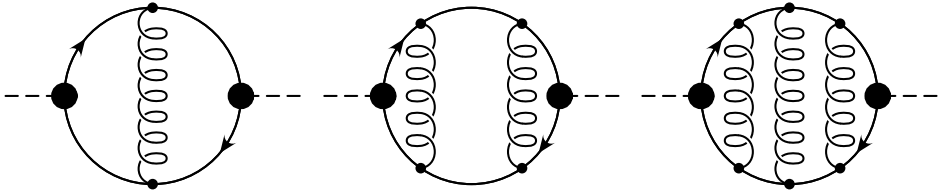}
 \end{center}
\caption{\label{fig:del22}
         Typical Feynman diagrams contributing to $\Delta_{22}^q$. 
         The solid circles represent the operator
         ${\cal O}_{2}$.}
\end{figure}

Typical diagrams contributing to the QCD corrections of
$\Delta_{22}^q$ are pictured in Fig.~\ref{fig:del22}. 
At one-loop order they have been evaluated 
in~\cite{Braaten:1980yq,Drees:1990dq} keeping the
full dependence of the quark mass.
The dominant corrections at order ${\cal O}(\alpha_s^2)$, 
i.e. those obtained 
keeping only the factor $m_q^2$ from the Yukawa coupling,  
have been evaluated in~\cite{GorKatLarSur90}.
The calculation has later on been improved and the correction 
terms proportional to 
$m_q^2/M_H^2$ became available~\cite{Sur94,CheKwi96}.
A naive expansion in $m_q$ is sufficient for their computation.
Beyond the quadratic term, however, one either has to adopt the method 
discussed in Section~\ref{sub:as3m4} or apply the large-momentum
expansion. The latter has been performed in~\cite{Harlander:1997xa}
and mass correction terms up to order $(m_b^2/M_H^2)^8$ 
have been evaluated.
For a Higgs boson in the intermediate-mass 
range considered in this Subsection the corrections
beyond the quadratic term are quite small and can safely be neglected.
In~\cite{Harlander:1997xa} the higher order terms
have been considered in the context of a heavy Higgs boson which can
also decay into top quarks. In this case it turned out that even the
quartic terms are important and only 
an expansion up to order $(m_t^2/M_H^2)^8$ gives satisfactory results.
In Section~\ref{sub:veccor} we review the calculation and 
numerical results are presented in Tab.~\ref{tab:Higgs}.
For completeness we want to mention that
the imaginary part of the correlator 
$\langle {\cal O}_2 {\cal O}_2 \rangle$
has even been evaluated at four-loop order 
using the technique described in Section~\ref{sub:IRR}.
This leads to corrections of order $\alpha_s^3$~\cite{Che97_Higgs}.

Next to the pure QCD corrections there are also the contributions 
of order $\alpha$ and the mixed QED/QCD terms of order
$\alpha\alpha_s$~\cite{Kataev:1992fe,Kataev:1997cq} which can easily
be extracted form the QCD results.
In summary, the numerical result for the discussed terms read
\begin{eqnarray}
  \Delta_{22}^{q}(M_H) &=& 
    \frac{\alpha_s^{(5)}(M_H)}{\pi}
    \left[
    5.66667 
    + \left(35.9400 - 1.3587 n_l\right) 
      \frac{\alpha_s^{(5)}(M_H)}{\pi}
    \right.\nonumber\\&&\left.\mbox{}
    + \left(164.1392 - 25.7712 n_l + 0.2590 n_l^2\right)
      \left(\frac{\alpha_s^{(5)}(M_H)}{\pi}\right)^2
    \right]
    \nonumber\\&&\mbox{}
    + \frac{\bar{\alpha}(M_H)}{\pi}
      Q_q^2\, 
   \left(
      4.2500
    + 11.7097 \frac{\alpha_s^{(5)}(M_H)}{\pi}
   \right)
   \,,
\end{eqnarray}
where $n_l$ is the number of light quarks.

\begin{figure}[ht]
 \begin{center}
   \leavevmode
   \epsfxsize=15.0cm
   \epsffile[70 630 570 720]{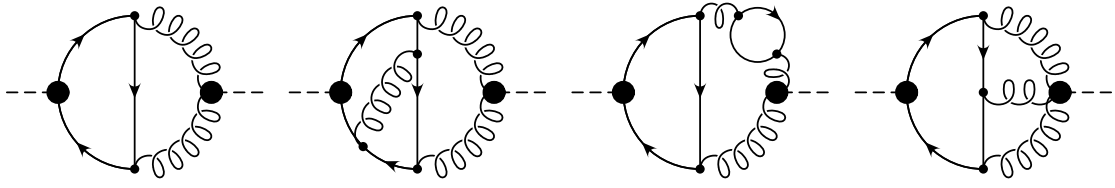}
 \end{center}
\caption{\label{fig:del12}Two- and some of the 
                         three-loop diagrams contributing
                         to $\Delta_{12}^q$. 
                         The solid circles represent the operators
                         ${\cal O}_1$ and ${\cal O}_{2}$, respectively.
                         }
\end{figure}

The imaginary part of the mixed correlator 
$\langle {\cal O}_1 {\cal O}_2 \rangle$ is denoted by $\Delta_{12}^q$. 
Next to contributions to the partial width into quarks it 
also involves purely gluonic final states as can be seen
from the diagrams shown in Fig.~\ref{fig:del12}.
The contributions at two- and three-loop order have been computed
in~\cite{CheKwi96,LarRitVer95_2,CheKniSte97hbb} and~\cite{CheSte97},
respectively. In numerical form $\Delta^q_{12}$ is given by
\begin{eqnarray}
  \Delta_{12}^q(M_H) &=&
    \frac{\alpha_s^{(5)}(M_H)}{\pi}
    \left[
       -30.667
       + \left(-524.853  + 20.647\,n_l \right) 
         \frac{\alpha_s^{(5)}(M_H)}{\pi}
    \right]  
  \,.
  \label{eq:del12}
\end{eqnarray}
In case one is only interested in final states involving quarks the
purely gluonic cuts have to be subtracted.
Currently they are only known at order $\alpha_s$. To this order the
subtracted result reads~\cite{CheKniSte97hbb}
\begin{eqnarray}
  \Delta_{12}^{q\prime}(M_H) &=&
    \frac{\alpha_s^{(5)}(M_H)}{\pi}
    \left[
       -\frac{76}{3} + 8\zeta_2 
       -\frac{4}{3} \ln^2\frac{m_q^2}{M_H^2}
       - 8 \ln\frac{\mu^2}{M_H^2}
    \right] 
       + \ldots
  \,,
  \label{eq:del12_2}
\end{eqnarray}
where the ellipses indicate terms of order $\alpha_s^2$.
Note the logarithmic singularity in the light-quark mass
which arises from the fact that 
only parts of the final state are contained in Eq.~(\ref{eq:del12_2}).

\begin{figure}[ht]
\leavevmode
\begin{center}
\epsfxsize=\textwidth
\epsffile[118 552 478 731]{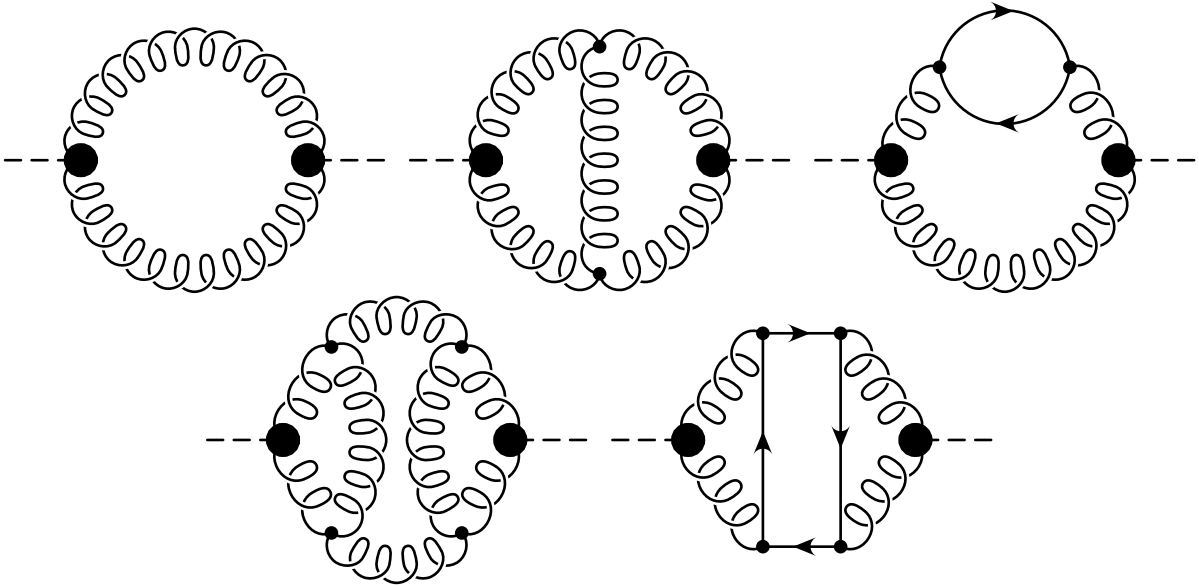}
\caption{Typical Feynman diagrams contributing to the correlator
  $\langle {\cal O}_1 {\cal O}_1\rangle$.
  Looped, solid, and dashed lines represent gluons, light quarks, and 
  Higgs bosons, respectively.
  Solid circles represent insertions of ${\cal O}_1$.}
\label{fig:O1O1}
\end{center}
\end{figure}

\begin{figure}[ht]
\leavevmode
\begin{center}
\epsfxsize=\textwidth
\epsffile[70 530 540 710]{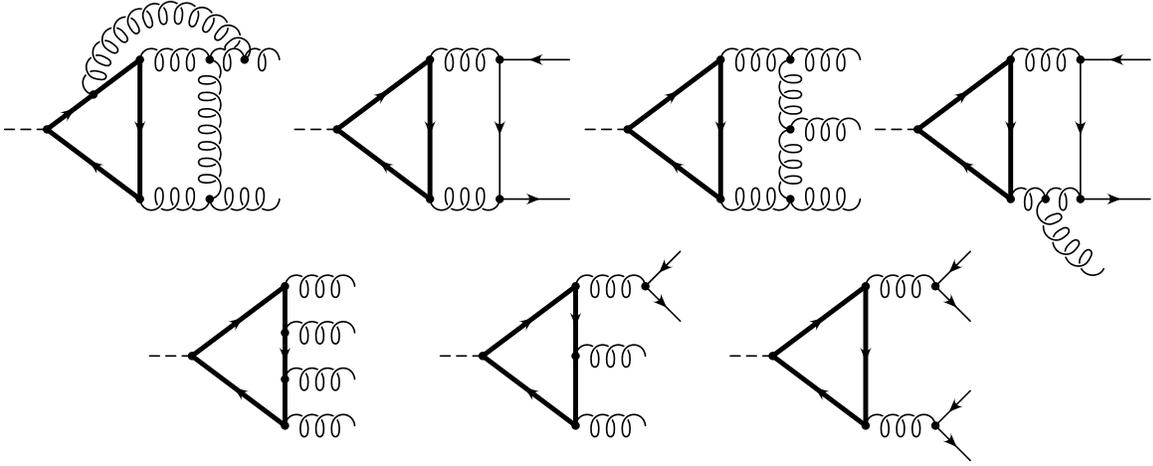}
\caption{Typical diagrams generating ${\cal O}(\alpha_s^2)$
corrections to $\Gamma(H\to gg)$.
Bold-faced (dashed) lines represent the top quark (Higgs boson).}
\label{fig:hggfig}
\end{center}
\end{figure}

The correlator formed by the operator ${\cal O}_1$ mainly contains
cuts arising from gluons. However, starting from two loops there are
also contributions from light quarks and at order $\alpha_s^2$ there
are even cuts involving no gluons at all (cf. second diagram in the
lower row of Fig.~\ref{fig:O1O1}).
In Fig.~\ref{fig:O1O1} some typical diagrams are pictured.
In particular, the combination $(C_1)^2 \Delta_{11}$ contains the
contribution from all the diagrams pictured in Figure~\ref{fig:hggfig}.
The two-loop contribution has been evaluated 
in~\cite{Inami:1983xt,Djouadi:1991tk} and the order $\alpha_s^2$ terms
can be found if~\cite{CheKniSte97}.
If we set $\mu^2=M_H^2$ and evaluate the correlator for $q^2=M_H^2$ we
obtain 
\begin{eqnarray}
  \Delta_{11}(M_H^2) &=& 
  1
  + \frac{\alpha_s^{(5)}(M_H)}{\pi} \left( 18.250 - 1.167 n_l \right)
  \nonumber\\&&\mbox{}
  + \left(\frac{\alpha_s^{(5)}(M_H)}{\pi}\right)^2
    \left(242.973 - 39.374\, n_l + 0.902\, n_l^2  \right)
  \,.
  \label{eq:del11}
\end{eqnarray}

The contributions to $\Delta^{\rm hdo}$ are not covered by 
Eq.~(\ref{eq:leff}). 
In the language of the effective Lagrangian 
it would require to deal with operators of dimension six and higher. 
However, up to now they have not been studied in detail.
The approach adopted in Refs.~\cite{CheKwi96,LarRitVer95_2} is based
on asymptotic expansion which is applied to the propagator-type
diagrams involving a top-quark loop.
At order $\alpha_s^2$ there are two classes of such diagrams
contributing to $\Delta^{\rm hdo}_q$, namely the
double-triangle (or singlet) and the double bubble diagrams.
The exact result for the imaginary part of the latter
with massless external quark lines and heavy virtual top quark 
can be found in~\cite{Kni95}.
The leading term contribution to $\Delta^{\rm hdo}_q$ reads
\begin{eqnarray}
  \Delta^{\rm hdo}_q 
  &=& 
  \left(\frac{\alpha_s^{(5)}}{\pi}\right)^2 \frac{M_H^2}{M_t^2} 
  \left( \frac{5863}{24300} - \frac{113}{1620} \ln\frac{M_H^2}{M_t^2} 
  \right)
  \nonumber\\
  &=& 
  \left(\frac{\alpha_s^{(5)}}{\pi}\right)^2 \frac{M_H^2}{M_t^2} 
  \left(0.241-0.070 \ln\frac{M_H^2}{M_t^2} \right)
  \,.
  \label{eq:hdoq}
\end{eqnarray}
The higher order contributions in $M_H^2/M_t^2$ are very
small and can safely be neglected --- even for $M_H=M_t$.
For $\Delta^{\rm hdo}_g$ one obtains~\cite{LarRitVer95_2}
\begin{eqnarray}
  \Delta^{\rm hdo}_g 
  &=& 
  \left(\frac{\alpha_s^{(5)}}{\pi}\right)^2 \frac{M_H^2}{M_t^2} 
  \,\frac{7}{60}
  + 
  \left(\frac{\alpha_s^{(5)}}{\pi}\right)^3 \frac{M_H^2}{M_t^2} 
  \left( \frac{2249}{1080} - \frac{7}{30} \ln\frac{M_H^2}{M_t^2} 
  \right)
  \nonumber\\
  &=& 
  \left(\frac{\alpha_s^{(5)}}{\pi}\right)^2 \frac{M_H^2}{M_t^2} 
  \, 0.11667
  + 
  \left(\frac{\alpha_s^{(5)}}{\pi}\right)^3 \frac{M_H^2}{M_t^2} 
  \left( 2.0824 - 0.23333 \ln\frac{M_H^2}{M_t^2} 
  \right)
  \,,
  \label{eq:hdog}
\end{eqnarray}
where again the higher order terms in $M_H^2/M_t^2$ are much smaller.
The contribution to $\Delta^{\rm hdo}_g$ is obtained from the
application of the hard-mass procedure to the propagator-type diagrams
where the external Higgs bosons are attached to  
top quarks and they subsequently couple to gluons. To leading order in
the expansion for large $m_t$ one obtains the one- and two-loop
diagrams of Fig.~\ref{fig:O1O1}.
The comparison of Eq.~(\ref{eq:hdog})
with Eq.~(\ref{eq:del11}) shows that also in this case
the contributions from the higher dimensional operators are small.
We want to mention
that for $H\to gg$ a numerical calculation of order
$\alpha_s^2$ exists which takes the complete mass
dependence into account \cite{Djouadi:1991tk}.

At one-loop the complete weak corrections have been computed in
analytical from~\cite{DabHol92,Kni92}. 
If we put $m_q=0$ and consider the limit $M_H\ll 2M_W$,
$\Delta_q^{\rm weak}|_{x_t=0}$ takes the form~\cite{Kni92}
\begin{eqnarray}
  \Delta_q^{\rm weak}|_{x_t=0} &=&
  \frac{G_FM_Z^2}{8\pi^2\sqrt2}
  \left[\frac{1}{2}-3\left(1-4s_w^2|Q_q|\right)^2
  +c_w^2\left(\frac{3}{s_w^2}\ln c_w^2-5\right)\right]
  \,.
\end{eqnarray}
The leading $m_t^2$ term is stripped off as it is already contained in
the universal factor $\delta_u$. 
Expressed in terms of the $\overline{\rm MS}$ top quark mass
the latter reads in numerical form~\cite{KwiSte94KniSpi94,KniSte95}
\begin{eqnarray}
  \delta_u &=&
  {7\over6}N_cx_t\left[1+
  \frac{\alpha_s^{(6)}(\mu)}{\pi}
  (2\,\ln\frac{\mu^2}{m_t^2}+0.869\,561)
\right.\nonumber\\&&\left.\mbox{}
  +\left(\frac{\alpha_s^{(6)}(\mu)}{\pi}\right)^2
   \left(3.750\,000\,\ln^2\frac{\mu^2}{m_t^2}
   +6.010\,856\,\ln\frac{\mu^2}{m_t^2}
   -2.742\,226\right)\right]
  \,.
\end{eqnarray}

We want to mention that also 
non-universal radiative corrections to $C_1$ and $C_{2q}$,  
which are enhanced by a factor $G_Fm_t^2$,
are available up to the three-loop
order~\cite{KniSte95,CheKniSte97hbb,Ste98_higgs}. 
They will be listed below 
in comparison with the terms of ${\cal O}(\alpha_s^3)$.

At this point we want to compare the relative size of the individual
terms. In particular we have in mind terms of order
${\cal O}(\alpha_s^3)$,
$\alpha_s^2m_b^2/M_H^2$,
$\alpha\alpha_s$,
$\alpha_s^2X_t$
and
$\alpha_s^2M_H^2/M_t^2$.
For this purpose we will consider all quarks with mass lighter than
$m_b$ as massless. This means that the sum in
Eq.~(\ref{eq:gamhad}) reduces to $q=b$
which is conveniently written in the form
\begin{eqnarray}
\Gamma(H\to\mbox{hadrons}) &=& 
A_{b\bar{b}}\left(
1+\Delta^b_l+\Delta_t
\right)
+
\frac{A_{gg}}{144}
\left(\frac{\alpha_s^{(5)}}{\pi}\right)^2
\,\Delta_g
\,,
\end{eqnarray}
where $\Delta_l^b$ contains only corrections from light 
degrees of freedom. All top-induced terms proportional to 
$A_{b\bar{b}}$ from 
Eq.~(\ref{eq:gamhad}) are contained in $\Delta_t$,
which we express in terms of $\alpha_s^{(5)}(\mu)$.
$\Delta_g$ contains the corrections from the gluonic final state.
Choosing $\mu^2=M_H^2$ and $n_l=5$ we find 
\begin{eqnarray}
\Delta_l^b &=&
- 6 \frac{(m_b^{(5)})^2}{M_H^2}
+ 0.472 \,\frac{\bar\alpha(M_H)}{\pi}
+ 1.301 \,\frac{\bar\alpha(M_H)}{\pi}a_H^{(5)}
+ a_H^{(5)}\left(
   5.667 
   - 40.000\frac{(m_b^{(5)})^2}{M_H^2}
\right)
\nonumber\\&&\mbox{}
+\left(a_H^{(5)}\right)^2\left( 
29.147 
- 87.725 \frac{(m_b^{(5)})^2}{M_H^2}
\right)
+ 41.758 \left(a_H^{(5)}\right)^3,
\label{eqdell}
\\
\Delta_t &=&
\left(a_H^{(5)}\right)^2
\left[
3.111
-0.667\,L_t
+\frac{(m_b^{(5)})^2}{M_H^2}
\left(
-10
+4\,L_t
+\frac{4}{3}\ln\frac{(m_b^{(5)})^2}{M_H^2}
\right)
\right] 
\nonumber\\&&\mbox{}
+\left(a_H^{(5)}\right)^3
\left(
50.474
-8.167\,L_t
-1.278\,L_t^2
\right)
+ \left(a_H^{(5)}\right)^2\frac{M_H^2}{M_t^2}
\left(
0.241 
- 0.070\, L_t
\right)
\nonumber\\&&\mbox{}
+X_t\left[1
- 4.913 a_H^{(5)}
+ \left(a_H^{(5)}\right)^2
\left(
-72.117
-20.945\,L_t
\right)
\right]
\,,
\label{eqdelt}
\\
\Delta_g &=&
  1 
+ X_t
+ a_H^{(5)}
  \left[
    17.917
   + 30.3369 \, X_t
  \right]
+ \left(a_H^{(5)}\right)^2\left[
       156.808
     + 5.708 \ln\frac{M_H^2}{m_t^2}
  \right]
\,,
\label{eq:delg}
\end{eqnarray}
with $a_H^{(5)}=\alpha_s^{(5)}(M_H)/\pi$, $L_t=\ln M_H^2/M_t^2$,
and $X_t = G_F M_t^2/(8\pi^2\sqrt{2})$.
In Eqs.~(\ref{eqdell}) and~(\ref{eqdelt}) also the quadratic mass
correction terms are listed.
In $\Delta_l$ they are obtained from the naive expansion of the diagrams.
The $(m_b^{(5)})^2/M_H^2$ corrections in $\Delta_t$ arise from
the singlet diagram with one top and one bottom quark triangle.
In this case a naive expansion fails as can be seen by the logarithmic
term in Eq.~(\ref{eqdelt}). 
Instead the asymptotic expansion has to be applied~\cite{CheKwi96}.
Both for $\Delta_l$ and $\Delta_t$
one observes that the ${\cal O}(\alpha_s^3)$ term
proves to be numerically more important than the power suppressed
contribution of ${\cal O}(\alpha^2_s m_b^2/M_H^2)$.
Note, that Eq.~(\ref{eqdelt}) contains 
contributions with pure gluonic final states 
which is due to diagrams of the type in Fig.~\ref{fig:del12}.

In the approximation considered here we have
$-2\lsim L_t<0$. This means that the logarithm needs not 
necessarily to be re-summed as in addition the coefficients
in front of $L_t$ are much smaller than the constant term.

A comparison of Eqs.~(\ref{eqdell}) and~(\ref{eqdelt}) shows
that the top-induced corrections in $\Delta_t$ of
${\cal O}(\alpha_s^3)$ are numerically of the same size as 
the ones arising from ``pure'' QCD.
Furthermore one should mention that the coefficient of the
$M_t$-suppressed terms are tiny and, as
$\alpha_s/X_t\approx30$, also the $\alpha_s^2X_t$ enhanced terms 
are less important than the cubic QCD corrections.
This is also the case for Eq.~(\ref{eq:delg}).
For comparison in Eq.~(\ref{eqdell}) also the two-loop
corrections of order $\alpha\alpha_s$ are listed.
In principle also higher order mass corrections are available 
\cite{Harlander:1997xa}. However, in the case of bottom quarks
it turns out that they are tiny.

In summary, we have shown that for an intermediate-mass Higgs boson
the application of the effective Lagrangian
(cf. Eq.~(\ref{eq:eff})) is quite successful and enables the
computation of the hadronic Higgs decay up to orders $\alpha_s^3$
and $\alpha_s^4$ for the quark and gluon final states, respectively.
The smallness of the higher dimensional operators 
(cf. Eq.~(\ref{eq:hdoq}) and~(\ref{eq:hdog})) justifies
this approach.
In conclusion we can state that the perturbative expansion 
of the hadronic width of the Higgs boson is well under control.


\subsection{\label{sub:as3m4}Quartic 
  mass corrections to $\sigma(e^+e^-\to\mbox{hadrons})$}

The total cross section for hadron production in
electron-positron annihilation, $\sigma(e^+e^-\to\mbox{hadrons})$,
is one of the most fundamental
observables in particle physics (for a review see~\cite{ckk96}). 
For energies sufficiently far above
threshold it can be predicted by perturbative QCD, and it is well
accessible experimentally from threshold up to the high energies of
LEP and a future linear collider.  It allows for a precise determination
of the strong coupling $\alpha_s$ and, once precision measurements at
different energies are available, for a test of its evolution dictated
by the renormalization group equation.

Often the center-of-mass energy is much larger than the quark masses
which then can safely be neglected. However, there are also many
situations where it is important to take into account the 
effect of finite quark masses~\cite{CheKueKwiPR}.
E.g., one can think of charm or bottom quark production 
not far above their production thresholds~\cite{CheKue95,CheKueTeu97}, 
or of top quark production at a future linear collider~\cite{HarSte98}.

The complete mass-dependence at order $\alpha_s$ to 
$R(s)\equiv\sigma(e^+e^-\to\mbox{hadrons})/\sigma(e^+e^-\to\mu^+\mu^-)$ 
has been evaluated quite
some time ago in analytical form~\cite{KalSab55} in the context of QED.
At order $\alpha_s^2$ this task
already becomes much less trivial.
The massless approximation became available quite some time 
ago~\cite{CheKatTka79DinSap79CelGon80}.
However, only for a certain 
class of diagrams --- the ones containing a second massless quark pair ---
the full quark mass dependence could be obtained in analytical form
using conventional methods~\cite{HoaKueTeu951}. 
For all other contributions different methods have to be applied
which are discussed in detail in Section~\ref{sec:pade}.
Here, we only want to mention that a crucial ingredient is 
the application of the large-momentum procedure which provides
an expansion in $m^2/s$.
At order $\alpha_s^3$ also this method fails 
as it would be necessary to evaluate massless
four-loop propagator-type diagrams. At the moment this is not
possible.
Thus a different strategy has to be employed which we will describe below.

In addition to the 
massless result, which has been obtained
in~\cite{GorKatLar91SurSam91,Che97_R}, 
the $m^2/s$ terms of
${\cal O}(\alpha_s^3)$ have been calculated
in~\cite{CheKue90}. They were obtained by reconstructing the
logarithmic $\alpha_s^3m^2/s$ terms for the polarization function
$\Pi(q^2)$ from the knowledge of the full three loop 
${\cal O}(\alpha_s^2m^2/s)$ result of~\cite{GorKatLar86} with the
help of the renormalization group equations. These are sufficient to
calculate the $m^2/s$ terms of the imaginary part in the time-like
region. 
A generalization of this
approach has been formulated for the quartic mass terms 
in~\cite{CheSpi87,CheKue94} 
and was originally adopted for the calculation
of $\alpha_s^2 m^4/s^2$ terms~\cite{CheKue94}.

The basic ingredients are the OPE~\cite{Wil69} and
the renormalization group equations (RGE).
The idea is to apply the OPE to the correlator of two currents and
compute its imaginary part which immediately leads to
corrections for $R(s)$.
The current correlator is expressed as a sum over
local operators multiplied by coefficient functions which represent
the short distance part of the process.
Afterwards one exploits the RGEs in order to relate different pieces in
the sum and to construct the logarithmic terms of the polarization
function.
In addition to the anomalous mass dimension and the $\beta$-function, the
anomalous dimensions of the operators of dimension four are required in
appropriate order.

Let us be more specific and consider the time-ordered product of two
currents. The application of the OPE leads to
\begin{eqnarray}
  T^j(q) &=& i\int{\rm d}^4 x e^{iqx} {\rm T}\,j(x)j(0)
         \,\,\stackrel{-q^2\to\infty}{\sim}\,\,\sum_n C_n(q) {\cal O}_n
  \,,
  \label{eq:ope}
\end{eqnarray}
where the dependence of the coefficient functions $C_n(q)$ on the
large scale $q$ is made explicit. In Eq.~(\ref{eq:ope})
we have to consider all operators ${\cal O}_n$ of dimension four
as exactly those contribute to the quartic mass corrections.
They can be found in
Eqs.~(\ref{eq:op1}) and~(\ref{eq:op3}).

Actually the contributing operators essentially coincide with the ones
of Section~\ref{sub:higgs}. 
There, however, the top quark
mass took over the role of the large scale and the external momentum was
supposed to be much smaller. As a consequence the coefficient
functions depend on $M_t$ and are computed with the help of vacuum
integrals. On the contrary, we will see below that the coefficient
functions of Eq.~(\ref{eq:ope}) are expressed through massless
propagator-type integrals.

The virtue of Eq.~(\ref{eq:ope}) becomes obvious if one takes the
vacuum expectation value. 
Then the left-hand side turns into the polarization
function which in the following we generically call $\Pi(q^2)$. 
On the right-hand side we obtain a sum of vacuum expectation values
of the local operators multiplied by the coefficient functions.
As already mentioned, the latter only depend on $q$ whereas the scale
of the vacuum expectation values is given by the quark mass. This
strongly resembles the large-momentum procedure. Also there a
factorization of the scales is achieved.
In fact, it can be shown that for a certain choice of the operator
basis an identification of the $C_n(q)$ with the hard subgraphs
and of the vacuum expectation values with the co-subgraphs is
possible. 

\begin{figure}[t]
  \begin{center}
  \leavevmode
  \epsfxsize=2.5cm
  \epsffile[150 260 420 450]{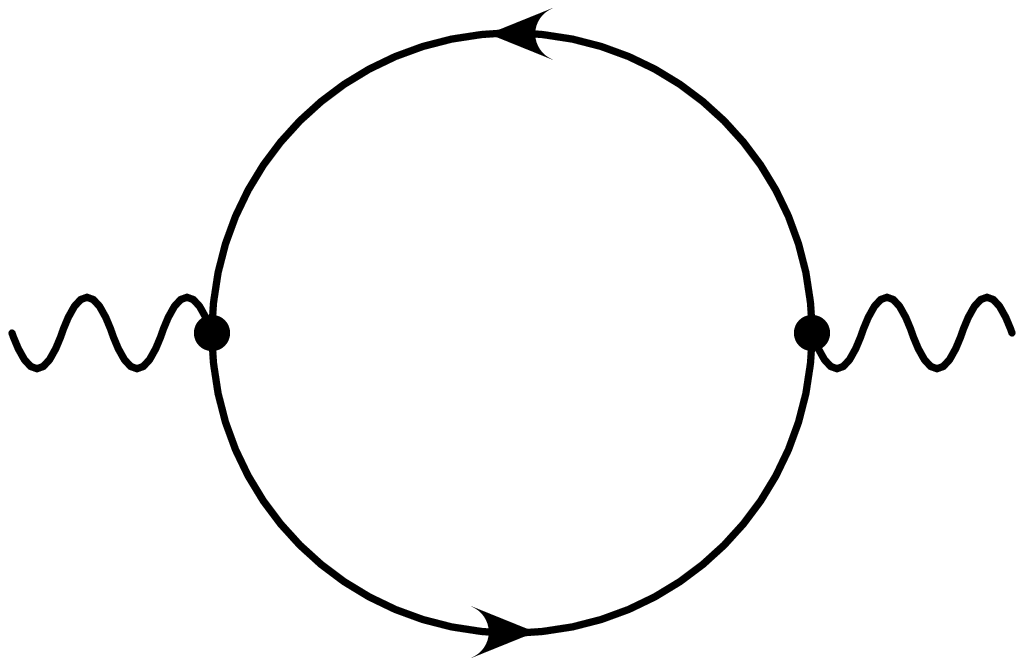}\hspace{1em}
  \raisebox{2.1em}
  {\Large $= \ \ 1\ \star\,\, $}
  \epsfxsize=2.5cm
  \epsffile[150 260 420 450]{figs/d1q.ps}\hspace{1em}
  \raisebox{2.1em}
  {\Large $+ \ \ 2\times \!\!\!\!$}
  \epsfxsize=2cm
  \raisebox{.5em}{\epsffile[150 260 420 450]{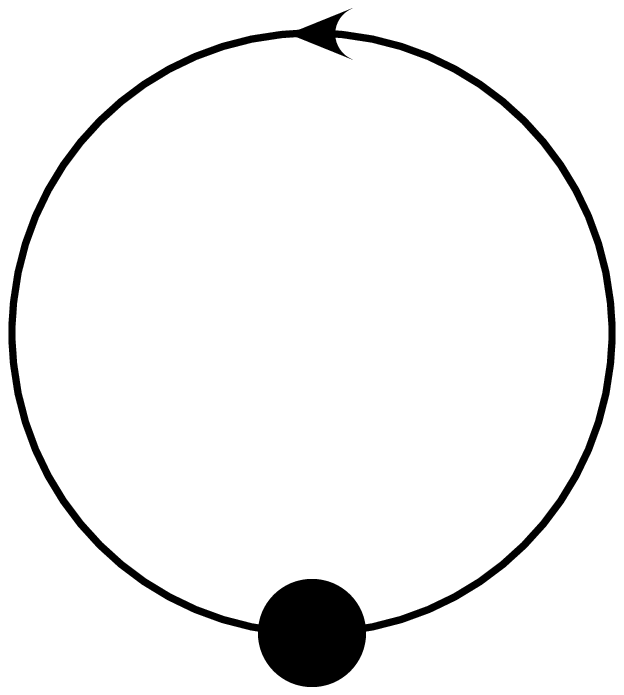}}\hspace{-.5em}
  \raisebox{2.1em}{\Large $\star\,\,$}
  \epsfxsize=2.5cm
  \epsffile[150 260 420 450]{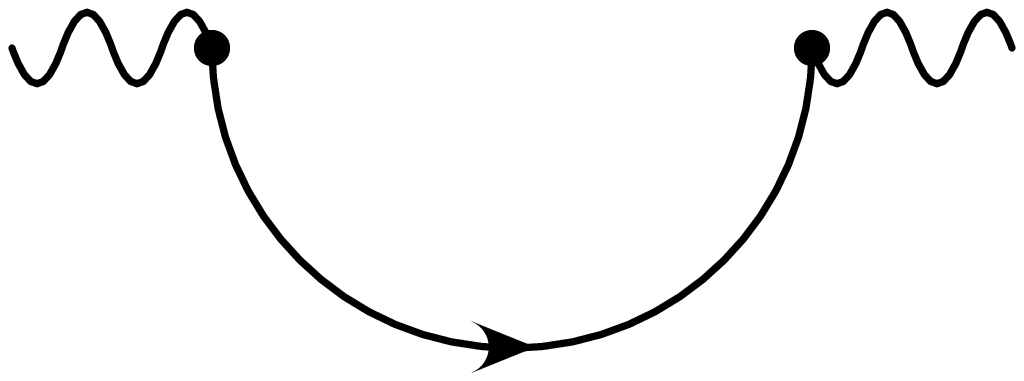}\hspace{1em}
      \caption[]{\label{fig:lmp1l}\sloppy
        Large-momentum procedure for the one-loop photon polarization
        function. The quark lines carry the mass $m$.
        }
  \end{center}
\end{figure}

For illustration we consider the correlator of two vector currents at
one-loop order. The application of the large-momentum procedure is
visualized in Fig.~\ref{fig:lmp1l}.
The prescription tells us that the first term on the right-hand side of
the equation has to be expanded in the quark mass, $m$, leading to
massless integrals. Also the subdiagram on the very right
in the second term
has to be expanded in $m$ and then has to be inserted in the blob of the 
loop-diagram. Thus one ends up with a vacuum integral.
Taking into account terms up to quartic order one 
gets\footnote{For a precise definition of the polarization function
  see Eqs.~(\ref{eqpivadef}) and~(\ref{eqpi2}) in Section~\ref{sec:pade}.}
(see, e.g., Ref.~\cite{Har:9910496,Harlander:1999dq})
\begin{eqnarray}
  \Pi^{(0)}_{\rm bare}(q^2)  
  &\stackrel{q^2\gg m^2}{=}&
  {3\over 16\pi^2}\,\Bigg\{  
  {4\over 3\varepsilon}
  + {20\over 9} - {4\over 3}\,\logqmums
  + 8 {m^2\over q^2}
  + \left({m^2\over q^2}\right)^2\,\bigg( 
    4 
    + 8\,\logqmms
  \bigg)
+\ldots
  \Bigg\}\,,
  \nonumber\\
  \label{eq:lmp1l}
\end{eqnarray}
with $\logqmums=\ln(-q^2/\mu^2)$ and $\logqmms=\ln(-q^2/m^2)$.
The ellipses in Eq.~(\ref{eq:lmp1l}) denote 
higher order mass correction terms.
On dimensional reasons it is clear that the $m^0$ and $m^2$ term only
comes from the first diagram on the right-hand side of
Fig.~\ref{fig:lmp1l}. Also the $m^4 \ln(-q^2)$ term can only arise 
from this diagram. On the other hand, the massive logarithm, 
$m^4 \ln m^2$, originates from the vacuum integral of Fig.~\ref{fig:lmp1l}.

Let us now look at the result provided by the OPE.
At lowest order in $\alpha_s$ only the two operators 
${\cal O}_2$ and ${\cal O}_6$ of 
Eq.~(\ref{eq:op1}) contribute.
The product of the corresponding vacuum expectation values and the
coefficient functions reads (see, e.g., Ref.~\cite{Har:9910496,Har:diss})
\begin{eqnarray}
  C_2^0 \langle {\cal O}_2^0 \rangle &=& 
  \frac{3}{16\pi^2} \left(\frac{m^2}{q^2}\right)^2 
        \left(\frac{8}{\varepsilon}+12+8\logmum\right)
  \,,
  \nonumber\\
  C_6^0 \langle {\cal O}_6^0 \rangle &=& 
  \frac{3}{16\pi^2} \left(\frac{m^2}{q^2}\right)^2 
        \left(-\frac{8}{\varepsilon}-8+8\logqmums\right)
  \,.
  \label{eq:C2C6}
\end{eqnarray}
Thus, $C_6^0\langle {\cal O}_6^0 \rangle$ exactly repoduces the 
$m^4\ln(-q^2)$ of the massless diagram of Fig.~\ref{fig:lmp1l},
whereas $\langle {\cal O}_2^0 \rangle$ provides the massive logarithm.
The sum of the two contributions in Eq.(\ref{eq:C2C6})
reproduces the $m^4$ terms of Eq.~(\ref{eq:lmp1l}).

Note that in the above consideration no normal-ordering prescription
has been used. Otherwise the vacuum expectation value of the operator 
${\cal O}_2$ would be zero. Furthermore, the coefficient function
$C_6$ would necessarily contain $\ln m^2$ terms in order to reproduce 
the quartic terms of Eq.~(\ref{eq:lmp1l}).
Thus, in case the normal-ordering prescription is applied, there is no
separation of the two scales $q$ and $m$.

This example, in particular Eq.~(\ref{eq:C2C6}) and
Fig.~\ref{fig:lmp1l}, shows that
in principle one could still use the large-momentum procedure for the
practical computation of $C_n(q)$ and $\langle{\cal
O}_n(0)\rangle$. However, in practice it turns out that this is
quite tedious.

The aim of the calculation is to obtain $R(s)$ up to
order $m^4 \alpha_s^3$, which means that due to the equation
\begin{eqnarray}
  R(s)\Big|_{m^4} &=& 12 \pi \, {\rm Im} \Pi(q^2=s+i\epsilon)\Big|_{m^4}
  \nonumber\\
                  &\sim& {\rm Im} \left[
    C_1 \langle {\cal O}_1 \rangle
   +C_2 \langle {\cal O}_2 \rangle
   +C_6 \langle {\cal O}_6 \rangle
                                  \right]
  \,,
\end{eqnarray}
one has to evaluate the coefficient functions and vacuum expectation
values up to sufficiently high order.
The imaginary parts can only arise from the coefficient functions as
by construction only they can develop logarithms of the form
$\ln(-s-i\epsilon)$. Note further that the information about the
considered current only enters into the coefficient functions; the
matrix elements of the operators are universal.
For the vector current correlator it turns out that $C_1$
develops an imaginary part starting from ${\cal O}(\alpha_s^3)$. As
$\langle {\cal O}_1 \rangle$ is proportional to $\alpha_s$
there is no contribution to order $\alpha_s^3$ from this term.
The lowest order of $\langle {\cal O}_2 \rangle$ is $\alpha_s^0$ 
and the imaginary part for $C_2$ starts at order $\alpha_s^2$.
This implies that $C_2$ is needed up to order $\alpha_s^3$ (three loops)
and $\langle {\cal O}_2 \rangle$ up to order $\alpha_s$ (two loops).
So far the occuring integrals are all available in the literature.
However, in the case of $C_6$ the logarithmic 
terms up to ${\cal O}(\alpha_s^3)$ are needed, which would require a
four-loop calculation. This can be avoided as we will see 
in the following~\cite{CheKue94}.

We consider the renormalization
group properties of the polarization function.
In general $\Pi(q^2)$ is not renormalization group invariant. Considering, 
however, only the quartic mass terms we have
\begin{eqnarray}
  0&=&\mu^2\frac{{\rm d}}{{\rm d}\mu^2} \Pi(q^2)\Bigg|_{m^4 {\rm terms}}
  \nonumber\\
   &=&\mu^2\frac{{\rm d}}{{\rm d}\mu^2} 
      \left(
        C_1 \left\langle {\cal O}_1 \right\rangle
      + C_2 \left\langle {\cal O}_2 \right\rangle
      + C_6 \left\langle {\cal O}_6 \right\rangle
      \right)
   \,,
\end{eqnarray}
as non-zero contributions may at most have mass dimension two.
Using Eq.~(\ref{eq:gam_mn}) and 
\begin{eqnarray}
  \mu^2\frac{{\rm d}}{{\rm d}\mu^2} C_6 &=& 
  \left(\mu^2 \frac{\partial}{\partial\mu^2}
   +\beta \pi \frac{\partial}{\partial\alpha_s}\right)C_6
\end{eqnarray}
one obtains
\begin{eqnarray}
  \frac{\partial}{\partial L} 
     C_6 \left\langle {\cal O}_6 \right\rangle &=& 
  - 4\gamma_m C_6 \left\langle {\cal O}_6 \right\rangle 
  - \beta \pi\frac{\partial}{\partial\alpha_s}
    C_6 \left\langle {\cal O}_6 \right\rangle 
  - C_1 4m^4\alpha_s \frac{\partial}{\partial\alpha_s} \gamma_0
  + C_2 4m^4\gamma_0
  \,,
\nonumber\\
  \label{eq:c6comp}
\end{eqnarray}
with $L=\ln(\mu^2/(-q^2))$. With the help of this equation the
logarithmic terms of $C_6$ at order $\alpha_s^3$ can be obtained
through two- and three-loop calculations. In particular $C_6$ itself
appears on the right-hand side of Eq.~(\ref{eq:c6comp}), however, only
at order $\alpha_s^2$ which corresponds to massless three-loop integrals.

As a simple example 
let us evaluate the order $\alpha_s^0$ term of Eq.~(\ref{eq:c6comp}).
In this limit there is only a contribution from 
the last term.
With the help of Eq.~(\ref{eq:gamma0}) and $C_2=2/q^4$ we obtain
\begin{eqnarray}
  \frac{\partial}{\partial L} C_6 \left\langle {\cal O}_6 \right\rangle 
  &=& 
  \frac{3}{16\pi^2} \left(\frac{m^2}{q^2}\right)^2 \left(-8\right)
  \,,
\end{eqnarray}
which after integration reproduces the logarithmic terms of the 
renormalized version of
Eq.~(\ref{eq:C2C6}):
\begin{eqnarray}
  C_6 \langle {\cal O}_6 \rangle &=& 
  \frac{3}{16\pi^2} \left(\frac{m^2}{q^2}\right)^2 
        \left(-4+8\logqmums\right)
  \,. 
\end{eqnarray}

At this point it is instructive to make again a comparison with the
large-momentum procedure.
Applied to the polarization function there is always one term where
the hard subgraph constitutes the complete diagram Taylor
expanded in the masses (cf. Appendix~\ref{sub:ae}). 
At $n$-loop order, i.e. considering QCD
corrections to order $\alpha_s^{n-1}$, this means that 
$n$-loop massless propagrator-type integrals have to be solved.
Actually, as we are only interested in the imaginary part of the
polarization function only the logarithmic parts of 
the integrals is needed.
This part exactly constitutes the left-hand side of
Eq.~(\ref{eq:c6comp}). 
The right-hand side of Eq.~(\ref{eq:c6comp}) contains 
lower-order terms of 
$C_6$\footnote{Note that $\gamma_m$ and $\beta$
  start at ${\cal O}(\alpha_s)$ and ${\cal O}(\alpha_s^2)$, respectively.}
and contributions of $(n-1)$-loop diagrams.
Thus the price one has to pay in order to avoid the computation of the
(imaginary part) of the $n$-loop diagrams is the construction of
appropriate operators, the computation of their anomalous dimension and
their coefficient functions.

For the practical computation of the coefficient functions the
so-called ``method of projectors''~\cite{GorLarTka83,GorLar87} is used.
For the projectors, $\pi_n$, 
an appropriate combination of initial and final
states, $|i\rangle$ and $|f\rangle$, 
and derivatives with respect to masses and momenta is chosen
in such a way that one has
\begin{eqnarray}
  \pi_n\left[{\cal O}^0_m\right] = \delta_{nm}
  \,.
  \label{eq:pro1}
\end{eqnarray}
Here ${\cal O}^0_n$ is one of the operators defined in
Eqs.~(\ref{eq:op1}) and~(\ref{eq:op3})
and $\pi_n$ has the form
\begin{eqnarray}
  \pi_n\left[X\right] &=& \sum_k P_k
  \left(
    \frac{\partial}{\partial p},\frac{\partial}{\partial m}
  \right)
  \left\langle f_k \left|X\right|i_k \right\rangle\Bigg|_{p=m=0}
  \,.
\end{eqnarray}
It is understood that the nullification of $p$ and $m$ happens before
the loop integrals are performed. Thus, in Eq.~(\ref{eq:pro1})
there is only the tree-level contribution; all loop corrections become
massless tadpoles which are set to zero in dimensional regularization.
The application of the projectors to Eq.~(\ref{eq:ope}) immediately
leads to
\begin{eqnarray}
  C_n^0(q) &=& \pi_n\left[T^j(q)\right]
  \,,
\end{eqnarray}
which relates the bare coefficient function $C_n^0$ to massless
two-point functions.
For $C_6^0$, e.g., the projector is quite simple. It essentially
consists of four derivatives 
of the polarization function with respect to $m$.
From the structure of the operators it is clear that the projectors for
$C_1^0$ and $C_2^0$ are more complicated, as the corresponding diagrams 
also involve external gluons and quarks,
respectively~\cite{SurTka90,Har:diss}.

There is quite some similarity between
the ``method of projectors'' 
and the procedure we have used for the
computation of the decoupling constants in Section~\ref{sec:dec}.
In fact, if one tries to construct a projector for $C_1^0$
one arrives at a similar system of equations as in
Eq.~(\ref{eq:C1C4C5})~\cite{SurTka90,Har:diss}. 

As already mentioned, the computation of the vacuum expectation values
reduces to the evaluation of vacuum integrals
which have been calculated up to three-loop 
order~\cite{Bro81,CheSpi87,BraNarPic92,CheKue94,Har:diss}.

We refrain from listing the individual results for the coefficient
functions and the vacuum expectation values of the operators but provide
directly the results for $R(s)$.
Thereby we want to list the results of those terms which 
contribute to the production of the heavy quark pair $Q\bar{Q}$
via the exchange of a photon
which will be denoted by $R_Q(s)$.
In this Subsection we want to list $R_Q(s)$ up to order
$\alpha_s^3 m^4/s^2$ and postpone the discussion of the remaining
terms (in particular the full mass dependence at order $\alpha_s^2$) to
Section~\ref{subsub:R}.

It is convenient to decompose the contributions to $R_Q(s)$ into three
parts
\begin{eqnarray}
  R_Q(s) &=& 3 \left( Q_Q^2 r_Q + \sum_q Q_q^2 r_{qQ} 
                    + r_{Q,\rm sing} \right)
  \,,
  \label{eq:RQ}
\end{eqnarray}
where the sum runs over all massless quark flavours $q$ and
$Q_q$ denotes the charge of quark $q$.
In Eq.~(\ref{eq:RQ}) we distinguish the contributions
where the massive quark $Q$ directly couples to photon ($r_Q$)
from the ones where in a first step a massless quark is produced
which subsequently splits into the massive quark $Q$.
Furthermore, the singlet contributions are displayed separately.
They arise from diagrams where the external current couples to a
closed quark line which is different from the one involving the 
final-state quarks.
In the following $r_{Q,\rm sing}$ will not be considered.
Its contribution is numerically small and can, e.g., be found
in~\cite{CheHarKue00}. 

Both $r_Q$ and $r_{qQ}$ are expanded in $m_q^2/s$
and can be written as
\begin{eqnarray}
  r_Q &=& r_0 + r_{Q,2} + r_{Q,4} + \ldots
  \,,
  \nonumber\\
  r_{qQ} &=& r_0 + r_{qQ,2} + r_{qQ,4} + \ldots
  \,,
  \label{eq:RQ2}
\end{eqnarray}
where $r_0$ belongs to the massless approximation, while
$r_{Q,n}$ and $r_{qQ,n}$ contain the mass terms of order $m_Q^n$.
A look to the contributing diagrams shows that 
the contributions to $r_{qQ,2}$ and $r_{qQ,4}$
arise for the first time at order $\alpha_s^2$.
However, it can be inferred from general renormalization group
considerations that the corresponding coefficient of $r_{qQ,2}$ has to
be zero~\cite{CheKue90,CheKue97}, 
which means that it starts out only at order $\alpha_s^3$.

The numerical result for the massless approximation 
reads~\cite{CheKatTka79DinSap79CelGon80,GorKatLar91SurSam91,Che97_R}
\begin{eqnarray}
  r_0 &=&
       1
      + {\alpha_s\over \pi} 
      + \left({\alpha_s\over \pi}\right)^2\,\Big(
         1.98571 - 0.115295\,n_f \Big)
         \nonumber\\&&\mbox{}
      + \left({\alpha_s\over \pi}\right)^3\,\Big(
         -6.63694 - 1.20013\,n_f - 0.00517836\,n_f^2\Big)
      + \ldots 
  \,,
  \label{eq:r0}
\end{eqnarray}
where the ellipses indicate higher orders in $\alpha_s$.
$n_f$ is the number of active quark flavours.
The quadratic mass corrections are given by~\cite{GorKatLar86,CheKue90}
\begin{eqnarray}
  r_{Q,2} &=&
    {m_Q^2\over s}\,{\alpha_s\over \pi}\,\bigg[
  12 + {\alpha_s\over \pi}\,\Big(
      126.5 - 4.33333\,n_f\Big)
\nonumber\\&&\mbox{}
 + \left({\alpha_s\over \pi}\right)^2\,\Big(
1032.14 - 104.167\,n_f + 1.21819\,n_f^2\Big)
\bigg]\,,
\nonumber\\
r_{qQ,2} &=&
  {m_Q^2\over s}\,\left({\alpha_s\over \pi}\right)^3\,
  \bigg[-7.87659 + 0.35007\, n_f\bigg]
  + \ldots
  \,,
  \label{eq:r2MS}
\end{eqnarray}
and, finally, for the quartic terms we
have~\cite{CheKue94,Har:diss,CheHarKue00}
\begin{eqnarray}
  r_{Q,4} &=&
 \left({m_Q^2\over s}\right)^2\,\bigg[-6 - 22\,{\alpha_s\over \pi}
\nonumber\\&&\mbox{\hspace{0em}}
 + \left({\alpha_s\over \pi}\right)^2\,(
     148.218 
     - 6.5\,\logmsms 
     + (-1.84078 + 0.333333\,\logmsms)\,n_f
)
 \nonumber\\&&\mbox{\hspace{0em}}
  + \left({\alpha_s\over \pi}\right)^3\,(
     4800.95 
     - 244.612\,\logmsms + 13\,\logmsms^2
 \nonumber\\&&\mbox{\hspace{1em}}
     + (-275.898 
        + 18.1861\,\logmsms 
        - 0.666667\,\logmsms^2 )\,n_f
 \nonumber\\&&\mbox{\hspace{1em}}
     + (4.97396 
        - 0.185185\,\logmsms)\,n_f^2
 ) + \ldots
\bigg]\,,
\nonumber\\
r_{qQ,4} &=&
      \left({m_Q^2\over s}\right)^2\,
      \left({\alpha_s\over \pi}\right)^2\,\bigg[
              -0.474894
              - \logmsms
\nonumber\\&&\mbox{\hspace{1em}}
          + {\alpha_s\over \pi}\,\Big(
            4.59784 
            - 22.8619\,\logmsms 
            + 2\,\logmsms^2  
\nonumber\\&&\mbox{\hspace{2em}}
            + (0.196497 + 0.88052\,\logmsms)\,n_f
              \Big)
  + \ldots
          \bigg]\,,
  \label{eq:r4MS}
\end{eqnarray}
with $\logmsms=\ln m_Q^2/s$.

In Eqs.~(\ref{eq:r2MS}) and~(\ref{eq:r4MS}) 
the $\overline{\rm MS}$ quark mass has been
chosen as a parameter. This is inherent to the method used for
the computation. Actually also the mass which is present in the
renormalized operators ${\cal O}_2$ and ${\cal O}_6$ is
defined in the modified minimal subtraction scheme.
In order to transform the expressions into the
on-shell scheme the 
two-~\cite{GraBroGraSch90} and
three-loop~\cite{CheSte99,CheSte00,MelRit99} (see also
Section~\ref{sub:msos}) relation between the masses is 
necessary\footnote{Due to the absence of a Born term 
  in Eq.~(\ref{eq:r2MS}) the two-loop relation between the 
  $\overline{\rm MS}$ and on-shell quark mass is sufficient in this case.}.
We obtain for the quadratic terms (see also
Ref.~\cite{CheHoaKueSteTeu97})
\begin{eqnarray}
  r_{Q,2}^{\rm OS} &=&
    {M_Q^2\over s}\,{\alpha_s\over \pi}\,\bigg[
    12 + {\alpha_s\over \pi} 
      \left(94.5000 + 24\lMs - 4.33333 n_f\right) 
    + \left({\alpha_s\over \pi}\right)^2\,\left( 
        347.168
      + 378\lMs 
  \right.\nonumber\\&&\left.\mbox{}
      - 9\lMs^2 
      + n_f\left(
         -67.6190 - 17.3333 \lMs + 2 \lMs^2
            \right) 
      + 1.21819 \, n_f^2
      \right)
    +\ldots
    \bigg]
    \,,
\nonumber\\
r_{qQ,2}^{\rm OS} &=&
  {M_Q^2\over s}\,\left({\alpha_s\over \pi}\right)^3\,
  \bigg[-7.87659 + 0.35007\, n_f\bigg]
  + \ldots
  \,,
  \label{eq:r2OS}
\end{eqnarray}
where $\lMs=\ln (M_Q^2/s)$.
Note the presence of mass logarithms which are introduced via
the transition to the on-shell scheme.

The quartic terms read in the on-shell scheme
\begin{eqnarray}
  r_{Q,4}^{\rm OS} &=&
    \left({M_Q^2\over s}\right)^2\,\bigg[
     -6 
      + {\alpha_s\over \pi}\left(
        10 - 24\lMs
      \right) 
      + \left({\alpha_s\over \pi}\right)^2\left(
          570.519
        - 155.5\lMs 
        - 15 \lMs^2 
  \right.\nonumber\\&&\left.\mbox{}
        + n_f\left(-26.8336
                   + 9 \lMs 
                   - 2 \lMs^2
              \right)
      \right)
      + \left({\alpha_s\over \pi}\right)^3\left(
         9157.82
       - 444.899 \lMs 
  \right.\nonumber\\&&\left.\mbox{}
       - 147.750 \lMs^2 
       + 7.50000 \lMs^3 
       + n_f\left(-936.140
         + 243.009 \lMs 
         - 26.1667 \lMs^2 
  \right.\right.\nonumber\\&&\left.\left.\mbox{}
         - 0.666667 \lMs^3
         \right)
       + n_f^2\left(
           20.6385
         - 7.86797 \lMs 
         + 1.44444 \lMs^2 
  \right.\right.\nonumber\\&&\left.\left.\mbox{}
         - 0.222222 \lMs^3
         \right)
       \right)
  \ldots
  \bigg]
  \,,
  \nonumber\\
   r_{qQ,4}^{\rm OS} &=&
  \left({M_Q^2\over s}\right)^2\,
  \left({\alpha_s\over \pi}\right)^2\,\bigg[
    -0.474894 - \lMs 
   + {\alpha_s\over \pi}\left(
         9.79728
       - 21.4282 \lMs 
       - 2\lMs^2 
  \right.\nonumber\\&&\left.\mbox{}
       + n_f\left(
              0.196497
              + 0.880520 \lMs
            \right)
      \right)
  \ldots
  \bigg]
  \,.
  \label{eq:Rm4OS}
\end{eqnarray}
Note that in this case the transition to the on-shell scheme 
even introduces cubic mass-logarithms.

\begin{figure}[ht]
  \begin{center}
    \leavevmode
    \begin{tabular}{cc}
      (a) & (b) \\[-.5em]
      \epsfxsize=18em
      \epsffile[110 265 465 560]{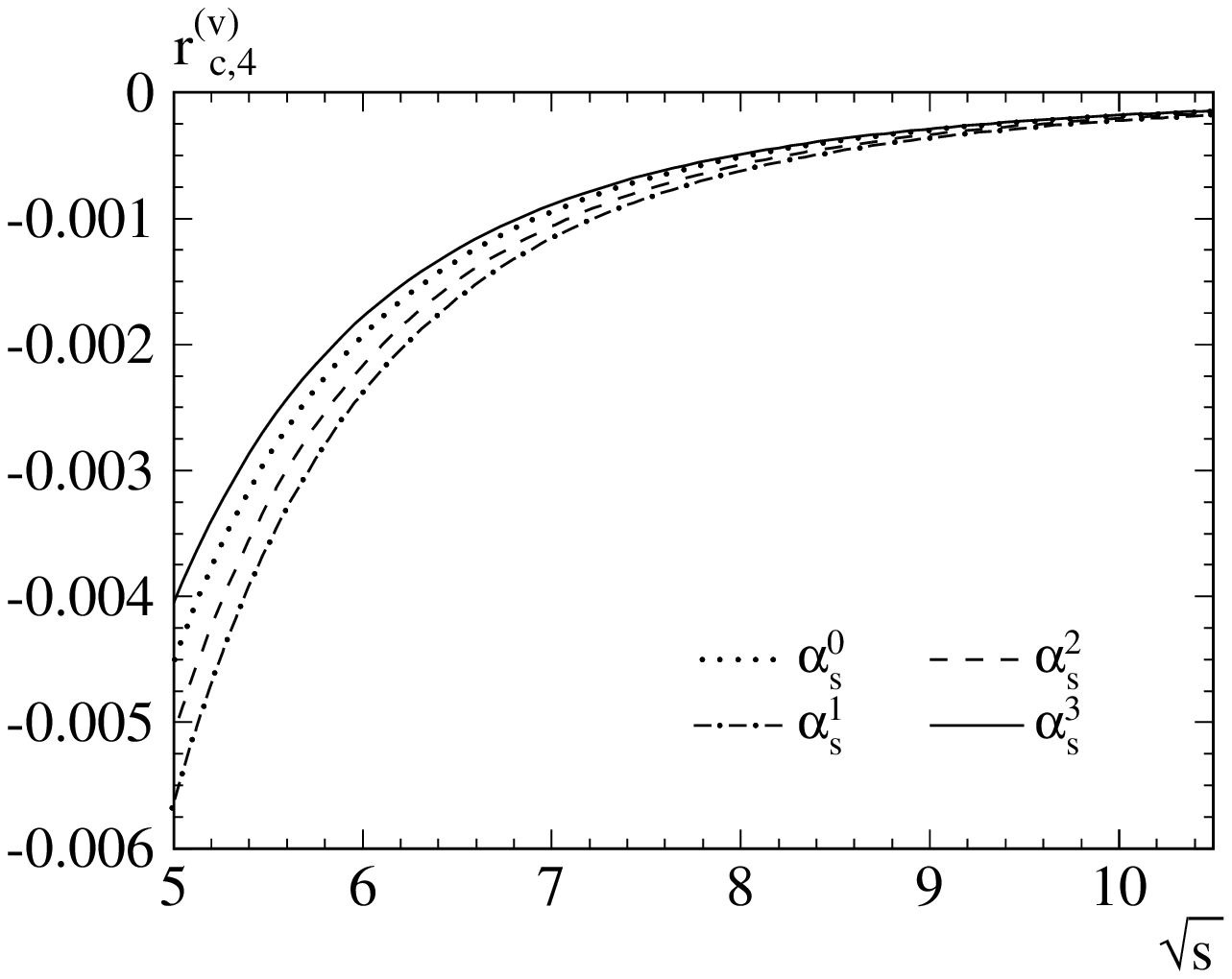} &
      \epsfxsize=18em 
      \epsffile[110 265 465 560]{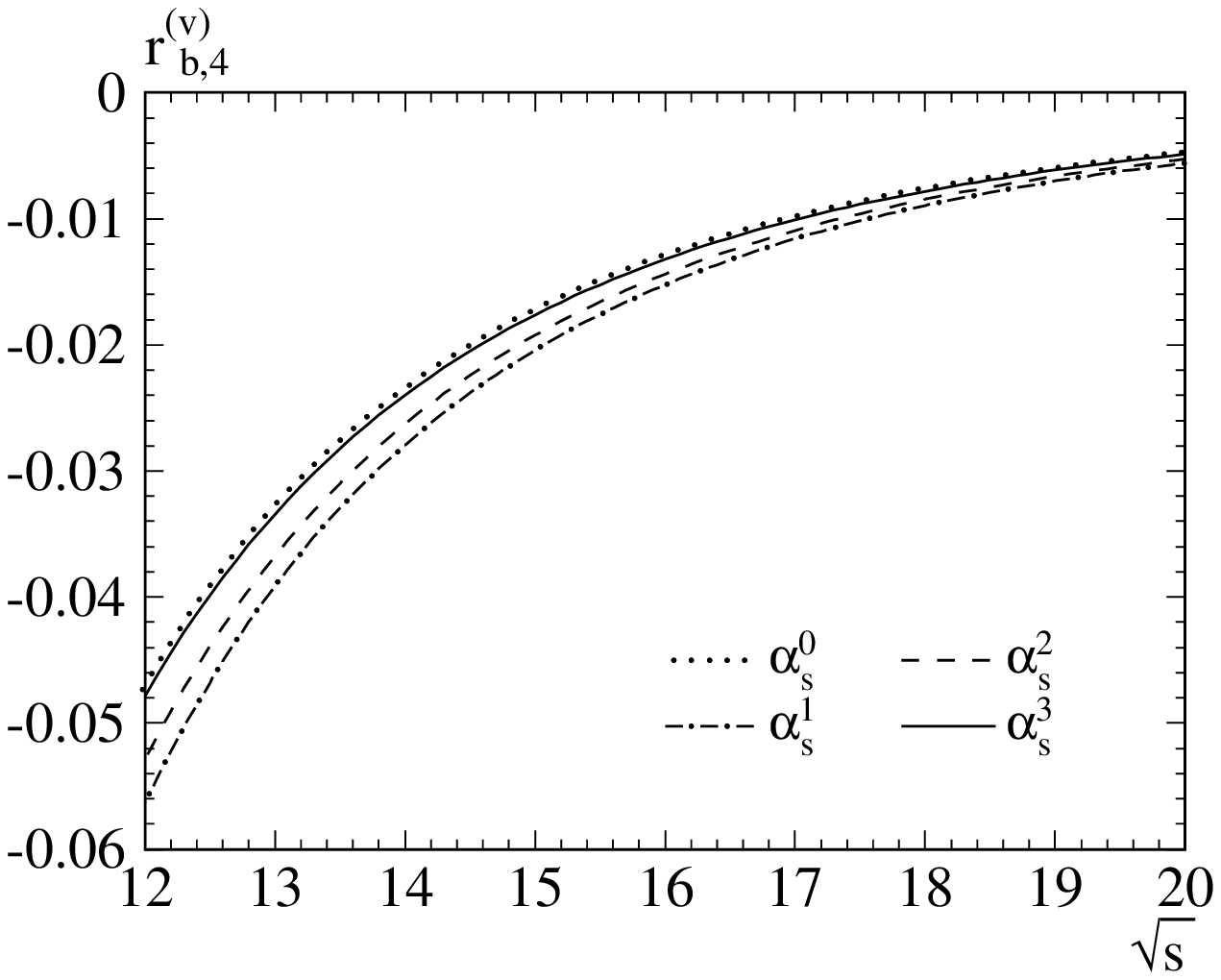} \\[.5em]
      \multicolumn{2}{c}{(c)} \\[-.5em]
      \multicolumn{2}{c}{
      \epsfxsize=18em
      \epsffile[110 265 465 560]{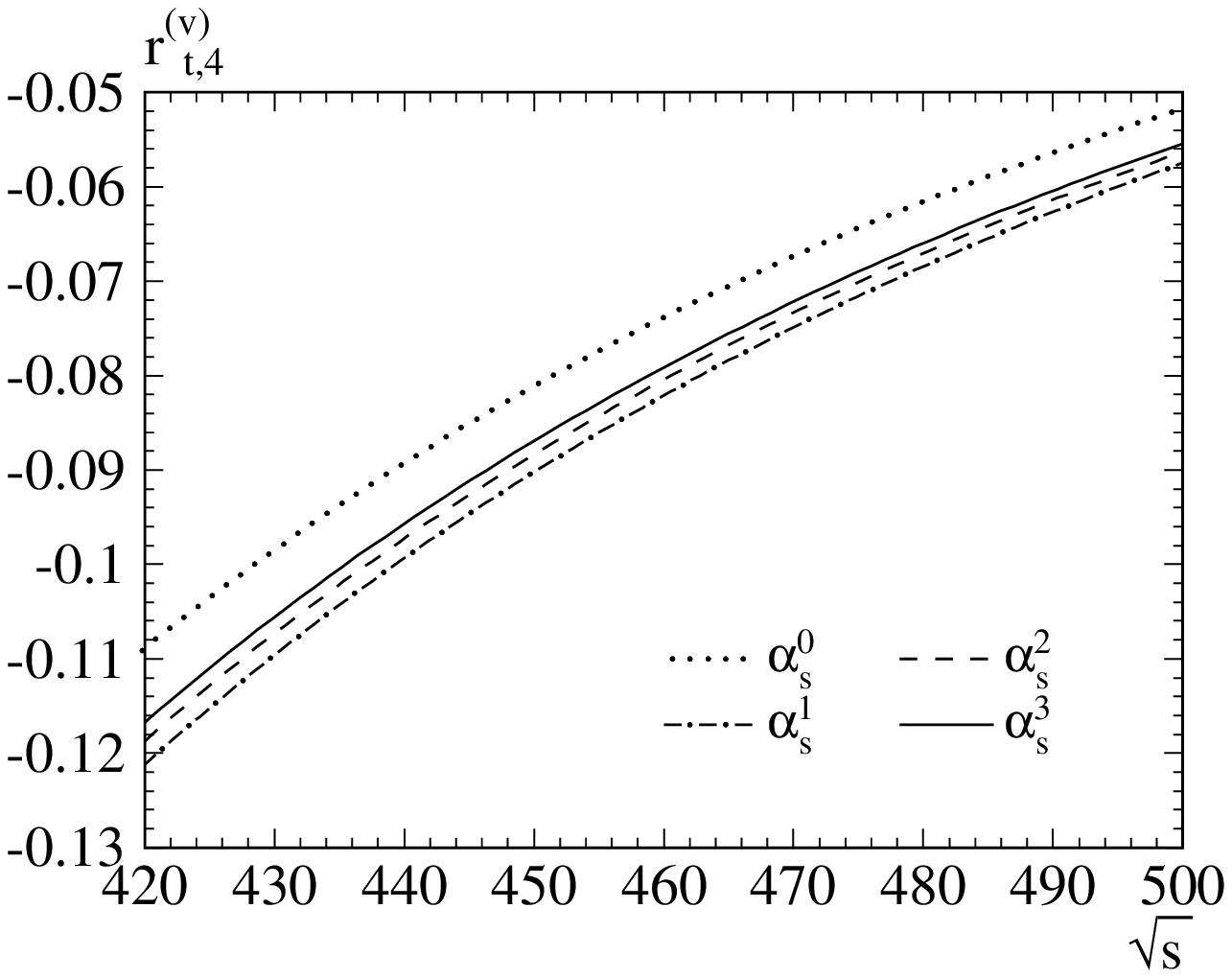}
      }
    \end{tabular}
      \caption[]{\label{fig:rQ4} 
        Quartic mass corrections ($\propto m^4$) to the non-singlet
        contribution of $r_c$ (a), $r_b$ (b), and  
        $r_t$ (c) arising from diagrams where
        the external current couples directly to the massive quark.
      }
  \end{center}
\end{figure}

In Fig.~\ref{fig:rQ4} $r_{c,4}$, $r_{b,4}$ and $r_{t,4}$ 
are shown as a function of
the center-of-mass energy, $\sqrt{s}$, where successively higher
orders in $\alpha_s$ are taken into account.
It can be seen that the major part of the result is given by the Born
approximation. The correction terms of order 
$\alpha_s$, $\alpha_s^2$ and $\alpha_s^3$
are significantly smaller than the leading terms. 
However, it can be observed that with increasing order 
they remain roughly comparable in magnitude which could indicate a bad
behaviour of the perturbative expansion.
Nevertheless, the higher orders are small compared to the $m_Q^4/s^2$ Born
terms. It was shown in~\cite{CheHarKue00} that the overall prediction for
$R(s)$ is stable and a variation of $\mu$ between $\sqrt{s}/2$ and
$2\sqrt{s}$ for $r_{c,4}$ at $6$~GeV varies by $\pm 0.0005$ and
for $r_{b,4}$ at $14$~GeV by $\pm 0.0016$.
In the case of $r_{t,4}$ the variation is negligible.

Thus a prediction for $R_Q(s)$ up to order $\alpha_s^3$ is available.
It includes mass terms in an expansion up to the quartic order.

\section{\label{sec:pade}Asymptotic expansion and Pad\'e approximation}
\setcounter{equation}{0} 
\setcounter{figure}{0} 
\setcounter{table}{0} 

Using the techniques currently available, diagrams beyond two loops 
can only be computed for special cases.
In particular, only diagrams depending on one scale, like an external
momentum or an internal mass, have been studied systematically.
This section is devoted to a method 
which allows for the numerical reconstruction of a function 
depending on two dimensionful parameters like
an external momentum $q$ and a mass $M$.
It constitutes a powerful combination of asymptotic expansion and
the analytic structure of the function to be approximated.

The basic idea is as follows.
Let us consider a Feynman diagram depending on an external momentum $q$
and one mass parameter, $M$, which can occur in some of the internal lines.
Then the final result is a function of $M^2/q^2$, which in general 
is quite involved and at three-loop order --- at least with the current
techniques --- not computable in an analytical form.
On the other hand, it is 
straightforward\footnote{With ``straightforward'' we mean that program
  packages exist which allow the computation 
  of the corresponding expressions with the help of computers.
  In this context see also the 
  Appendices~\ref{sub:single} and~\ref{sub:aut}.}
to evaluate the diagram in the limits $q^2\ll M^2$ and $q^2\gg M^2$.
The information from the different kinematical regions
is combined and
a semi-numerical function of $M^2/q^2$ is constructed. Below we will
demonstrate on typical examples that it
provides a very good approximation to the exact result.

In Subsection~\ref{sub:method}
we start with a detailed description of the method
and present explicit results for a two-loop example.
The physical processes discussed afterwards
in Sections~\ref{sub:veccor},~\ref{sub:mudec} and~\ref{sub:msos}
point out different aspects and fields of application.
In the first application we consider current correlators at three
loops both for a diagonal and a non-diagonal coupling to quarks.
In the latter case we assume that one of the quarks is massless.
The correlators depend on the external momentum $q$ and the quark mass
$M$. The main interest is in the imaginary part which represents a
physical observable.

Also the two-point function considered in Subsection~\ref{sub:msos}
--- proper combinations of the quark selfenergy ---
depends on an external momentum and one mass parameter. However,
the main interest is not on the functional behaviour but on the value
at threshold, i.e. $q^2=M^2$. 
The successful application of our method in that case is
not obvious, especially as only the information for $q^2\ll M^2$ and
$M^2\ll q^2$ are incorporated.

In Subsection~\ref{sub:mudec} four-loop integrals are computed in
order to obtain the order $\alpha^2$ QED corrections to the muon decay.
As we are only interested in the imaginary part they can be reduced to
three-loop integrals using asymptotic expansions.
From the technical point of view we want to obtain the value of a
function for $q^2=M^2$ using only expansion terms for $q^2\ll M^2$.
Also here the Pad\'e approximation turns out to be quite successful.
We want to mention already here that the example of the muon decay
does not fit completely into the philosophy which is developed in
Section~\ref{sub:method} as no high-energy
information can be incorporated. This leads to worse approximations.
Furthermore, in a first step the imaginary part is taken and
afterwards the Pad\'e approximation is applied.


\subsection{\label{sub:method}The Method}

The basic ingredients and tools 
of our method are moments of the function to be considered,
conformal mapping and Pad\'e approximation. 
In this Section we will explain the role of each of them.

The aim is to obtain 
an approximation to the function $f(z)$
which can not be computed directly.
The analytical properties of $f(z)$ are exploited
and expansions of $f(z)$ for small and large argument are used
in combination with a conformal mapping and
Pad\'e approximation.
In what follows we have in mind the computation of approximations to a 
physical function, like a vector boson self energy,
at a given loop-order\footnote{This is in contrast to the
  considerations of Ref.~\cite{padeals} where the Pad\'e approximation
  has been performed in the coupling constant in order to 
  estimate higher order terms in $\alpha_s$.}.
In particular, we assume that $f(z)$
is analytical for $z\to0$. However, in the limit $1/z\to0$ we allow
for a non-analytical behaviour.

In general a Pad\'e approximation of a function $f(z)$ is defined
through
\begin{eqnarray}
  [n/m](z) &=& \frac{a_0 + a_1 z + \ldots a_n z^n}
                    {1   + b_1 z + \ldots b_m z^m}
  \,,
  \label{eq:PAdef}
\end{eqnarray}
where the coefficients $a_i$ and $b_j$ are determined from the
requirement that the Taylor expansion of Eq.~(\ref{eq:PAdef})
coincides with the first $n+m+1$ terms of the
Taylor expansion of $f(z)$ 
around\footnote{In general also Taylor expansion around $z_0\not=0$
  can be considered. However, for our purpose the choice $z_0=0$
  is sufficient.}  
$z=0$.
Thus, in case the Taylor expansion is known up to terms of order $z^k$
Pad\'e approximations $[n/m]$ fulfilling
the condition
\begin{eqnarray}
  k - 1 &\ge& n+m
  \,,
\end{eqnarray}
can be computed.

In the approach discussed in the remaining part of this Section 
the considerations of the previous paragraph are improved with respect
to two points. First, we perform the Pad\'e approximation not only in $z$
but also in a new variable which is confined to the interior of the unit
circle and thus provides better convergence properties. 
Second, we want to include in the approximation for $f(z)$
not only Taylor coefficients around $z=0$ but also information from
other kinematical regions, in particular from $z\to\infty$.

Our procedure is restricted to the approximation of functions which
depend on one dimensionless variable, $z$. 
Concerning the physical 
applications we have in mind two-point functions where $z$
is given by the ratio of the squared external momentum and the square
of an internal mass. This motivates the following
form of $f(z)$ for small and large 
argument\footnote{It is advantageous to consider the space-like region
  of $z$ where no imaginary part occurs.}
\begin{eqnarray}
  f(z) &=& \left\{
  \begin{array}{ll}
    \displaystyle
    \sum_{k=1}^{n_{sma}} c_k z^k
    & \mbox{for $z\to0$}
    \\
    \displaystyle
    \sum_{k=0}^{n_{lar}} \sum_{i\ge0} d_{k,i} \left[\ln (-z)\right]^i z^{-k}
    & \mbox{for $z\to-\infty$}
  \end{array}
  \right.
  \,,
  \label{eq:deff}
\end{eqnarray}
where the normalization $f(0)=0$ has been chosen. 
As we will see below, in the computation of 
the coefficients $c_k$ and $d_{k,i}$
--- in the following also refered to as moments ---
one of the scales drops out and the integrals to be evaluated are much
simpler.
The moments $c_k$ and $d_{k,i}$ will serve as input for our procedure.
In addition we also admit information 
about the behaviour at the physical threshold, which we choose to be at
$z=1$, as input.

In order to obtain a semi-numerical approximation of the function $f(z)$
the following steps have to be performed:

\begin{enumerate}
\item
Compute as many moments as possible for small and large $z$.
As we require analyticity for $z\to0$ one gets in this limit
a simple Taylor
series of the Feynman diagrams in the external momentum.
The expansion in $q$ can be performed before the momentum integrations
are performed.
As a result the external momentum no longer appears
in the integrand and one ends up with vacuum diagrams. They are 
analytically known up to three-loop order in case of one internal mass
parameter (cf. Appendix~\ref{sub:single}).

However, for $z\to-\infty$ the rules of asymptotic 
expansion~\cite{Smi91} have to be applied. 
As a consequence the number of individual terms to be considered 
in the practical computation is larger. 
However, also here the number of scales in the individual
diagrams is reduced. One ends up with
either vacuum integrals or massless two-point functions.
The latter are responsible for the logarithmic terms in $z$.

\item
\label{item:thr}
Incorporate the information for $z\to1$ which we denote as $f^{thr}(z)$.
In the physical examples considered below
this information is either logarithmical, i.e. of the form
$\ln(1-z)$, or proportional to $1/\sqrt{1-z}$.
The latter occurs, for instance, in the abelian contribution to the
vector current correlator (cf. Section~\ref{sub:veccor})
and corresponds to the Coulomb singularity.
The further steps slightly depend on which case is present.

In case the leading threshold behaviour is logarithmic
one constructs $f^{thr}(z)$ in such a way that 
the singularity is reproduced for $z\to1$.
One has to take care that $f^{thr}(z)$ does not destroy the
behaviour of $f(z)$ for small and large $z$. In particular,
the expansion of $f^{thr}(z)$ has to be analytical for $z\to0$.
By construction the difference $f(z)-f^{thr}(z)$
is regular for $z\to1$ and has the same limiting behaviour as the one 
required in Eq.~(\ref{eq:deff}).
We should mention already here that 
due to the construction of the Pad\'e method the resulting function
has a vanishing imaginary part at
$z=1$. Thus it is crucial to implement the
leading threshold behaviour in this way.

Threshold singularities of the form $1/\sqrt{1-z}$ are not taken into
account at this step, i.e. formally $f^{thr}(z)=0$ is chosen. 
They are treated below. In contrast
to the logarithmic singularities they are removed
via multiplication and not by subtractions.

\item
\label{item:log}
Construct a function $f^{log}(z)$ in such a way that the
combination
\begin{eqnarray}
  \tilde{f}(z) &\equiv& f(z) - f^{thr}(z) - f^{log}(z)
  \,,
  \label{eq:ftilde}
\end{eqnarray}
is polynomial both in $z$ and $1/z$, i.e. in the small- and 
high-energy region. Furthermore no logarithmic singularities may be 
introduced for $z\to1$.

In this step a large part of information (e.g., the large high-energy
logarithms) which is known analytically is extracted and only a small
remainder 
$\tilde{f}(z)$ is left. It parameterizes the unknown part of $f(z)$.
 
\item
Perform a conformal mapping. The change of
variables~\cite{Fleischer:1994ef}
\begin{eqnarray}
  z &=& {4\omega\over (1+\omega)^2}
  \,,
  \label{eq:confmap}
\end{eqnarray}
maps the $z$ plane into the interior of the unit circle of the $\omega$
plane. Thereby the cut $[1,\infty)$ is mapped to the perimeter.
The conformal mapping is visualized in Fig.~\ref{fig:trafo}.

\begin{figure}[ht]
  \begin{center}
    \epsfxsize=12.0cm
    \leavevmode
    \epsffile[83 290 529 502]{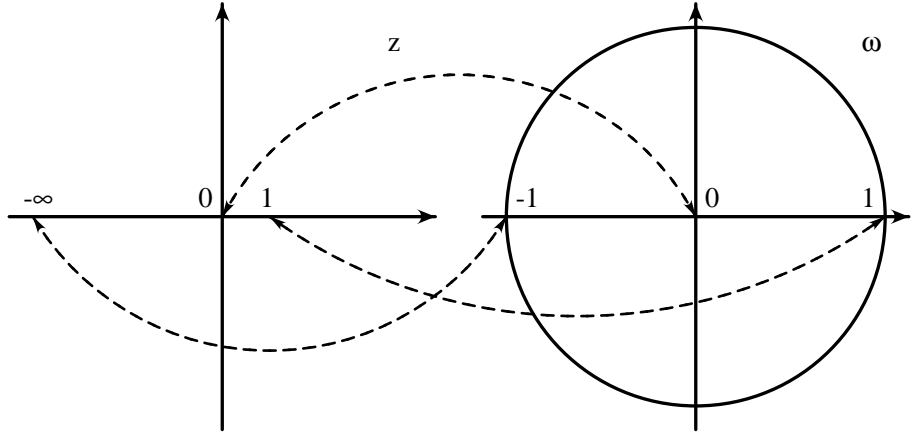}
    \caption[]{\label{fig:trafo}
      The conformal mapping (\ref{eq:confmap}) maps the $z$ plane into
    the interior of the unit circle in the $\omega$ plane.
      }
  \end{center}
\end{figure}

\item
In a next step a function is defined for which finally the Pad\'e
approximation is performed. Due to the discussion in the context of
Eq.~(\ref{eq:PAdef}) we are interested to shift the information
available for $\omega\to-1$
to $\omega\to0$. Furthermore we have to take care of possible
power-like threshold singularities.

Following Ref.~\cite{CheHarSte98} we define
\begin{eqnarray}
  P_{n_{lar}}(\omega) &=&  
  {p^{thr}(\omega)(4\omega)^{n_{lar}-1}\over (1+\omega)^{2n_{lar}}}\left(
  \tilde f(z) - 
  \sum_{j=0}^{n_{lar}-1}{1\over j!}\left(
  {d^j\over d(1/z)^j}\tilde f(z)\bigg|_{z =
  -\infty}\right) {(1+\omega)^{2j}\over (4\omega)^j}\right)
  \,,
  \nonumber\\
  \label{eq:Pnlar}
\end{eqnarray}
where for $\tilde{f}$ the moments up to order $1/z^{n_{lar}}$ must be known.
The function $p^{thr}(\omega)$ is equal to $1$ in case 
$f(z)$ has logarithmic divergences at threshold
and $p^{thr}(\omega)=(1-\omega)$ if $f(z)$ is proportional to 
$1/\sqrt{1-z}$ for $z\to1$.
The available information from the moments transforms into
$P_{n_{lar}}(-1)$ and $P_{n_{lar}}^{(k)}(0)$, 
$(k = 0,1,\ldots,n_{lar}+n_{sma}-1)$, 
where $n_{sma}$ is the number of moments for $z\to0$.
Whereas for a logarithmic threshold behaviour the corresponding
information is already
taken into account in step~\ref{item:thr}, the
$1/\sqrt{1-z}$ behaviour is treated with the factor $p^{thr}(\omega)$
and in addition $P_{n_{lar}}(\omega)$ is known for $\omega=1$.

\item
In the last step a Pad\'e approximation is performed 
for the function $P_{n_{lar}}(\omega)$. This means that 
$P_{n_{lar}}(\omega)$ is identified with a function
$[n/m](\omega)$ as defined in Eq.~(\ref{eq:PAdef}), where
the number of coefficients on the right-hand side depends on the amount of
information available for $P_{n_{lar}}(\omega)$.
In particular, one has $n+m = n_{lar}+n_{sma}+1$ if $P_{n_{lar}}(1)$
is available and otherwise $n+m = n_{lar}+n_{sma}$.
This leads to a system of (non-linear) equations which can be solved for
the coefficients $a_i$ and $b_j$ in Eq.~(\ref{eq:PAdef}).
For large values of $n+m$ the analytical solution becomes quite 
lengthy and time consuming. Thus it is preferable to solve the
equations numerically (using high precision).

\item
Finally, Eq.~(\ref{eq:Pnlar}) has to be solved for $\tilde{f}(z)$ and
from Eq.~(\ref{eq:ftilde}) an approximation for the function
$f(z)$ is obtained.

\end{enumerate}

Due to the structure of Eq.~(\ref{eq:PAdef}) 
some Pad\'e approximants develop poles inside the unit circle
($|\omega|\le1$). In general we will
discard such results as they would induce unphysical
poles in the $z$-plane.
In some cases, however, a
pole coincides with a zero of the numerator up to several digits
accuracy. These Pad\'e approximations 
will be taken into account in constructing our results.
If not stated otherwise we will, in addition
to the Pad\'e results without any poles inside the unit circle,
also use the ones where the poles are accompanied by zeros within a
circle of radius 0.01, and the distance between the pole and the
physically relevant point $q^2/M^2=1$ is larger than $0.1$.

There are situations where the information for $z\to-\infty$ 
can not be used as this would lead to physically 
not allowed scenarios (cf. Subsection~\ref{sub:mudec}).
In this case it is not necessary (and even not possible) to 
define the function $P_{n_{lar}}(\omega)$.
Instead one can directly perform 
a Pad\'e approximation either in $z$ or in $\omega$.

At this point we should spend some words on the estimation of the
errors to be assigned to the final results. It is difficult to provide
general rules for their determination as it very much depends on the problem
under consideration. Experience on the estimation of the error can be
gained from the comparison with known results at lower order or
for other colour structures. A reasonable choice for the 
systematic error due to the used method is to take the
spread of the individual Pad\'e results.

Concerning the above list some comments to point~\ref{item:log} are in
order. In principle there are many ways to subtract the high-energy
logarithms.
However, one has to keep in mind that the subtraction 
must not spoil
the polynomial behaviour for small $z$.
Furthermore, no divergences may be introduced. In particular, 
$f^{log}(z)$ has to be regular for $z=1$.

For the construction of $f^{log}(z)$ it is convenient to use the
function 
\begin{eqnarray}
  G(z) &=&
  \frac{2u\ln u}{u^2-1}
  \,,
\end{eqnarray}
with
\begin{eqnarray}
  u &=& \frac{\sqrt{1-\frac{1}{z}}-1}{\sqrt{1-\frac{1}{z}}+1}
  \,,
\end{eqnarray}
which naturally occurs in the result of the one-loop photon
polarization function, as a building block.
That this is possible in a systematic way can best be seen
by looking at the expansion of $G(z)$ in the different kinematical
regions
\begin{eqnarray}
  G(z) &=& \left\{
  \begin{array}{ll}
    \displaystyle
    1+\frac{2}{3}z+\frac{8}{15}z^2 + {\cal O}\left(z^3\right)
    &
    \displaystyle
    z\to0
    \,,
    \\
    \displaystyle
    \frac{1}{2z}\ln(-1/4z) + \frac{1}{4z^2}\left(1+\ln(-1/4z)\right)
    + {\cal O}\left(\frac{1}{z^3}\right)
    &
    \displaystyle
    z\to-\infty
    \,,
    \\
    \displaystyle
    \frac{\pi}{2\sqrt{1-z}} - 1 + {\cal O}\left(\sqrt{1-z}\right)
    &
    \displaystyle
    z\to1
    \,.
  \end{array}
  \right.
\end{eqnarray}
From this equation one can see that
$f^{log}(z)$ can be chosen as a linear combination of terms
$(1/z)^j(1-z)^l z^m (G(z))^n$ ($j,l,m,n\ge0$)
where the corresponding coefficients
are determined as follows:
\begin{itemize}
\item
  Consider the term in the second equality of Eq.~(\ref{eq:deff})
  which has the lowest value of $k$ and the highest value
  of $i\not=0$ and fix the index $n$
  such that the powers of the logarithms coincide.
\item
  Determine $l$ in such a way that there is no singular behaviour
  for $z\to1$. 
\item
  As $G(z)$ starts with order $1/z$ for $z\to-\infty$ one eventually
  has to correct for it with the help of the index $m$.
\item
  In a similar way the index $j$ is used in order to subtract the
  logarithms suppressed by higher powers in $1/z$.
\item
  Finally, terms involving $l=m=n=0$ and $j\ge0$ are added 
  in order to restore the behaviour for $z\to0$.
\item
  Repeat the procedure with the next values of $k$ and $i$, i.e. lower
  $i$ by one unit until $i=0$ is reached; then increase $k$ by one unit.
\end{itemize}
By construction this algorithm terminates once the linear
logarithm of the largest high-energy moment is treated.

To our knowledge a simple version of 
the method was first applied in~\cite{BaiBro95}
for the evaluation of certain four-loop contributions 
to the anomalous magnetic
moment of the muon. It was obtained by a convolution over the photon
polarization function. For the latter an approximation formulae
was obtained with the help of the Pad\'e-method. One should stress
that for this application only the integral over the space-like
momenta of the approximation was used. In the applications which
are discussed in this review the approximated function itself and in
particular its analyticity properties are of interest.
A brief introduction to the Pad\'e-method 
and the discussion of some results 
can also be found in~\cite{Harlander:2001sa}.


\subsubsection{Explicit example at two loops}

For clarity let us present an example where
in all steps explicit results are given.
We consider two-loop QCD corrections to the correlator
of two vector currents 
(for a precise definition see Eq.~(\ref{eqpivadef}) below).

\begin{enumerate}
\item
The expansion of the diagrams for small external momentum leads to
two-loop vacuum integrals which, e.g., can be computed with the help
of {\tt MATAD}~\cite{matad}.
After renormalizing the quark mass in the on-shell scheme and
subtracting the constant, the result
for the first three expansion terms reads
\begin{eqnarray}
  \Pi^{(1),v}(q^2) = \frac{3}{16\pi^2}\left(
    \frac{328}{81}z 
  + \frac{1796}{675} z^2 
  + \frac{999664}{496125} z^3
  + \ldots
  \right)
  \,. 
\end{eqnarray}
In the high-energy region we restrict ourselves to the
first two terms which can be obtained by solving 
massless integrals with the help of {\tt MINCER}~\cite{mincer}.
Using again the on-shell quark mass definition and
taking into account the condition $\Pi(0)=0$
gives
\begin{eqnarray}
 \Pi^{(1),v}(z) &=& \frac{3}{16\pi^2}\left( 
     \frac{5}{6}
   - 4\zeta_3
   - \ln(-4z) 
   - \frac{3}{z} \ln(-4z)
   + \ldots
   \right)
   \,.
\end{eqnarray}

\item
At threshold $\Pi(z)$ has a logarithmic singularity
which can be cast in the form
\begin{eqnarray}
  \Pi^{(1),v,thr}(z) &=& \frac{3}{16} \ln\left(\frac{1}{1-z}\right)
  \,.
\end{eqnarray}
Thus, the combination $\Pi(z)-\Pi^{thr}(z)$ is constant for $z=1$, has 
a polynomial behaviour for $z\to0$ and at most logarithmic
singularities for $z\to-\infty$.

\item
The high-energy logarithms are subtracted with the help of the
function
\begin{eqnarray}
  \Pi^{(1),v,log}(z) &=&
   \frac{3}{16\pi^2} 
   \frac{1}{3z}\Bigg[
   - 21 + z + \pi^2(3 + 5z) 
   \nonumber\\&&\mbox{}
   + 3(-1 + z)\left(-7 - 2z + \pi^2(1 + 2z)\right) G(z)
   \Bigg]
   \,,
\end{eqnarray}
which is constructed using the algorithm outlined above.
The resulting function
$\tilde{\Pi}^{(1),v}(z)=\Pi^{(1),v}(z)-\Pi^{(1),v,thr}(z)-\Pi^{(1),v,log}(z)$ 
is analytical for $z\to0$ and free of logarithms in the 
first two high-energy terms.
\end{enumerate}

After the conformal mapping (cf. Eq.~(\ref{eq:confmap})) and the $[2/2]$ 
Pad\'e approximation in $\omega$
are performed one obtains for $\tilde{\Pi}^{(1),v}(z)$
\begin{eqnarray}
  \tilde{\Pi}^{(1),v}(z) &=&
  -0.874397 + 
  \frac{(0.874397 + 0.905702\, \omega + 0.165184\, \omega^2 ) (1+\omega)^2}
       { 1 + 0.860764\, \omega + 0.069525\, \omega^2 }
   \,,
   \label{eq:pi2pade}
\end{eqnarray}
which finally 
leads to $\Pi^{(1),v}_{appr}(z) = \tilde{\Pi}^{(1),v}(z) 
  + \Pi^{(1),v,thr}(z) + \Pi^{(1),v,log}(z)$.
In Eq.~(\ref{eq:pi2pade}) the numbers are truncated. Usually 
high numerical precision is needed in order not to loose
significant digits in the final result.

By construction, $\Pi^{(1),v}_{appr}(z)$
has the same analyticity properties as the
exact function. As in this case the latter is known one can also
check the quality of the approximation.
It turns out that even for the relatively small amount of input used here
there is a perfect agreement between $\Pi^{(1),v}_{appr}(z)$ and
$\Pi^{(1),v}(z)$. E.g., it is not possible to 
distinguish the imaginary parts plotted in the range
$0 < 2m/\sqrt{s} < 1$~\cite{CheKueSte96,CheKueSte97}.


\subsection{\label{sub:veccor}Current correlators in QCD}

A variety of important observables can be described by the correlators
of two currents. If the
coupling of the currents to quarks is diagonal quantities like 
$e^+ e^-$ annihilation into hadrons and the decay of the $Z$ boson are
covered by the vector and axial-vector current correlators. 
Total decay rates of CP even or CP odd Higgs bosons can be obtained by
the scalar and pseudo-scalar current densities, respectively.
For these cases the full mass dependence at order $\alpha_s^2$ has
been computed in~\cite{CheKueSte96} for the non-singlet and
in~\cite{CheHarSte98} for the singlet correlators.
In Subsection~\ref{subsub:R} these results will be briefly reviewed.

On the other hand the correlators involving different quarks describe,
e.g., properties of a charged gauge or Higgs boson.
In particular a 
certain (gauge invariant) class of corrections to the 
single-top-quark
production via the process $q\bar{q}\to t\bar{b}$ becomes available.
As an application of the (pseudo-)scalar current correlator we want to
mention the decay of a charged Higgs boson, which occurs in extensions
of the SM, into a massive and a massless quark.
Another important application of the non-diagonal current correlator 
is connected to the meson decay constant. 
Within the heavy quark effective QCD it is related
to the spectral density, evaluated near threshold.  
The latter can be obtained from the correlator of the
full theory which is considered below.

Let us in a first step introduce the polarization functions for the four
cases of interest. 
The vector and axial-vector ($\delta=v,a$) correlators are defined as
\begin{eqnarray}
  \left(-q^2g_{\mu\nu}+q_\mu q_\nu\right)\,\Pi^\delta(q^2)
  +q_\mu q_\nu\,\Pi^\delta_L(q^2)
  &=&
  i\int {\rm d}x\,{\rm e}^{iqx}
  \langle 0|Tj^\delta_\mu(x) j^{\delta\dagger}_\nu(0)|0 \rangle
  \,,
  \label{eqpivadef}
\end{eqnarray}
and the scalar and pseudo-scalar ones ($\delta=s,p$) read
\begin{eqnarray}
  q^2\,\Pi^\delta(q^2)
  &=&
  i\int {\rm d}x\,{\rm e}^{iqx}
  \langle 0|Tj^\delta(x)j^{\delta\dagger}(0)|0 \rangle
  \,.
\label{eqpispdef}
\end{eqnarray}
The currents are given by
\begin{eqnarray}
  j_\mu^v = \bar{\psi}_1\gamma_\mu \psi_2,\qquad
  j_\mu^a = \bar{\psi}_1\gamma_\mu\gamma_5 \psi_2,\qquad
  j^s = \frac{m(\mu)}{M} \bar{\psi}_1\psi_2,\qquad
  j^p = i \frac{m(\mu)}{M} \bar{\psi}_1\gamma_5 \psi_2
  \,.
\label{eq:currents}
\end{eqnarray}
Here $m$ is the $\overline{\rm MS}$ and $M$ the on-shell quark mass.
In Eqs.~(\ref{eqpivadef}) and (\ref{eqpispdef}) two powers
of $q$ are factored out in order to end up with dimensionless
quantities $\Pi^\delta(q^2)$. 
As we are mainly interested in the imaginary part, the overall 
renormalization can be performed in such a way that this is 
possible.
Furthermore it is advantageous to adopt the QED-like 
renormalization $\Pi^\delta(0)=0$.

The physical observable $R(s)$ is related to $\Pi(q^2)$ by
\begin{eqnarray}
R^\delta (s)&=&12\pi\,\mbox{Im}\,\Pi^\delta(q^2=s+i\epsilon)
\qquad\qquad \mbox{for  } \delta=v,a\,,
\\
R^\delta (s)&=&8\pi\,\,\,\,\mbox{Im}\,\Pi^\delta(q^2=s+i\epsilon)
\qquad\qquad \mbox{for  } \delta=s,p\,.
\label{eq:rtopisp}
\end{eqnarray}
It is convenient to define 
\begin{eqnarray}
\Pi^\delta(q^2) &=& \Pi^{(0),\delta}(q^2) 
         + \frac{\alpha_s(\mu^2)}{\pi} C_F \Pi^{(1),\delta}(q^2)
         + \left(\frac{\alpha_s(\mu^2)}{\pi}\right)^2\Pi^{(2),\delta}(q^2)
         + \ldots\,,
\nonumber
\\
\Pi^{(2),\delta} &=&
                C_F^2       \Pi_A^{(2),\delta}
              + C_A C_F     \Pi_{\it NA}^{(2),\delta}
              + C_F T   n_l \Pi_l^{(2),\delta}
              + C_F T       \Pi_F^{(2),\delta}
              + C_F T       \Pi_S^{(2),\delta}\,,
\label{eqpi2}
\end{eqnarray}
and similarly for $R^\delta(s)$.
The abelian contribution $\Pi_A^{(2),\delta}$ is already present in 
(quenched) QED and $\Pi_{NA}^{(2),\delta}$ originates from the 
non-abelian structure
specific for QCD. The polarization functions containing a second
massless or massive quark loop are denoted 
by $\Pi_l^{(2),\delta}$ and $\Pi_F^{(2),\delta}$, respectively.
$\Pi_S^{(2),\delta}$ represents the double-triangle contribution.

Actually, we are mainly interested in the imaginary part $R^\delta(s)$ 
which in principle could be obtained from tree diagrams with five
external legs, from one-loop four-point integrals and from
two-loop three-point integrals.
However, in particular the latter can not be evaluated analytically
using current methods.
Also numerically the treatment of these integrals is inconvenient.
On the other hand, if one has to rely on approximations 
like small or large external momenta it is much more advantageous
to stick to two-point functions simply because the resulting 
integrals are easier to solve and the corresponding techniques are
much more advanced.


\subsubsection{\label{subsub:R}Diagonal current correlators}

The diagonal correlators with all their applications 
have extensively been discussed in the
literature.
Therefore we will be brief in this subsection and mainly
refer to the original literature where in most cases also the
analytical results can be found.

It is useful to define dimensionless variables
\begin{eqnarray}
  z\,\,=\,\,\frac{q^2}{4m^2},
  &&
  x\,\,=\,\,\frac{2m}{\sqrt{s}},
\end{eqnarray}
where $q$ is the external momentum of the polarization function
and $s$ corresponds to the center-of-mass energy in the 
process $e^+e^-\to\mbox{hadrons}$ or the mass of the boson in case
decay processes are considered.
Then the velocity, $v$, of one of the produced quarks reads
\begin{eqnarray}
  v&=&\sqrt{1-x^2}.
\end{eqnarray}
Every time the generic index $\delta$ appears without further explanation
it is understood that
$\delta$ represents one of the letters $a,v,s$ or $p$.

{\bf Vector and axial-vector correlators.}
The vector correlator certainly plays the most important role,
mostly because it covers the processes induced by the photon.
Already in 1979 the massless $\alpha_s^2$ corrections have been
evaluated~\cite{CheKatTka79DinSap79CelGon80} and roughly ten years
later even the order~$\alpha_s^3$ corrections became
available~\cite{GorKatLar91SurSam91,Che97_R}.
However, due to the impressive experimental precision 
the massless approximations are not sufficient for a reliable
comparison as we will see below in the case of the hadronic
contribution of $\Delta\alpha$ (cf. Section~\ref{subsub:delal}).
However,
a complete analytical computation of $\Pi^v(q^2)$
at three-loop order is currently not feasible.
In Ref.~\cite{CheKueSte96,CheKueSte97} 
the Pad\'e method described above has been
applied and semi-numerical results have been obtained.
At that time only the mass corrections of order 
$m^2/q^2$~\cite{GorKatLar86}, for which no asymptotic expansion
has to be applied, have been
available in the high-energy region. 
They have been combined with the 
terms up to order $(q^2/m^2)^8$~\cite{CheKueSte97} 
in the small-$q^2$ expansion
in order to get semi-analytical results for the individual colour
structures.

\begin{figure}[ht]
\begin{center}
\begin{tabular}{cc}
    \leavevmode
    \epsfxsize=5.5cm
    \epsffile[110 265 465 560]{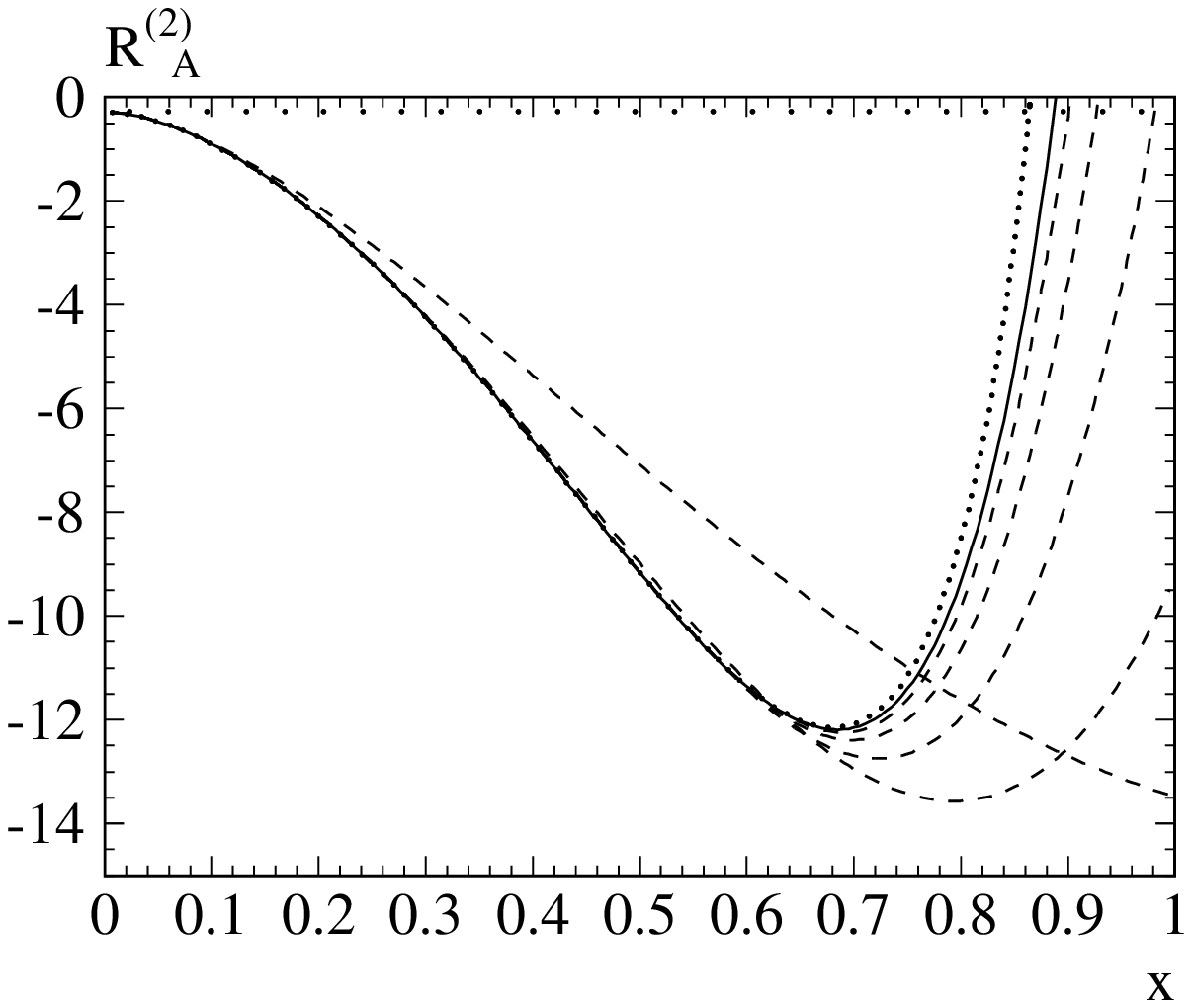}&
    \epsfxsize=5.5cm
    \epsffile[110 265 465 560]{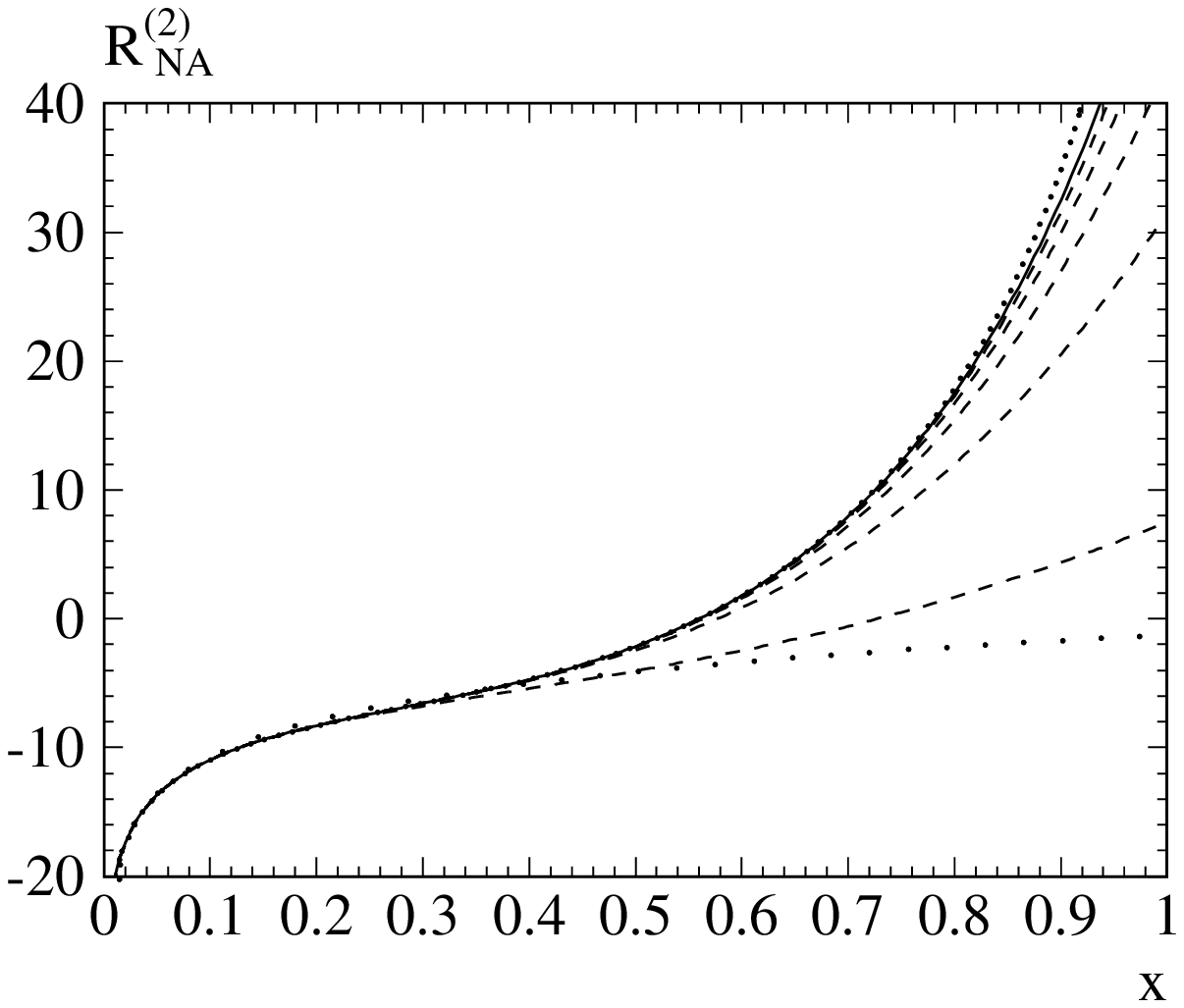}\\
    \epsfxsize=5.5cm
    \epsffile[110 265 465 560]{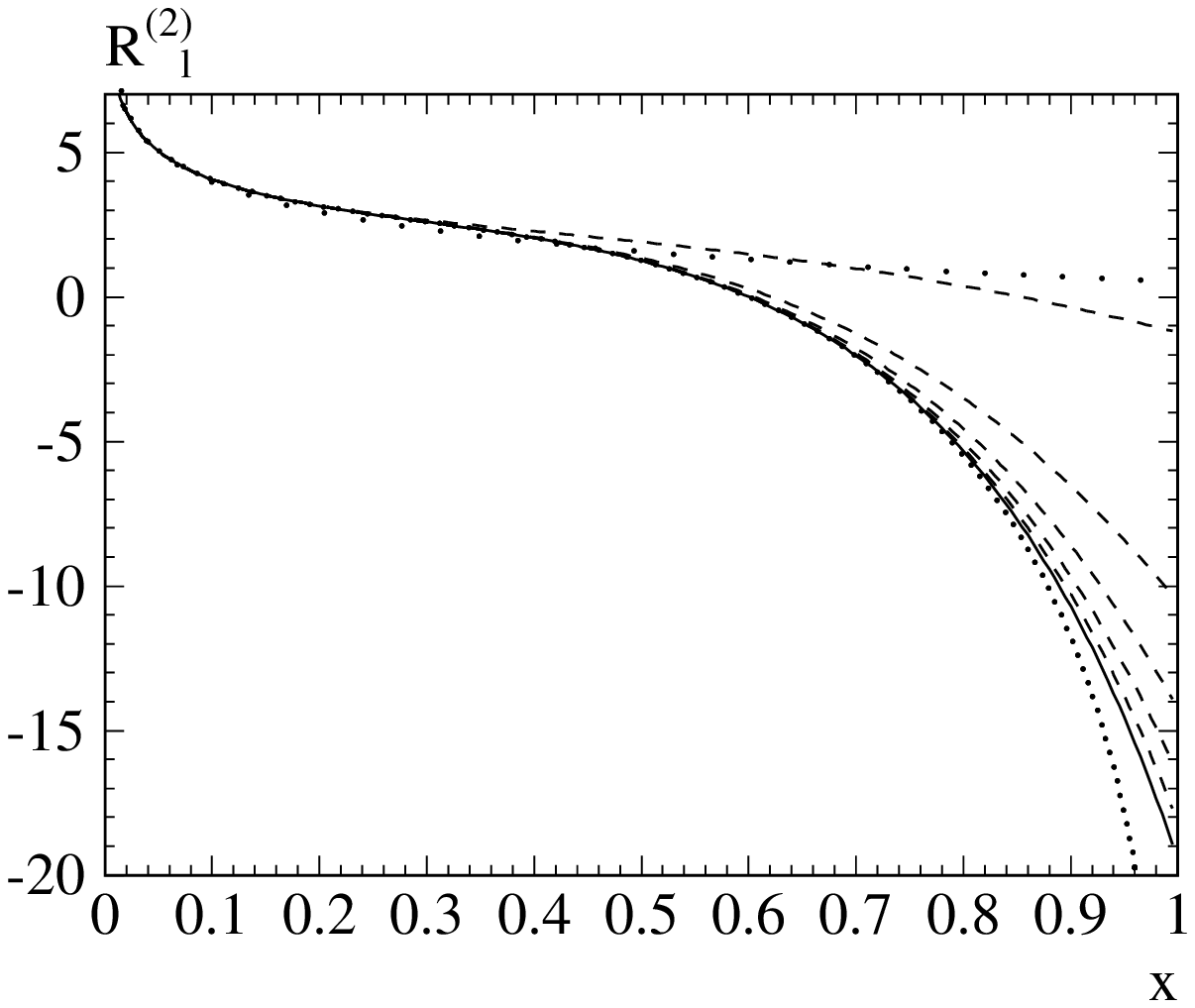}&
    \epsfxsize=5.5cm
    \epsffile[110 265 465 560]{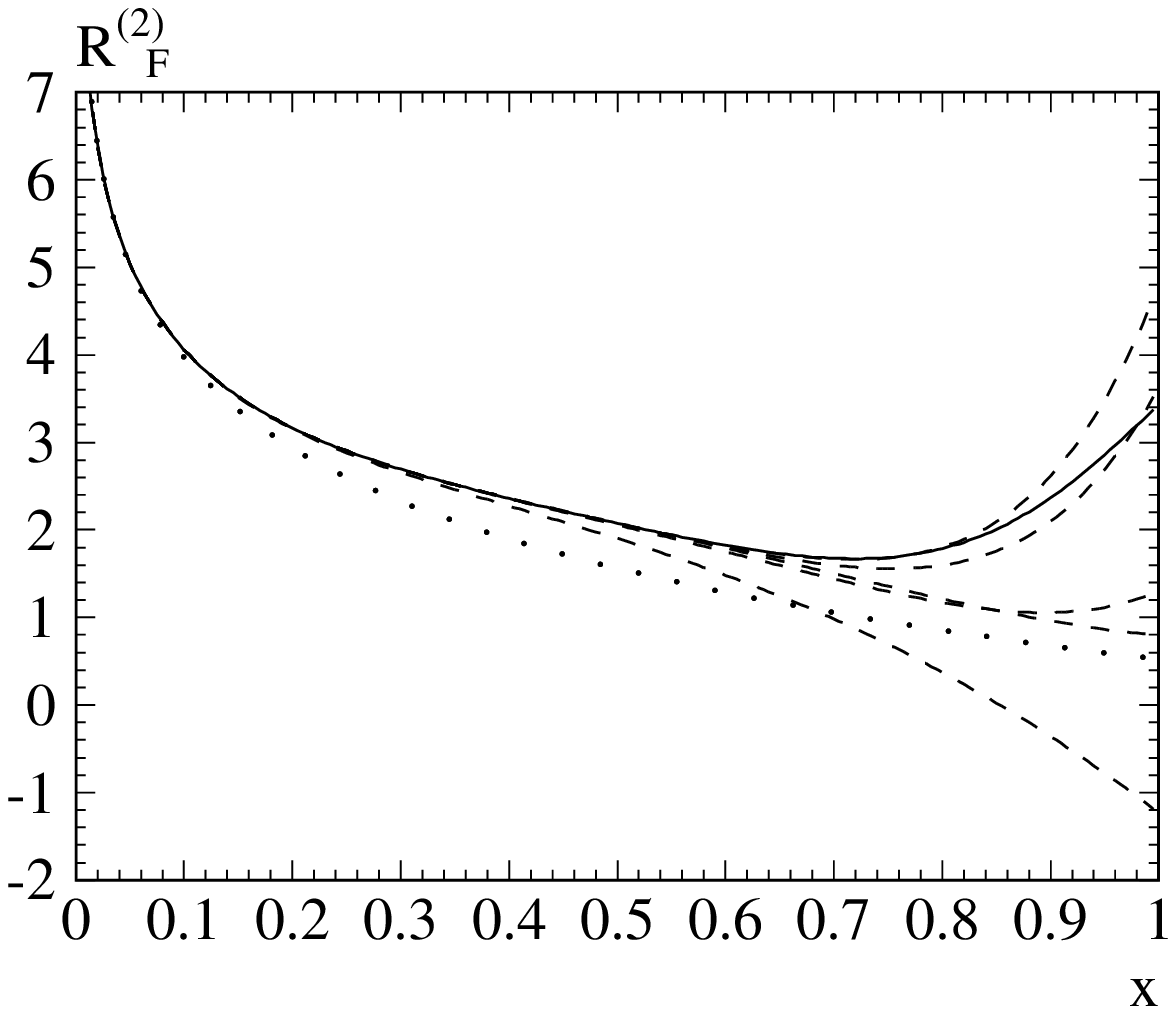}
\end{tabular}
    \caption[]{\label{fig:rvlar}\sloppy\rm
      The abelian contribution $R_A^{(2),v}$, 
      the non-abelian piece $R_{\it NA}^{(2),v}$,
      the contribution from light internal quark loops $R_l^{(2),v}$ 
      and the contribution $R_F^{(2),v}$ from the double-bubble diagram 
      with the heavy fermion in both the inner and outer loop
      as functions of $x = 2m/\sqrt{s}$.
      Wide dots: no mass terms; 
      dashed lines: including mass terms $(m^2/s)^n$ up to $n=5$; 
      solid line: including mass terms up to $(m^2/s)^6$;
      narrow dots: semi-analytical result (except for $R_F^{(2),v}$). The
      scale $\mu^2 = m^2$ has been adopted.
      }
\end{center}
\end{figure}

As the main interest is in the imaginary part, one could also adopt the
attitude to compute as many terms in the high-energy expansion as
possible. Going, however, beyond the $m^2$ terms a naive expansion
fails and the large-momentum
procedure has to be applied. The calculation becomes very cumbersome 
if it has to be applied by hand.
For this reason the large-momentum procedure has been automated
and the program ${\tt lmp}$~\cite{Har:diss} has been developed.
As a first application correction terms up to order $(m^2/q^2)^6$
have been evaluated~\cite{CheHarKueSte96,CheHarKueSte97} 
for the vector correlator.
In Fig.~\ref{fig:rvlar} the comparison of the individual expansion
terms with the Pad\'e result from Ref.~\cite{CheKueSte96} is shown.

For all three functions $R^{(2),v}_{A}$, $R^{(2),v}_{\it NA}$ and
$R^{(2),v}_l$  
and values between
$x=0$ and $x=0.6$ ($x=2m/\sqrt{s}$)
the expansions including terms of order $(m^2/s)^3$ (or
more) are in perfect agreement with the semi-analytical Pad\'e result.
Conversely this provides a completely independent test of the method of
\cite{CheKueSte96} which did rely mainly on low energy information. Including
more terms in the expansion, one obtains an improved
approximation even in the low energy region. However, the quality of the
``convergence'' is significantly better for $R^{(2),v}_l$ and
$R^{(2),v}_{\it NA}$ 
than for $R^{(2),v}_A$. Two reasons may be responsible for this difference: 
{\rm (i)} In a high energy expansion it is presumably more difficult to
approximate the $1/v$ Coulomb singularity in $R^{(2),v}_A$ than the mild $\ln
v$ singularity in $R^{(2),v}_{\it NA}$ and $R^{(2),v}_l$.
{\rm (ii)} The function $R^{(2),v}_l$ can be approximated in the whole
energy region 
$2m<\sqrt{s}<\infty$ by an increasing number of terms with arbitrary
accuracy. This is evident from the known analytical form of 
$R^{(2),v}_l$~\cite{HoaKueTeu951}, a
consequence of the absence of thresholds above $2m$ in this piece. In
contrast the functions $R^{(2),v}_A$, $R^{(2),v}_{\it NA}$ 
and $R^{(2),v}_F$ exhibit a four particle
threshold at $\sqrt{s} = 4m$. The high energy expansion is, therefore,
not expected to converge to the correct answer in the interval between
$2m$ and $4m$.
In particular, in the case of $R^{(2),v}_F$ it can be seen that for $x>0.5$
no convergence is observed.

\begin{figure}[ht]
\begin{center}
\begin{tabular}{cc}
    \leavevmode
    \epsfxsize=5.5cm
    \epsffile[110 265 465 560]{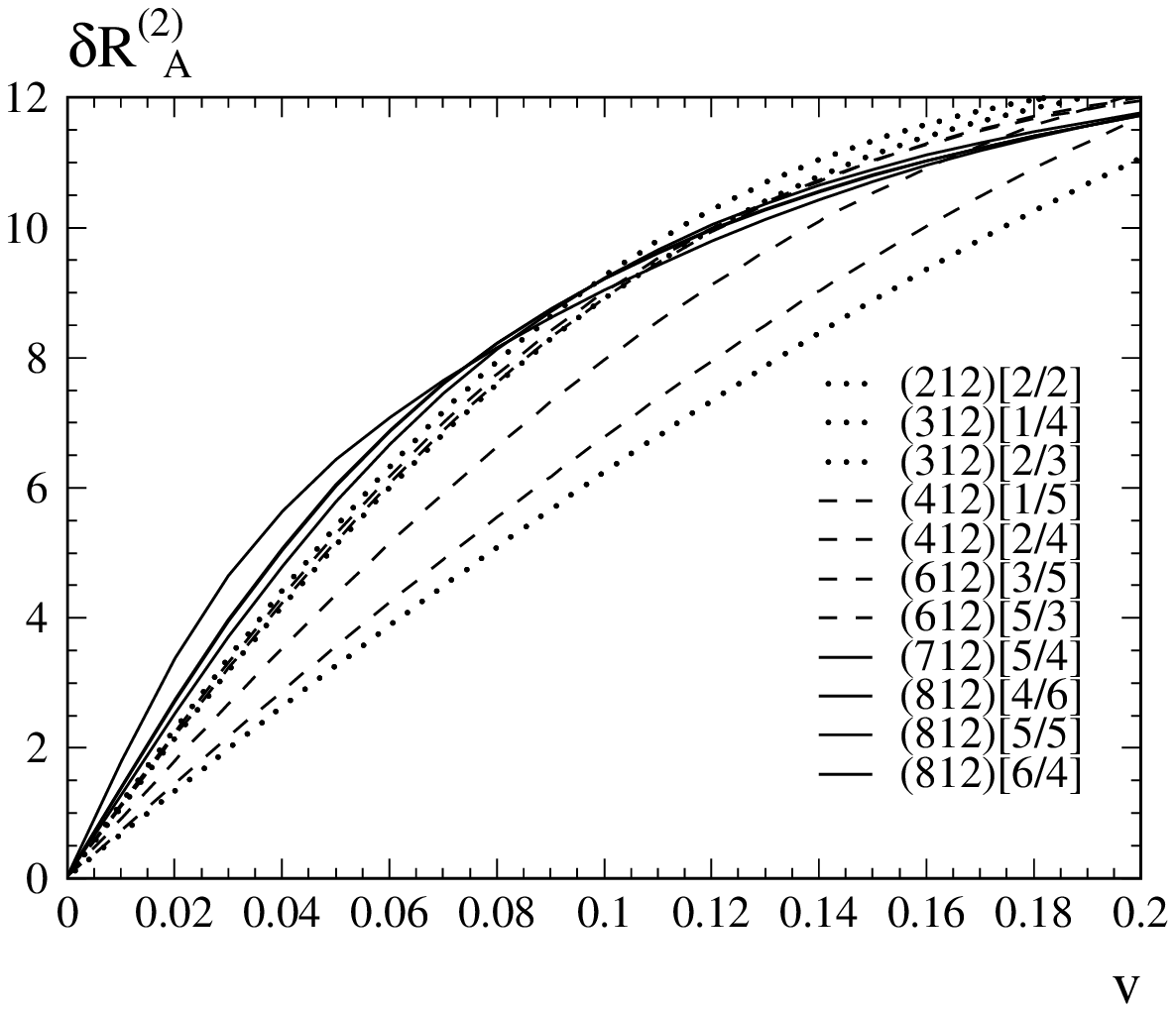}&
    \epsfxsize=5.5cm
    \epsffile[110 265 465 560]{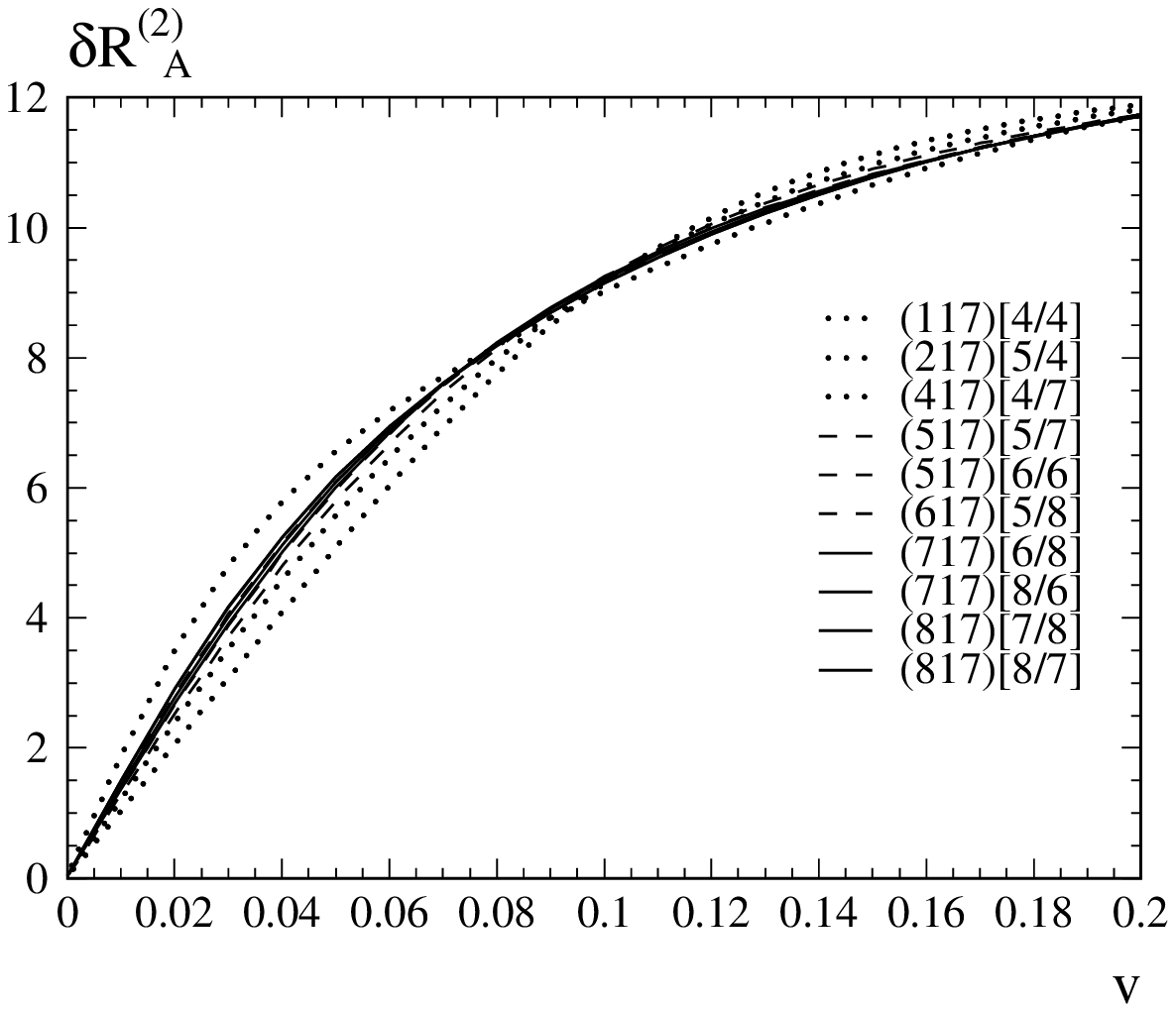}\\
    (a) & (b) \\
    \epsfxsize=5.5cm
    \epsffile[110 265 465 560]{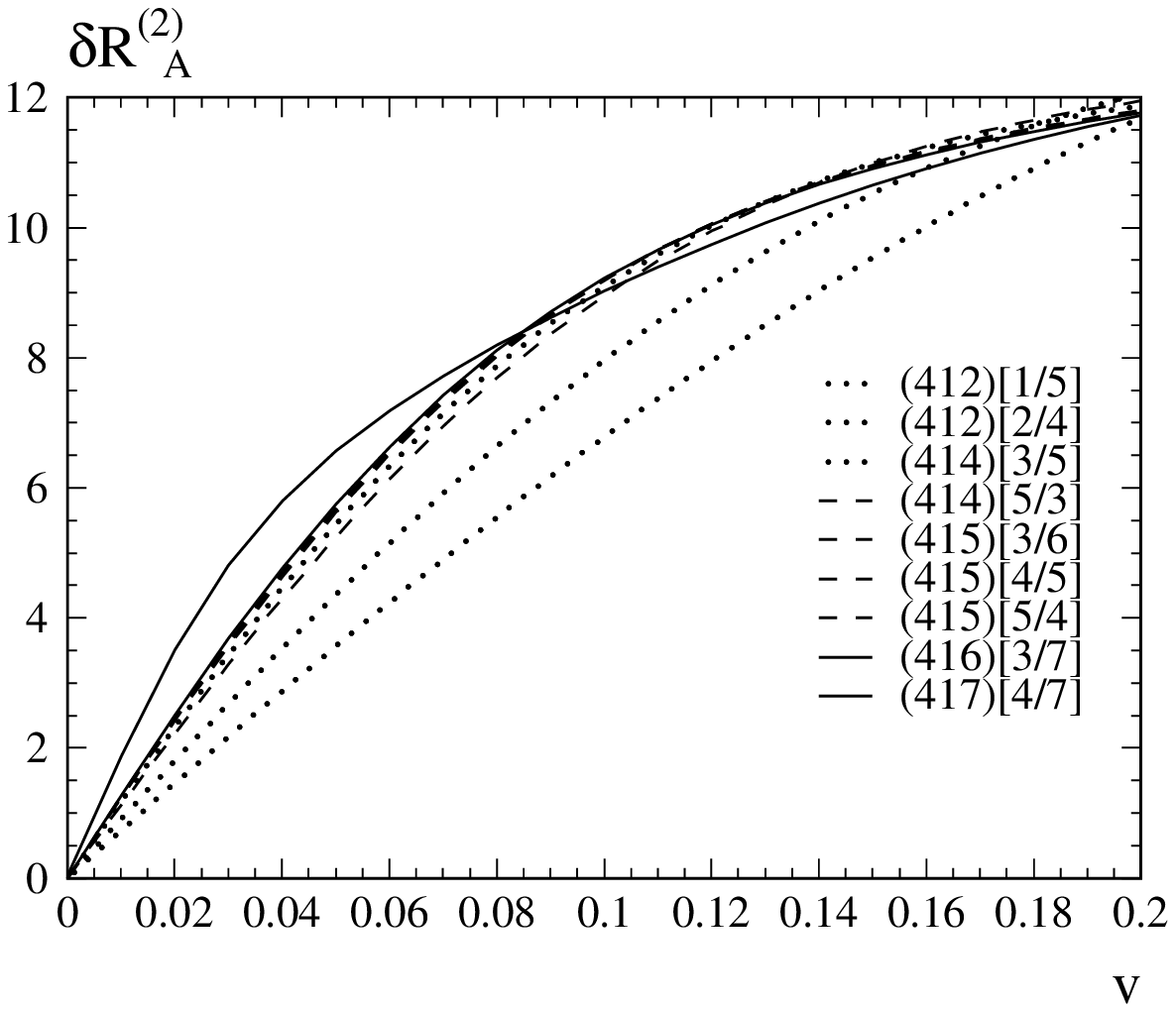}&
    \epsfxsize=5.5cm
    \epsffile[110 265 465 560]{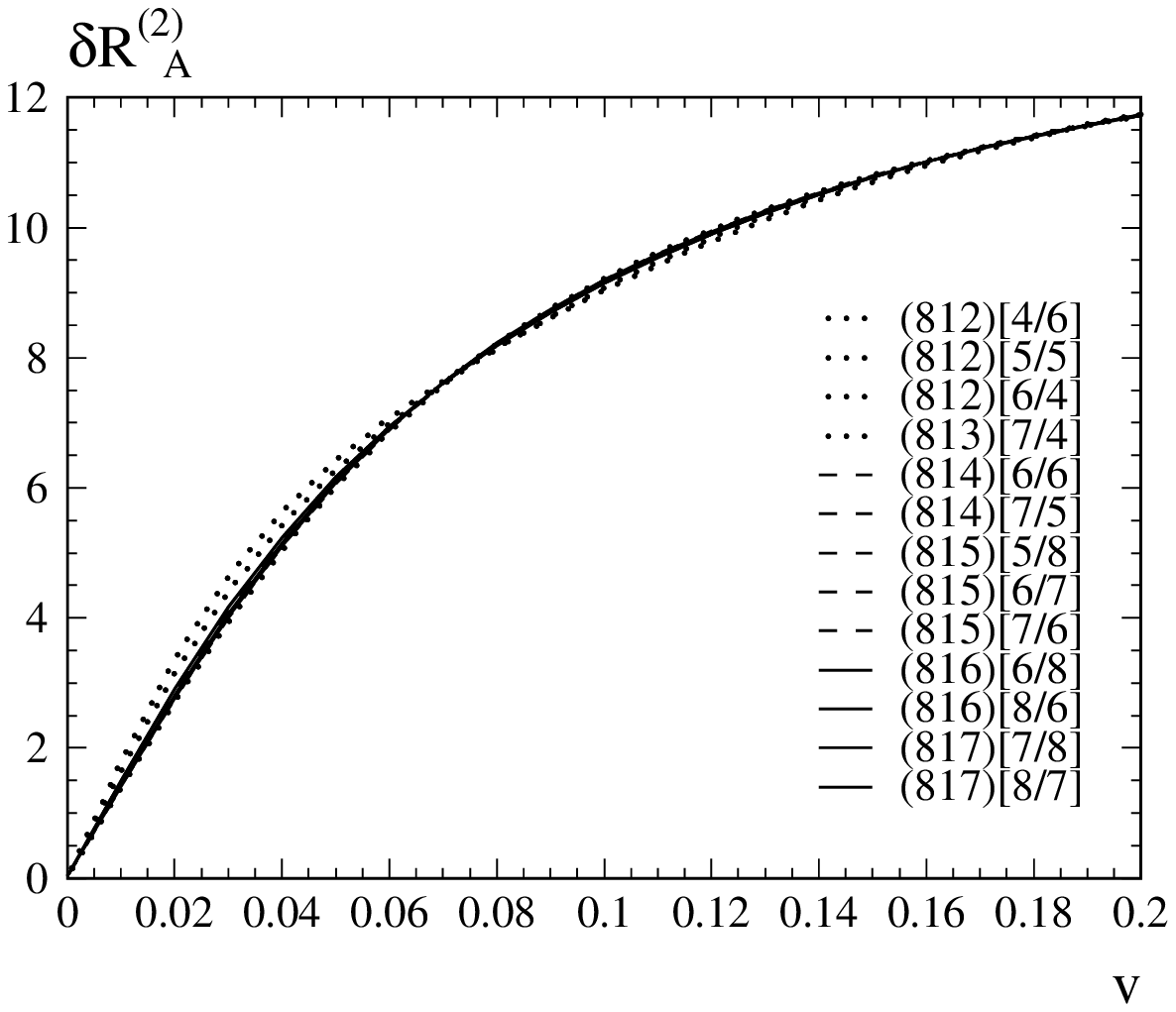}\\
    (c) & (d)
\end{tabular}
    \caption[]{\label{fig:padehigh1}Abelian part of the vector
      correlator as a function of $v$. The leading threshold term is
      subtracted as described in the text. In (a) and (b) the number
      of low-energy moments are varied and the high-energy terms are
      fixed to two and seven, respectively. In (c) and (d) the
      low-energy moments are fixed to four and eight, respectively, and
      the high-energy terms are varied. The notation $(l1k)[i/j]$
      means that terms up to order $z^l$ and $1/z^{k-1}$ are taken
      into account in order to construct the Pad\'e approximation $[i/j]$.
      }
\end{center}
\end{figure}

At this point it is tempting to combine both approaches 
--- asymptotic expansion in the high-energy region and the Pad\'e
method --- and evaluate high-order Pad\'e approximants.
A detailed study in the case of the vector correlator can be found
in~\cite{Har:diss} from which Fig.~\ref{fig:padehigh1}
is taken. In Fig.~\ref{fig:padehigh1}
the influence of the number of 
low- and high-energy input data is studied. 
In all four plots the quantity
$\delta R_A^{(2)}=R_A^{(2),v}-3(\pi^4/(8v)-3\pi^2)$
is shown, i.e. the leading threshold term is subtracted. Otherwise it
would not be possible to detect any difference between the individual
Pad\'e results.
Moreover, the abscissa only extends to $v=0.2$ as for $v\gsim0.5$
all curves coincide.
In Figs.~\ref{fig:padehigh1}(a) and~\ref{fig:padehigh1}(b)
the higher-energy terms are fixed to two and seven, respectively,
whereas an increasing number of low-energy moments are considered.
On the other hand, on the plots in the lower row the number of moments
is fixed to four and eight, respectively, with varying high-energy input.

From Figs.~\ref{fig:padehigh1}(a) and~\ref{fig:padehigh1}(c)
a clear stablilization of the results 
can be observed with increasing degree of the
Pad\'e approximation. 
The same is true for Figs.~\ref{fig:padehigh1}(b)
and~\ref{fig:padehigh1}(d): although the degree of the Pad\'e
approximants is higher from the very beginning a further stabilization
is visible.

From these considerations one can conclude that both the small- and
high-energy expansion terms are crucial as input for the Pad\'e
procedure. A significant stabilization of $R(s)$ in the
threshold region is observed
if more terms are taken into account.
However, for practical purposes it is probably more than enough
to consider, e.g., only the quadratic terms in the high-energy region.
The reason for this is that also the leading threshold behaviour is
incorporated into the analysis. 
The situation is different in those cases where no information about
the threshold is available or one even wants to determine the 
value of the (real part of the) considered function for $z=1$.
We will come back to this point in Section~\ref{sub:msos}.

At this point we refrain from listing the results of the individual
terms contributing to $R(s)$ as all of them are available in the
literature. In particular we want to refer to the Appendix of
Ref.~\cite{CheKueKwiPR} where detailed results 
up to order $\alpha_s^2 m^4/s^2$ and $\alpha_s^3 m^2/s$ are listed.
Concerning the full mass dependence at order $\alpha_s^2$
a complete discussion and a detailed
compilation of the individual terms in the on-shell scheme
can be found in Ref.~\cite{CheHoaKueSteTeu97} (see
also~\cite{Chetyrkin:1996yp}).
Together with the quartic on-shell terms at order $\alpha_s^3$
given in Eq.~(\ref{eq:Rm4OS})
this constitutes the current state-of-the-art radiative corrections
for $R(s)$.

In~\cite{CheKueSte97} next to the vector case also the axial-vector,
scalar and pseudo-scalar correlators have been considered.
Moments up to order $(q^2)^8$ have been combined with quadratic mass
terms and the leading threshold behaviour in order to obtain 
semi-numerical approximations for the correlators.

\begin{figure}[ht]
 \begin{center}
 \begin{tabular}{cc}
   \leavevmode
   \epsfxsize=6.5cm
   \epsffile[110 270 480 560]{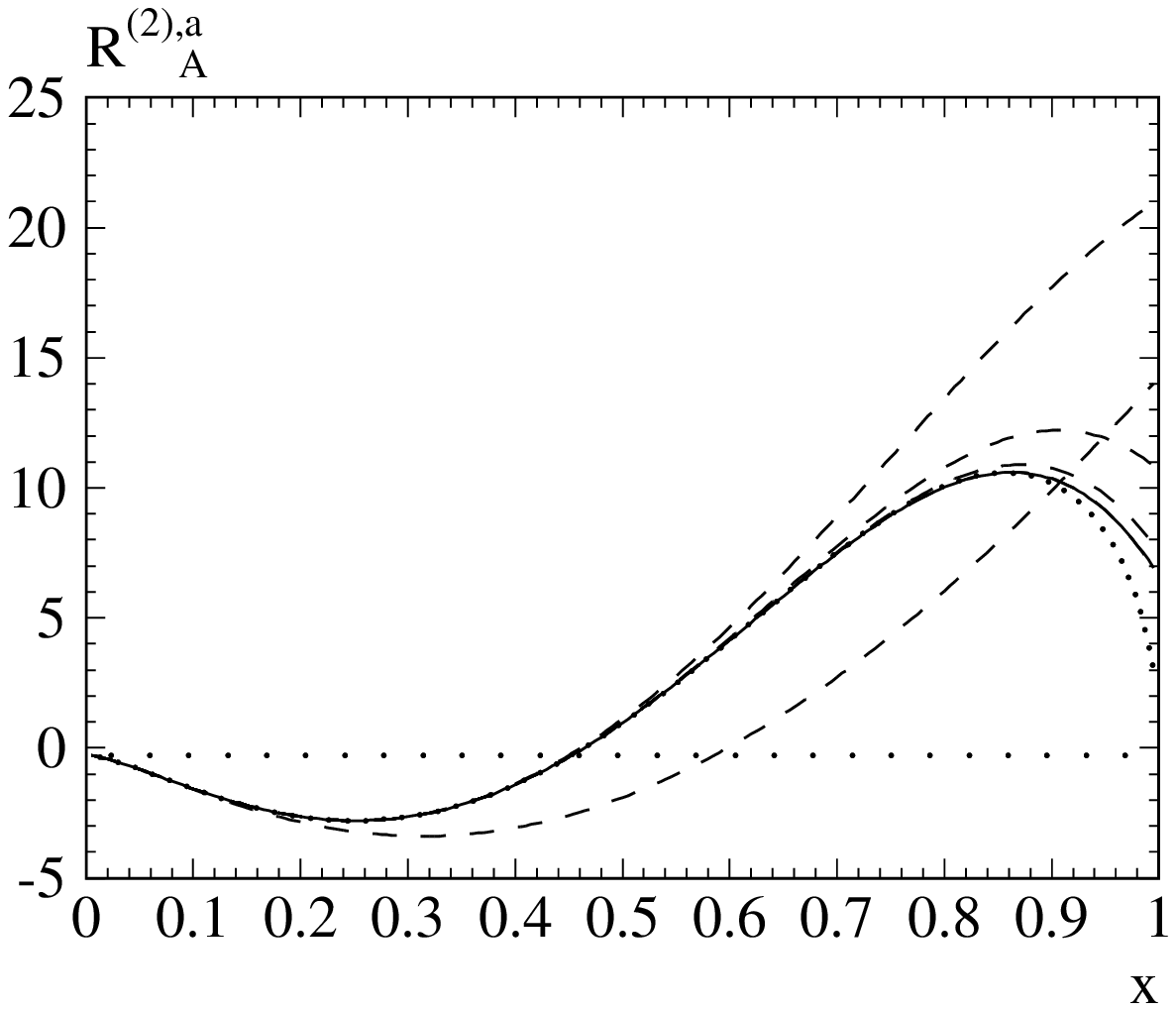}
   &
   \epsfxsize=6.5cm
   \epsffile[110 270 480 560]{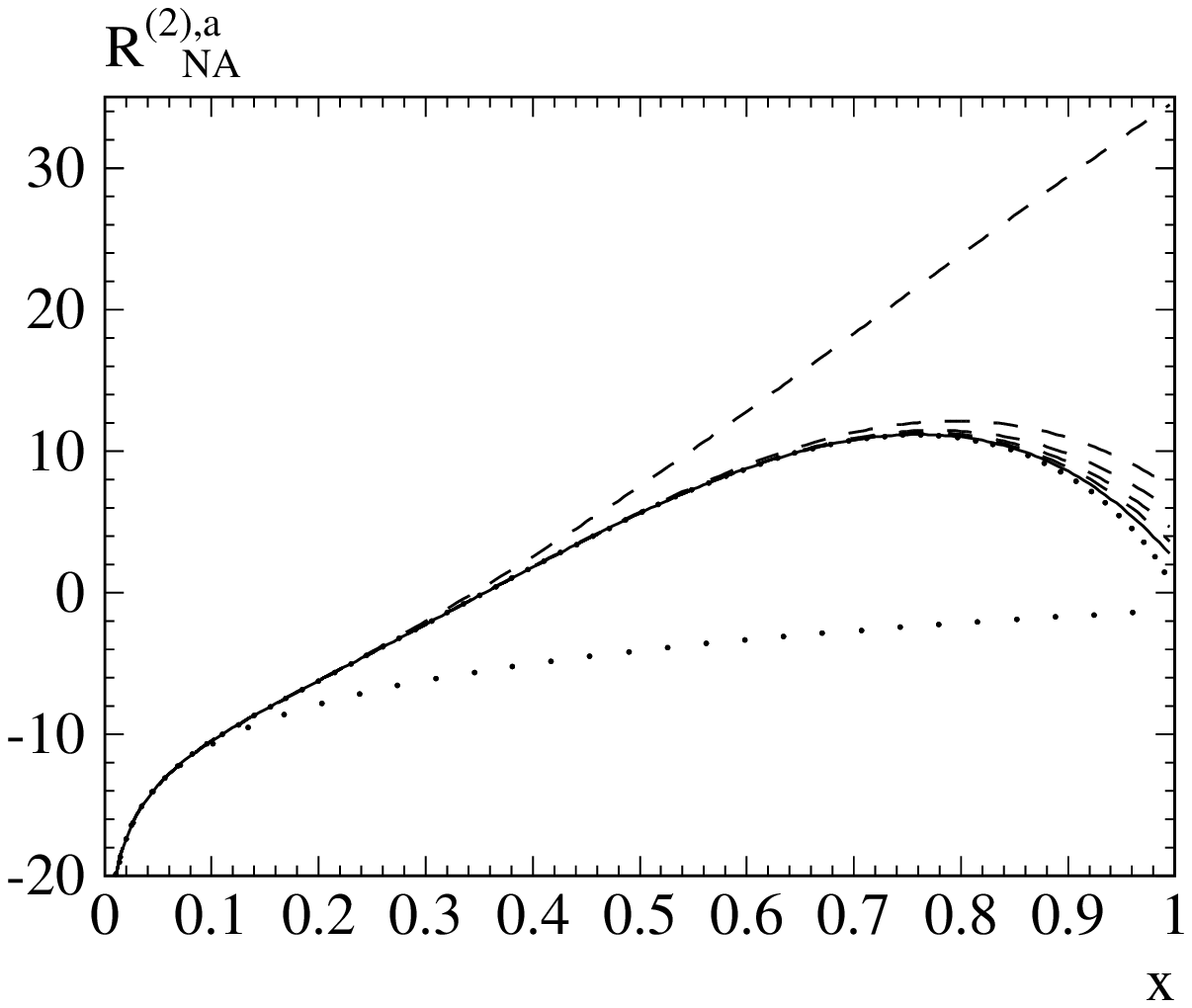} 
\\
   \epsfxsize=6.5cm
   \epsffile[110 270 480 560]{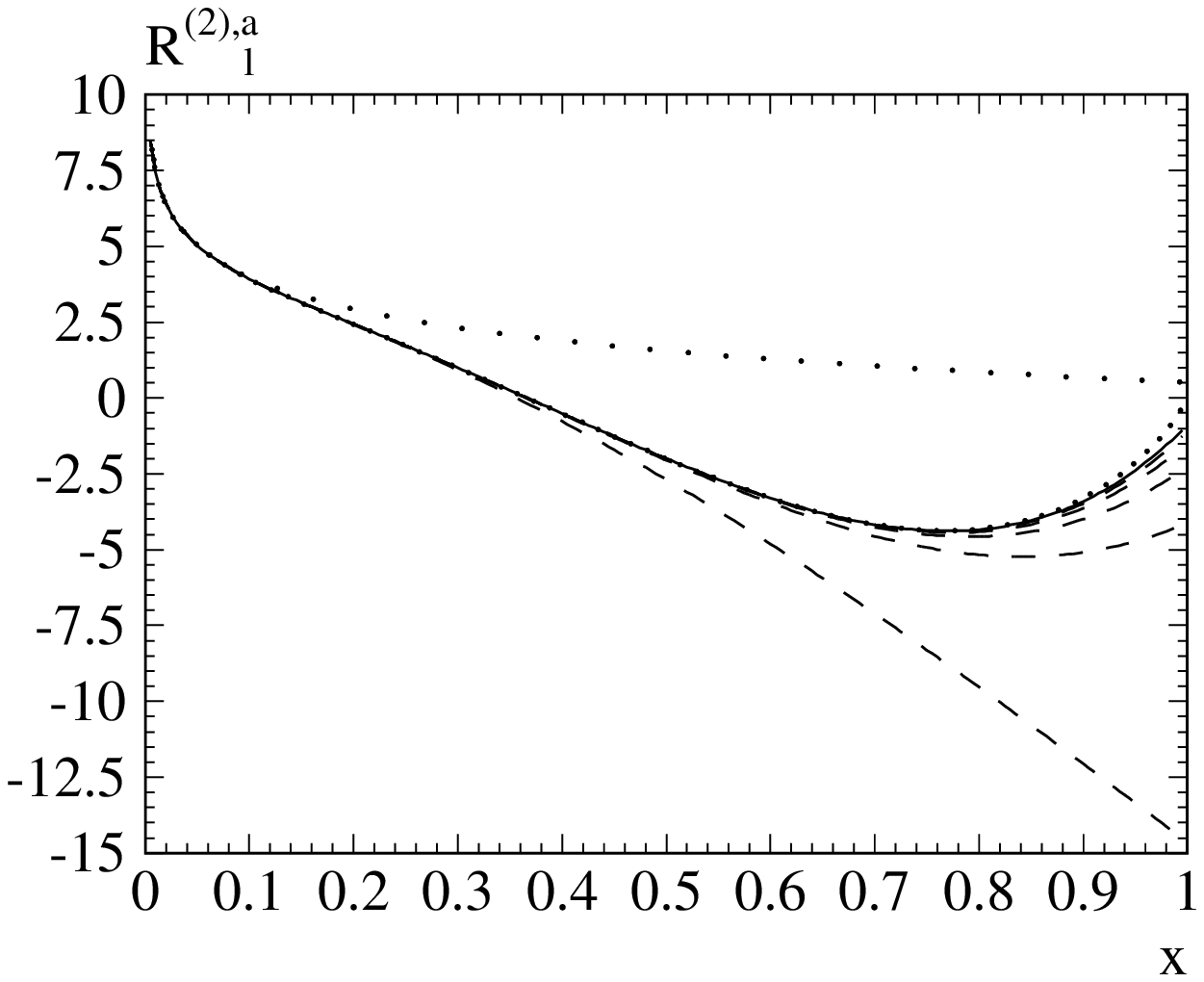}
   &
   \epsfxsize=6.5cm
   \epsffile[110 270 480 560]{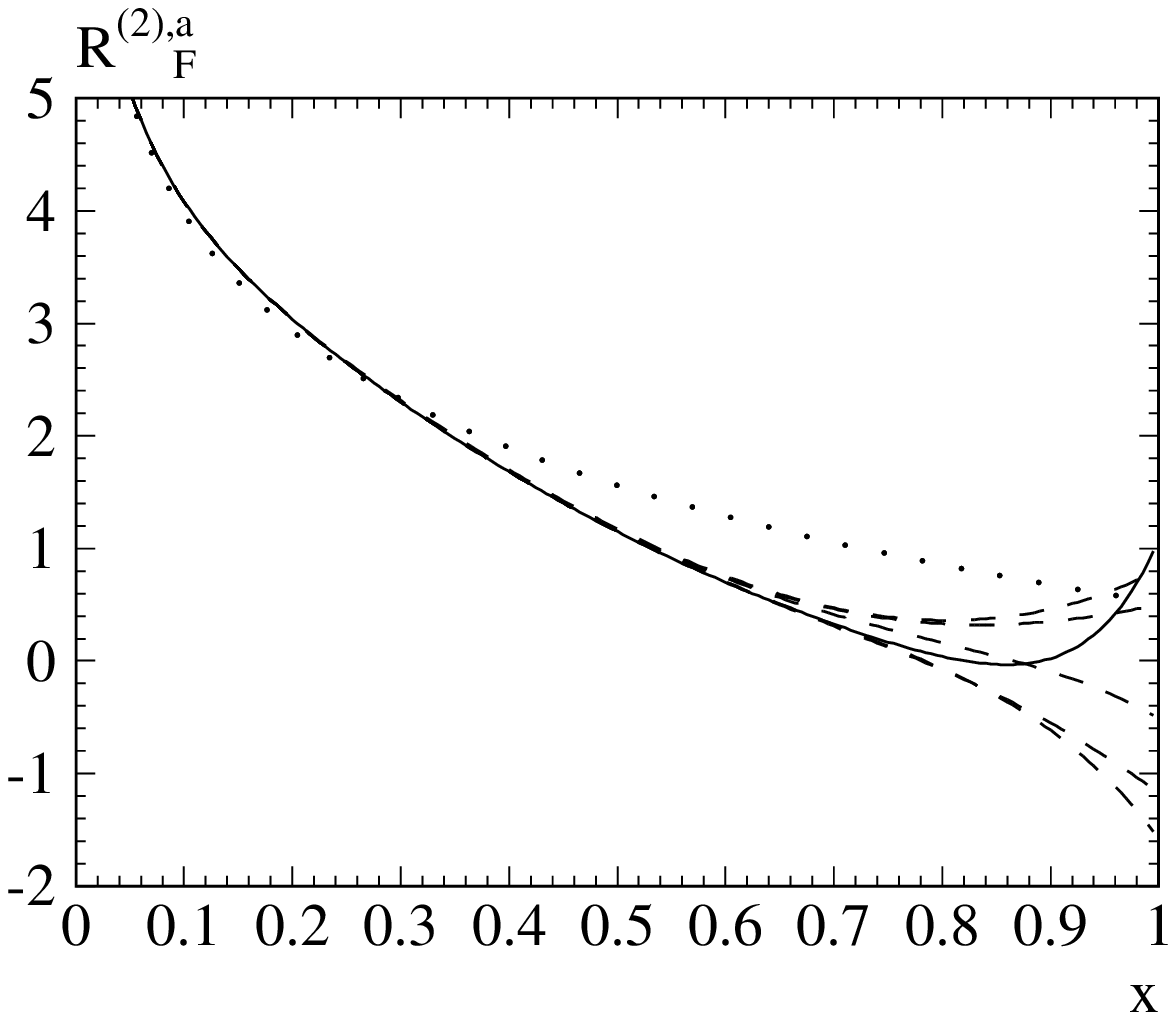}
\\
   \epsfxsize=6.5cm
   \epsffile[110 270 480 560]{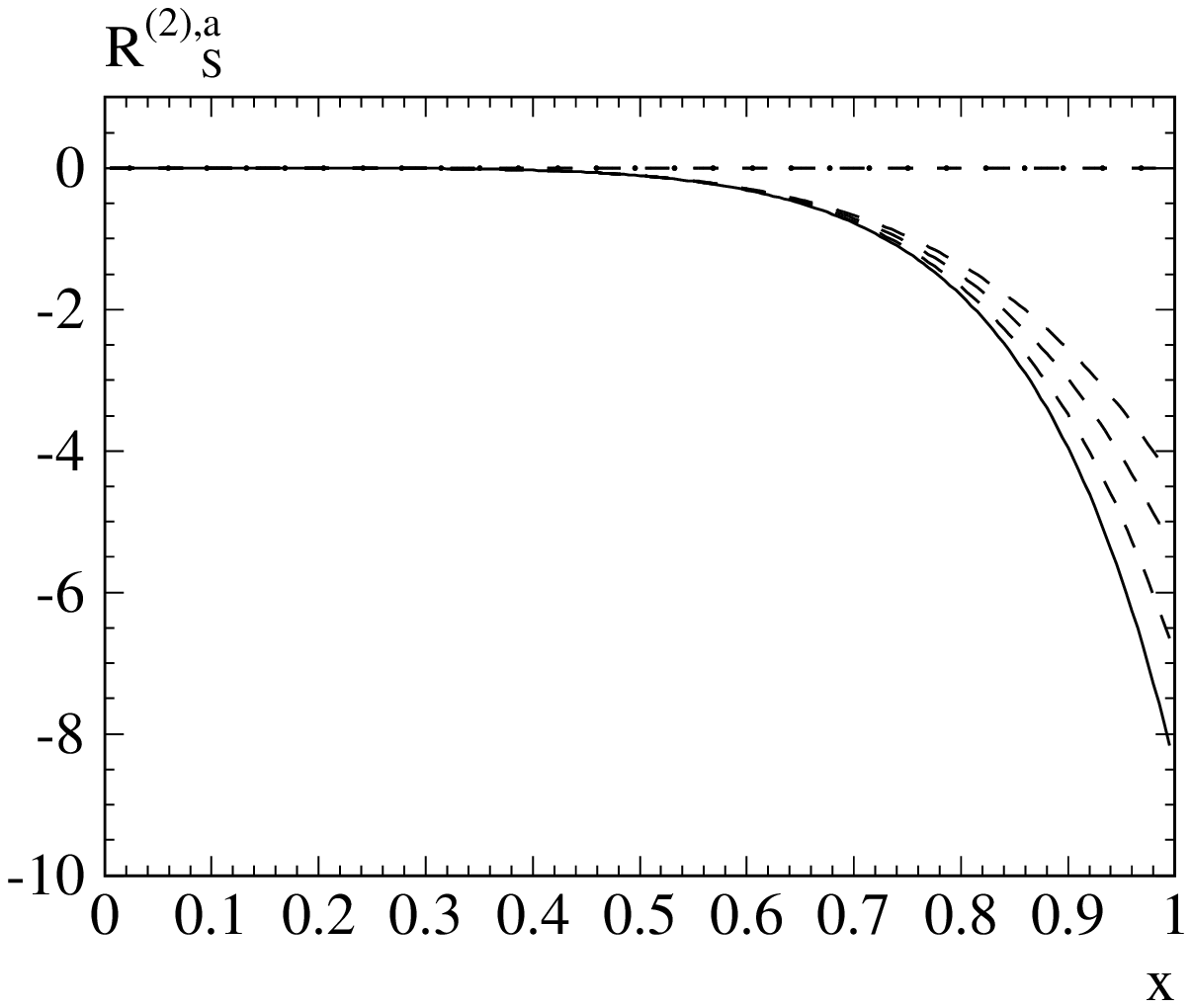}
   &
   \epsfxsize=6.5cm
   \epsffile[110 270 480 560]{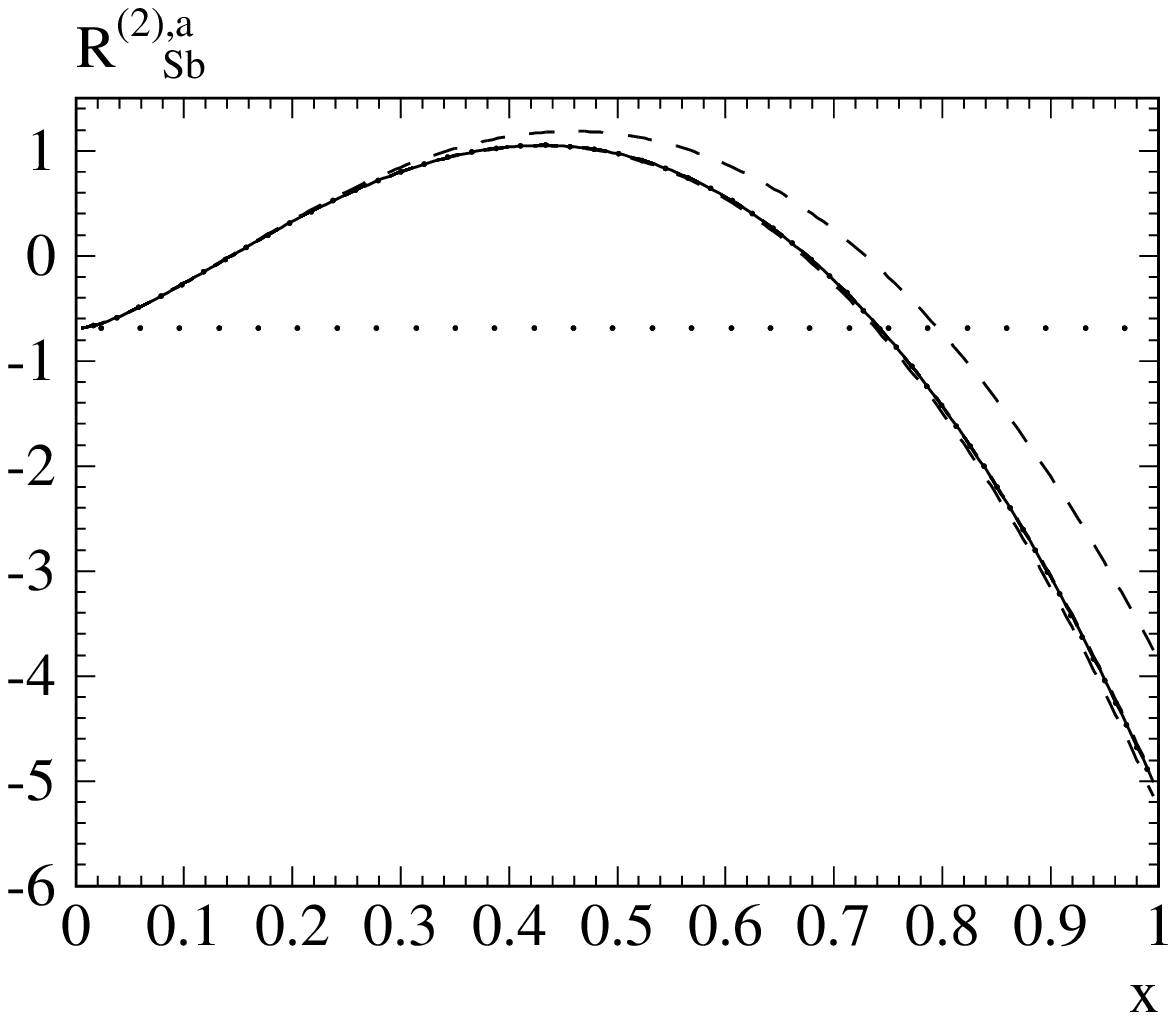}
 \end{tabular}
 \caption{\label{fig:ra} 
   $R^{(2),a}_i$, $i={\it A, NA, l, F, S, Sb}$ as functions of $x =
   2M_t/\protect\sqrt{s}$ at $\mu^2 = M_t^2$. Successively higher order
   terms in $(M_t^2/s)^n$ are included: 
   Dotted: $n=0$; dashed: $n=1,\ldots,5$; solid:
   $n=6$. Narrow dots: exact result ($R_l^{(2),a}$, $R_{Sb}^{(2),a}$) or 
   semi-analytical results ($R_A^{(2),a}$, $R_{NA}^{(2),a}$).
   In the case of the top-bottom doublet $R_{Sb}$ contains only the
 imaginary parts of the singlet diagrams which arise from the gluon
 and bottom quark cuts (see also Fig.~\ref{fig:pademxfig}).}
 \end{center}
\end{figure}

QCD corrections to the axial-vector correlator have also been
evaluated in~\cite{HarSte98}. In the limit
$q^2\gg M^2$ the first seven terms could be evaluated using automated 
asymptotic expansion. In Fig.~\ref{fig:ra} the 
expansion terms are compared with the Pad\'e results~\cite{CheKueSte97}
for the individual colour factors.
Again perfect agreement is found in the region where the 
asymptotic expansion is expected to converge to the exact result.

\begin{figure}[ht]
  \includegraphics[width=\linewidth]{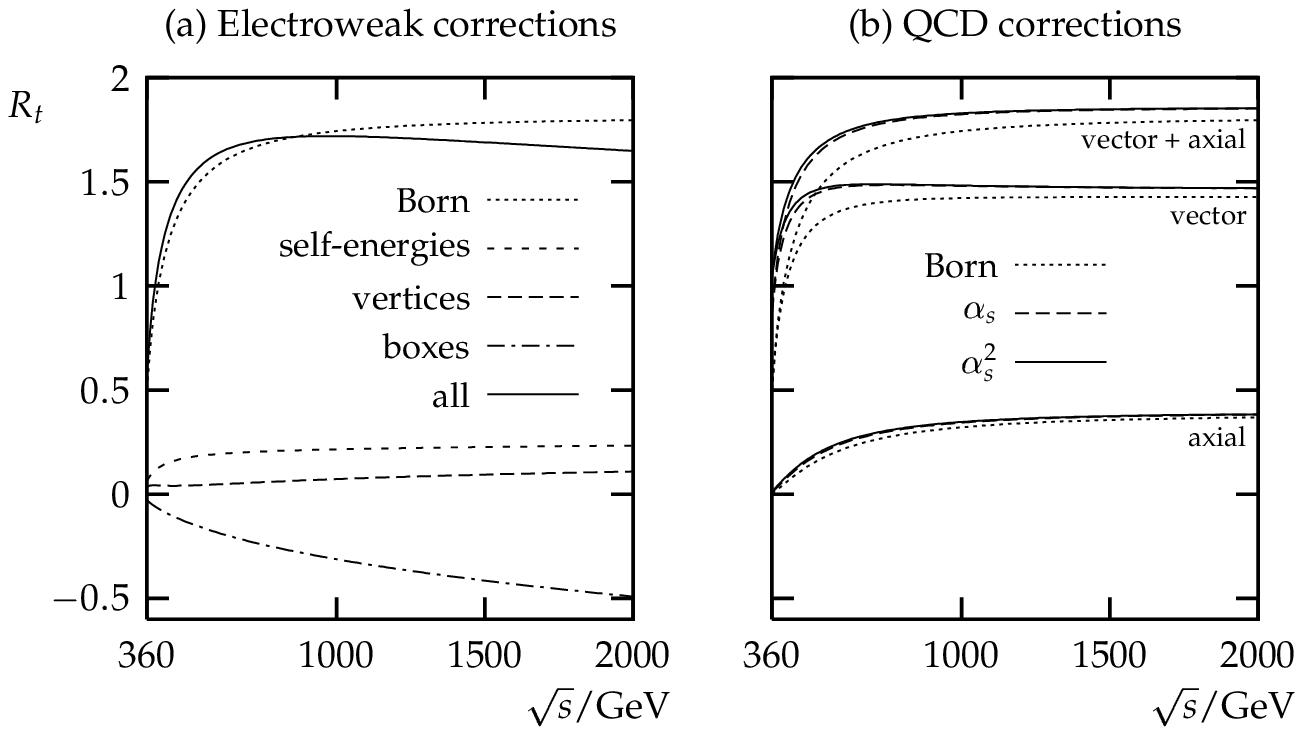}
  \caption{\label{fig:ewqcdalone}The electroweak (a) and QCD (b)
  corrections to $e^+e^-\to t\bar{t}$.
  In this Figure the normalized cross section $R_t=\sigma(e^+e^-\to
  t\bar{t})/\sigma_{pt}$ is shown. $\sigma_{pt}=4\pi\alpha^2/3s$.}
\end{figure}

In~\cite{HarSte98}
the results have been used in order to obtain in combination with 
the vector correlator, order $\alpha_s^2$
corrections to the top quark production in $e^+ e^-$ annihilation
above the threshold. The results for the
cross section are shown in Fig.~\ref{fig:ewqcdalone} 
where also the electro-weak corrections have been 
included~\cite{KueHahHar99}.

{\bf Scalar and Pseudo-scalar correlators.} 
For completeness we want to mention that 
the high-energy expansions 
of the scalar and pseudo-scalar correlators have been
considered in~\cite{Harlander:1997xa}. The results have been applied to
the decay of a scalar and pseudo-scalar Higgs boson into top quarks
where higher order mass effects are important. This can be seen in 
Tab.~\ref{tab:Higgs}
where a (pseudo-)scalar Higgs boson mass of $450$~GeV has been
considerd.
Both the results for the individual mass-correction terms 
$(M_t^2/M_H^2)^i$ ($i=0,\ldots,4$) and their
proper sum is listed up to order $\alpha_s^2$.

\begin{table}[th]
{\footnotesize
\renewcommand{\arraystretch}{1.4}
\begin{center}
\begin{tabular}{|l||r|r|r|r|r||l||l|}
 \hline 
   \mbox{} & ${(M_t^2)^0}$ &  ${(M_t^2)^1}$ & 
      ${(M_t^2)^2}$ &  ${(M_t^2)^3}$ & 
      ${(M_t^2)^4}$ &  $\Sigma$   &  exact \\ 
 \hline \hline
 \mbox{} & \multicolumn{5}{|c||}{scalar, on-shell} & &  \\ 
  \hline 
$R^{(0),s}/3$ & $
               1.000$ & $
              -0.907$ & $
               0.137$ & $
               0.014$ & $
               0.003$ & $
       0.247$ & $       0.248$ \\ 
 \hline 
$C_FR^{(1),s}/3$ & $
              -0.778$ & $
               5.646$ & $
              -3.080$ & $
               0.010$ & $
               0.004$ & $
       1.802$ & $       1.802$ \\ 
 \hline 
$R^{(2),s}/3$ & $
             -35.803$ & $
              77.056$ & $
             -17.792$ & $
              -5.347$ & $
              -0.680$ & $
      17.435$ & $-$ \\ 
 \hline 
 $\Sigma_i (\alpha_s/\pi)^i$ 
&$  0.943$
&$ -0.663$
&$  0.026$
&$  0.009$
&$  0.003$
&$  0.318$ & \mbox{}\\ \hline \hline
 \mbox{} & \multicolumn{5}{|c||}{scalar, $\overline{\rm MS}$} & &  \\ 
  \hline 
$\frac{m_t^2}{M_t^2}\bar{R}^{(0),s}/3$ & $
 0.784$
&$ -0.558$
&$  0.066$
&$  0.005$
&$  0.001$
&$  0.298$
&$  0.299  $ 
\\
 \hline 
$\frac{m_t^2}{M_t^2}C_F\bar{R}^{(1),s}/3$ & $
  4.445$
&$ -3.723$
&$ -0.238$
&$  0.143$
&$  0.032$
&$  0.659$
&$  0.673  $
\\
 \hline 
$\frac{m_t^2}{M_t^2}\bar{R}^{(2),s}/3$ & $
 21.799$
&$ -7.606$
&$-15.334$
&$  0.680$
&$  0.546$
&$  0.086$
&$  -  $
\\
 \hline 
 $\Sigma_i (\alpha_s/\pi)^i$ 
&$  0.941$
&$ -0.679$
&$  0.045$
&$  0.010$
&$  0.002$
&$  0.319$
& \mbox{}\\ \hline \hline
 \mbox{} & \multicolumn{5}{|c||}{pseudo-scalar, on-shell} & &  \\ 
  \hline 
$R^{(0),p}/3$ & $
               1.000$ & $
              -0.302$ & $
              -0.046$ & $
              -0.014$ & $
              -0.005$ & $
       0.633$ & $       0.629$ \\ 
 \hline 
$C_FR^{(1),p}/3$ & $
              -0.778$ & $
               3.495$ & $
               0.417$ & $
               0.047$ & $
               0.027$ & $
       3.208$ & $       3.238$ \\ 
 \hline 
$R^{(2),p}/3$ & $
             -35.803$ & $
              25.024$ & $
              12.173$ & $
               3.780$ & $
               1.293$ & $
       6.467$ & $-$ \\ 
 \hline 
 $\Sigma_i (\alpha_s/\pi)^i$ 
&$  0.943$
&$ -0.172$
&$ -0.022$
&$ -0.009$
&$ -0.003$
&$  0.737$ & \mbox{}\\ \hline \hline
 \mbox{} & \multicolumn{5}{|c||}{pseudo-scalar, $\overline{\rm MS}$} & &  \\ 
  \hline 
$\frac{m_t^2}{M_t^2}\bar{R}^{(0),p}/3$ & $
  0.784$
&$ -0.186$
&$ -0.022$
&$ -0.005$
&$ -0.002$
&$  0.569$
&$  0.569  $
\\
 \hline 
$\frac{m_t^2}{M_t^2}C_F\bar{R}^{(1),p}/3$ & $
  4.445$
&$ -0.248$
&$ -0.215$
&$ -0.119$
&$ -0.043$
&$  3.821$
&$  3.791  $
\\
 \hline 
$\frac{m_t^2}{M_t^2}\bar{R}^{(2),p}/3$ & $
 21.799$
&$  9.854$
&$  2.455$
&$ -0.781$
&$ -0.520$
&$ 32.807$
&$ -  $
\\
 \hline 
 $\Sigma_i (\alpha_s/\pi)^i$
&$  0.941$
&$ -0.185$
&$ -0.026$
&$ -0.010$
&$ -0.003$
&$  0.717$
&\mbox{}\\\hline 
\end{tabular}
\end{center}
}
\caption{\label{tab:Higgs}
  Numerical results for $R^s$ and $R^p$ both in the on-shell and
  $\overline{\mbox{MS}}$ schemes.  The contributions from the mass terms
  $(M_t^2)^i$, their sum ($\Sigma$) and, where available, the exact
  results are shown.  $\Sigma_i(\alpha_s/\pi)^i$ is the sum of the 1-,
  2- and 3-loop terms.  The numbers correspond to $M_t=175$~GeV and
  $M_{H/A}=450$~GeV.  The renormalization scale $\mu^2$ is set to
  $s=M_{H/A}^2$.  }
\end{table}

\begin{figure}[t]
 \begin{center}
 \begin{tabular}{cc}
   \leavevmode
   \epsfxsize=6.5cm
   \epsffile[131 314 481 478]{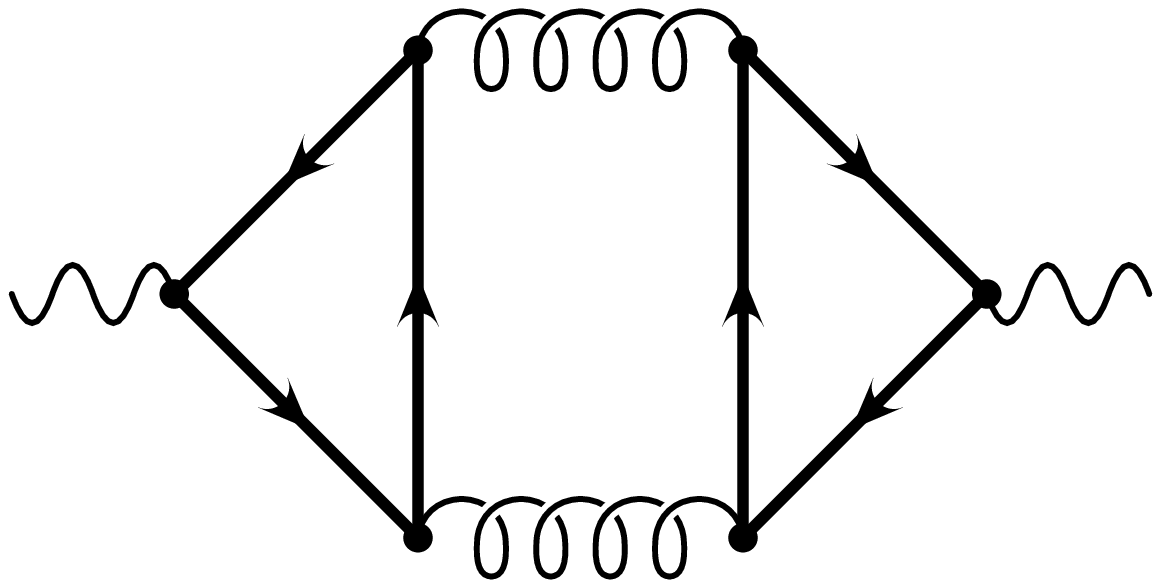}
   &
   \epsfxsize=6.5cm
   \epsffile[131 314 481 478]{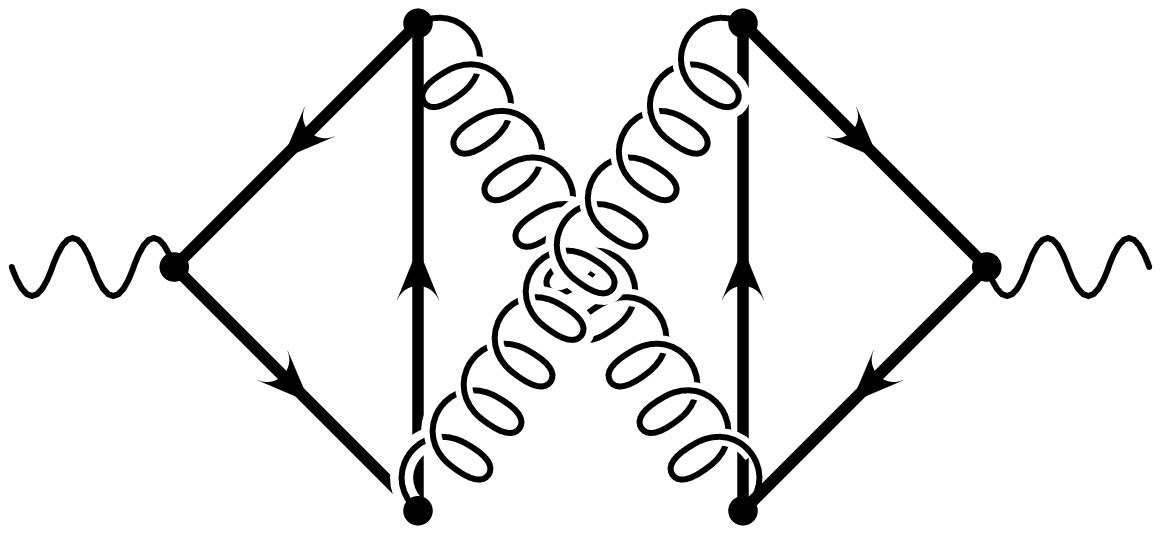}
 \end{tabular}
 \caption{\label{fig:diasing}Singlet or double-triangle diagrams.
   In the fermion lines either the quark $\psi$ or its isospin partner
   $\chi$ may be present.}
 \end{center}
\end{figure}

{\bf Singlet contribution.}
A special feature of the diagonal current correlator is the
occurence of so-called singlet diagrams as pictured in
Fig.~\ref{fig:diasing}. They are often also
referred to as double-triangle diagrams as the external currents are not
connected through the same fermion line.  
Note that for the vector correlator there are no singlet diagrams 
at three-loop level according to Furry's theorem~\cite{Fur37}.
In case of axial-vector couplings one has to take both members
of a weak isospin doublet into account 
in order for the axial anomaly to cancel.
It is therefore convenient to
replace the current $j^a_\mu$ in Eq.~(\ref{eqpivadef}) by 
$
j_{S,\mu}^a = \bar{\psi}\gamma_\mu\gamma_5 \psi - 
\bar{\chi}\gamma_\mu\gamma_5 \chi ,
$
where $\psi$ and $\chi$ are isospin partners.
The diagrams contributing to the singlet part, $\Pi^{(2),a}_S(q^2)$, are
depicted in Fig.~\ref{fig:diasing}
where in the fermion triangles either $\psi$ or $\chi$ may be present.
Note that for a degenerate quark doublet $\Pi_S^{(2),a}(q^2)$ vanishes.
Having furthermore in mind the physical case $(\psi,\chi) = (t,b)$, 
it is justified to set $m_\psi = M$ and $m_{\chi} = 0$.
In the case of the scalar and pseudo-scalar currents 
there is no anomaly. Furthermore,
as the coupling is proportional to the quark mass,
only one diagram
has to be considered, namely the one where in both quark lines the same
heavy quark flavour is present.

As mentioned in Section~\ref{sub:method} it is essential that the
expansion for $q^2\to0$ is analytical.
However, in the case of the singlet diagrams this is not fulfilled
as there exist massless cuts containing gluons and light fermions.
Thus the method of Section~\ref{sub:method} cannot directly be applied
to $\Pi_S^{(2),\delta}(q^2)$. Rather it is applied to
\begin{eqnarray}
  \Pi_{S,mod}^{(2),a}(q^2) &=& \Pi_S^{(2),a}(q^2) -
  {1\over 12\pi^2}\int_0^1 {\rm d} r {R_{Sb}^{(2),a}(s)\over
  r-z}
  \,,
  \nonumber\\
  \Pi_{S,mod}^{(2),\kappa}(q^2) &=& \Pi_S^{(2),\kappa}(q^2) -
  {1\over 8\pi^2}\int_0^1 {\rm d} r {R_{gg}^{(2),\kappa}(s)\over
  r-z}
  \,,\,\,\,\,\kappa=s,p
  \,,
\label{eq:piamod}
\end{eqnarray}
where $R_{Sb}^{(2),a}$ and $R_{gg}^{(2),\kappa}$ denote 
the contribution of these massless cuts to $R_S^{(2),a}$ and 
$R_S^{(2),\kappa}$, respectively.
Thus by definition, $\Pi_{S,mod}^{(2),\delta}(q^2)$ 
has the same analytical
properties as the non-singlet polarization functions.
The notation already suggests that in the scalar and pseudo-scalar
case there is only the cut through the two gluons. On the contrary,
this cut is zero in the axial-vector case
according to the Landau-Yang-Theorem~\cite{LanYan} and only cuts
involving the massless quark contribute. 
The analytical expressions for $R^{(2),a}_{Sb}$ 
and $R^{(2),\kappa}_{gg}$ can be
found in~\cite{KniKue89} and~\cite{CheHarSte98}, respectively. 
Expansions of the former are also listed
in~\cite{CheHarSte98}.

\begin{figure}[ht]
 \begin{center}
 \begin{tabular}{cc}
   \small (a) & \small (b) \\[-2ex]
   \leavevmode
   \epsfxsize=6.5cm
   \epsffile[110 270 480 560]{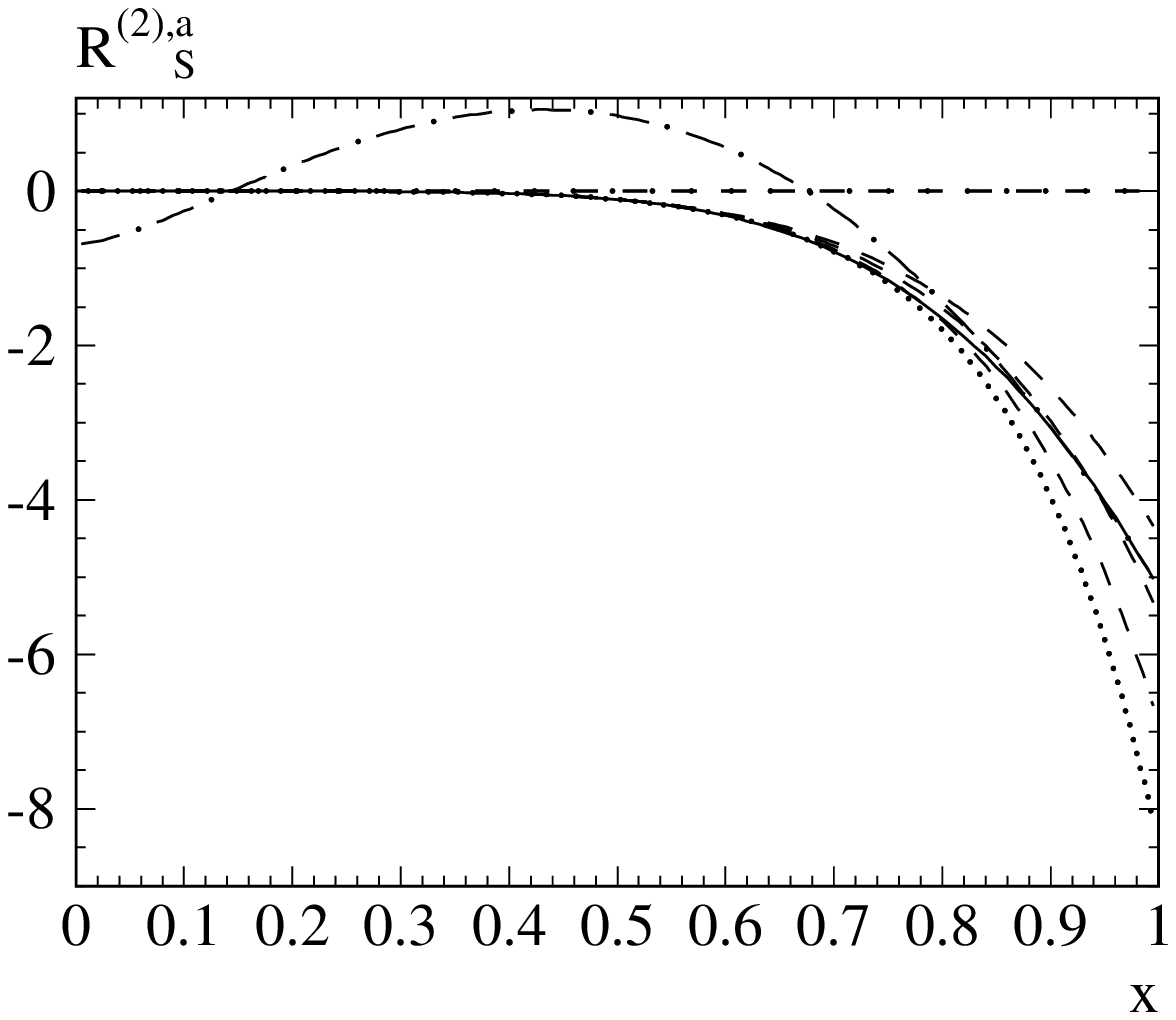}
&
   \epsfxsize=6.5cm
   \epsffile[110 270 480 560]{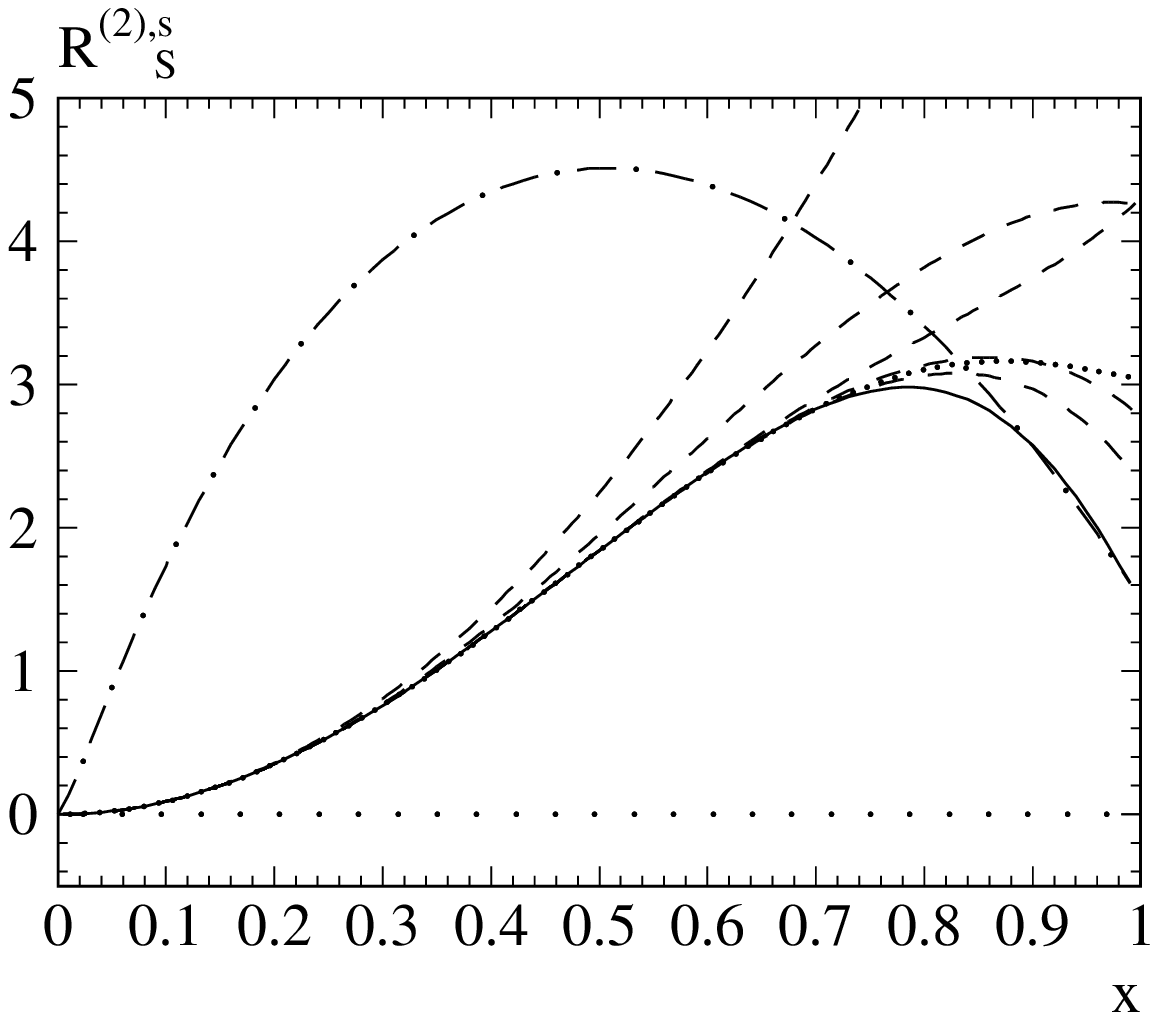}
\\
   \small (c) & \small (d)\\[-2ex]
   \epsfxsize=6.5cm
   \epsffile[110 270 480 560]{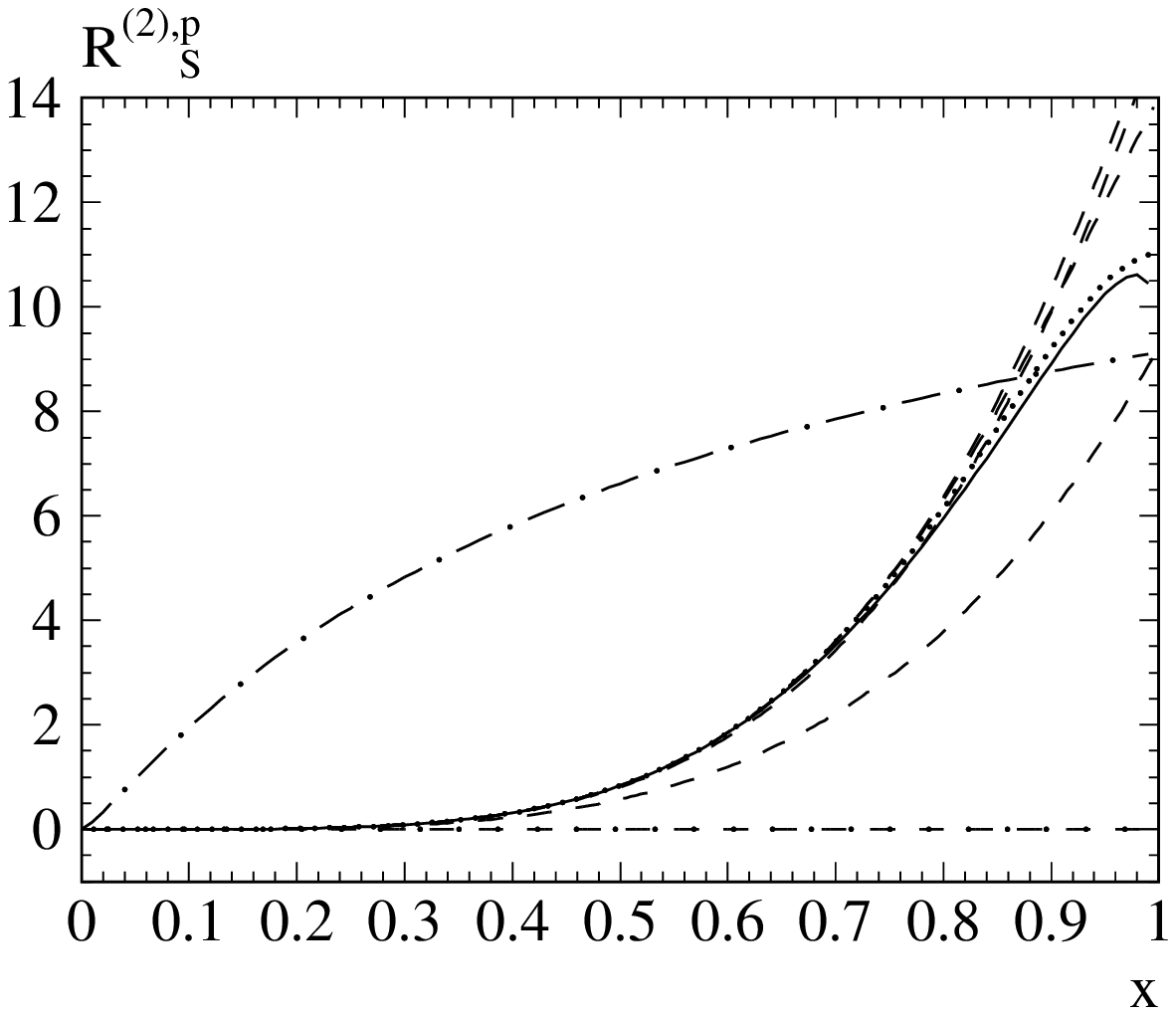}
&
   \epsfxsize=6.5cm
   \epsffile[110 270 480 560]{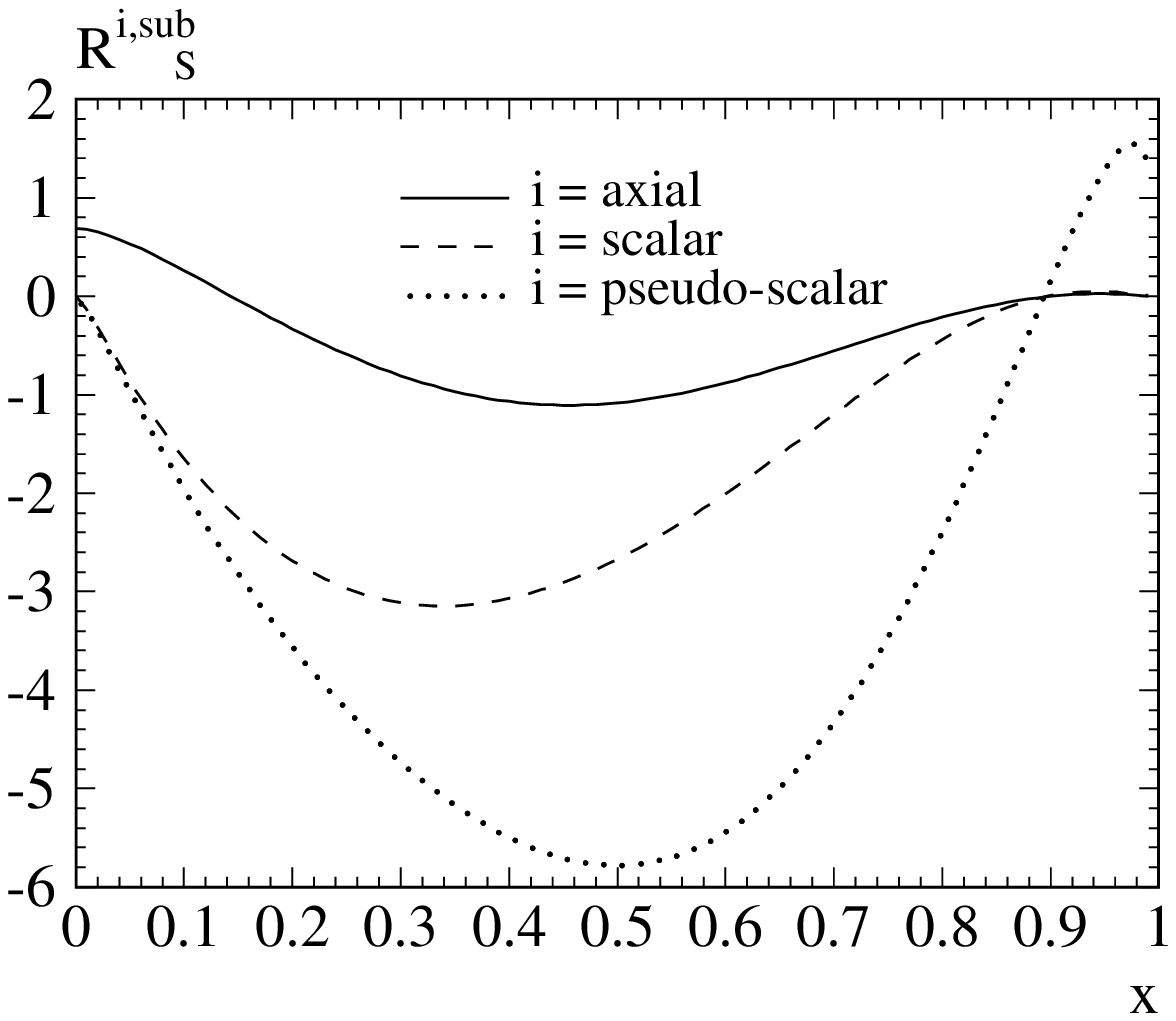}
 \end{tabular}
 \caption{\label{fig:pademxfig} The imaginary parts $R_S^{(2),a}$,
   $R_S^{(2),s}$ and $R_S^{(2),p}$ of (a) the axial-vector, (b)
   scalar and (c) pseudo-scalar singlet diagrams, respectively.
   For the abscissa the variable $x=2m/\sqrt s$ is chosen.
   Solid line: Pad\'e result; wide dots, dashes and narrow dots:
   $(m^2/q^2)^n$-expansion for $n=0$, $n=1,\ldots,5$ and $n=6$,
   respectively; dash-dotted line: purely massless cuts $R_{Sb}^{(2),a}$,
   $R_{gg}^{(2),s}$ and $R_{gg}^{(2),p}$. Fig.~(d) shows the difference
   between the solid and the dash-dotted line (i.e., the contribution
   of the massive quarks) of Figs.~(a), (b) and (c) as
   solid, dashed and dotted line, respectively.
   The curves in (a) can also be found in the last row of Fig.~\ref{fig:ra}.}
 \end{center}
\end{figure}

In~\cite{CheHarSte98} the asymptotic expansion has been applied to
the singlet diagrams and terms up to $(M^2/q^2)^6$ in the high-energy
region have been evaluated.
For small external momentum moments up to order
$z^7$ (axial-vector) and $z^8$ (scalar and pseudo-scalar) are
available.
This input was used to compute roughly 30 Pad\'e approximants for each
correlator.
In Fig.~\ref{fig:pademxfig}~(a)--(c) the results for the imaginary part of
$\Pi_{S}^{(2),a}$, $\Pi_{S}^{(2),s}$ and $\Pi_{S}^{(2),p}$ (solid
lines), together with the first seven terms of the high energy expansion
(dashed and dotted lines) are shown as functions of $x=2m/\sqrt s$.
Note, that in the displayed region, $0<x<1$, ${\rm Im}\Pi_{S}^{(2)} =
{\rm Im}\Pi_{S,mod}^{(2)}$.  Therefore, if one is interested, e.g., in
(inclusive) production of the heavy quarks only, the corresponding
massless cuts (depicted as dash-dotted lines)
have to be subtracted. 
The resulting curves are shown in Fig.~\ref{fig:pademxfig}~(d).

The requirement that no logarithmic terms may appear in the expansion
for $z\to0$ (cf. Eq.~(\ref{eq:deff})) seems to be quite
restrictive. However, the example of the singlet diagrams shows that
also problems which at first sight do not
match the definition of $f(z)$ in Eq.~(\ref{eq:deff}) can be treated.

\vspace{1em}

From the systematic evaluation of the polarization function at low and
high energies and the information about the leading threshold
behaviour it is possible to construct --- with the help of Pad\'e
approximation and conformal mapping --- an approximation to $\Pi(z)$
taking into account the complete quark mass dependence. The imaginary
part immediately leads to important physical quantities like $R(s)$,
top quark pair production above threshold and the total decay rate of
scalar and pseudo-scalar Higgs bosons. 


\subsubsection{\label{subsub:delal}An 
  application of $R(s)$: $\Delta\alpha_{\rm had}(M_Z)$}

An important application of the vector current correlator, 
as discussed in Sections~\ref{sub:as3m4} and~\ref{subsub:R},
is the evaluation of the fine-structure constant at
the scale of the $Z$ boson, $\alpha(M_Z^2)$.
It plays a crucial role in the indirect determination of the
Higgs boson mass which constitutes one of the most important
goals of precision experiments.
In particular, the error on $\alpha(M_Z^2)$ induces one of the largest
uncertainties.

In Ref.~\cite{EidJeg95} a conservative analysis has been performed
which exclusively
relies on data below a center-of-mass energy, $\sqrt{s}$, of 40~GeV.
Consequently it suffers from sizable experimental errors.
Only for $\sqrt{s}\ge40$~GeV perturbative QCD has been used.
Recently several suggestions
have been made, which significantly reduce the uncertainties
in $\alpha(M_Z^2)$.
Most of them were actually triggered from the knowledge of the 
complete mass dependence at order
$\alpha_s^2$ (see Section~\ref{sub:veccor}).
There is a number of
so-called theory-driven 
analyses~\cite{DavHoe98_1,KueSte98,
  GroKoeSchNas98,Erl98,DavHoe98_2,JegRADCOR98,MarOutRys00}  
which replace to smaller and larger extent unprecise data by
the results of perturbative QCD (pQCD). They will be discussed below.

Let us start with some basic definitions.
Re-summation of the leading logarithms leads to
\begin{eqnarray}
\alpha(s) &=& \frac{\alpha(0)}
      {1-\Delta\alpha_{\rm lep}(s)
        -\Delta\alpha^{(5)}_{\rm had}(s)
        -\Delta\alpha_{\rm top}(s)}
\,,
\label{eq:delaldef}
\end{eqnarray}
with $\alpha=\alpha(0)=1/137.0359895$.
$\Delta\alpha_{\rm lep}$ denotes the leptonic contribution and is
known up to the three-loop order~\cite{Ste98}
\begin{eqnarray}
\Delta\alpha_{\rm lep}(M_Z^2) &=& 314.97686 \times 10^{-4}
\,.
\label{eq:delallep}
\end{eqnarray}
Perturbation theory is also applicable to treat the contribution arising from
the top quark. Including three-loop QCD corrections one gets~\cite{KueSte98}
\begin{eqnarray}
\Delta\alpha_{\rm top}(M_Z^2) &=& (-0.70 \pm 0.05) \times 10^{-4}
\,.
\label{eq:delaltop}
\end{eqnarray}
The contribution from the remaining five quarks has to be taken into account
using the dispersion integral
\begin{eqnarray}
\Delta\alpha_{\rm had}^{(5)}(M_Z^2)
&=&
-\frac{\alpha M_Z^2}{3\pi}\,\mbox{Re}\,
\int_{4m_\pi^2}^\infty\,{\rm d}
s\,\frac{R(s)}{s\left(s-M_Z^2-i\epsilon\right)}
\,,
\label{eq:delaldisp}
\end{eqnarray}
with
$R(s)=\sigma(e^+e^-\to\mbox{hadrons})/\sigma(e^+e^-\to\mu^+\mu^-)$
(cf. Eq.~(\ref{eq:rtopisp})).
It is not possible to use perturbation theory for $R(s)$ in the whole energy
region. Thus one has to rely to some extent on
experimental results.
The results of the recent evaluations can be found in
Tab.~\ref{tab:cmp}
where also the resulting values for $\alpha^{-1}(M_Z^2)$ are listed. The
latter has been obtained with the help of 
Eqs.~(\ref{eq:delaldef}),~(\ref{eq:delallep}) and~(\ref{eq:delaltop}).
In the fourth column of Tab.~\ref{tab:cmp} some keywords are given
which shall indicate the method used for the analysis. In the
following we briefly describe the abbreviations. For more details we
refer to the original papers.

\begin{table}[t]
  \renewcommand{\arraystretch}{1.3}
  \begin{center}
    {\small
      \begin{tabular}{|l|l|l|l|}
        \hline\hline
        $\Delta\alpha^{(5)}_{\rm had}(M_Z^2)$ &
        $\alpha^{-1}(M_Z)$ & Reference 
        & comment \\
        $\times 10^4$ &&&\\
        \hline
        $273.2(4.2)$   &128.985(58)& \cite{MarZep95}, Martin et al. `95 
        & (low-order) pQCD \\
        $280(7)$       &128.892(95)& \cite{EidJeg95}, Eidelman et al. `95
        & data \\
        $280(7)$       &128.892(95)& \cite{BurPie95}, Burkhardt et al. `95
        & data \\
        $275.2(4.6)$   &128.958(63)& \cite{Swa95},    Swartz `96
        & --- \\
        $281.7(6.2)$   &128.869(85)& \cite{AleDavHoe97}, Alemany et al. `97
        & $\tau$ data\\
        $278.4(2.6)^*$ &128.914(35)& \cite{DavHoe98_1}, Davier et al. `97
        & + pQCD \\
        $277.5(1.7)$   &128.927(23)& \cite{KueSte98}, K\"uhn et al. `98
        & + CT \\
        $277.6(4.1)$   &128.925(56)& \cite{GroKoeSchNas98}, Groote et al. `98
        & SR (pQCD) \\
        $277.3(2.0)^{**}$ &128.929(27)& \cite{Erl98}, Erler `98
        & $\tau$ data + UDR \\
        $277.0(1.6)^*$   &128.933(22)& \cite{DavHoe98_2}, Davier et al. `98
        & $\tau$ data + pQCD + SR\\
        $277.8(2.5)$   &128.922(34)& \cite{JegRADCOR98}, Jegerlehner `99
        & pQCD (ER)\\
        $274.3(1.9)$   &128.970(26)& \cite{MarOutRys00}, Martin et al. `00
        & new data + pQCD (ER)\\
        $276.1(3.6)$   &128.946(49)& \cite{BurPie01}, Burkhardt et al. `01
        & new data \\
        $277.3(2.1)$   &128.930(29)& \cite{Jeg01}, Jegerlehner `01
        & pQCD (ER) + new data \\
        \hline\hline
      \end{tabular}
      }
    \caption{\label{tab:cmp}
      Comparison of the different evaluations of
      $\Delta\alpha^{(5)}_{\rm had}(M_Z^2)$.
      The column ``comment'' reminds on the different methods used in the
      analysis as described in the text.
      (${}^*\Delta\alpha_{\rm top}(M_Z^2)$ subtracted;
      ${}^{**}$ value corresponding to $\alpha_s(M_Z^2)=0.118$ adopted.)
      }
  \end{center}
\end{table}

{\bf $\tau$ data}~\cite{AleDavHoe97}.
$\tau$ data from ALEPH have been used in order to
get more information about $R(s)$ for energies below roughly 1.8~GeV.
The hypothesis of conserved vector current (CVC) in combination with isospin
invariance relates, e.g.,  the vector part of the two-pion $\tau$ spectral
function to the corresponding part of the isovector $e^+e^-$ cross section
through the following relation
\begin{eqnarray}
\sigma^{I=1}\left(e^+e^-\to\pi^+\pi^-\right) &=&
\frac{4\pi\alpha^2(0)}{s}v_{J=1}\left(\tau\to\pi\pi^0\nu_\tau\right)
\,.
\end{eqnarray}
A similar equation holds for the four-pion final state.
Their incorporation into the analysis has been performed
in~\cite{AleDavHoe97} leading to a slight reduction of the error on
$\Delta\alpha^{(5)}_{\rm had}$\footnote{On the contrary, the inclusion
  of the $\tau$ data leads to a 
  significant reduction of the error of the anomalous magnetic moment
  of the muon as it is more sensitive to the low-energy region.}.

{\bf Perturbative QCD (pQCD)}.
The first attempt to replace unprecise data by pQCD can be found
in~\cite{MarZep95}. At that time, however, mass effects were barely
known. Thus pQCD could only be applied far above the particle thresholds.
Meanwhile $R(s)$ can be calculated in the framework of
pQCD up to order $\alpha_s^3$
if quark masses are neglected~\cite{GorKatLar91SurSam91,Che97_R} and up to
${\cal O}(\alpha_s^2)$ with full quark mass 
dependence~\cite{HoaKueTeu951,CheKueSte96,CheKueSte97,CheHoaKueSteTeu97}.
In~\cite{DavHoe98_1,KueSte98} pQCD has been used down to an energy scale of 
$\sqrt{s}=1.8$~GeV and it has been shown that the non perturbative
contributions are small. This leads to a further reduction of the error of
about a factor two. For convenience we list in Tab.~\ref{tab:pert}
the perturbative hadronic contributions for a variety of energy intervals.
As our default values we adopt $\mu^2 = s$, 
$\alpha^{(5)}_s(M_Z^2)=0.118$~\cite{Kniehl:2000cr},
$M_c=1.6$~GeV and $M_b=4.7$~GeV. In separate columns we list the
variations with a change in the renormalization scale, the strong
coupling constant and the quark masses:
\begin{eqnarray}
\delta\alpha_s\,\,=\,\,\pm0.003,\quad
\delta M_c\,\,=\,\,\pm0.2~{\mbox GeV},\quad
\delta M_b\,\,=\,\,\pm0.3~{\mbox GeV}.\quad
\label{eqdelta}
\end{eqnarray}

\begin{table}[t]
\begin{center}
{\small
\begin{tabular}{|l|r|r|r|r|r|}
\hline\hline
Energy range (GeV) & central value &$\delta\mu$ &
$\delta\alpha_s$ & $\delta M_c$ & $\delta M_b$\\ \hline
$  1.800-  2.125$&$    5.67$&$    0.22$&$    0.04$&$    0.00$&$    0.00$\\
$  2.125-  3.000$&$   11.66$&$    0.21$&$    0.06$&$    0.01$&$    0.00$\\
$  3.000-  3.700$&$    7.03$&$    0.06$&$    0.03$&$    0.00$&$    0.00$\\
$  1.800-  3.700$&$   24.36$&$    0.48$&$    0.13$&$    0.01$&$    0.01$\\
\hline
$  5.000-  5.500$&$    5.44$&$    0.03$&$    0.03$&$    0.06$&$    0.00$\\
$  5.500-  6.000$&$    4.93$&$    0.03$&$    0.02$&$    0.04$&$    0.00$\\
$  6.000-  9.460$&$   25.45$&$    0.11$&$    0.08$&$    0.10$&$    0.00$\\
$  9.460- 10.520$&$    5.90$&$    0.02$&$    0.01$&$    0.01$&$    0.00$\\
$ 10.520- 11.200$&$    3.48$&$    0.01$&$    0.01$&$    0.00$&$    0.00$\\
$  5.000- 11.200$&$   45.20$&$    0.19$&$    0.15$&$    0.21$&$    0.01$\\
(without $b\bar{b}$) &&&&&\\
\hline
$ 11.200- 11.500$&$    1.63$&$    0.00$&$    0.01$&$    0.00$&$    0.00$\\
$ 11.500- 12.000$&$    2.62$&$    0.00$&$    0.01$&$    0.00$&$    0.00$\\
$ 12.000- 13.000$&$    4.93$&$    0.01$&$    0.01$&$    0.00$&$    0.00$\\
$ 13.000- 40.000$&$   72.92$&$    0.08$&$    0.12$&$    0.02$&$    0.02$\\
$ 12.000- 40.000$&$   77.85$&$    0.09$&$    0.14$&$    0.02$&$    0.02$\\
$ 40.000-\infty$&$   42.67$&$    0.03$&$    0.06$&$    0.00$&$    0.00$\\
$ 11.200-\infty$&$  124.77$&$    0.12$&$    0.21$&$    0.03$&$    0.02$\\
\hline
$1.8-\infty$ (pQCD) & $  194.33$&$    0.79$&$    0.49$&$    0.24$&$    0.03$\\
\hline
QED & $    0.11$ & -- & -- & -- & -- \\
\hline\hline
\end{tabular}
}
\caption{\label{tab:pert}
Contributions to $\Delta\alpha^{(5)}_{\rm had}(M_Z^2)$ 
(in units of $10^{-4}$) from the energy
regions where pQCD is used (adopted from~\protect\cite{CheHoaKueSteTeu97}).
For the QED corrections the same intervals
have been chosen. For the variation of $\alpha_s(M_Z^2)$, $M_c$ and $M_b$
(cf. Eqs.~(\ref{eqdelta})) have been used. $\mu$ has been varied between 
$\protect\sqrt{s}/2$ and $2\protect\sqrt{s}$.
}
\end{center}
\end{table}

The typical contributions which have to be taken into account 
look as follows~\cite{KueSte98}.
In the perturbative regions one receives contributions from light ($u$,
$d$ and $s$) quarks whose masses are neglected throughout, and from
massive quarks which demand a more refined treatment. Below the charm
threshold the light quark contributions are evaluated in order
$\alpha_s^3$ plus terms of order
$\alpha_s^2\, s/(4M^2_c)$
from virtual massive quark loops. Above $5$~GeV the full $M_c$ dependence is
taken into account up to order $\alpha^2_s$, and in addition the
dominant cubic terms in the strong coupling are incorporated, as well as
the corrections from virtual bottom quark loops of order
$\alpha_s^2\, s/(4M^2_b)$. Above $11.2$~GeV the same formalism
is applied to the massive bottom quarks and charm quark mass effects are
taken into account through their leading contributions in an $M^2_c/s$
expansion.
All formulae are available for arbitrary renormalization scale $\mu$
which allows to test the scale dependence of the final answer. This
was used to estimate the theoretical uncertainties
from uncalculated higher orders. Matching of $\alpha_s$
between the treatment with
$n_f=3$, $4$ and $5$ flavours is performed at the respective threshold
values (cf. Section~\ref{sec:dec}). 
The influence of
the small ${\cal O}(\alpha^3_s)$ singlet piece which prevents a
clear separation of contributions from different quark species can safely be
ignored for the present purpose.

{\bf Charm threshold region (CT)}~\cite{KueSte98}.
Perturbative QCD is clearly inapplicable in the charm
threshold region
between $3.7$ and $5$~GeV
where rapid variations of the cross section are
observed. Data have been taken more than $15$ years ago by the 
PLUTO~\cite{PLUTO},
DASP~\cite{DASP}, and MARK~I collaborations~\cite{MARK1}. 
The systematic errors of $10$ to
$20$~\% exceed the statistical errors significantly and are reflected in a
sizeable spread of the experimental results.
In~\cite{KueSte98} the experimental data are normalized to match the
predictions of perturbative QCD both below $3.7$ and above $5.0$~GeV.
Two models have been constructed which describe the differences of the
normalization factors below and above the considered energy interval.

Recently, the BES collaboration has measured $R(s)$ in the energy 
range between 2 and 5~GeV with substantially improved
precision~\cite{Bai:2000pk}.
We applied the method of~\cite{KueSte98} and obtained perfect
agreement with the results of PLUTO, DASP and MARK~I.

{\bf QCD sum rules (SR)}~\cite{GroKoeSchNas98,DavHoe98_2}.
Global parton-hadron duality is used in order to reduce the influence of
the data in the different intervals.
This is achieved by choosing a proper polynomial, $Q_N(s)$,
which is supposed to 
approximate the weight function $M_Z^2/s(s-M_Z^2)$ 
in Eq.~(\ref{eq:delaldisp}) as good as possible.
Adding and subtracting $Q_N(s)$ in Eq.~(\ref{eq:delaldisp})
and exploiting the analycity of the
subtracted term leads to
\begin{eqnarray}
  \int_{s_0}^{s_1}\,{\rm d}s\,
  \frac{R(s)\,M_Z^2}{s\left(s-M_Z^2\right)}
  &=&
  \int_{s_0}^{s_1}\,{\rm d}s\,R(s)
  \left(
    \frac{M_Z^2}{s\left(s-M_Z^2\right)} - Q_N(s)
  \right)
  \nonumber\\&&\qquad\qquad\qquad
  +
  6\pi i\oint_{|s|=s_1} \,{\rm d}s\, \Pi^{\rm QCD}(s) Q_N(s)
  \,.
  \label{eq:delalsub}
\end{eqnarray}
Thus the influence of the experimental data is significantly reduced in the
first term of the right-hand side and pQCD only has to be used for 
$|s|=s_1$ which is indicated by the superscript QCD.

{\bf Unsubtracted dispersion relations (UDR)}~\cite{Erl98}.
They are used in order to evaluate the electromagnetic coupling in the
$\overline{\rm MS}$ scheme. Four-loop running is
accompanied by three-loop matching in order to arrive at $\bar\alpha(M_Z^2)$,
which subsequently has to be transformed to the on-shell quantity
$\alpha(M_Z^2)$. Via this method no complications in connection with the
$J/\Psi$ or $\Upsilon$ resonances occur. However, one encounters a much
stronger dependence on the quark masses.

{\bf Perturbative QCD in Euclidian region
(ER)}~\cite{JegRADCOR98,MarOutRys00,Jeg01}.
The authors of Ref.~\cite{EidJeg95} also re-evaluated
$\Delta\alpha^{(5)}_{\rm had}(M_Z)$ using pQCD.
In a first step $\Delta\alpha^{(5)}_{\rm had}$ was calculated at the
large negative scale $s=-M_Z^2$ and then analytically continued
to $s=M_Z^2$. Thus pQCD has only been applied in the Eulidian region
where it
is supposed to work best as one is far from resonances and thresholds.
Furthermore pQCD has been used down to $-(2.5~\mbox{GeV})^2$.

{\bf New data in the low-energy region.}
In the meantime new experimental data for $R(s)$ in the low-energy region have
become available. 
Besides improvements in the energy interval below 1.4~GeV by the 
CMD-2 detector at the VEPP-2M collider in Novosibirsk a
measurement of $R(s)$ in the range
$2~\mbox{GeV}<\sqrt{s}<5$~GeV has been performed by the experiment BES~II
at Beijing. In Ref.~\cite{MarOutRys00} these data have been 
incorporated and accompanied with pQCD in the regions
$3~\mbox{GeV}<\sqrt{s}<3.74$~GeV
and $\sqrt{s}>5$~GeV in order to evaluate
$\Delta\alpha^{(5)}_{\rm had}(M_Z)$. The result shown in
Tab.~\ref{tab:cmp} has been obtained using only inclusive measurements of
$R(s)$ for $\sqrt{s}\lsim1.9$~GeV.
Based on the comparision of time-like and space-like (i.e. in
the Euclidian region) evaluations of $\Delta\alpha^{(5)}_{\rm had}(M_Z)$
it has been argued in~\cite{MarOutRys00}
that this is preferred to the exclusive measurements of $R(s)$.

More recently the data-based analysis of~\cite{BurPie95} has been 
updated~\cite{BurPie01} using pQCD only above $\sqrt{s}=12$~GeV. 
The main improvements are due to the new BES measurements.

\vspace{1em}

Tab.~\ref{tab:cmp} shows that the inclusion of pQCD leads to 
a significant reduction of the error 
in $\Delta\alpha^{(5)}_{\rm had}(M_Z^2)$.
The new analyses (with the exception 
of~\cite{MarOutRys00}\footnote{Actually, in~\cite{MarOutRys00} also a
  result based on exclusive data is given
  ($\Delta\alpha^{(5)}_{\rm had}(M_Z^2)=(276.49\pm2.14)\times10^{-4}$)
  which is in better agreement with the other values.}) 
agree well both in
their central values and in their quoted errors.
These promising developments suggest to use
the new values in the interpretation of the electroweak measurements.
Once more precise experimental results on $R$ are
available it can replace the corresponding parts in the 
theory-motivated analyses.
Certainly these measurements would be extremely valuable for a cross check of
the theory-driven results. 


\subsubsection{\label{subsub:nondiag}Heavy-light
  current correlators}

The method of Section~\ref{sub:method} has also been applied to the
non-diagonal correlators where one of the quark fields in 
Eq.~(\ref{eq:currents}) has mass $M$ 
and the other one is massless.
In this limit the vector (scalar) and
axial-vector (pseudo-scalar) correlators coincide.
Furthermore
it is convenient to work with the variables
\begin{eqnarray}
  z\,\,=\,\,\frac{q^2}{M^2}\,,\qquad
  x\,\,=\,\,\frac{M}{\sqrt{s}}\,,\qquad
  v\,\,=\,\, \frac{1-x^2}{1+x^2}\,.
\end{eqnarray}

For the application of the Pad\'e method 
the polarization function $\Pi^\delta(z)$ has to be considered in the
different kinematical regions.
In~\cite{CheSte01,CheSte01_2} seven terms for small and eight terms
for large external momentum have been computed
both for the vector and scalar correlators
of Eqs.~(\ref{eqpivadef}) and~(\ref{eqpispdef}), respectively.
This means expansion coefficients up to order $z^6$, respectively, $1/z^7$
are available for the Pad\'e procedure.

The threshold behaviour constitutes the main difference as compared to
the diagonal correlator discussed in the Section~\ref{subsub:R}.
The Born result for $R^\delta$
of the diagonal vector and pseudo-scalar correlators
are proportional to $v$ for small velocities. 
At order $\alpha_s$ $R^{(1),\delta}$ approaches a constant and at order
$\alpha_s^2$ one either has a $1/v$ or a $\ln v$ behaviour ---
depending on the colour structures~\cite{CheKueSte96,CheKueSte97}.
The axial-vetor and scalar correlators show the same pattern with a
additional factor $v^2$ at each order.

On the contrary the imaginary part of the non-diagonal correlators are
proportional to $v^2$ (possibly accompanied with $\ln v$ terms)
independent of the order in $\alpha_s$ and of the colour structure.
This is valid in every order in $\alpha_s$
as follows from  Heavy Quark Effective Theory (HQET)~\cite{HQET}.
Actually the latter can be used 
to obtain the leading threshold behaviour of $R^v(s)$
and $R^s(s)$ at ${\cal O}(\alpha_s^2)$
from the corresponding correlators in HQET. In
particular, the renormalization group equation in the effective theory
is used to get the leading logarithmic behaviour at order
$\alpha_s^2$~\cite{CheSte01_2}. Afterwards the decoupling
relation between the currents in the full
and effective theory~\cite{JiMus91,BroGro95,Gro98} is exploited
to get the information about $R^v(s)$ and $R^s(s)$~\cite{CheSte01,CheSte01_2}.
It turns out that linear and quadratic logarithms occur.
This translates into quadratic and cubic logarithms of the
corresponding polarization function which are incorporated into the
Pad\'e method as descibed in step~\ref{item:log} of 
Section~\ref{sub:method}.

Before discussing the results in the full theory we want to spend time
on the spectral function in HQET.
In numerical form it reads~\cite{CheSte01_2}
\begin{eqnarray}
   \tilde{R}^\prime(\omega) &=& N_c \omega^2 
    \Bigg[ 1 + \frac{\alpha_s^{(n_l)}(\mu)}{\pi}\left(8.667 +
               \Lw\right)
             + \left(\frac{\alpha_s^{(n_l)}(\mu)}{\pi}\right)^2
               \left(
                 \tilde{c}_{n_l} + 35.54 \Lw 
\right.\nonumber\\&&\left.\mbox{}
                 + 1.875 \Lw^2
                 +n_l\left( - 1.583 \Lw - 0.08333 \Lw^2\right)
               \right)     
    \Bigg]
  \,,
  \label{eq:rtilfin}
\end{eqnarray}
where $\omega=\sqrt{s}-M$ is the only dimensionful quantity in the effective
theory and $\Lw=\ln(\mu^2/\omega^2)$.
The tilde and the prime remind 
that the quantity on the left-hand side of Eq.~(\ref{eq:rtilfin})
is defined in the effective theory
where the heavy quark is decoupled~\cite{BroGro95,Gro98}.
In Eq.~(\ref{eq:rtilfin})
the corrections of order $\alpha_s$ are known since quite some
time~\cite{BroGro92} whereas 
the constant $\tilde{c}_{n_l}$, which is not accessible using renormalization
group techniques, has been determined in~\cite{CheSte01_2}
with the help of the Pad\'e results in the full theory.
Its dependence on the number of massless quarks is given 
by
\begin{eqnarray}
  \tilde{c}_{n_l} &=& 46(15) - 1.2(4) \, n_l
  \,.
  \label{eq:ctilnl}
\end{eqnarray}
In the meantime the coefficient $\tilde{c}_{n_l}$ has been computed
analytically~\cite{CzaMel01} 
with the result $\tilde{c}_{n_l} = 58 - 1.7 \,n_l$. 
Agreement with Eq.~(\ref{eq:ctilnl}) can be observed for 
all physically reasonable values of $n_l$.

We want to mention that $\tilde{R}^\prime$ enters as a building block
into the sum rules which are used to determine the meson decay
constants like, e.g., $f_B$ (see, e.g., Ref.~\cite{BroGro92}).
The typical scale where Eq.~(\ref{eq:rtilfin}) has to be evaluated
is 1~GeV. Thus the first order QCD corrections amount to 
about 100\% and the terms of order $\alpha_s^2$ contribute
with additional 60(20)\%
where the sign is the same as for the LO correction.

\begin{figure}[t]
  \begin{center}
    \begin{tabular}{cc}
      \leavevmode
      \epsfxsize=7.cm
      \epsffile[110 280 460 560]{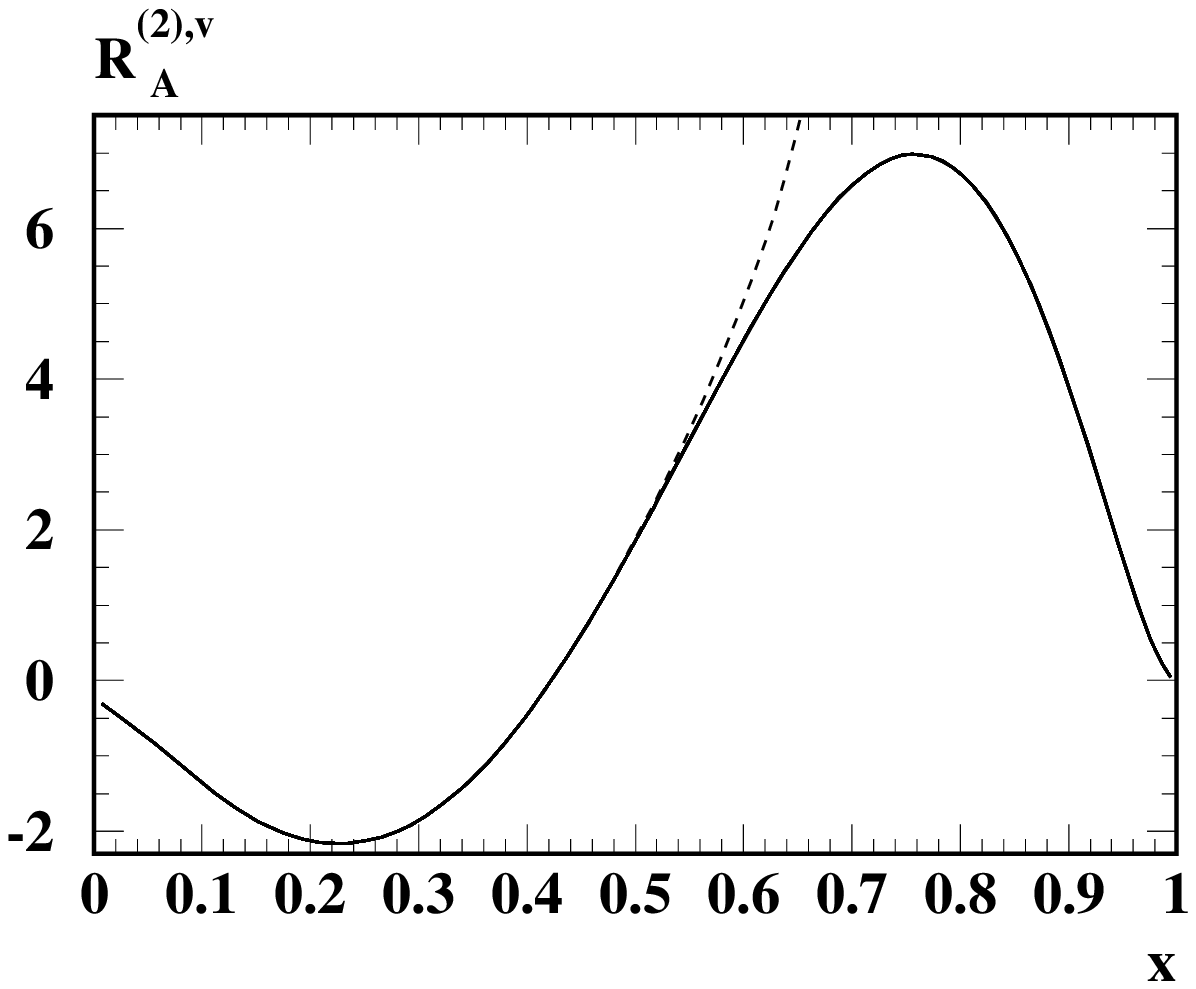}
      &
      \epsfxsize=7.cm
      \epsffile[110 280 460 560]{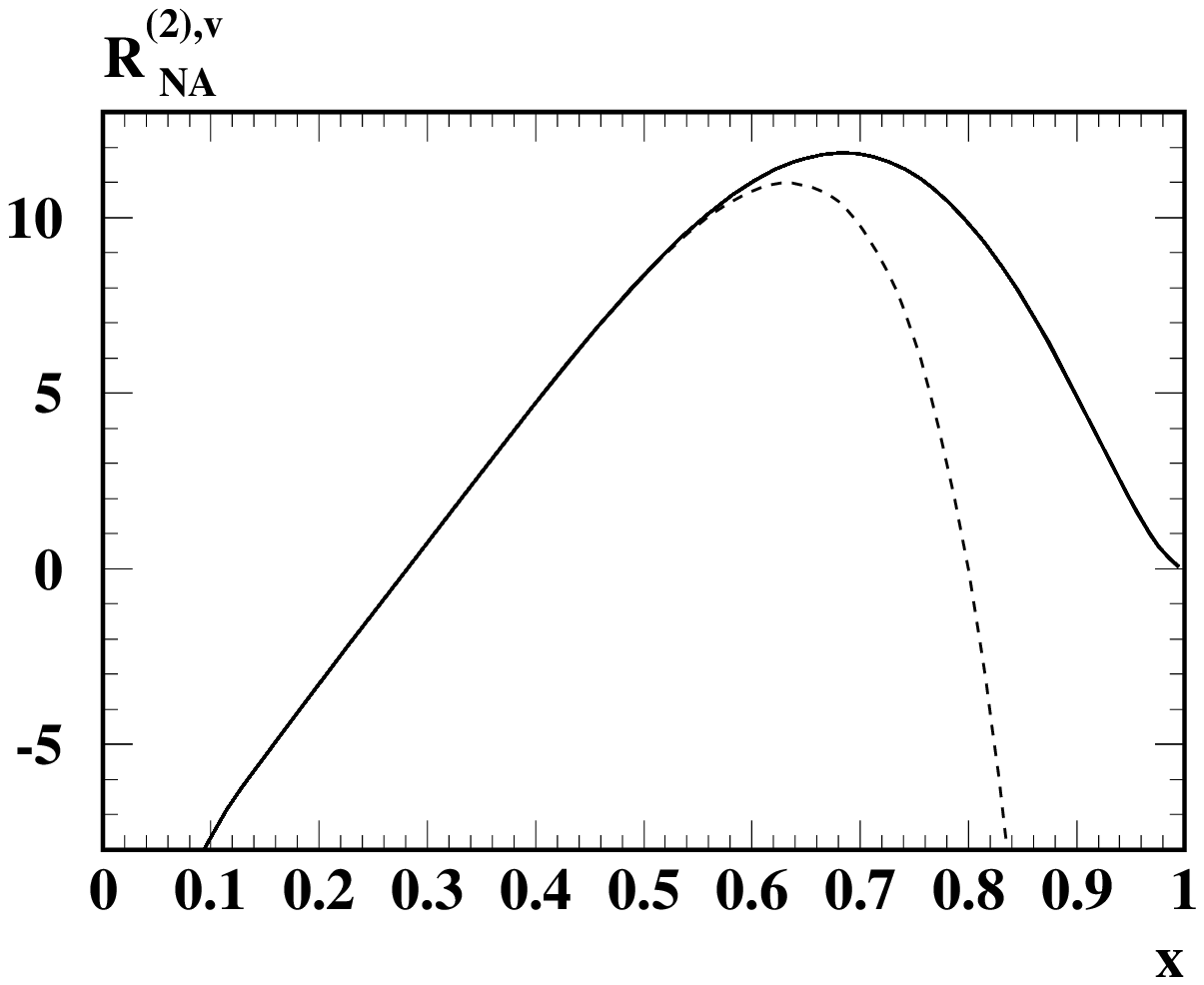}
      \\
      \epsfxsize=7.cm
      \epsffile[110 280 460 560]{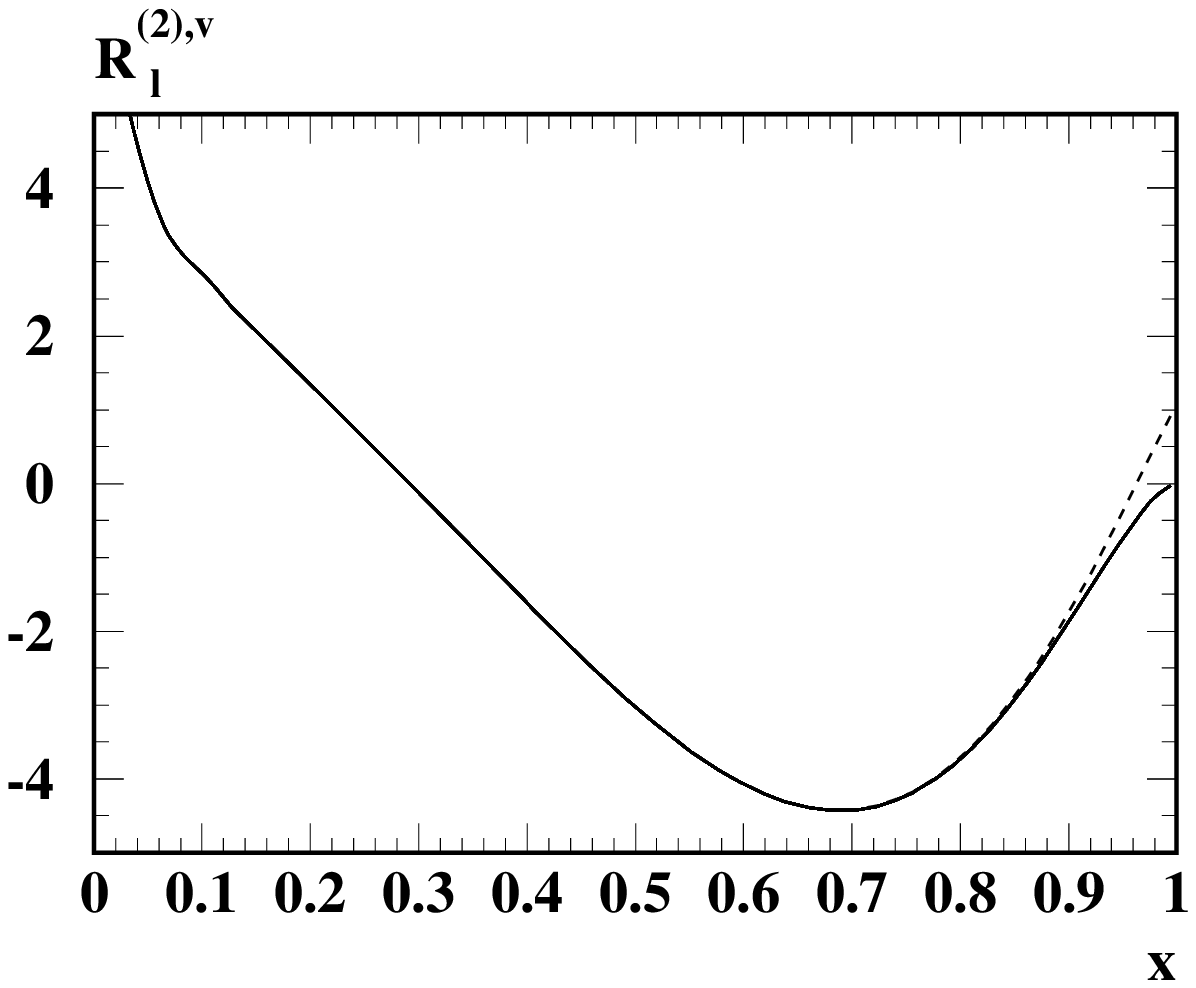}
      &
      \epsfxsize=7.cm
      \epsffile[110 280 460 560]{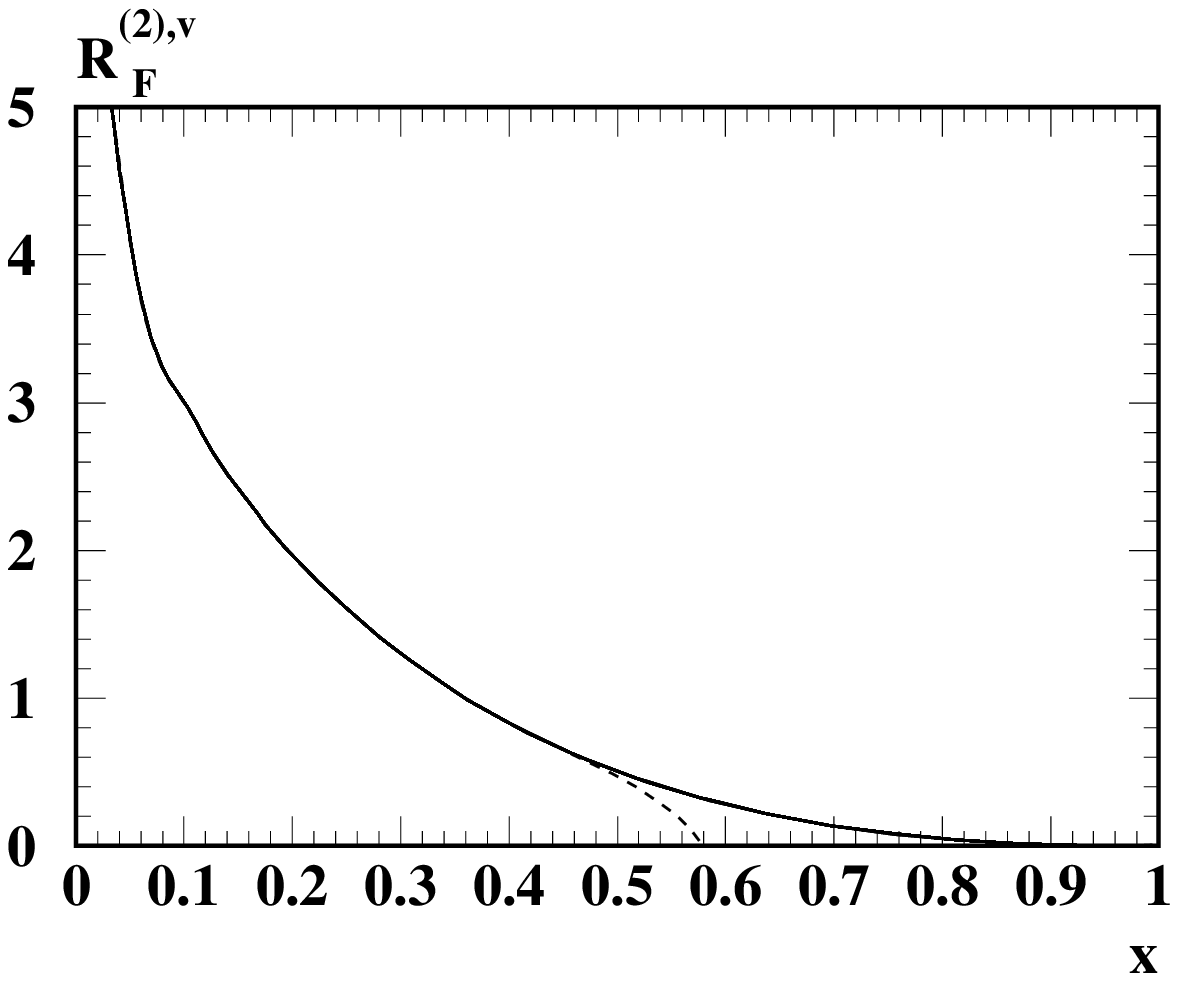}
    \end{tabular}
  \end{center}
  \caption{\label{fig:Rvx}$R^{(2),v}_A(s)$, $R^{(2),v}_{NA}(s)$,
    $R^{(2),v}_l(s)$  and $R^{(2),v}_F(s)$ as a
    function of $x$. The dashed curves correspond to the analytical 
    expressions obtained via asymptotic expansion containing
    the terms up to order $1/z^7$.
          }
\end{figure}

\begin{figure}[t]
  \begin{center}
    \begin{tabular}{cc}
      \leavevmode
      \epsfxsize=7.cm
      \epsffile[110 280 460 560]{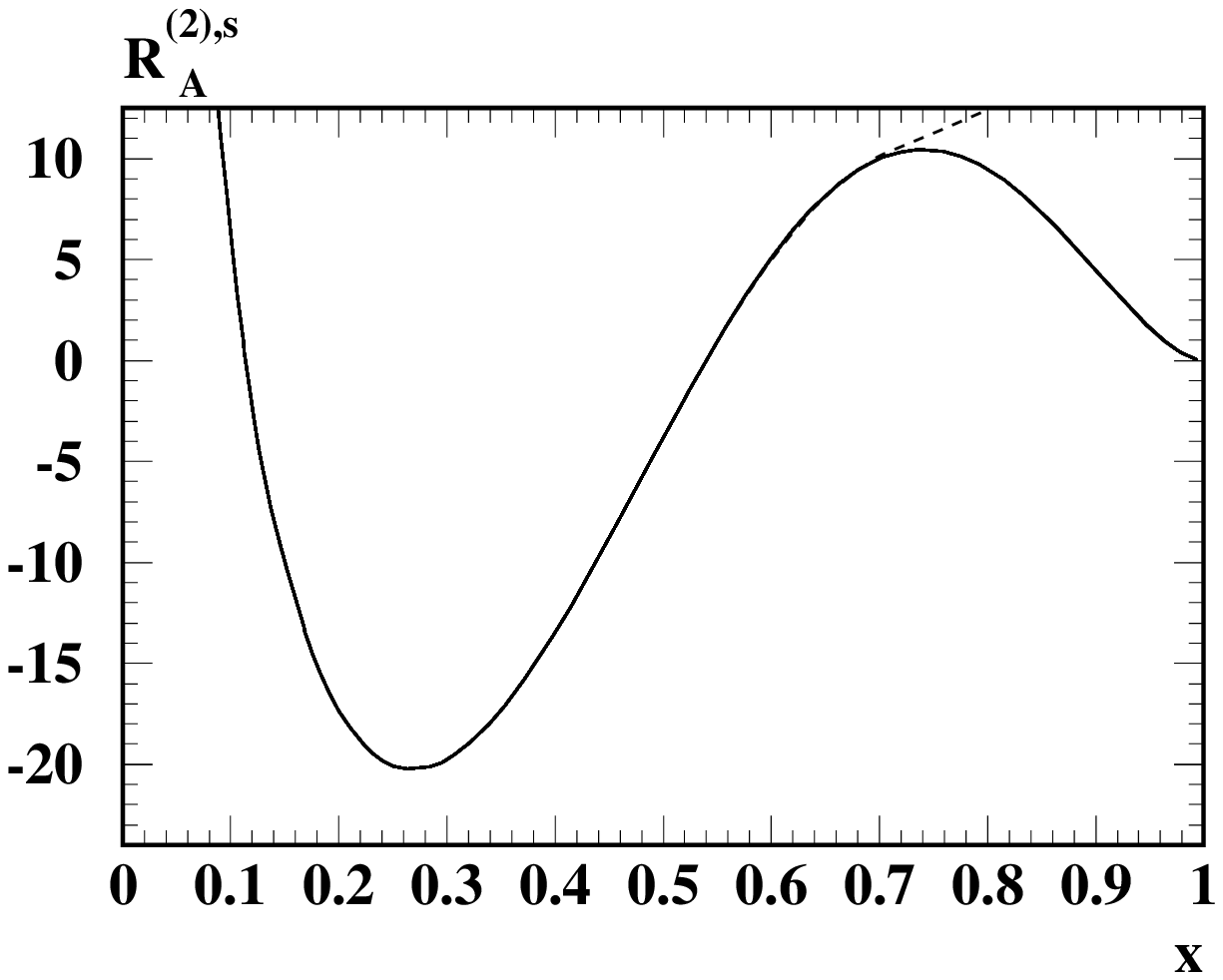}
      &
      \epsfxsize=7.cm
      \epsffile[110 280 460 560]{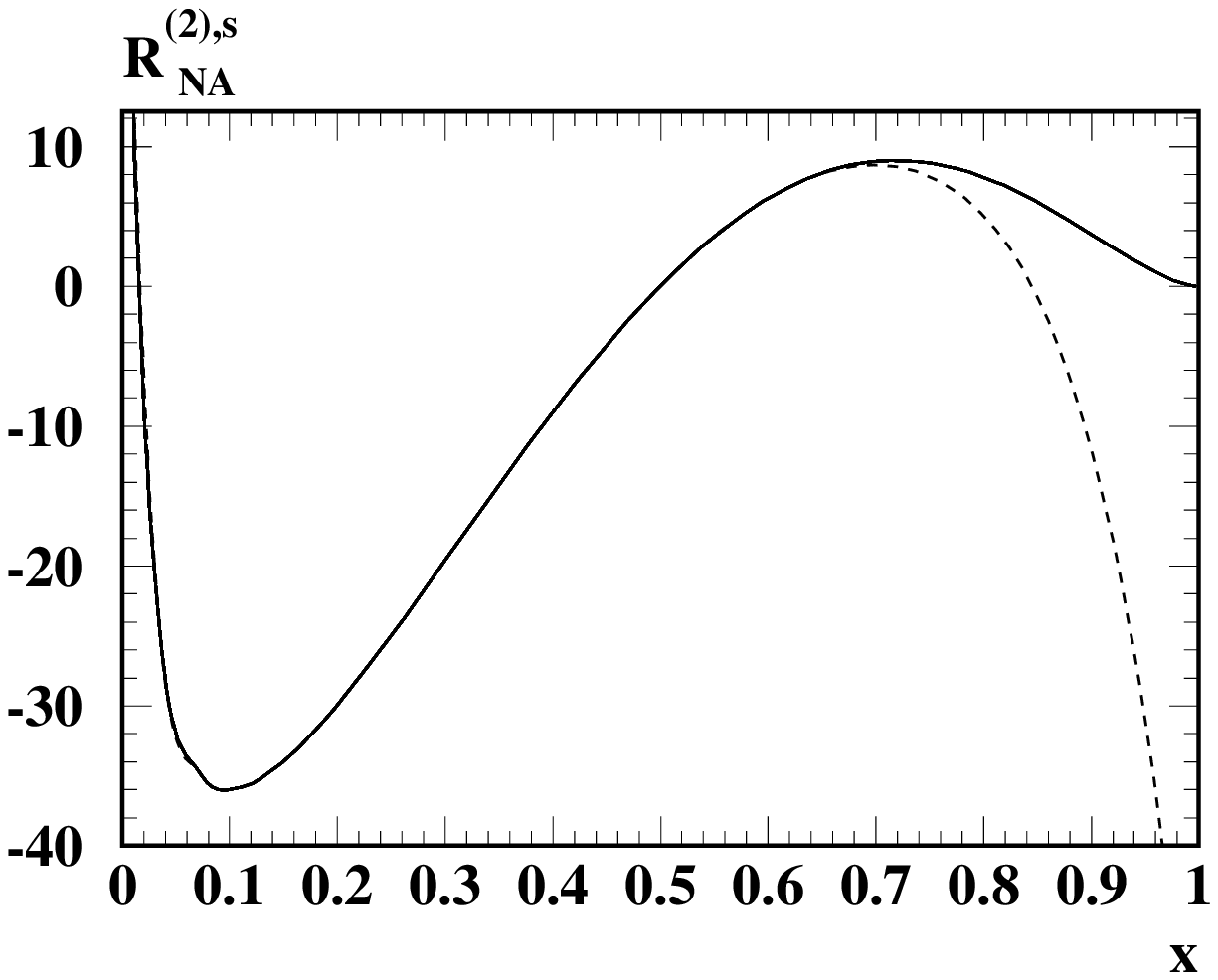}
      \\
      \epsfxsize=7.cm
      \epsffile[110 280 460 560]{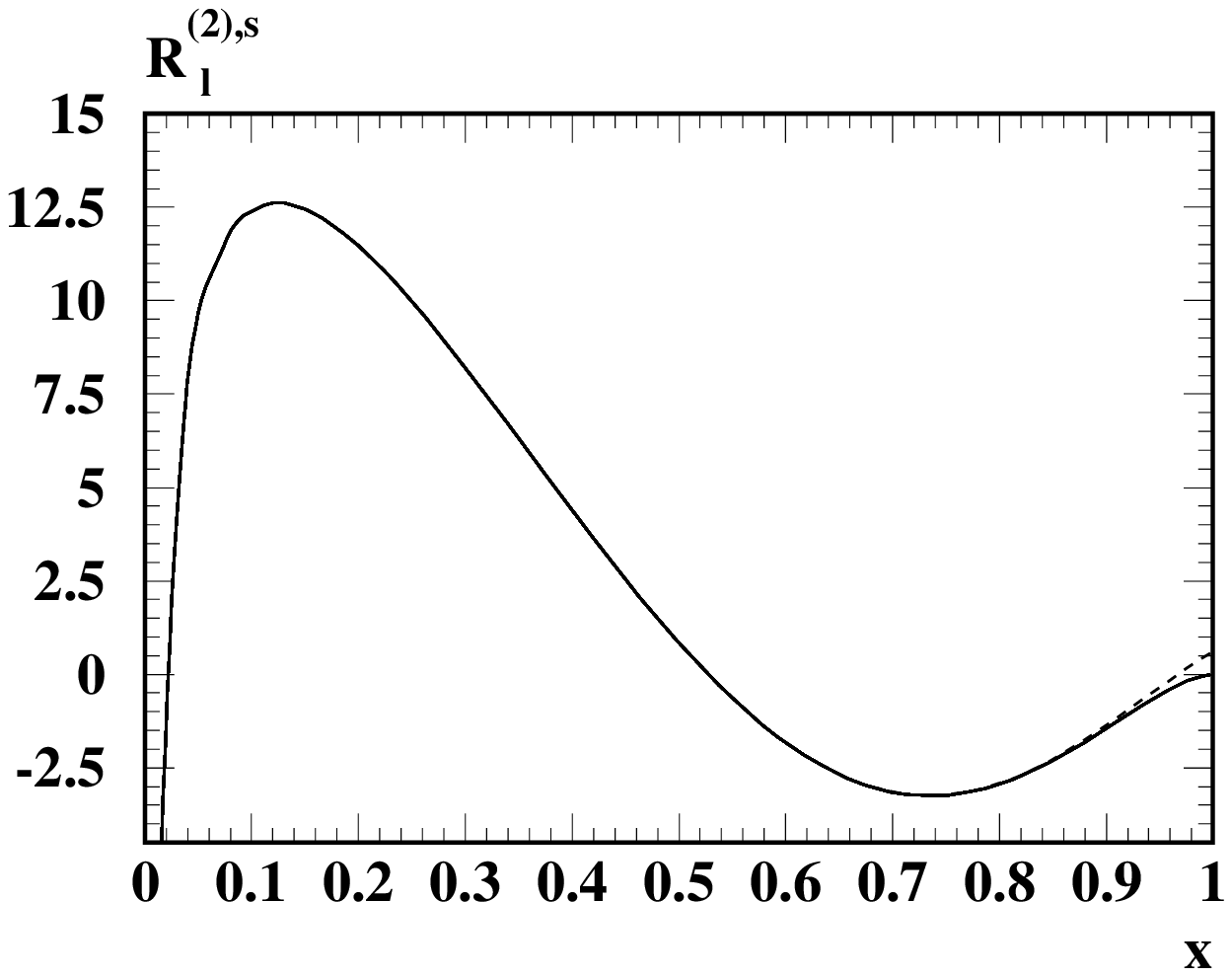}
      &
      \epsfxsize=7.cm
      \epsffile[110 280 460 560]{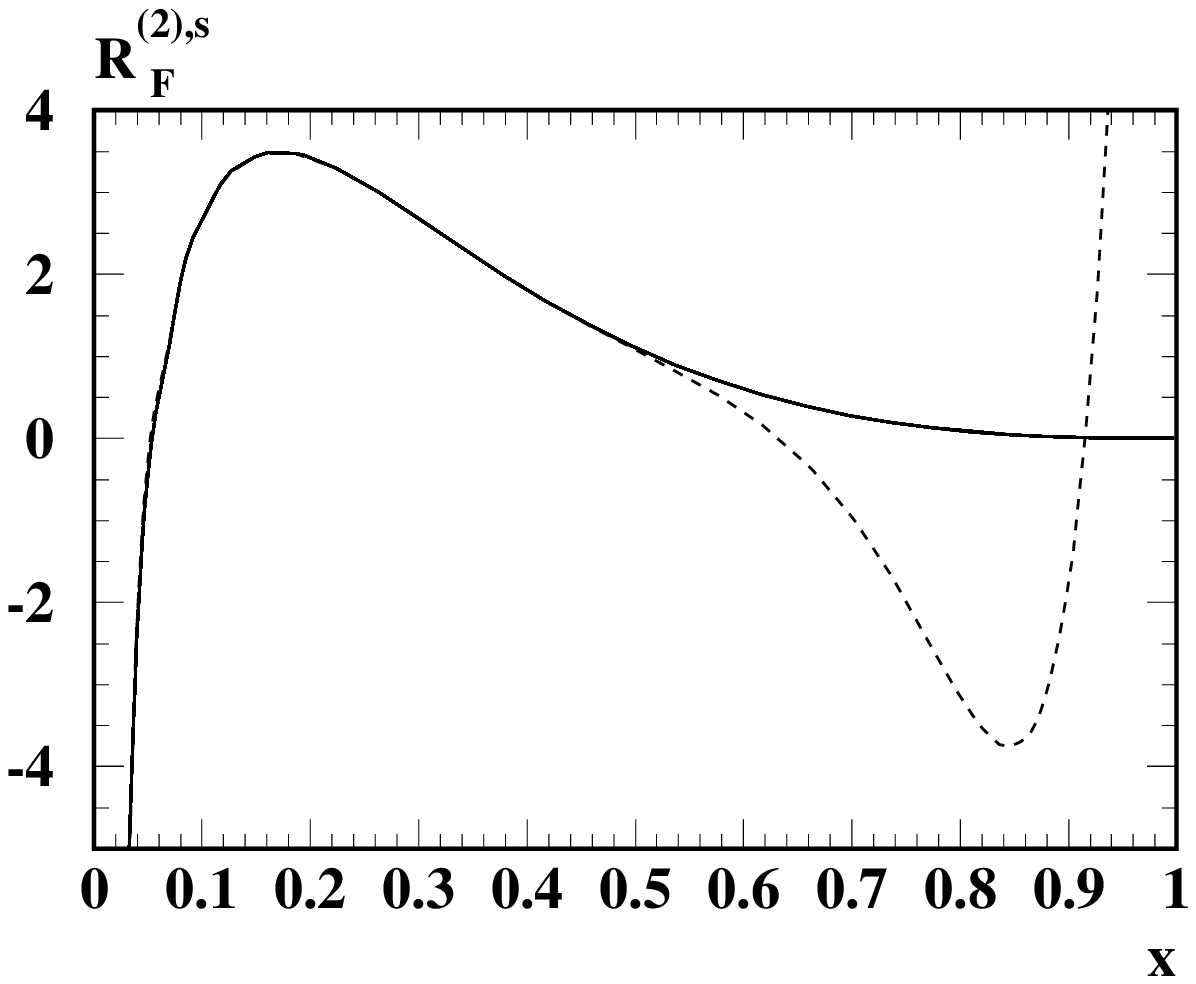}
    \end{tabular}
  \end{center}
  \caption{\label{fig:Rsx}$R^{(2),s}_A(s)$, $R^{(2),s}_{NA}(s)$,
    $R^{(2),s}_l(s)$  and $R^{(2),s}_F(s)$ as a
    function of $x$. The dashed curves correspond to the analytical 
    expressions obtained via asymptotic expansion containing
    the terms up to order $1/z^7$.
          }
\end{figure}

In Figs.~\ref{fig:Rvx} and~\ref{fig:Rsx}
the imaginary parts of the individual three-loop colour structures
for the vector and scalar correlators
are plotted as a function of the variable $x$.
Each solid line contains of the order of 15 Pad\'e approximants
which show perfect agreement among each other.
For comparison also the high-energy expansion terms including order
$1/z^7$ are shown as dashed curves.
Excellent agreement with the
semi-numerical Pad\'e results is observed up to 
$x\approx 0.5$ which corresponds to $v\approx 0.60$.
Some colour structures 
($R^{(2),v}_l(s)$, 
$R^{(2),s}_A(s)$, $R^{(2),s}_{NA}(s)$, $R^{(2),s}_l(s)$)
show even an agreement up to $x\approx 0.7$ ($v\approx 0.34$)
which is already fairly close to threshold.
We want to remind that the functions which exhibit next to the
cut at $\sqrt{s}=m$ also a cut at $\sqrt{s}=3m$ 
are not expected to converge to the correct answer
in the interval between $m$ and $3m$. This explains the somewhat crazy
behaviour of $R^{(2),s}_F(s)$ for larger values of $x$.

The non-diagonal correlators describe properties connected to
the $W$ boson. In particular,
a certain (gauge invariant) class of corrections to the Drell-Yan
process, i.e. to
the production of a quark pair through the decay of a virtual $W$
boson generated in $p\bar{p}$ collisions,
are covered by the vector and axial-vector correlator.
The absorptive part
is directly related to the decay width of the (highly virtual)
$W$ bosons into quark pairs and gluons.
Of particular interest in this connection is the single-top-quark
production via the process $q\bar{q}\to t\bar{b}$. The imaginary part
of the transversal $W$ boson polarization function constitutes a
finite and gauge invariant contribution of ${\cal O}(\alpha_s^2)$.

The corrections of order $\alpha_s$  to the (total) single-top-quark 
production rate are quite large. They amount to
about  54\% and 50\% for Tevatron and LHC energies,
respectively~\cite{SmithWillen96}, where 18\%, respectively, 17\% arise
from the final state corrections.
This calls for a complete ${\cal O}(\alpha_s^2)$ calculation.

If one considers the leading term of the large-$N_c$ limit
it is possible to use the results for $R^v$ 
to perform a theoretical analysis of
order $\alpha_s^2$ to the single-top-quark production. 
The production cross-section of the virtual $W^*$ boson is identical to
that of the Drell-Yan process $q\bar{q}\to e\bar{\nu}_e$.
The latter is known to
${\cal{O}}(\alpha^2_s)$ from Ref.~\cite{Drell_Yan}.
Thus we can take proper ratios to make predictions 
in the large-$N_c$ limit 
at NNLO free from any  dependence on parton distribution functions.  
As an example, we consider
\begin{eqnarray}
  \frac{\frac{{\rm d}\sigma}{{\rm d}q^2}\left(pp\to W^*\to tb\right)}
       {\frac{{\rm d}\sigma}{{\rm d}q^2}\left(pp\to W^*\to
       e\nu_e\right)}
  &=& \frac{\mbox{Im}\left[\Pi_{tb}(q^2)\right]}
           {\mbox{Im}\left[\Pi_{e\nu}(q^2)\right]} 
  \nonumber\\
  &=& N_c |V_{tb}|^2 R^v(s)
  \,.
  \label{eq:ratio}
\end{eqnarray}

\begin{figure}[t]
  \begin{center}
    \begin{tabular}{c}
      \leavevmode
      \epsfxsize=14.cm
      \epsffile[110 280 470 560]{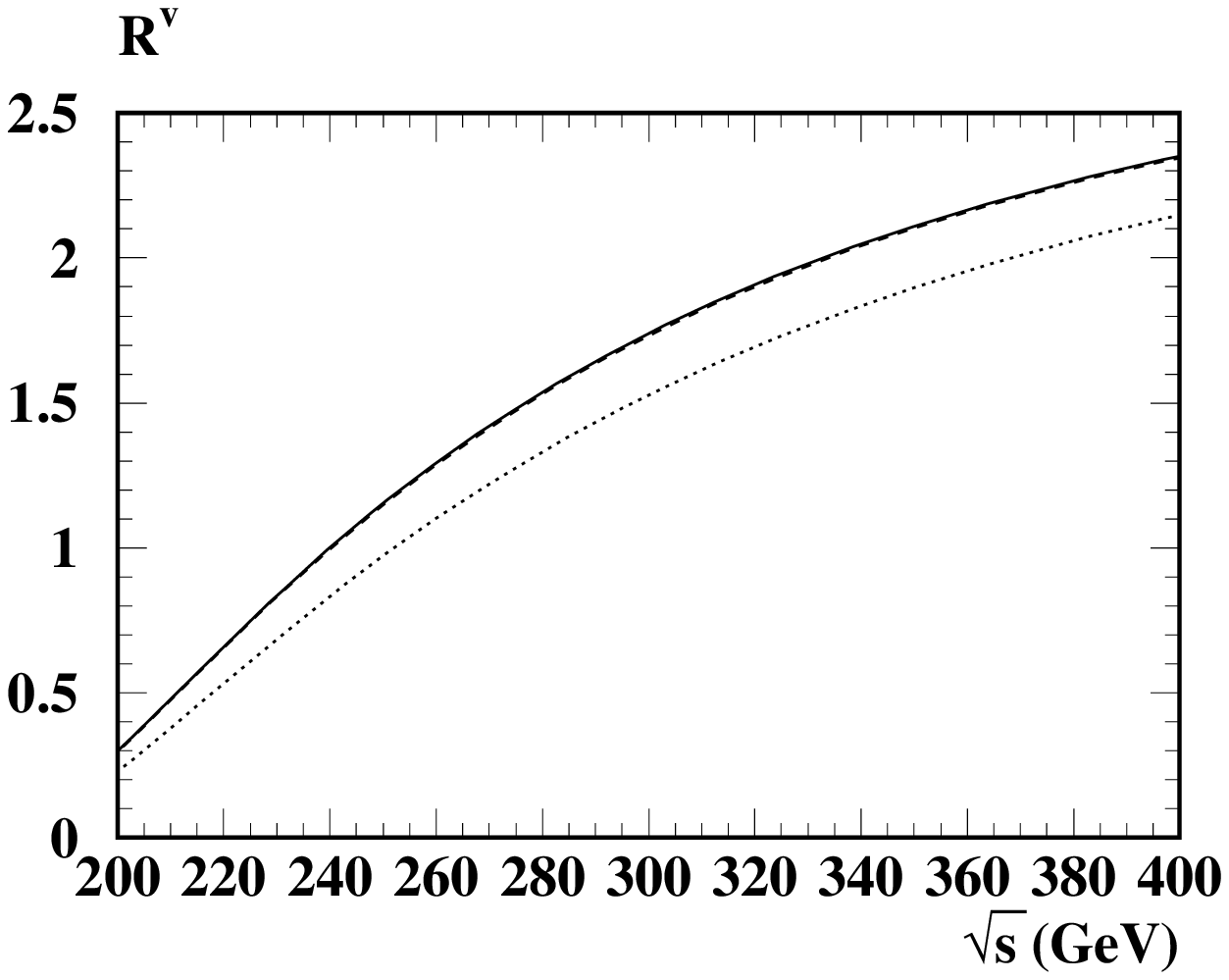}
    \end{tabular}
  \end{center}
  \caption{\label{fig:dsigma_qqtb}LO (dotted), 
    NLO (dashed) and NNLO (solid) results of $R^v(s)$.
          }
\end{figure}

The numerical significance of the order $\alpha_s^2$ corrections
is shown in Fig.~\ref{fig:dsigma_qqtb} where the LO, NLO and NNLO result
of $R^v(s)$ is plotted in the range $\sqrt{s} = 200 \ldots 400$ GeV. 
For the numerical values $M_t=175$~GeV and 
$\alpha_s(M_Z)=0.118$ has been chosen.
Whereas the ${\cal O}(\alpha_s)$ corrections are significant
there is only a moderate contribution from the order $\alpha_s^2$
terms. In the range in $q^2$ shown in Fig.~\ref{fig:dsigma_qqtb} they are
below 1\% of the Born result.
Note that (at least for $\mu^2=M_Z^2$)
the NNLO corrections to the Drell-Yan process are also 
small and amount to at most a few percent (see e.g.~\cite{Mar00}).

As an application of the scalar and pseudo-scalar
current correlator we want to
mention the decay of a charged Higgs boson which occurs in extensions
of the Standard Model. The corrections to $R^s$
describe the total decay rate into a massive and a massless quark.
To be more precise, the hadronic decay rate of the charged Higgs boson
takes the form
\begin{eqnarray}
  \Gamma(H^+ \to U \bar{D})
  & = &\frac{\sqrt{2}G_F}{8\pi}M_{H^+} (a^2+b^2) R^s(M_{H^+}^2)
  \,,
\end{eqnarray}
where $a$ and $b$ parametrize the coupling of the Higgs boson to the
massive quark $U$ and the massless quark $D$
\begin{eqnarray}
  J_{H^+} &=& 
  \frac{m_U}{\sqrt{2}} \,  
  \bar{U} \left[ a \, (1-\gamma_5) + b \,  (1+\gamma_5) \right] D
  \label{eq:Higgs_current} 
  \,.
\end{eqnarray}

\begin{figure}[t]
  \begin{center}
    \begin{tabular}{c}
      \leavevmode
      \epsfxsize=14.cm
      \epsffile[110 280 470 560]{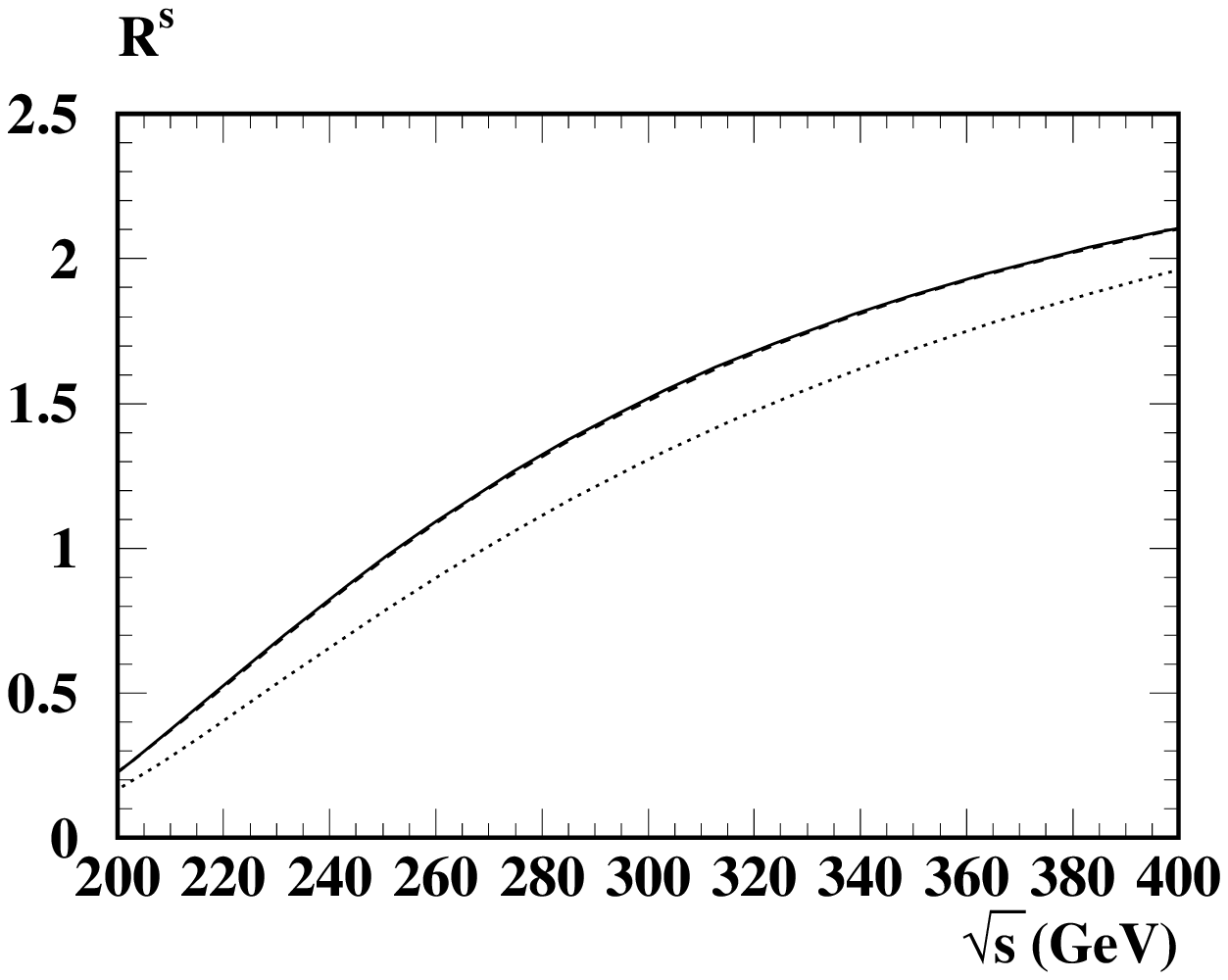}
    \end{tabular}
  \end{center}
  \caption{\label{fig:higgs}LO (dotted), 
    NLO (dashed) and NNLO (solid) results of $R^s(s)$, $M=M_t=175$~GeV.
          }
\end{figure}

In Fig.~\ref{fig:higgs} $R^s(s)$ is 
plotted at LO, NLO and NNLO~\cite{CheSte01,CheSte01_2}.
Again it turns out that the radiative corrections are well under
control as order $\alpha_s^2$ terms contribute at most of the order of 1\%.

\subsection{\label{sub:mudec}QED corrections to muon decay}

The Fermi coupling constant, $G_F$, constitutes together with 
the electromagnetic coupling constant and the mass of the $Z$ boson
the most precise input parameters of the SM of elementary
particle physics. 
$G_F$ is defined through the muon lifetime
which in turn is obtained from the decay rate.
The one-loop corrections of order $\alpha$ have been evaluated more
than 40 years ago in~\cite{KinSir59Ber58}.
Only recently the two-loop corrections of order $\alpha^2$ have been 
computed by two independent groups~\cite{RitStu99,SeiSte99}. 
The inclusion of the new terms leads to the following value for the
Fermi coupling constant~\cite{Gro00}
\begin{eqnarray}
  G_F &=& 1.16639(1) \times 10^{-5}~\mbox{GeV}^{-2}
  \,,
\end{eqnarray}
where the error is reduced by a factor two. It is now entirely
experimental.
In this Subsection we want to concentrate on the method used 
in~\cite{SeiSte99} which is based on Pad\'e approximation and
conformal mapping.
We also want to mention that
in Ref.~\cite{FerOssSir99} optimization methods have been used in order to  
estimate the coefficient of order $\alpha^3$.

It is common to both calculations of the order $\alpha^2$ corrections
that the muon propagator is considered in the framework
of the effective theory where the $W$ 
boson is integrated out. The QED corrections to the resulting Fermi contact 
interaction were shown to be finite to all orders~\cite{BerSir62}. It is quite
advantageous to perform a Fierz transformation which for a pure $V-A$ theory
has the consequence that afterwards the two neutrino lines appear in the same 
fermion trace. Thus the QED corrections only affect the fermion trace 
involving
the muon and the electron. This also provides some simplifications in the
treatment of $\gamma_5$ since in the case of vanishing electron mass a fully 
anticommuting prescription can be used. 

\begin{figure}[t]
\begin{center}
      \leavevmode
      \epsfxsize=\textwidth
      \epsffile[70 700 510 740]{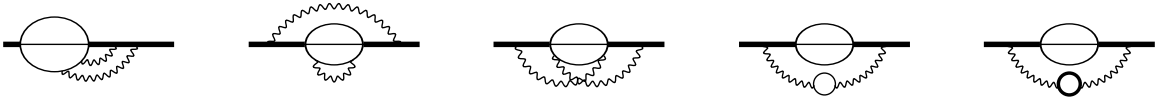}
\end{center}
\caption{\label{fig:bdec}Sample diagrams for the muon (thick line) 
self energy. Two of the thin
lines represent the electron and the corresponding neutrino
and the third one the muon neutrino. All
(one-particle-irreducible) diagrams involving the coupling of the
photon (wavy lines) to the muon and electron have to be taken into
account.
}
\end{figure}

The decay rate can be written in the 
form\footnote{Some discussion of Eq.~(\ref{eq:gam}) --- in particular
  the additional factor $z$ in front of $S_V^{\rm OS}$ --- can be found 
  in~\cite{CheHarSeiSte99}.}
\begin{eqnarray}
  \Gamma &=& 
  2 M \, \mbox{Im} \left[ z \, S_V^{\rm OS} - S_S^{\rm OS}
  \right]\bigg|_{z=1}, 
\label{eq:gam}
\end{eqnarray}
where
\begin{eqnarray}
  S_S^{\rm OS} 
  \,\,=\,\, Z_2^{\rm OS} Z_m^{\rm OS} \left( 1 - \Sigma_S^0 \right)
  \,,&&
  \qquad
  \label{eq:sssv}
  S_V^{\rm OS} \,\,=\,\, Z_2^{\rm OS}\left(1+\Sigma_V^0\right)
  \,,
\end{eqnarray}
are functions of the variable
\begin{equation}
  z = {q^2\over M_\mu^2}\,.
\end{equation}
$\Sigma_S^0$ and $\Sigma_V^0$ represent the scalar and vector part of the 
muon propagator. They are functions of the 
external momentum $q$ and the bare mass $m^0$.
In our case they further depend on the bare electromagnetic 
coupling $\alpha^0$ 
and are proportional to the square of the Fermi coupling 
constant, $G_F^2$.
Typical diagrams contributing to $\Sigma_S^0$ and $\Sigma_V^0$
are depicted in Fig.~\ref{fig:bdec}.
$Z_2^{\rm OS}$ and $Z_m^{\rm OS}$ represent the wave function and mass
renormalization in the on-shell scheme.

It is convenient to parameterize the QED corrections for the
muon decay in the following form
\begin{eqnarray}
  \Gamma(\mu\to \nu_\mu e \bar{\nu}_e) &=& \Gamma^0_\mu\left[
    A_\mu^{(0)}
    +\frac{\bar\alpha}{\pi} A_\mu^{(1)}
    +\left(\frac{\bar\alpha}{\pi}\right)^2 
        \left( A_\mu^{(2)} + A_{\mu,\tau}^{(2)} 
      + A_{\mu,{\rm had}}^{(2)} \right)
    +\ldots
  \right]
  \,,
  \label{eq:gamparmu}
\end{eqnarray}
with $\Gamma^0_\mu=G_F^2 M_\mu^5 /(192\pi^3)$.
$\bar{\alpha}=\bar{\alpha}(\mu)$ represents the electromagnetic coupling in
the $\overline{\rm MS}$ scheme
and $A_{\mu,\tau}^{(2)}$ corresponds to the 
contribution involving a virtual $\tau$ loop.
It is suppressed by $M_\mu^2/M_\tau^2$ and
almost four orders of magnitudes smaller than the other
terms~\cite{RitStu98}.
The hadronic contribution is denoted by 
$A_{\mu,{\rm had}}^{(2)}$~\cite{RitStu98}. 

To lowest order the result is known for finite electron mass, $M_e$
\begin{eqnarray}
  A_\mu^{(0)} &=& 1 - 8 \, \frac{M_e^2}{M_\mu^2} 
  - 12 \, \frac{M_e^4}{M_\mu^4} \, \ln \frac{M_e^2}{M_\mu^2} 
  + 8 \, \frac{M_e^6}{M_\mu^6} - \frac{M_e^8}{M_\mu^8}
  \,,
\end{eqnarray}
and in the limit $M_e=0$ we obtain for the one-loop corrections
\begin{eqnarray}
  A_\mu^{(1)} &=& \frac{25}{8}-\frac{\pi^2}{2} \,\,\approx\,\,-1.810
  \,.
\end{eqnarray}
The approach chosen in~\cite{RitStu99} to evaluate $A_\mu^{(2)}$
has some similarity to the computation of the four-loop contribution to the
$\beta$ function~\cite{RitVerLar97_bet} (see also
Section~\ref{sub:massint}). In fact, since only the imaginary part
has to be calculated one is only interested in the pole part
like in the case of a $\overline{\rm MS}$ renormalization constant.
However, only the pole part arising from the
cuts through the electron and neutrino lines have to be taken into
account and not the ones through the muon
line. This is ensured by considering the on-shell muon propagator
which requires the evaluation of (the imaginary part of)
four-loop on-shell integrals with external momentum $q^2=M_\mu^2$.
Integration-by-parts relations~\cite{CheTka81} are applied to reduce
the number of occuring integrals to a small set of so-called master
integrals. Only for the latter a hard calculation is necessary.
Usually a large number of terms is generated in intermediate steps
while using integration-by-parts relations. However, 
those four-loop integrals that have no imaginary part can immediately
be discarded. This includes four-loop vacuum graphs and diagrams
with a through-going on-shell line.
A comprehensive discussion and lots of intermediate results can be
found in Ref.~\cite{RitStu99}.

The approach of~\cite{SeiSte99} used
for the computation of the QED corrections of order
$\alpha^2$ is based on an expansion of
the full fermion propagator in the limit
$M_\mu^2 \gg q^2$,
where $q$ is the external momentum and $M_\mu$ is the on-shell
propagator mass of the muon. The on-shell 
limit $q^2 \to M^2$ will be performed afterwards with the help of Pad\'e 
approximations. This, of course, only provides an approximation to the exact
result. However, the integrals to be evaluated are simplified
considerably.
Furthermore, the accuracy obtained with this method is sufficient to 
check the result of~\cite{RitStu99}
and enables the same reduction of the theoretical 
error on $G_F$. 
Good convergence to the exact result is also expected from the
analysis performed in~\cite{FleSmiTar97} where the hard-mass procedure
has been applied to scalar two-loop integrals involving massless
thresholds.

Due to the Fierz transformation the loop integration connected to
the two neutrino lines can be performed immediately as it constitutes a 
massless two-point function. This is also the case after allowing for 
additional photonic corrections. As a result one encounters in the resulting 
diagram a propagator with one of the momenta raised to power $\varepsilon$
where $D=4-2\varepsilon$ is the space-time dimension. This slightly increases
the difficulty of the computation of the resulting
diagrams. Especially for the 
order $\alpha^2$ corrections, where the original four-loop diagrams
are reduced  
to three-loop ones with non-integer powers of denominators, it is a priori not
clear that these integrals can be solved analytically. However, it turns out 
that for the topologies needed in our case this is indeed possible. For the 
computation of the massless two-point functions we have used the package 
{\tt MINCER}~\cite{mincer}. Only slight modifications are necessary in
order to use
this package also for the computation of the new type of integrals.

In contrast to the current correlators considered in
Section~\ref{sub:veccor} 
the expansion terms for $z\to-\infty$ can not be used as they
describe the unphysical process $\mu\to\mu+\gamma$.
Another difference is the
presence of massless cuts in the limit $q^2\to0$.
Thus a naive expansion is not possible and rather the asymptotic
expansion has to be applied, which generates 
from the 44 contributing four-loop diagrams 72 sub- and
cosub-diagrams that have to be evaluated.
The analytical results are rather lengthly and cannot be listed.
Instead the results are presented in numerical form~\cite{SeiSte99}.

In order to get reliable results it is necessary to compute as many terms as
possible in the expansion parameter $z$. Subsequently a Pad\'e approximation 
is applied as described in Section~\ref{sub:method}. We want
to recall that before the Pad\'e procedure a conformal mapping can be used 
which maps the complex $z$-plane into the interiour of the unit circle. 
Following Ref.~\cite{CheHarSeiSte99} we denote those results by 
$\omega$-Pad\'es and the ones obtained without conformal mapping by 
$z$-Pad\'es.

The calculation is performed with the help of the package 
{\tt GEFICOM}~\cite{geficom}. It uses {\tt QGRAF}~\cite{qgraf} for the 
generation of the diagrams and {\tt EXP}~\cite{Sei:dipl} for the
application of 
the asymptotic expansion procedures.

\begin{table}[t]
\begin{center}
\begin{tabular}{|l|l||c|c|}
\hline
input & P.A. & $z$ & $\omega$ \\
\hline
6 & [3/2] & 5.836 & 7.249 \\
6 & [2/3] & 5.836 & 7.057 \\
\hline
7 & [4/2] & 5.935 & 7.040 \\
7 & [3/3] & 5.833 & 7.076 \\
7 & [2/4] & 5.938 & 7.080 \\
\hline
8 & [4/3] & 6.110 & 6.873 \\
8 & [3/4] & 6.113  & 7.060 \\
\hline
\multicolumn{2}{|c||}{exact:} & \multicolumn{2}{|c|}{6.743} \\
\hline
\end{tabular} 
\end{center}
\caption{\label{tab:Oal2_2}Pad\'e results for the corrections of ${\cal
    O}(\alpha^2)$ to the muon decay, $A^{(2)}_\mu$.
    The first row indicates the order in $z$ which has been
    used to construct the Pad\'e approximations.}
\end{table}

The method has been successfully tested at Born level and  
at order $\alpha$ where a large number of moments 
can be evaluated.
This gives a hint on 
how of many terms are necessary at ${\cal O}(\alpha^2)$ in order to obtain a
reliable answer.
The results for $A_\mu^{(2)}$ of the individual Pad\'e approximations 
are shown in Tab.~\ref{tab:Oal2_2}.
It leads to the final answer~\cite{SeiSte99}
\begin{eqnarray}
  A_\mu^{(2)}  &=& 6.5(7) \,,
\end{eqnarray}
where the  
deviation of the central value from the exact result of 6.743 is less
than 3\% and well covered by the extracted error of roughly 10\%. Thus
the sole 
knowledge of our results would also reduce the theoretical error on
$G_F$.
Finally, the total decay rate of the muon takes the form
\begin{eqnarray}
  \Gamma(\mu\to \nu_\mu e \bar{\nu}_e) &=& \Gamma^0_\mu\left[
    0.9998
    - 1.810 \frac{\bar\alpha(M_\mu)}{\pi}  
    + 6.700(2) \left(\frac{\bar\alpha(M_\mu)}{\pi}\right)^2 
    +\ldots
  \right]
  \,,
  \label{eq:gammu}
\end{eqnarray}
where $A_{\mu,{\rm had}}^{(2)}=-0.042(2)$~\cite{RitStu98} and
$A_{\mu,\tau}^{(2)}=-0.00058$~\cite{RitStu98} has been used. Furthermore 
$\mu^2=M_\mu^2$ has been adopted.

As already noted in~\cite{FerOssSir99} the numerical coefficient in front of 
the second order corrections becomes very 
small\footnote{Instead of ``6.7'' one has ``0.27'' in Eq.~(\ref{eq:gammu}).} 
if one uses the on-shell 
scheme for the definition of the coupling constant 
$\alpha$. Then the
$\overline{\mbox{MS}}$ coupling is given by 
$\bar\alpha(M_\mu)=\alpha(1+\alpha/(3\pi) \ln(M_\mu^2/M_e^2))$ and
there is an accidental cancellation between the constant and the
logarithm in the second order corrections. 

A similar kinematical situation as in the $\mu$ decay is also given
for the semileptonic decay of a bottom quark, $b\to ue\nu_e$.
From the technical point of view the difference is only due to the
non-abelian structure of QCD. The two methods described above have
been applied to the total rate
\begin{eqnarray}
  \Gamma(b\to ue\bar{\nu}_e) &=& \Gamma^0_b\left[
    1
    - 2.413 \frac{\alpha_s(M_b)}{\pi} 
    + A_b^{(2)} \left(\frac{\alpha_s(M_b)}{\pi}\right)^2 
    +\ldots
  \right]
  \,,
\end{eqnarray}
where $\Gamma^0_b=G_F^2 M_b^5 |V_{ub}|^2/(192\pi^3)$.
The order $\alpha_s^2$ results read
$A_b^{(2)}=-21.296$~\cite{Rit99} 
and
$A_b^{(2)}=-21.1(6)$~\cite{SeiSte99}.
Again perfect agreement between the two methods is observed.

At the end of this Subsection we want to mention that similar methods
have been used to compute the decay rate of a top quark into a $W$
boson and a bottom quark~\cite{CheHarSeiSte99}. 
Contrary to the case of the muon the $W$
boson is not integrated out from the Lagrangian. Thus, at order
$\alpha_s^2$ three-loop diagrams contributing to the top quark
propagator have to be considered.
In~\cite{CheHarSeiSte99} a double expansion in $q^2/M_t^2$ and 
$M_W^2/M_t^2$ has been performed, where $q$ is the external momentum
of the top quark propagator, which leads to a reliable prediction for
$\Gamma(t\to Wb)$ including finite $W$-mass effects.


\subsection{\label{sub:msos}The 
  relation between the $\overline{\rm MS}$ and on-shell quark mass}

In this Subsection we consider the relation between the 
on-shell and the $\overline{\rm MS}$ quark mass at three-loop order in QCD.
The result has been obtained for the first time
with the help of the Pad\'e method~\cite{CheSte99}
and has been confirmed half a year later by a completely
independent calculation~\cite{MelRit99}.

Here, we want to describe the approach of~\cite{CheSte99}
as it constitutes an other facet of applications of the 
method as described in Section~\ref{sub:method}.
In contrast to the examples described before one is
directly interested in the real part of the considered function.
Furthermore, in the relation between the $\overline{\rm MS}$ and
the on-shell value of the masses we want to know
the function $f(z)$ as defined in~(\ref{eq:deff}) 
for $z=1$, whereas it is only available for small and large values of $z$.
Nevertheless, the method is powerful enough to get the
value for $f(1)$ with an error of $2-3\%$ (see below).

The basic object entering the mass relation is the fermion
propagator, $\Sigma(q)$. However, the Pad\'e method cannot be applied
directly to $\Sigma(q)$ as it contains (unknown) singularities
at threshold. Thus, proper combinations have to be considered which
are regular for $z=1$. They are obtained from the requirement that
the inverse fermion propagator has a zero at the position of the
on-shell mass.

In the following three different types of masses will appear:
the bare mass, $m_0$, the $\overline{\rm MS}$, $m(\mu)$ and the
on-shell mass $M$. The relation between them is given by
\begin{eqnarray}
  m(\mu) &=& Z_m m^0 \,\,=\,\,z_m(\mu) M
  \,,
  \label{eq:mmsos}
\end{eqnarray}
where $z_m$ is finite and has an explicit dependence on the
renormalization scale $\mu$. It is the purpose of this Section to
describe its calculation at order $\alpha_s^3$.

As already mention above, in order to obtain the mass relation we have
to consider the fermion propagator as shown in Eq.~(\ref{eq:sfinv0}).
The renormalized version can be cast in the form
\begin{eqnarray}
  \left(S_F(q)\right)^{-1} &=& i \left[
    \left(M-\qsla\right) S_V(z) + M \left(z_m(\mu) S_S(z) -S_V(z) \right)
  \right]
    \,,
\label{eq:sfinv}
\end{eqnarray}
with\footnote{Note that in contrast to the quantities defined
  in Eq.~(\ref{eq:sssv}) (see also Ref.~\cite{CheHarSeiSte99})
  the wave function renormalization for functions $S_{S/V}$
  is still defined in the $\overline{\rm MS}$ scheme.}
\begin{eqnarray}
  S_V(z) &=& Z_2(1+\Sigma^0_V)
  \,,
  \nonumber\\
  S_S(z) &=& Z_2Z_m(1-\Sigma^0_S)
  \,.
  \label{eq:SvSs}
\end{eqnarray}
$Z_2$ denotes the wave function renormalization in the $\overline{\rm MS}$
scheme which is sufficient for our considerations.
Note that the functions $S_S$ and $S_V$ are $\overline{\rm MS}$ quantities
which later on are expressed in terms of the on-shell mass.
The two-loop relation between $m$ and $M$ is enough to
do this at order $\alpha_s^3$.
It is convenient to write the functions $S_{S/V}$ in the following way
\begin{eqnarray}
S_{S/V} = 1 
+ \sum_{n\ge1} S_{S/V}^{(n)} \left(\frac{\alpha_s}{\pi}\right)^n
\label{eq:decas}
\,,
\end{eqnarray}
where the quantities $S_{S/V}^{(n)}$ exhibit the following colour
structures
(the indices $S$ and $V$ are omitted in the following):
\begin{eqnarray}
S^{(1)} &=& C_F S_F 
\,,
\nonumber\\
S^{(2)} &=& C_F^2 S_{FF} + C_FC_A S_{FA} + C_FTn_l S_{FL} + C_FT S_{FH}
\,,
\nonumber\\
S^{(3)} &=& C_F^3 S_{FFF} + C_F^2C_A S_{FFA} + C_FC_A^2 S_{FAA} 
          + C_F^2Tn_l S_{FFL}   + C_F^2T S_{FFH} 
\nonumber\\&&\mbox{}
          + C_FC_ATn_l S_{FAL}  + C_FC_AT S_{FAH}
          + C_FT^2n_l^2 S_{FLL} + C_FT^2n_l S_{FLH} 
\nonumber\\&&\mbox{}
          + C_FT^2 S_{FHH}
\,.
\label{eq:deccf}
\end{eqnarray}
The same decomposition also holds for the function $z_m$.
In~(\ref{eq:deccf}) $n_l$ represents the number of light (massless) quark
flavours. $C_F$ and $C_A$ are the Casimir operators of the fundamental and
adjoint representation. In the case of $SU(N_c)$ they are given by
$C_F=(N_c^2-1)/(2N_c)$ and $C_A=N_c$. The trace normalization of the
fundamental representation is $T=1/2$.
The subscripts $F$, $A$ and $L$ in Eq.~(\ref{eq:deccf})
shall remind us on the colour factors $C_F$,
$C_A$ and $Tn_l$, respectively. $H$ simply stands for the
colour factor $T$.

A formula which allows for the computation 
of the $\overline{\rm MS}$--on-shell relation for the quark
mass is obtained from the requirement that the inverse fermion propagator has
a zero at the position of the on-shell mass:
\begin{eqnarray}
\left(S_F(q)\right)^{-1}\bigg|_{q^2=M^2} &=& 0
\,.
\label{eq:oscond}
\end{eqnarray}
In the literature there are two different approaches to compute the
occuring Feynman diagrams. In the first evaluation of the three-loop
$\overline{\rm MS}$--on-shell relation the method of
Section~\ref{sub:method} has been applied which we will discuss
below.
On the contrary, a subsequent analysis~\cite{MelRit99} 
has chosen $q^2=M^2$ from the
very beginning which makes it necessary to solve three-loop on-shell
integrals. This can effectively be done using the 
integration-by-parts method within dimensional
regularization~\cite{CheTka81}. It enables the derivation of
recurrence relations which express complicated integrals in terms of
simpler ones. At the end one arrives at a small set of integrals ---
so-called master integrals --- which
actually have to be evaluated.
In~\cite{MelRit99,Melnikov:2000zc} 
the considerations of~\cite{LapRem96} have been
extended and the missing master integrals have been evaluated.
The technique used for the computation is based on 
the hard-mass procedure for large $M$ 
which represents the on-shell integrals in terms
of a power series in $q^2/M^2$.
The coefficients contain nested harmonic sums which in the on-shell
limit, i.e. for $q^2=M^2$,
can be reduced to known mathematical constants.
For explicit examples we refer to~\cite{Melnikov:2000zc}.
At the end of this section we will list the analytical result 
for $z_m(\mu)$ obtained in~\cite{MelRit99}.

The starting point for the approach of~\cite{CheSte99,CheSte00}
is Eq.~(\ref{eq:oscond}).
Applying it to Eq.~(\ref{eq:sfinv}) leads to the
condition
\begin{eqnarray}
h(z) &\equiv& z_m(\mu)\,S_S(z) - S_V(z) \,\,=\,\, 0 \qquad \mbox{for}
\qquad z=1
\,.
\label{eq:f}
\end{eqnarray}
At a given loop-order $L$, Eqs.~(\ref{eq:SvSs})
are inserted and the resulting equation is solved for $z_m^{(L)}$.
Thus Eq.~(\ref{eq:f}) can be cast in the form
\begin{eqnarray}
h(z) &=& f(z) + z_m^{(L)} \left(\frac{\alpha_s}{\pi}\right)^L
\,.
\label{eq:f2}
\end{eqnarray}
Our aim is the computation of $f(1)$.
Note that the individual self energies $\Sigma_S$ and $\Sigma_V$
develop infra-red singularities when they are evaluated on-shell.
The proper combination which leads to the relation between the 
$\overline{\rm MS}$ and on-shell mass is, however, free of infra-red
problems.

The explained procedure has been applied to each colour structure occuring
in $f(z)$ separately as outlined in~\cite{CheSte00}.
E.g., collecting all terms proportional to $C_F^3$ in
Eq.~(\ref{eq:f2}) leads to
\begin{eqnarray}
  f_{FFF}(z) &=& S_{S,FFF}(z) - S_{V,FFF}(z)
  + z_m^F S_{S,FF} + z_m^{FF}S_{S,F}
  \,,
  \label{eq:gFFF}
\end{eqnarray}
which has to be evaluated for $z=1$.
The individual terms on the right-hand side 
develop a (unknown) singular behaviour 
which is encoded in the corresponding moments.
In case the Pad\'e method is (naively) perfomed with the individual pieces,
the Pad\'e approximants try to imitate the threshold singularity.
However, due to the very construction of the method 
the analytical structure of the Pad\'e result is polynomial for
$z\to1$ and the typical threshold logarithms can not be reproduced.
Thus, the results show instabilities in the vicinity of $z=1$.
On the contrary, the proper combination as given in Eq.~(\ref{eq:gFFF})
has to be regular for $z\to1$, as the on-shell mass does not contain
any infra-red singularity~\cite{Kro98,Gambino:2000ai}. The corresponding
Pad\'e results demonstrate great stability.

To summarize, although superficially
only information about small and large momenta enter the 
Pad\'e procedure, it is sensitive to the analytical structure
at threshold 
as the information about the singularity is,
to some extend, also contained in the moments of the analytical
function $f(z)$.

At this place we will refrain from the discussion of the individual
colour structures which can be found in~\cite{CheSte00}
but only present the results for the sum where the numerical values
for $C_A$, $C_F$, $T$ and $n_l$ have been inserted.
Still care has to be taken because of the diagrams involving a closed
heavy quark loop.

It was already realized in~\cite{CheKueSte96} that the Pad\'e procedure
shows less stability as soon as diagrams are involved which
exhibit more than one particle threshold. In our case the interest is
in the lowest particle cut which happens to be for $q^2=M^2$.
The Pad\'e method heavily relies on the combination of expansions in the small
and large momentum region. The large momentum expansion, however, is
essentially sensitive to the highest particle threshold. Thus, if this
threshold numerically dominates the lower-lying ones it
cannot be expected that the Pad\'e approximation leads to stable results.
In such cases a promising alternative to the above method is the one where
only the expansion terms for $q^2\to0$ are taken into account in order to
obtain a numerical value at $q^2=M^2$.
This significantly reduces the calculational effort as
the construction of the Pad\'e approximation from low-energy moments alone is
much simpler.
In practice this approach will be applied if the Pad\'e results involving
also the high-energy data look ill-behaved.

In the present analysis diagrams with other cuts than for $q^2=M^2$
are already present at the two-loop level (see Fig.~\ref{fig:fpdiags})
which allows us to test these ideas.
Indeed, taking into account terms up to order $z^5$ and performing
a Pad\'e approximation there is an agreement of four digits with the
exact result~\cite{CheSte00}.
Also at three-loop order either $q^2=M^2$ or $q^2=9M^2$ cuts
appear. Cuts involving five or more fermion lines are first possible starting
from four-loop order. Note that cuts involving an even number of fermions
cannot occur.

Concerning the colour structures introduced in Eq.~(\ref{eq:deccf})
we use for the sum of the structures
$FFF$, $FFA$, $FFL$, $FAA$, $FAL$ and $FLL$
both the low- and high-energy moments whereas for the sum of the structures
$FFH$, $FAH$, $FLH$ and $FHH$
only the expansion for $z\to0$ is used.
In Tabs.~\ref{tab:TOTL} and~\ref{tab:TOTH} 
the results for different Pad\'e approximations
are listed. $n$ indicates the number of low-energy moments involved in the
analysis, i.e. $n=6$ implies the inclusion of terms of ${\cal O}(z^6)$.
The number of high-energy terms can be obtained in combination with
the order of the Pad\'e approximant ($[x/y]$) and is given by
$x+y+1-n$.

\begin{table}[t]
{\footnotesize
\begin{center}
\begin{tabular}{|l|l||r|r|r|r|r|r|} 
\hline
$n$ & P.A.& $ n_l=0 $& $ n_l=1 $& $ n_l=2 $& $ n_l=3 $& $ n_l=4 $& $ n_l=5 $\\
\hline
$5$ & $[4/5]$ &  $-200.2787$ &  $-173.6663$ &  $-148.3787$ &  $-124.4156$ &  $-101.7771$ &  $-80.4628$ \\
$5$ & $[4/6]$ &  $-201.6419$ &  $-174.8844$ &  $-149.4553$ &  $-125.3553$ &  $-102.5862$ &  $-81.1625$ \\
$5$ & $[5/4]$ &  $-203.9394$ &  $-176.7290$ &  $-150.8970$ &  $-126.4411$ &  $-103.3591$ &  $-81.6482$ \\
$5$ & $[5/5]$ &  $-201.4721$ &  $-174.7445$ &  $-149.3422$ &  $-125.2644$ &  $-102.5104$ &  $-81.0786$ \\
$5$ & $[5/6]$ &  $-198.7799$ &  $-172.8884$ &  $-148.2001$ & --- &  $-102.2739$ &  $-81.0336$ \\
$5$ & $[6/4]$ &  $-202.8435$ &  $-175.8651$ &  $-150.2298$ &  $-125.9387$ &  $-102.9929$ &  $-81.3939$ \\
\hline 
$6$ & $[4/6]$ &  $-201.0906$ &  $-174.3880$ &  $-149.0165$ &  $-124.9749$ &  $-102.2619$ &  $-80.8758$ \\
$6$ & $[5/5]$ &  $-200.9265$ &  $-174.2458$ &  $-148.8927$ &  $-124.8668$ &  $-102.1673$ &  $-80.7929$ \\
$6$ & $[5/6]$ &  $-200.4927$ &  $-173.9600$ &  $-148.7433$ &
$-124.8358$ &  $-102.2290$ &  $-80.9131$ \\
$6$ & $[5/7]$ &  $-200.3603$ &  $-173.7018$ &  $-148.3764$ &  $-124.3940$ &  $-101.7753$ &  $-80.5533$ \\
$6$ & $[6/4]$ &  $-201.6970$ &  $-174.9293$ &  $-149.4861$ &  $-125.3673$ &  $-102.5725$ &  $-81.1016$ \\
$6$ & $[6/6]$ &  $-200.3195$ &  $-173.6857$ &  $-148.3751$ &  $-124.3879$ &  $-101.7244$ &  $-80.3848$ \\
$6$ & $[7/5]$ &  $-202.1300$ &  $-175.2569$ &  $-149.7173$ &  $-125.5125$ &  $-102.6443$ &  $-81.1143$ \\
\hline 
\end{tabular}
\caption{\label{tab:TOTL}
  Pad\'e results for the sum of those contributions which don't have a closed
  heavy fermion loop. $n_l$ has been varied from 0 to 5.
}
\end{center}
}
\end{table}

\begin{table}[t]
{\footnotesize
\begin{center}
\begin{tabular}{|l|l||r|r|r|r|r|r|} 
\hline
$n$ & P.A.&$n_l=0 $& $n_l=1  $& $n_l=2  $& $n_l=3  $& $n_l=4  $& $n_l=5  $\\
\hline
$4$ & $[1/3]$ &  $-0.9345$ &  $-0.9572$ &  $-0.9798$ &  $-1.0024$ &  $-1.0249$ &  $-1.0475$ \\
$4$ & $[2/2]$ &  $-0.9321$ &  $-0.9546$ &  $-0.9770$ &  $-0.9995$ &  $-1.0218$ &  $-1.0442$ \\
$4$ & $[3/1]$ &  $-0.9324$ &  $-0.9551$ &  $-0.9777$ &  $-1.0003$ &  $-1.0229$ &  $-1.0455$ \\
$4$ & $[4/0]$ &  $-0.9604$ &  $-0.9828$ &  $-1.0053$ &  $-1.0277$ &  $-1.0501$ &  $-1.0725$ \\
\hline 
$5$ & $[1/4]$ &  $-0.9271$ &  $-0.9495$ &  $-0.9720$ &  $-0.9944$ &  $-1.0169$ &  $-1.0393$ \\
$5$ & $[2/3]$ &  $-0.9219$ &  $-0.9440$ &  $-0.9661$ &  $-0.9882$ &  $-1.0103$ &  $-1.0324$ \\
$5$ & $[3/2]$ &  $-0.9086$ &  $-0.9347$ &  $-0.9591$ &  $-0.9827$ &  $-1.0060$ &  $-1.0290$ \\
$5$ & $[4/1]$ &  $-0.9254$ &  $-0.9478$ &  $-0.9703$ &  $-0.9927$ &  $-1.0151$ &  $-1.0375$ \\
$5$ & $[5/0]$ &  $-0.9495$ &  $-0.9719$ &  $-0.9942$ &  $-1.0166$ &  $-1.0389$ &  $-1.0613$ \\
\hline 
$6$ & $[1/5]$ &  $-0.9217$ &  $-0.9441$ &  $-0.9665$ &  $-0.9888$ &  $-1.0112$ &  $-1.0336$ \\
$6$ & $[2/4]$ &  $-0.9140$ &  $-0.9364$ &  $-0.9589$ &  $-0.9813$ &  $-1.0037$ &  $-1.0261$ \\
$6$ & $[3/3]$ &  $-0.9125$ &  $-0.9352$ &  $-0.9578$ &  $-0.9803$ &  $-1.0028$ &  $-1.0252$ \\
$6$ & $[4/2]$ &  $-0.9126$ &  $-0.9352$ &  $-0.9578$ &  $-0.9803$ &  $-1.0028$ &  $-1.0253$ \\
$6$ & $[5/1]$ &  $-0.9202$ &  $-0.9425$ &  $-0.9649$ &  $-0.9872$ &  $-1.0096$ &  $-1.0319$ \\
$6$ & $[6/0]$ &  $-0.9416$ &  $-0.9639$ &  $-0.9862$ &  $-1.0085$ &  $-1.0308$ &  $-1.0532$ \\
\hline 
\end{tabular}
\caption{\label{tab:TOTH}
  Pad\'e approximations performed in the variable $z$. No high-energy results
  have been used. Again $n_l$ has been varied from 0 to 5, the dependence,
  however, is very weak.
}
\end{center}
}
\end{table}

The final result of~\cite{CheSte99,CheSte00} 
for the mass relation can be found in 
Tab.~\ref{tab:nl} where a comparison with the results of~\cite{MelRit99}
is performed.
Note that there is perfect agreement for all values of $n_l$.
At this point we want to mention that the result
of~\cite{CheSte99,CheSte00} is more general as the function $f(z)$ in
Eq.~(\ref{eq:f2}) has been computed for all values of $z$ and not only
for the special point $z=1$. 
This opens the possibility to obtain the fermion propagator in QCD at
three-loop order for arbitrary external momentum extending the
considerations of~\cite{FleJegTarVer99} by one more loop.

  \begin{table}[t]
{\tiny
    \begin{center}
      \begin{tabular}{|l||r|r|r|r|r|r|r|r|r|} 
        \hline
        & \multicolumn{3}{c|}{$z_m(M)=m(M)/M$}
        & \multicolumn{3}{c|}{$z_m^{SI}(M)=\mu_m/M$}
        & \multicolumn{3}{c|}{$z_m^{inv}(m)=M/\mu_m$}
        \\
        \hline
        $n_l$ 
        & ${\cal O}(\alpha_s^2)$ 
        & ${\cal O}(\alpha_s^3)$~\cite{CheSte99}
        & ${\cal O}(\alpha_s^3)$~\cite{MelRit99}
        & ${\cal O}(\alpha_s^2)$ 
        & ${\cal O}(\alpha_s^3)$~\cite{CheSte99}
        & ${\cal O}(\alpha_s^3)$~\cite{MelRit99}
        & ${\cal O}(\alpha_s^2)$ 
        & ${\cal O}(\alpha_s^3)$~\cite{CheSte99}
        & ${\cal O}(\alpha_s^3)$~\cite{MelRit99}
        \\
        \hline
$0$ &
$    -14.33$ & $   -202(5)$ & $-198.7$ &
$    -11.67$ & $   -170(5)$ & $-166.3$ &
$     13.44$ & $    194(5)$ & $190.6$ \\
$1$ &
$    -13.29$ & $   -176(4)$ & $-172.4$ &
$    -10.62$ & $   -146(4)$ & $-142.5$ &
$     12.40$ & $    168(4)$ & $164.6$ \\
$2$ &
$    -12.25$ & $   -150(3)$ & $-147.5$ &
$     -9.58$ & $   -123(3)$ & $-120.0$ &
$     11.36$ & $    143(3)$ & $139.9$ \\
$3$ &
$    -11.21$ & $   -126(3)$ & $-123.8$ &
$     -8.54$ & $   -101(3)$ & $-98.76$ &
$     10.32$ & $    119(3)$ & $116.5$ \\
$4$ &
$    -10.17$ & $   -103(2)$ & $-101.5$ &
$     -7.50$ & $    -81(2)$ & $-78.86$ &
$      9.28$ & $     96(2)$ & $94.42$ \\
$5$ &
$     -9.13$ & $    -82(2)$ & $-80.40$ &
$     -6.46$ & $    -62(2)$ & $-60.27$ &
$      8.24$ & $     75(2)$ & $73.64$ \\
        \hline
      \end{tabular}
      \caption{\label{tab:nl}
        Two- and three-loop coefficients of the relation between on-shell and
        $\overline{\rm MS}$ mass.
        The choice $\mu^2=M^2$, respectively, $\mu^2=m^2$
        has been adopted.
        }
    \end{center}
}
  \end{table}

For the relation between the $\overline{\rm MS}$ and on-shell quark
mass one finds up to three 
loops~\cite{Tar81,GraBroGraSch90,CheSte99,CheSte00,MelRit99}
\begin{eqnarray}
  \frac{m(\mu)}{M} &=&
  1 
  + \frac{\alpha_s^{(n_f)}(\mu)}{\pi}
  \left[-\frac{4}{3}-\lmM\right]
  + \left(\frac{\alpha_s^{(n_f)}(\mu)}{\pi}\right)^2 
  \Bigg[
  -\frac{3019}{288}
  - 2\zeta_2 
  - \frac{2}{3}\zeta_2\ln2
  + \frac{1}{6}\zeta_3
  \nonumber\\&&\mbox{}
  - \frac{445}{72}\lmM
  - \frac{19}{24}\lmM^2
  + \left(\frac{71}{144} 
    + \frac{1}{3}\zeta_2
    + \frac{13}{36}\lmM
    + \frac{1}{12}\lmM^2
  \right)n_l
  - \frac{4}{3}\sum_{1\le i\le n_l} \Delta\left(\frac{\smM_i}{M}\right)
  \Bigg]
  \nonumber\\&&\mbox{}
  + \left(\frac{\alpha_s^{(n_f)}(\mu)}{\pi}\right)^3
  \Bigg[z_m^{(3)}(M)
  + \left(
    -\frac{165635}{2592} 
    - \frac{25}{3}\zeta_2
    - \frac{25}{9}\zeta_2\ln2
    + \frac{55}{36}\zeta_3
  \right)  \lmM
  \nonumber\\&&\mbox{}
  -\frac{11779}{864}\lmM^2
  -\frac{475}{432}\lmM^3
  + n_l\left(
    \left(
      \frac{10051}{1296} 
      + \frac{37}{18}\zeta_2
      + \frac{2}{9}\zeta_2\ln2
      + \frac{7}{9}\zeta_3
    \right)\lmM
    +\frac{911}{432}\lmM^2
    \right.\nonumber\\&&\left.\mbox{}
    +\frac{11}{54}\lmM^3
  \right)
  + n_l^2\left(
    \left(
      -\frac{89}{648} 
      -\frac{1}{9}\zeta_2
    \right)\lmM^2
    -\frac{13}{216}\lmM^2
    -\frac{1}{108}\lmM^3
  \right)
  \Bigg]
  \,,
  \label{eq:zmlog}
\end{eqnarray}
where $\zeta_2=\pi^2/6$ and $\lmM=\ln\mu^2/M^2$.
The constant $z_m^{(3)}(M)$ is given by~\cite{MelRit99}
\begin{eqnarray}
  z_m^{(3)}(M) &=&
  - \frac {9478333}{93312} 
  + \frac {55}{162}\ln^4 2 
  +\left( - \frac {644201}{6480}
    + \frac {587}{27}\ln 2 
    + \frac {44}{27} \ln^2 2 
  \right)\zeta_2
  - \frac {61}{27}\zeta_3 
  \nonumber\\&&\mbox{}
  + \frac {3475}{432} \zeta_4
  + \frac {1439}{72}\zeta_2\zeta_3
  - \frac {1975}{216}\zeta_5 
  + \frac {220}{27} a_4 
  + n_l  \left[ \frac {246643}{23328} 
    - \frac {1}{81}\ln^4 2  
  \nonumber\right.\\&&\left.\mbox{}
    +\left(
        \frac {967}{108}
      + \frac {22}{27}\ln 2 
      - \frac {4}{27} \ln^2 2 
    \right)\zeta_2
    + \frac {241}{72}\zeta_3
    - \frac {305}{108}\zeta_4
    - \frac {8}{27}a_4 
  \right]
  \nonumber\\&&\mbox{}
  + n_l^2  \left[  - \frac {2353}{23328} 
    - \frac {13}{54}\zeta_2
    - \frac {7}{54}\zeta_3 
  \right]
  \,,
  \label{eq:zm3}
\end{eqnarray}
where $a_4=\mbox{Li}_4(1/2)\approx 0.517\,479$.

The function $\Delta(r)$ in Eq.~(\ref{eq:zmlog}) 
takes into account the effects
of secondary light quarks.
If $0 \le r \le 1$  then the function
$\Delta(r)$ may be conveniently approximated as
follows~\cite{GraBroGraSch90} 
\begin{equation}
  \Delta(r) = \frac{\pi^2}{8}~r - 0.597~r^2 + 0.230~r^3
  \,,
  \label{eq:K-appr}
\end{equation}
which is accurate to $1$\%.
Up to now there is no calculation available
taking into account the complete mass dependence of the light quarks
at order $\alpha_s^3$.
The subclass of diagrams containing a one-loop light quark vacuum
polarization insertion has been considered in~\cite{Hoang:2000fm},
where it was observed that the dominant contribution is provided by
the linear mass corrections as at order $\alpha_s^2$
(cf. Eq.~(\ref{eq:K-appr})).

At this point it is worthwhile to compare the results 
of~\cite{CheSte99,CheSte00,MelRit99} with
estimations for the ${\cal O}(\alpha_s^3)$ terms obtained with the 
help of different optimization procedures.
In~\cite{CheKniSir97} the fastest apparent convergence (FAC)~\cite{FAC}
and the principle of minimal sensitivity (PMS)~\cite{PMS}
have been used in order to predict the three-loop coefficient of
$M/m(m)$. For $n_l=2$ one observes a discrepancy of 9\%. 
It even reduces to below 1\%
for $n_l=5$, i.e. in the case of the top quark.
The results obtained in the large
$\beta_0$-limit~\cite{BenBra95},
where $\beta_0$ is the first coefficient of the QCD $\beta$ function,
agree to 2\% for $n_l=3$,
roughly 8\% for $n_l=4$ and 17\% for $n_l=5$.

Among the various applications of the order $\alpha_s^3$ term in the
$\overline{\rm MS}$--on-shell relation we only want to mention the
improvement in the determination of the top quark mass
to be measured at a future $e^+e^-$ linear collider.
To be specific, let us consider the production of top quarks in
$e^+e^-$ collisions.
The corresponding physical observables
expressed in terms of $M_t$ show in general a bad
convergence behaviour. In the case of the total cross section, e.g., the
next-to-next-to-leading order corrections partly exceed the next-to-leading
ones. Furthermore the peak position which is the most striking feature of the
total cross section and from which finally the mass value
can be extracted depends very much on the number of terms one includes into
the analysis.
The commonly accepted explanation for this is that
the pole mass is sensitive to long-distance effects which result in
intrinsic uncertainties of order $\Lambda_{QCD}$~\cite{BenBra94,Big94}.
In other words, it is not possible to determine the pole mass from the
analysis of the cross section at threshold with an accuracy better than
$\Lambda_{QCD}$. 

Several strategies have been proposed to circumvent this
problem~\cite{Ben98,HoaSmiSteWil98,HoaTeu99}.
They are based on the observation that
the same kind of ambiguities also appear in the static quark
potential, $V(r)$. In the combination $2M_t + V(r)$, however, the
infra-red sensitivity drops out. Thus a definition of a short-distance
mass extracted from threshold quantities should be possible.
The relation of the new mass parameter
to the pole mass is used in order to re-parameterize the threshold phenomena.
On the other hand a relation of the new quark mass to the $\overline{\rm MS}$
mass must be established as
it is commonly used for the parameterization of those quantities which are not
related to the threshold.
In order to do this consistently
the three-loop relation between the
$\overline{\rm MS}$ and the on-shell mass is needed.

In~\cite{Ben98} the concept of the so-called potential mass, $m_{t,PS}$, has
been introduced.
Its relation to the $\overline{\rm MS}$ mass, $m_t(m_t)$, reads
\begin{eqnarray}
  m_{t,PS}(20~{\rm GeV}) &=& \left( 165.0 + 6.7 + 1.2 + 0.28
  \right)~\mbox{GeV}
  \,,
  \label{eq:mtPSmtmt}
\end{eqnarray}
where the different terms represent the contributions from order $\alpha_s^0$
to $\alpha_s^3$. For the numerical values
$m_t(m_t)=165.0$~GeV and $\alpha_s^{(6)}(m_t(m_t))=0.1085$
have been used.
The comparison of Eq.~(\ref{eq:mtPSmtmt}) with the analogous expansion
for $M_t$,
\begin{eqnarray}
  M_t &=& ( 165.0 + 7.6 + 1.6 + 0.51 )\mbox{~GeV}
  \,,
\end{eqnarray}
shows that the potential mass can be 
more accurately related to the $\overline{\rm MS}$ mass than $M_t$.

Further details and more examples can be found in~\cite{CheSte00}.


\section*{Acknowledgments}

I would like to thank S.~Casalbuoni, K.G.~Chetyrkin, R.~Harlander and
B.A.~Kniehl for carefully reading the manuscript and
for valuable suggestions and discussions. 
Furthermore I would like to thank J.H.~K\"uhn and B.A.~Kniehl 
for encouraging me to complete this work.


\begin{appendix}
\renewcommand {\theequation}{\Alph{section}.\arabic{equation}}
\renewcommand {\thefigure}{\Alph{section}.\arabic{figure}}
\renewcommand {\thetable}{\Alph{section}.\arabic{table}}


\section{Technical remarks}
\setcounter{equation}{0} 
\setcounter{figure}{0} 
\setcounter{table}{0} 


\subsection{\label{sub:ae}Asymptotic expansion}

A promising approach to compute --- at least in certain kinematical
limits --- multi-loop diagrams is
based on asymptotic expansion.
An asymptotic expansion can be considered as a generalization of a
Taylor expansion. In both cases one obtains an expansion in powers of
a small quantity. However, in case of the asympotitic expansion the
corresponding coefficients are not constant but contain non-analytical
functions of the small parameter. As a simple example let us consider
the function $f(x)=\mbox{Li}_2(1-x)$ for which the Taylor expansion
for $x\to0$ does not exist beyond the leading order. Nevertheless
there is an asymptotic expansion which reads
\begin{eqnarray}
  f(x) &=& \frac{\pi^2}{6} + x\left(\ln x-1\right) 
  + {\cal O}\left(x^2\ln x\right)
  \,,
\end{eqnarray}
with the non-analytic (for $x\to0$)
function $\ln x$ in the coefficient of the
linear term in $x$.

In the case of Feynman diagrams the situation is similar.
Systematic procedures have been developed in the case of 
a large external mometum (``large-momentum
procedure'') and a large internal mass (``hard-mass
procedure'')~\cite{Smi91}.
Both procedures apply to problems which can be formulated in 
Euclidean space. This is the case for all calculations presented in
this review. In contrast to that there are also phenomena which are
tightly connected to the Minkowskian space-time. Also in such cases
rules for an asymptotic expansion have been developed. In particular,
two-loop on-shell two-point diagrams~\cite{CzaSmi97}
and two-loop vertex diagrams in the Sudakov limit~\cite{Smi97}
have been considered. Also the threshold expansion~\cite{Beneke:1998zp} 
belongs to this class of phenomena.

One may treat the large-momentum and hard-mass procedures on the same
footing. Thus, in what follows we only present the general formulae in
the case of large external momenta --- the transition to the hard-mass
procedure is straightforward.  The prescription for the large-momentum
procedure is summarized by the following 
formula\footnote{In the case of the hard-mass procedure one
essentially has to replace $Q$ by the large mass $M$.}:
\begin{eqnarray}
\Gamma(Q,m,q) & \stackrel{Q\to \infty}{\simeq} &
\sum_\gamma \Gamma/\gamma(q,m)
\,\,\star\,\, 
T_{\{q_\gamma,m_\gamma\}}\gamma(Q,m_\gamma,q_\gamma)
\,.
\label{eqasexp}
\end{eqnarray}
Here, $\Gamma$ is the Feynman diagram under consideration, $\{Q\}$
($\{m,q\}$) is the collection of the large (small) parameters, and the
sum goes over all subgraphs $\gamma$ of $\Gamma$ with masses $m_\gamma$
and external momenta $q_\gamma$, subject to certain conditions to be
described below.  $T_{\{q,m\}}$ is an operator performing a Taylor
expansion in $\{q,m\}$ {\em before} any integration is carried out.  The
notation $\Gamma/\gamma\star T_{\{q,m\}}\gamma$ indicates that the
subgraph $\gamma$ of $\Gamma$ is replaced by its Taylor expansion which
should be performed in all masses and external momenta of $\gamma$ that
do not belong to the set $\{Q\}$.  In particular, also those external
momenta of $\gamma$ that appear to be integration momenta in $\Gamma$
have to be considered as small. Only after the Taylor expansions have
been carried out, the loop integrations are performed.  In the following
we will refer to the set $\{\gamma\}$ as {\em hard subgraphs} or simply {\em
  subgraphs} and to $\{\Gamma/\gamma\}$ as {\em co-subgraphs}.

The conditions for the subgraphs $\gamma$ are different for
the hard-mass and large-momentum procedures\footnote{
  Actually they are very similar and it is certainly possible to merge
  them into one condition using a more abstract language. For our
  purpose, however, it is more convenient to distinguish the two procedures.}.
For the large-momentum procedure, $\gamma$ must
\begin{itemize}
\item contain all vertices where a large momentum enters or leaves the
  graph
\item be one-particle irreducible after identifying these
  vertices.
\end{itemize}
From these requirements it is clear that the hard subgraphs become
massless integrals where the scales are given by the large momenta. In
the simplest case of one large momentum one ends up with propagator-type
integrals.  The co-subgraph, on the other hand, may still contain small
external momenta and masses. However, the resulting integrals are
typically much simpler than the original one.

In the case of hard-mass procedure, $\gamma$ has to
\begin{itemize}
\item contain all the propagators carrying a large mass
\item be one-particle irreducible in its connected parts after
  contracting the heavy lines.
\end{itemize}
Here, the hard subgraphs reduce to tadpole integrals with the large masses
setting the scales. The co-subgraphs are again simpler to
evaluate than the initial diagram.

\begin{figure}[t]
\begin{center}
\begin{tabular}{ccc}
\leavevmode
\epsfxsize=5.0cm
\epsffile[142 267 470 525]{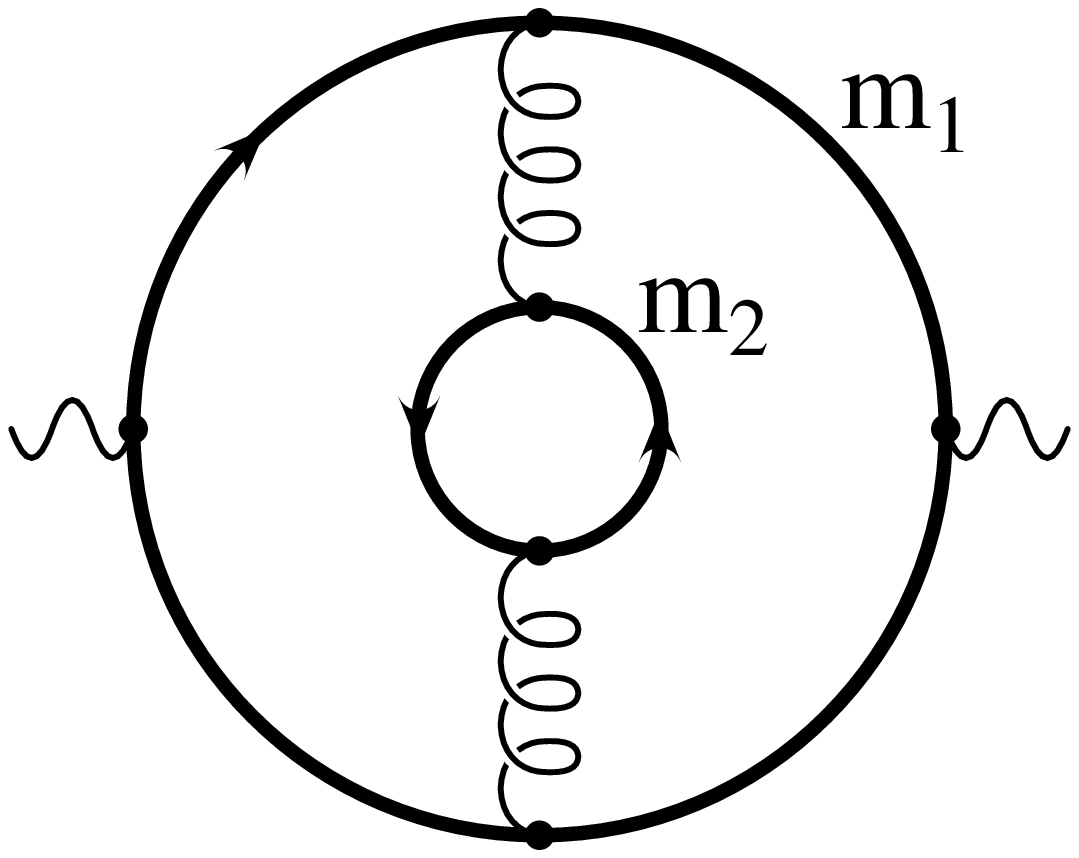}
&\hspace{2em}&
\leavevmode
\epsfxsize=5.0cm
\epsffile[142 267 470 525]{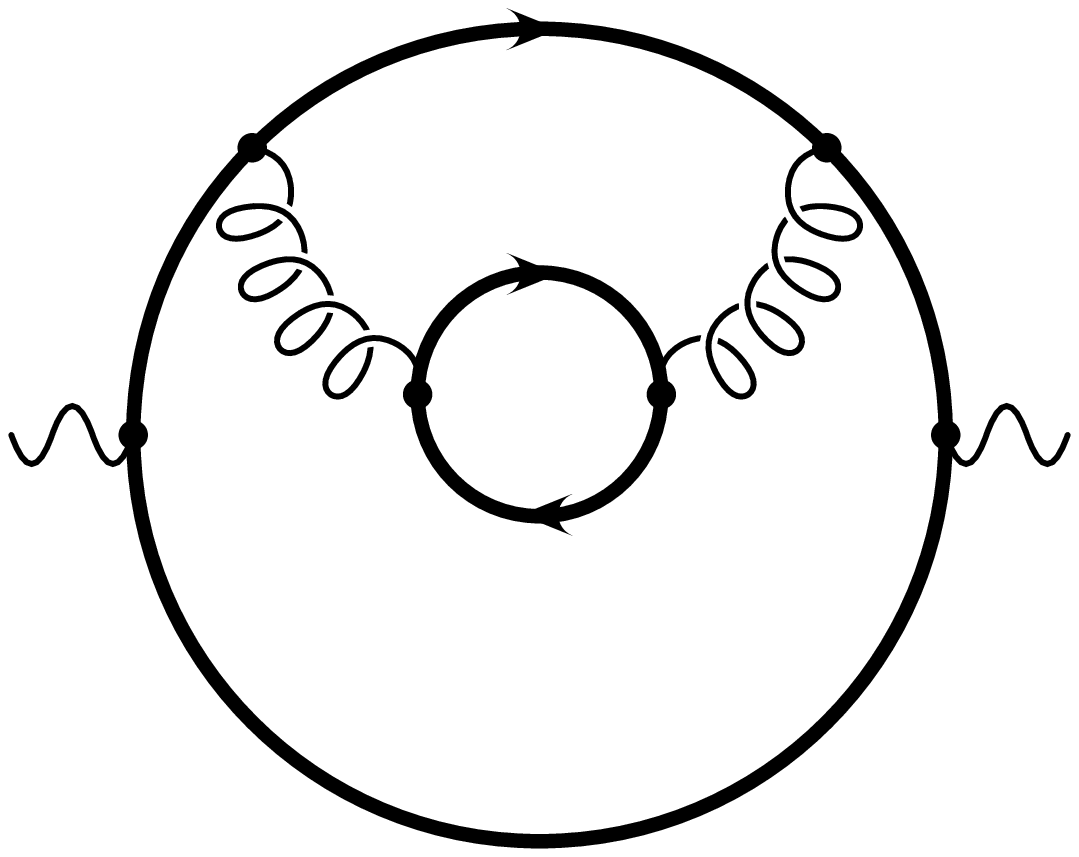} \\
(a) &\hspace{2cm}& (b)
\end{tabular}
\caption[]{\label{fig:db} 
Fermionic double-bubble diagrams with generic masses $m_1$ and $m_2$.
}
\end{center}
\end{figure}

An example demonstrating the practical application of the
large-momentum expansion was already presented in
Section~\ref{sec:dim4} (cf. Fig.~\ref{fig:lmp1l};
see also~\cite{Harlander:1999dq}).
As an application of the hard-mass procedure let us consider 
the double-bubble diagram of Fig.~\ref{fig:db}
with the hierarchy 
$m_1^2\ll q^2\ll m_2^2$. The imaginary part leads to contributions
for the total cross section $\sigma(e^+e^-\to\mbox{hadrons})$.
One may think of charm quark production ($m_1=M_c$)
in the presence of a virtual bottom quark ($m_2=M_b$). It turns out that
already the first term provides a very good approximation almost up
to the threshold $\sqrt{s}=2M_b$~\cite{Che93,HoaJezKueTeu94,Teu:diss}.
For simplicity we set $m_1=0$ and $m_2=m$ in the following.

\begin{figure}[t]
  \begin{center}
  \leavevmode
   \epsfxsize=3cm
   \epsffile[150 260 420 450]{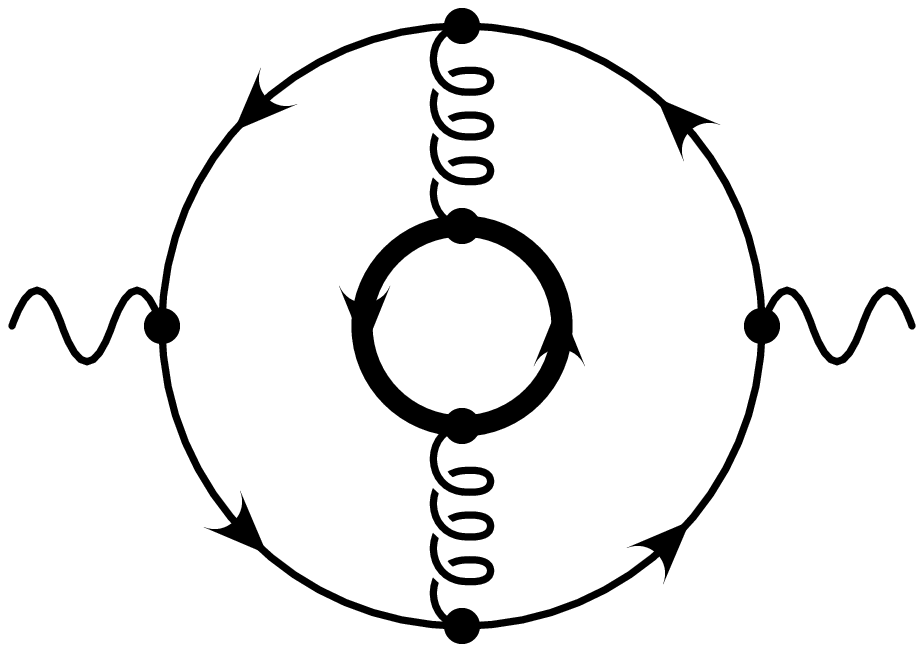}\hspace{1em}
   \raisebox{2.8em}
   {\Large $=$}
   \epsfxsize=2.cm
   \raisebox{1.em}{\epsffile[150 260 420 450]{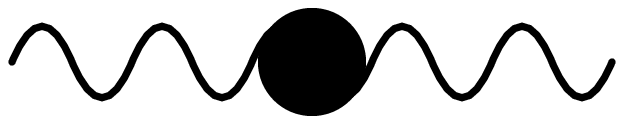}}\hspace{-1em}
   \raisebox{2.8em}{\Large $\ \ \star \!\!$}
   \epsfxsize=3.cm
   \epsffile[150 260 420 450]{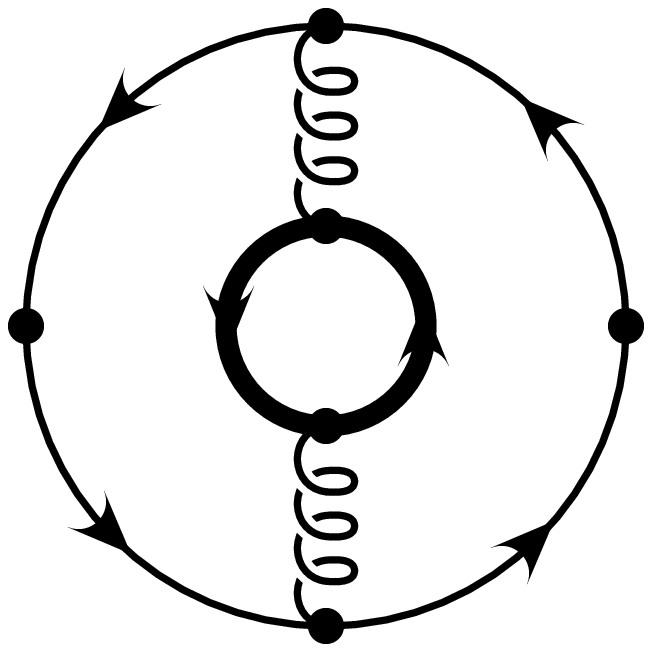}\hspace{0em}\\[1em]
   \mbox{\hspace{1em}}
   \raisebox{2.8em}
   {\Large $+ \ \ 2\times \ $}
   \epsfxsize=2.cm
   \raisebox{1.em}{\epsffile[150 260 420 450]{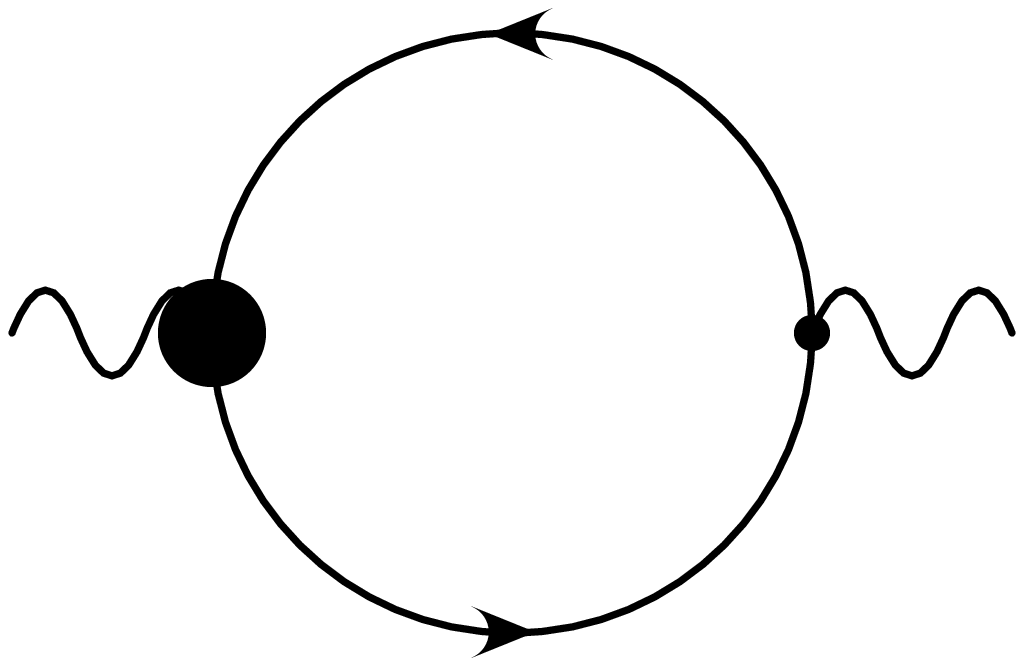}}\hspace{0em}
   \raisebox{2.8em}{\Large $\ \ \star \!\!$}
   \epsfxsize=3.cm
   \epsffile[150 260 420 450]{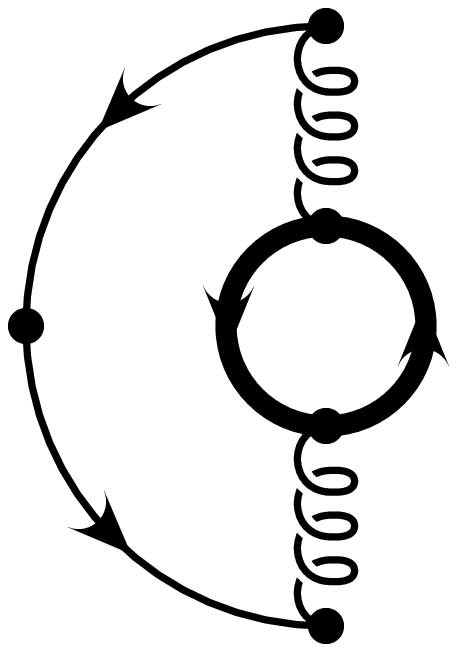}
   \raisebox{2.8em}
   {\Large $\!\!\!\!+\ \ $}
   \epsfxsize=3cm
   \raisebox{0em}{\epsffile[150 260 420 450]{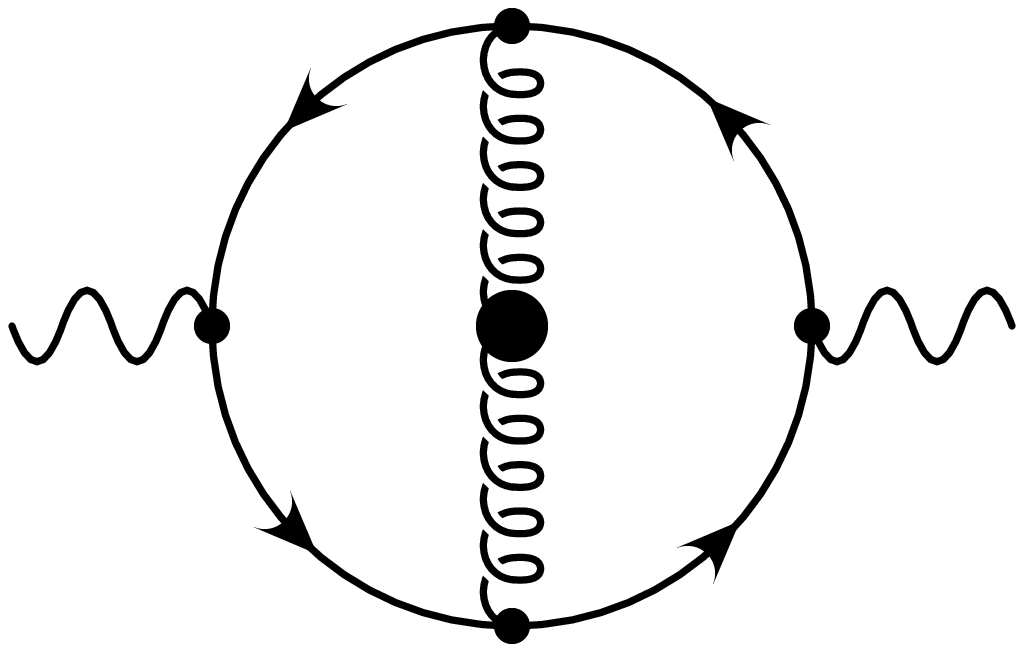}}
   \epsfxsize=3.cm
   \raisebox{2.8em}{\Large $\ \ \star \!\!\!\!\!\!$}
   \epsffile[150 260 420 450]{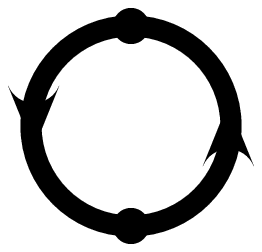}\hspace{0em}
   \caption[]{\label{fig:dbhmp}\sloppy
     Hard-mass procedure for the double-bubble diagram. The
     hierarchy $q^2 \ll m^2$ is considered where $m$ is the mass of the
     inner line.  The hard subdiagrams (right of ``$\star$'') are to be
     expanded in all external momenta including $q$ and reinserted into
     the fat vertex dots of the co-subgraphs (left of~``$\star$'').}
 \end{center}
\end{figure}

The corresponding diagrammatic representation
is shown in Fig.~\ref{fig:dbhmp}. There are three subdiagrams, where one
again corresponds to the naive Taylor expansion of the integrand
in the external momentum $q$.
After Taylor expansion, the subdiagrams are reduced to tadpole
integrals with mass scale $m$. The scale of the co-subgraphs is given
by $q$, thus leading to massless propagator-type integrals.
The result for the first three terms
reads~\cite{Ste:diss,Harlander:1999dq}
\begin{eqnarray}
\bar{\Pi}_{gs}(q^2) &\stackrel{q^2\ll m^2}{=}&
   \frac{3}{16\pi^2}
  \left({\alpha_s\over \pi}\right)^2\,C_{\rm F}\,T\,\Bigg[
        \frac{295}{648} 
      + \frac{11}{6}\logqmums
      - \frac{1}{6}\logqmums^2
      - \frac{11}{6}\logqmms 
      + \frac{1}{6}\logqmms^2
\nonumber\\
&&\mbox{}
      - \frac{4}{3}\zeta_3\logqmums 
      + \frac{4}{3}\zeta_3\logqmms
      + \frac{q^2}{m^2}\left(
         \frac{3503}{10125} 
       - \frac{88}{675}\logqmms
       + \frac{2}{135}\logqmms^2
                       \right)
\nonumber\\
&&
       + \left(\frac{q^2}{m^2}\right)^2\left(
        - \frac{2047}{514500} 
        + \frac{1303}{529200}\logqmms
        - \frac{1}{2520}\logqmms^2
                                        \right)
\Bigg] + \cdots\,,
\label{eq:dbhmpres}
\end{eqnarray}
with $\logqmums=\ln(-q^2/\mu^2)$ and $\logqmms=\ln(-q^2/m^2)$.

We want to stress that the main simplification, which 
is common to all kinds of asymtotic expansions, comes from the fact
that the expansions in the small parameters are done before any momentum
integration is performed.
The proof that this leads to correct results is based on the
so-called strategy of regions~\cite{Smirnov:1999bz}.
There different regions of each loop momentum are selected and in each
of them Taylor expansions with respect to the small parameters are
performed. 
In the limits of the hard-mass and large-momentum procedures
an interpretation of the different regions in terms
subgraphs and cosub-graphs is possible (see above).
This is different in the case of the threshold
expansion~\cite{Beneke:1998zp} where a graphical representation
becomes much less transparent. However, the application of the 
strategy of regions~\cite{Smirnov:1999bz} leads to correct results.


\subsection{\label{sub:single}Single-scale Feynman diagrams up to
three loops}

The problem of evaluating one-loop Feynman diagrams is --- at least in
principle --- solved (see, e.g.,
Refs.~\cite{'tHooft:1979xw,Passarino:1979jh,
vanOldenborgh:1990wn,Denner:1993kt}).   
However, in case many
legs and lots of different masses appear also one-loop computations can
become very tedious, in particular if degenerate momentum configurations
are involved.

At two-loop order the class of Feynman diagrams which have been
studied in detail is much more restricted.
There is a good understanding of two-point 
functions (see, e.g.,~\cite{Weiglein:1994hd})
which also has found important physical applications~\cite{Weiglein:2001ci}.
Concerning three- and four-point functions one is essentially
restricted to the massless case.
In this context we want to draw the attention to 
the recent activity in the computation of
the two-loop box diagrams (for a brief overview
see~\cite{Gehrmann:2001ih}).
Within the last two years the basis has been established to compute
two-loop virtual corrections to the four-point Feynman amplitudes
where all internal lines are massless and at most one external leg is
off-shell. 
This opens the door to investigate next-to-next-to-leading order
processes like the two-jet
production at hadron colliders or three-jet production
in $e^+e^-$ annihilation.

It is obvious that the complexity of the computation 
of a Feynman diagram strongly depends on the number of different
scales involved.
There is one class of diagrams which is studied in great detail up to
three loops, namely diagrams which only depend on one dimensionful
scale.
Next to massless propagator-type diagrams with one external momentum,
$q$, we have in mind vacuum
integrals with one non-zero mass, $M$ and so-called on-shell integrals
for which the condition $q^2=M^2$ is fulfilled.

The basic idea for the computation of the integrals is
common to all three types: after the numerator is decomposed in terms
of the denominator recurrence relations are applied which express the
diagram as a linear combination of so-called master integrals.
Only for the latter a hard computation is necessary. However, since in
the case of single-scale diagrams the master integrals are essentially
pure numbers it is also possible to use high-precision numerical
methods in case an analytical calculation is not possible.
We want to mention that for the massless propagator-type and 
the massive vacuum integrals two and nine
master integrals are needed, respectively.
In the case of the three-loop on-shell integrals a list
of all master integrals can be found in the Appendix of
Ref.~\cite{Melnikov:2000zc}.
Counting also those integrals which are composed of products of
lower-order diagrams they amount to 18.

For convenience we want to provide the analytical results for
the massless
one-loop two-point functions, $P_{ab}(Q)$, 
the on-shell two-point functions, $O_{ab}(Q)$,
and 
the one- ($V_a$) and two-loop ($V_{abc}$) vacuum integrals
in Euclidian space.
\begin{eqnarray}
P_{ab}(Q) &=& \int\frac{{\rm d}^Dp}{\left(2\pi\right)^D}
              \frac{1}{p^{2a}\left(p+Q\right)^{2b}}
\nonumber\\
          &=& 
\frac{\left(Q^2\right)^{D/2-a-b}}{\left(4\pi\right)^{D/2}}
\frac{
  \Gamma(a+b-D/2)
  \Gamma(D/2-a)
  \Gamma(D/2-b)
}{
  \Gamma(a)
  \Gamma(b)
  \Gamma(D-a-b)
}  
\,,
\label{eq:Pab}
\\
O_{ab}(Q) &=& \int\frac{{\rm d}^Dp}{\left(2\pi\right)^D}
              \frac{1}{p^{2a}\left(p^2+2p\cdot Q\right)^{b}}
\nonumber\\
          &=& 
\frac{\left(Q^2\right)^{D/2-a-b}}{\left(4\pi\right)^{D/2}}
\frac{
  \Gamma(a+b-D/2)
  \Gamma(D-2a-b)
}{
  \Gamma(b)
  \Gamma(D-a-b)
}  
\,,
\label{eq:Oab}
\\
V_{a}&=& \int\frac{{\rm d}^Dp}{\left(2\pi\right)^D}
              \frac{1}{\left(p^2+M^2\right)^{a}}
\,\,=\,\,
\frac{\left(M^2\right)^{D/2-a}}{\left(4\pi\right)^{D/2}}
\frac{
  \Gamma(a-D/2)
}{
  \Gamma(a)
}  
\,,
\label{eq:Va}
\\
V_{abc} &=&
  \int\frac{{\rm d}^D p}{(2\pi)^D}\frac{{\rm d}^D k}{(2\pi)^D}
  \frac{1}{(p^2+M^2)^{a}(k^2+M^2)^{b}(\left(p+k\right)^2)^{c}}
\nonumber\\
          &=& 
\frac{\left(M^2\right)^{D-a-b-c}}{\left(4\pi\right)^{D}}
\frac{
  \Gamma(a+b+c-D)
  \Gamma(a+c-D/2)
  \Gamma(b+c-D/2)
  \Gamma(D/2-c)
}{
  \Gamma(a)
  \Gamma(b)
  \Gamma(a+b+2c-D)
  \Gamma(D/2)
}
\,.
\nonumber\\
\label{eq:Vabc}
\end{eqnarray}

The results of a general three-loop diagram can not be expressed in
terms of $\Gamma$ functions. Moreover, 
due to the large number of contributing diagrams and the 
complexity of intermediate expressions it is absolutely
necessary to use computer algebra programs for the computation of
multi-loop diagrams.
It is thus also hardly possible to provide intermediate results which
eventually could be used in other calculations.
Therefore, on one side one is left with the description of the method
used for the evaluation of the diagrams.
On the other hand it is possible to provide the program 
code which was used for
the computation. Thus everybody can repeat the calculation
or apply it to own problems.

Both for the massless propagator-type integrals and the massive vacuum
integrals {\tt FORM}~\cite{form} packages have been published:
massless integrals up to three loops can be computed using  
{\tt MINCER}~\cite{mincer}; the package for the 
massive integrals is called {\tt MATAD}~\cite{matad}.
In the following we will present an example which demonstrates the use
of {\tt MATAD}.
The use of {\tt MINCER} is very similar.
Actually, in the package {\tt GEFICOM}~\cite{geficom} {\tt MATAD} and
{\tt MINCER} are used in parallel using the same notation
for the input.

\begin{figure}[ht]
  \begin{center}
  \leavevmode
      \epsfxsize=5cm
      \epsffile[189 314 481 478]{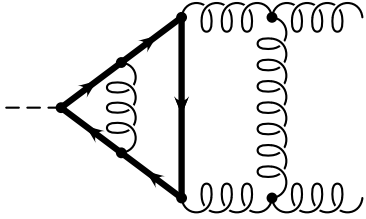}\\
      \caption[]{\label{fig:hgg}Sample diagram contributing to the
  decay of the Higgs boson. Solid and looped lines represent quarks
  and gluons, respectively.
        }
  \end{center}
\end{figure}

Let us consider the triangle diagram as pictured in Fig.~\ref{fig:hgg}.
It is one of the 657 diagrams which contribute to the coefficient
function $C_1^0$ appearing in the effective Lagrangian of
Eq.~(\ref{eq:eff}).
According to the Lorentz structure the result can be written as
\begin{eqnarray}
  K(M_t) \, \left( q_1^\nu q_2^\mu - q_1 q_2 g^{\mu\nu} \right)
  \,,
  \label{eq:k1}
\end{eqnarray}
where $q_1$ and $q_2$ are the momenta of the gluons with polarization
vectors $\epsilon^\mu(q_1)$ and $\epsilon^\nu(q_2)$.
Thus the vertex diagrams have to be expanded up to linear order both
in $q_1$ and $q_2$, and an appropriate projector has to be applied 
in order to get $K(M_t)$.

{\tt MATAD} requires one file containing the diagrams and the
projectors which have to be applied.
In our case it could look as follows

{\footnotesize
\begin{verbatim}
*--#[ TREAT0:
multiply, (
  a*deno(2,-2)*(q1.q2*d_(mu,nu)-q2(nu)*q1(mu)-q2(mu)*q1(nu))
 +b*deno(2,-2)*(-q1.q2*d_(mu,nu)+(3-2*ep)*q2(nu)*q1(mu)+q2(mu)*q1(nu))
);
.sort
*--#] TREAT0:

*--#[ TREAT1:
*--#] TREAT1:

*--#[ TREAT2:
*--#] TREAT2:

*--#[ TREATMAIN:
*--#] TREATMAIN:

*--#[ d3l335:
        ((-1)
        *M
        *Dg(nu1,nu2,p1)
        *Dg(nu7,nu8,-p4)
        *Dg(nu3,nu4,q1,-p1)
        *Dg(nu5,nu6,q2,p1)
        *S(-q1,-p3m,nu7,-q1,p5m,nu4,-p2m,nu6,q2,p5m,nu8,q2,-p3m)
        *V3g(mu,q1,nu1,-p1,nu3,p1-q1)
        *V3g(nu,q2,nu2,p1,nu5,-p1-q2)
        *1);

        #define TOPOLOGY "O4"
*--#] d3l335:
\end{verbatim}}

\noindent
where the diagram {\tt d3l335} corresponds to the one shown in
Fig.~\ref{fig:hgg} (for details concerning the nomenclature of the
momenta see~\cite{matad}).
The fold {\tt TREAT0} contains (up to an overall factor
$(q_1\cdot q_2)^{-2}$) the projector on the coefficients in front of
the structures $g^{\mu\nu}$ and $q^\nu_1 q^\mu_2$ of Eq.~(\ref{eq:k1}).
They are marked by 
the symbols \verb|a| and \verb|b|, respectively.
Thus the transversality of Eq.~(\ref{eq:k1}) can be explicitly checked
in the sum of all contributing diagrams (the result of a single
diagram does in general not have a transverse structure).

A second file which is required, the so-called {\tt main}-file,
reads

{\footnotesize
\begin{verbatim}
#define PRB "hgg"
#define PROBLEM0 "1"
#define DALA12 "1"
#define GAUGE "0"
#define POWER "2"
#define CUT "0"
#define FOLDER "hgg"
#define DIAGRAM "d3l335"
#-
#include main.gen
\end{verbatim}}

\noindent
The fifth line
ensures that an expansion of the integrand up to the second order in the 
external momenta is performed and the third line sets
$q_1^2$ and $q_2^2$ to zero and factors out
the scalar product $q_1\cdot q_2$.
\verb|#define CUT "0"| sets $\varepsilon$ to zero in the final result.
In this example we choose Feynman gauge which is achieved with 
\verb|#define GAUGE "0"|.

After calling {\tt MATAD} it takes of the order of a minute 
to obtain the result:

{\footnotesize
\begin{verbatim}
   d3l335 =
       + ep^-2 * ( 40*Q1.Q2*M^2*a + 344/9*Q1.Q2^2*a - 232/9*Q1.Q2^2*b )

       + ep^-1 * (  - 308/3*Q1.Q2*M^2*a - 3530/27*Q1.Q2^2*a + 1786/27*Q1.Q2^2*
         b )

       + 60*Q1.Q2*M^2*z2*a + 734/3*Q1.Q2*M^2*a - 1936/9*Q1.Q2^2*z3*a + 1136/9*
         Q1.Q2^2*z3*b + 172/3*Q1.Q2^2*z2*a - 116/3*Q1.Q2^2*z2*b + 46817/81*
         Q1.Q2^2*a - 26239/81*Q1.Q2^2*b;
\end{verbatim}}

\noindent
Note that the terms proportional to \verb|Q1.Q2*M^2| cancel after
adding all contributing diagrams.


\subsection{\label{sub:aut}Automation of Feynman diagram computation}

In this section we briefly want to mention the program packages which
have been used to obtain most of the results discussed in this review.
For a general overview concerning the automation of Feynman diagram
computation we refer to~\cite{Harlander:1999dq}.

The large number of diagrams which occurs in particular if one
considers higher loop orders makes it necessary to generate the diagrams
automatically. The {\tt Fortran} program {\tt QGRAF}~\cite{qgraf}
provides the possiblility to implement own models in a simple
way. Furthermore it is quite fast and generates several thousand diagrams in
a few seconds. One of the disadvantages of {\tt QGRAF} is that the
user has to put the Feynman rules himself. On the other hand, this
provides quite some flexibility in the choices of the vertices. 
E.g., it is straightforward to implement the vertex involving the
coupling of the operator 
${\cal O}_1$ to four gluons (cf. Eq.~(\ref{eq:op1})).

In general,
the application of asymptotic expansions, in particular if serveral of
them are applied successively, generates many subdiagram which have
to be expanded in several small quantities. Even for a single
multi-loop diagram this becomes very tedious if it has to be performed
by hand. For this reason the programs {\tt LMP}~\cite{Har:diss} 
and {\tt EXP}~\cite{Sei:dipl} have been developed. 
{\tt LMP} was especially developed in order to apply the
large-momentum procedure to the diagonal current
correlators~\cite{CheHarKueSte96,CheHarKueSte97,Harlander:1997xa,HarSte98}.
In some sense 
{\tt EXP} can be considered as the successor of {\tt LMP}. Next to the
hard-mass procedure also the succesive application of 
large-mometum and/or hard-mass procedure is possible.
This broadens the area of applications. Here we just want to mention
as examples
the correction of ${\cal O}(\alpha\alpha_s)$ to the $Z$ boson
decay~\cite{Harlander:1998zb}, ${\cal O}(\alpha_s^2)$ corrections to
the top quark decay~\cite{CheHarSeiSte99} or two-loop QED corrections
to the muon decay~\cite{SeiSte99} (cf. Section~\ref{sub:mudec}).

The very computation of the integrals is performed with the program 
packages {\tt MINCER}~\cite{mincer} and 
{\tt MATAD}~\cite{matad} (see Appendix~\ref{sub:single}).
The former deals with massless propagator-type integrals up to three
loops and {\tt MATAD} was written to deal with vacuum diagrams 
at one-, two- and three-loop order where several of the internal lines
may have a common mass. The area of application for each of the
individual packages seems to be quite restrictive. However, in
particular the combined application offers a quite flexible use.

In order to handle problems where a large number of diagrams are
involved and where eventually an asymptotic expansion has to be
applied in a convenient way the program package {\tt GEFICOM} has been
written. A very limited number of small input files allows the user to
rule the flow of the computation. 
{\tt Qgraf} is called to generate the diagrams. A {\tt
Mathematica}~\cite{math} script determines the toplogy of each
individual diagram and provides input which either can be directly
read from {\tt MINCER} and/or {\tt MATAD} or can be passed to {\tt
EXP} or {\tt LMP}.
At the end the results of the individual diagrams are summed and the
bare result is stored.
Moreover a convenient environment is provided which, e.g., makes
sure that all result files are up-to-date. 
Thus, processes involving a large number of (sub-)diagrams
can be treated without taking care of each individual result.


\section{\label{app:decconst}Decoupling constants and coefficient
functions}
\setcounter{equation}{0} 
\setcounter{figure}{0} 
\setcounter{table}{0} 

Transforming the decoupling constants of Eqs.~(\ref{eq:zetamOS}) 
and~(\ref{eq:zetagOS}) to the $\overline{\rm MS}$ scheme one obtains
\begin{eqnarray}
\zeta_m^{\rm MS}&=&1
+\left(\frac{\alpha_s^{(n_f)}(\mu)}{\pi}\right)^2
\left(\frac{89}{432} 
-\frac{5}{36}\ln\frac{\mu^2}{m_h^2}
+\frac{1}{12}\ln^2\frac{\mu^2}{m_h^2}\right)
+\left(\frac{\alpha_s^{(n_f)}(\mu)}{\pi}\right)^3
\left[\frac{2951}{2916} 
\right.
\nonumber\\&&\left.\mbox{}
-\frac{407}{864}\zeta_3
+\frac{5}{4}\zeta_4
-\frac{1}{36}B_4
+\left(-\frac{311}{2592}
-\frac{5}{6}\zeta_3\right)\ln\frac{\mu^2}{m_h^2}
+\frac{175}{432}\ln^2\frac{\mu^2}{m_h^2}
\right.
\nonumber\\&&\left.\mbox{}
+\frac{29}{216}\ln^3\frac{\mu^2}{m_h^2}
+n_l\left(
\frac{1327}{11664}
-\frac{2}{27}\zeta_3
-\frac{53}{432}\ln\frac{\mu^2}{m_h^2}
-\frac{1}{108}\ln^3\frac{\mu^2}{m_h^2}\right)\right]
\nonumber\\
&\approx&1
+0.2060\left(\frac{\alpha_s^{(n_f)}(\mu_h)}{\pi}\right)^2
+\left(1.8476+0.0247\,n_l\right)
\left(\frac{\alpha_s^{(n_f)}(\mu_h)}{\pi}\right)^3,
\label{eq:zetamMS}
\end{eqnarray}
\begin{eqnarray}
\left(\zeta_g^{\rm MS}\right)^2&=&1
+\frac{\alpha_s^{(n_f)}(\mu)}{\pi}
\left(
-\frac{1}{6}\ln\frac{\mu^2}{m_h^2}
\right)
+\left(\frac{\alpha_s^{(n_f)}(\mu)}{\pi}\right)^2
\left(
\frac{11}{72} 
-\frac{11}{24}\ln\frac{\mu^2}{m_h^2}
+\frac{1}{36}\ln^2\frac{\mu^2}{m_h^2}
\right)
\nonumber\\
&&\mbox{}+\left(\frac{\alpha_s^{(n_f)}(\mu)}{\pi}\right)^3
\left[
\frac{564731}{124416} 
-\frac{82043}{27648}\zeta_3
-\frac{955}{576}\ln\frac{\mu^2}{m_h^2}
+\frac{53}{576}\ln^2\frac{\mu^2}{m_h^2}
\right.
\nonumber\\
&&\left.\mbox{}
-\frac{1}{216}\ln^3\frac{\mu^2}{m_h^2} 
+n_l\left(
-\frac{2633}{31104}
+\frac{67}{576}\ln\frac{\mu^2}{m_h^2} 
-\frac{1}{36}\ln^2\frac{\mu^2}{m_h^2}
\right)
\right]
\nonumber\\
&\approx&1
+0.1528\left(\frac{\alpha_s^{(n_f)}(\mu_h)}{\pi}\right)^2
+\left(0.9721-0.0847\,n_l\right)
\left(\frac{\alpha_s^{(n_f)}(\mu_h)}{\pi}\right)^3
\,.
\label{eq:zetagMS}
\end{eqnarray}

In the following, we list the decoupling constants $\zeta_m$ and $\zeta_g$
appropriate for the general gauge group SU($N_c$).
The results read 
\begin{eqnarray}
\zeta_m^{\rm MS} &=&1
+\left(\frac{\alpha_s^{(n_f)}(\mu)}{\pi}\right)^2\left(\frac{1}{N_c}-N_c\right)
\left(-\frac{89}{1152}+\frac{5}{96}\ln\frac{\mu^2}{m_h^2}
-\frac{1}{32}\ln^2\frac{\mu^2}{m_h^2}\right)
\nonumber\\
&&\mbox{}
+\left(\frac{\alpha_s^{(n_f)}(\mu)}{\pi}\right)^3
\Bigg\{
\frac{1}{N_c^2}\left(
-\frac{683}{4608}
+\frac{57}{256}\zeta_3
-\frac{9}{64}\zeta_4
+\frac{1}{32}B_4\right)
\nonumber\\
&&\mbox{}
+\frac{1}{N_c}\left(
\frac{1685}{62208}
-\frac{7}{144}\zeta_3\right)
+\frac{907}{31104}
-\frac{397}{2304}\zeta_3
-\frac{1}{32}B_4
\nonumber\\
&&\mbox{}
+N_c\left(
-\frac{1685}{62208}
+\frac{7}{144}\zeta_3\right)
+N_c^2\left(
\frac{14813}{124416}
-\frac{29}{576}\zeta_3
+\frac{9}{64}\zeta_4\right)
\nonumber\\
&&\mbox{}
+\left[\frac{1}{N_c^2}\left(
-\frac{13}{512}
+\frac{3}{32}\zeta_3\right)
+\frac{31}{864N_c}
+\frac{1}{32}
-\frac{31}{864}N_c
-N_c^2\left(
\frac{3}{512}
\right.\right.\nonumber\\&&\left.\left.\mbox{}
+\frac{3}{32}\zeta_3\right)
\right]\ln\frac{\mu^2}{m_h^2} 
+\left(-\frac{1}{32N_c^2}
-\frac{5}{576N_c}
-\frac{5}{384}
+\frac{5}{576}N_c
+\frac{17}{384}N_c^2\right)
\ln^2\frac{\mu^2}{m_h^2}
\nonumber\\
&&\mbox{}
+\left(\frac{1}{144N_c}
-\frac{11}{576}
-\frac{1}{144}N_c
+\frac{11}{576}N_c^2\right) 
\ln^3\frac{\mu^2}{m_h^2}
\nonumber\\
&&\mbox{}
+n_l\left(\frac{1}{N_c}-N_c\right)\left(
-\frac{1327}{31104}
+\frac{1}{36}\zeta_3
+\frac{53}{1152}\ln\frac{\mu^2}{m_h^2}
+\frac{1}{288}\ln^3\frac{\mu^2}{m_h^2}\right)
\Bigg\}
\,,
\label{eq:zetamMSnc}
\end{eqnarray}

\begin{eqnarray}
\left(\zeta_g^{\rm MS}\right)^2&=&1
+\frac{\alpha_s^{(n_f)}(\mu)}{\pi}
\left(-\frac{1}{6}\ln\frac{\mu^2}{m_h^2}\right)
\nonumber\\
&&\mbox{}
+\left(\frac{\alpha_s^{(n_f)}(\mu)}{\pi}\right)^2
\left[
\frac{13}{192N_c}
+\frac{25}{576}N_c
-\left(\frac{1}{16N_c}
+\frac{7}{48}N_c\right)
\ln\frac{\mu^2}{m_h^2}
+\frac{1}{36}\ln^2\frac{\mu^2}{m_h^2}\right]
\nonumber\\
&&\mbox{}
+\left(\frac{\alpha_s^{(n_f)}(\mu)}{\pi}\right)^3
\Bigg\{
\frac{1}{N_c^2}\left(
-\frac{97}{2304}
+\frac{95}{1536}\zeta_3\right)
+\frac{1}{N_c}\left(
-\frac{103}{10368}
+\frac{7}{512}\zeta_3\right)
\nonumber\\
&&\mbox{}
-\frac{1063}{5184}
+\frac{893}{3072}\zeta_3
+N_c\left(
\frac{451}{20736}
-\frac{7}{256}\zeta_3\right)
+N_c^2\left(
\frac{7199}{13824}
-\frac{17}{48}\zeta_3
\right)
\nonumber\\
&&\mbox{}
+\left(-\frac{9}{256N_c^2}
-\frac{5}{192N_c}
-\frac{119}{1152}
-\frac{23}{3456}N_c
-\frac{1169}{6912}N_c^2\right)
\ln\frac{\mu^2}{m_h^2}
\nonumber\\
&&\mbox{}
+\left(\frac{5}{192N_c}
-\frac{11}{384}
+\frac{35}{576}N_c
-\frac{1}{128}N_c^2\right)
\ln^2\frac{\mu^2}{m_h^2}
-\frac{1}{216}\ln^3\frac{\mu^2}{m_h^2}
\nonumber\\
&&\mbox{}
+n_l\Bigg[
\frac{41}{1296N_c}
-\frac{329}{10368}N_c
+\left(-\frac{5}{384N_c}
+\frac{139}{3456}N_c\right)
\ln\frac{\mu^2}{m_h^2}
\nonumber\\
&&\mbox{}
+\left(\frac{1}{96N_c}
-\frac{1}{96}N_c\right)
\ln^2\frac{\mu^2}{m_h^2}\Bigg]\Bigg\}
\,.
\label{eq:zetagMSnc}
\end{eqnarray}
For $N_c=3$, we recover Eqs.~(\ref{eq:zetamMS}) and (\ref{eq:zetagMS}).

The renormalized
decoupling constants $\zeta_2$ and $\zeta_3$ for the quark and gluon fields, 
respectively, arise from 
Eqs.~(\ref{eq:zeta20}) and~(\ref{eq:zeta30}).
Of course, $\zeta_2$ and $\zeta_3$ are both gauge dependent.
Restricting ourselves to the case $N_c=3$, we find in the covariant
gauge~(\ref{eq:gluprop})
\begin{eqnarray}
\zeta_2^{\rm MS}&=&1
+\left(\frac{\alpha_s^{(n_f)}(\mu)}{\pi}\right)^2
\Bigg({5\over 144}
-{1\over 12}\ln\frac{\mu^2}{m_h^2}\Bigg) 
+\left(\frac{\alpha_s^{(n_f)}(\mu)}{\pi}\right)^3\Bigg[
{42811\over 62208}
+{1\over 18}\zeta_3
\nonumber\\
&&\mbox{}
-{155\over 192}\ln\frac{\mu^2}{m_h^2}
+{49\over 576}\ln^2\frac{\mu^2}{m_h^2}
-{1\over 96}\ln^3\frac{\mu^2}{m_h^2}
+n_l\Bigg(
{35\over 3888}
+{5\over 432}\ln\frac{\mu^2}{m_h^2}\Bigg) 
\nonumber\\
&&\mbox{}
+\xi\Bigg(
-{2387\over 6912}
+{1\over 12}\zeta_3
+{121\over 576}\ln\frac{\mu^2}{m_h^2}
-{13\over 192}\ln^2\frac{\mu^2}{m_h^2}
+{1\over 96}\ln^3\frac{\mu^2}{m_h^2}\Bigg)\Bigg]
\,,
\label{eq:zeta2res}
\\
\zeta_3^{\rm MS}&=&1
+\frac{\alpha_s^{(n_f)}(\mu)}{\pi}\Bigg(
{1\over 6}\ln\frac{\mu^2}{m_h^2}\Bigg) 
+ \left(\frac{\alpha_s^{(n_f)}(\mu)}{\pi}\right)^2\Bigg(
{91\over 1152}
+{29\over 96}\ln\frac{\mu^2}{m_h^2}
+{3\over 32}\ln^{2}\frac{\mu^2}{m_h^2}\Bigg) 
\nonumber\\
&&\mbox{}
+\left(\frac{\alpha_s^{(n_f)}(\mu)}{\pi}\right)^3\Bigg[
-{284023\over 62208}
+{86183\over 27648}\zeta_3
+{99\over 128}\zeta_4
-{1\over 32}B_4
\nonumber\\
&&\mbox{}
+\left({52433\over 27648}
-{33\over 64}\zeta_3\right)\ln\frac{\mu^2}{m_h^2}
+{383\over 2304}\ln^2\frac{\mu^2}{m_h^2}
+{119\over 768}\ln^3\frac{\mu^2}{m_h^2}
\nonumber\\
&&\mbox{}
+n_l\Bigg(
{3307\over 15552}
-{1\over 12}\zeta_3
-{293\over 1152}\ln\frac{\mu^2}{m_h^2}
+{1\over 36}\ln^2\frac{\mu^2}{m_h^2}
-{1\over 96}\ln^3\frac{\mu^2}{m_h^2}\Bigg) 
\nonumber\\
&&\mbox{}
+\xi\Bigg(
-{677\over 1536}
+{3\over 32}\zeta_3
+{233\over 1024}\ln\frac{\mu^2}{m_h^2}
-{3\over 32}\ln^2\frac{\mu^2}{m_h^2}
+{3\over 256}\ln^3\frac{\mu^2}{m_h^2}\Bigg)\Bigg]
\,.
\end{eqnarray}

\begin{figure}[t]
 \begin{center}
 \begin{tabular}{c}
   \leavevmode
   \epsfxsize=14cm
   \epsffile[77 570 560 780]{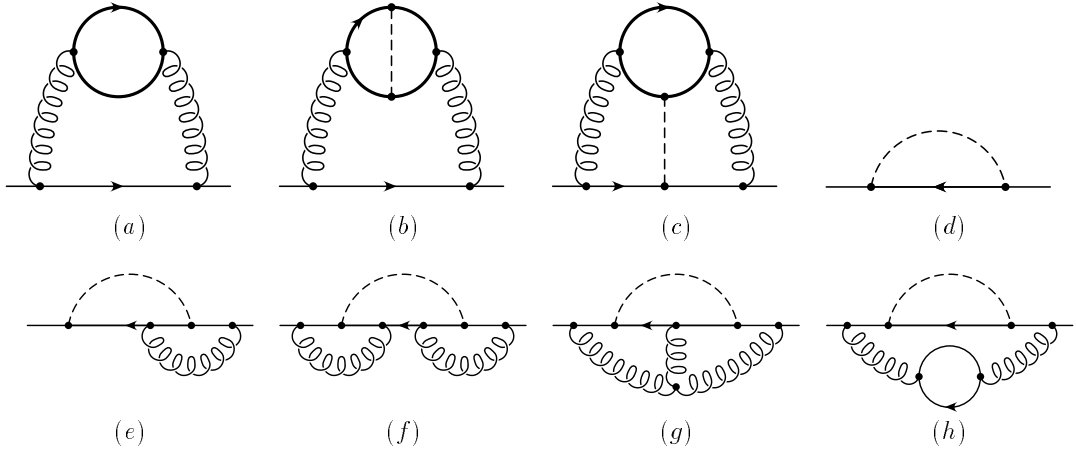}
 \end{tabular}
\caption{\label{fig:sig}
  Feynman diagrams contribution to $Z_{2}$, $Z_{m}$ and
  $\zeta_{m_q}^0$. The
  dashed line either represents the Higgs boson ($h$) or the neutral ($\chi$)
  or charged ($\phi^\pm$) Goldstone boson.
}
 \end{center}
\end{figure}

\begin{figure}[t]
 \begin{center}
 \begin{tabular}{c}
   \leavevmode
   \epsfxsize=14cm
   \epsffile[110 640 490 730]{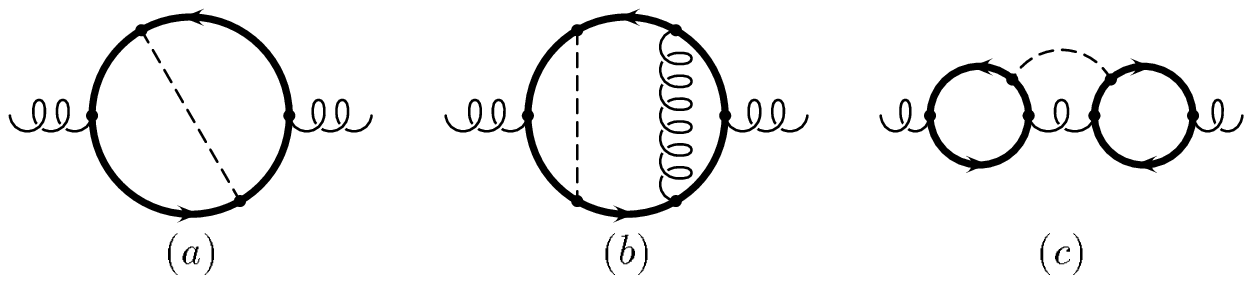}
 \end{tabular}
\caption{\label{fig:Z3}
  Feynman diagrams contribution to $Z_3$ and
  $\zeta_g^0$. The
  dashed line either represents the Higgs boson ($h$) or the neutral
  ($\chi$) or
  charged ($\phi^\pm$) Goldstone boson.
}
 \end{center}
\end{figure}

In Ref.~\cite{Ste98_higgs} the Yukawa corrections to $\zeta_m$ and 
$\zeta_g$ enhanced by the top quark mass have been
evaluated. They are conveniently expressed in terms of the variable
\begin{eqnarray}
  x_t &=& \frac{G_F m_t^2}{8\pi^2\sqrt{2}}\,,
  \label{eq:xt}
\end{eqnarray}
where $m_t$ is the top quark mass
defined in the $\overline{\rm MS}$ scheme.
Corrections proportional to $x_t$ arise if in addition to the pure QCD
Lagrangian also the couplings of the Higgs boson ($h$) and
the neutral ($\chi$)
and charged ($\phi^\pm$) Goldstone boson to the top
quark are considered.
Sample diagrams contributing to $\zeta_m$ and $\zeta_g$
are shown in Figs.~\ref{fig:sig} and~\ref{fig:Z3}, respectively.

The decoupling constant for the $u, d, s$ and $c$ quark mass 
reads\footnote{For convenience, the corrections of order $\alpha_s^2$
  are repeated from Eq.~(\ref{eq:zetamMS}).}
\begin{eqnarray}
\zeta_{m_l}^{{\rm MS},x_t} &=& 1
+
  \left(\frac{\alpha_s^{(6)}(\mu)}{\pi}\right)^2\Bigg\{
         \frac{89}{432} 
       - \frac{5}{36}\ln\frac{\mu^2}{m_t^2} 
       + \frac{1}{12}\ln^2\frac{\mu^2}{m_t^2}
\nonumber\\&&\mbox{}
+ x_t\left[
         \frac{101}{144} 
       - \frac{5}{12}\zeta_2 
       + \frac{73}{12}\zeta_3 
       - 9\zeta_4 
       - \frac{7}{6}\ln\frac{\mu^2}{m_t^2}
       + 6\zeta_3\ln\frac{\mu^2}{m_t^2} 
\right.\nonumber\\&&\mbox{}\left.
       + 2I_{3l}\left(
         - \frac{37}{18} 
         - \frac{19}{3}\zeta_3 
         + 9\zeta_4 
         - \ln\frac{\mu^2}{m_t^2} 
         - 6\zeta_3\ln\frac{\mu^2}{m_t^2} 
       \right)
  \right]
\Bigg\}
\,,
\label{eq:zetaml}
\end{eqnarray}
where $I_{3l}$ is the third component of the weak isospin, i.e. $I_{3l}=+1/2$
for up-type quarks and $I_{3l}=-1/2$ for down-type quark flavours.
For the bottom quark one receives
\begin{eqnarray}
\zeta_{m_b}^{{\rm MS},x_t} &=& \zeta_{m_d} + x_t\Bigg\{
    \frac{5}{4} 
  + \frac{3}{2}\ln\frac{\mu^2}{m_t^2}
+
  \frac{\alpha_s^{(6)}(\mu)}{\pi} \left[
      \frac{16}{3} 
    - 4\zeta_2 
    + \frac{7}{2}\ln\frac{\mu^2}{m_t^2} 
    + \frac{3}{2}\ln^2\frac{\mu^2}{m_t^2}
  \right]
\nonumber\\&&\mbox{}
+
  \left(\frac{\alpha_s^{(6)}(\mu)}{\pi}\right)^2 \left[
      \frac{472933}{12096} 
    - \frac{6133}{168}\zeta_2 
    + \frac{905}{72}\zeta_3 
    + \frac{383}{18}\zeta_4 
    + \frac{1251}{112}S_2 
    + \frac{19}{72}D_3
\right.\nonumber\\&&\mbox{}\left.
    - \frac{7}{9}B_4
    + \left(
        \frac{763}{18}
      - \frac{55}{3}\zeta_2
      - \frac{43}{4}\zeta_3
    \right)\ln\frac{\mu^2}{m_t^2}
    + \frac{529}{48}\ln^2\frac{\mu^2}{m_t^2} 
    + \frac{29}{12}\ln^3\frac{\mu^2}{m_t^2} 
\right.\nonumber\\&&\mbox{}\left.
    + n_l\left(
      - \frac{23}{24}
      + \frac{31}{36}\zeta_2
      - 2\zeta_3
      + \left(
        - \frac{241}{144}
        + \frac{2}{3}\zeta_2
      \right)\ln\frac{\mu^2}{m_t^2}
      - \frac{1}{2}\ln^2\frac{\mu^2}{m_t^2}
\right.\right.\nonumber\\&&\mbox{}\left.\left.
      - \frac{1}{12}\ln^3\frac{\mu^2}{m_t^2}
    \right)
  \right]
\Bigg\}
\,,
\label{eq:zetamb}
\end{eqnarray}
where we have used $C_F=4/3$, $C_A=3$ and $T=1/2$.
The constants
\begin{eqnarray}
S_2&=&{4\over9\sqrt3}\mbox{Cl}_2\left({\pi\over3}\right)
\,\,\approx\,\,0.260\,434
\,,
\nonumber\\
D_3&=&6\zeta_3-\frac{15}{4}\zeta_4
     -6\left(\mbox{Cl}_2\left({\pi\over3}\right)\right)^2
\,\,\approx\,\,-3.027\,009
\,,
\nonumber\\
B_4&=&16\li\left({1\over2}\right)-{13\over2}\zeta_4-4\zeta_2\ln^22
+{2\over3}\ln^42
\,\,\approx\,\,-1.762\,800
\,,
\end{eqnarray}
where $\zeta_4=\pi^4/90$, $\mbox{Cl}_2$ is Clausen's function and 
$\mbox{Li}_4$ is the quadrilogarithm,
occur in the evaluation of the three-loop master
diagrams~\cite{Bro92,Avdrho,CheKueSte95rho,Bro98}.

Finally, for $\zeta_g$ we obtain the following result:
\begin{eqnarray}
\left(\zeta_g^{{\rm MS},x_t}\right)^2 &=& 1 +
  \frac{\alpha_s^{(6)}(\mu)}{\pi}\,T\,\Bigg\{
       - \frac{1}{3} \ln\frac{\mu^2}{m_t^2}
+
  \frac{\alpha_s^{(6)}(\mu)}{\pi} \left[
    C_F\,\left(
       - \frac{13}{48}
       + \frac{1}{4} \ln\frac{\mu^2}{m_t^2} 
    \right)
\right.\nonumber\\&&\left.\mbox{}
    +C_A\,\left(
         \frac{2}{9}
       - \frac{5}{12} \ln\frac{\mu^2}{m_t^2}
    \right)
       + T\,\frac{1}{9} \ln^2\frac{\mu^2}{m_t^2}
  \right]
+ x_t\Bigg\{
       - \frac{2}{3}
       + \ln\frac{\mu^2}{m_t^2} 
\nonumber\\&&\mbox{}
+
  \frac{\alpha_s^{(6)}(\mu)}{\pi} \left[
     C_F\,\left(
       - \frac{17}{16}
       + \frac{5}{4}\zeta_2
       + \frac{25}{8}\zeta_3
       - 3 \ln\frac{\mu^2}{m_t^2} 
       + \frac{3}{4} \ln^2\frac{\mu^2}{m_t^2}
     \right)
\right.\nonumber\\&&\left.\mbox{}
     +C_A\,\left(
       - \frac{5}{4} 
       + \frac{3}{8}\zeta_2
       - \frac{95}{64}\zeta_3
       + \frac{7}{4} \ln\frac{\mu^2}{m_t^2} 
     \right)
     +T\,\left( 
         \frac{5}{4}
       + \frac{7}{8}\zeta_3
       + \frac{4}{9} \ln\frac{\mu^2}{m_t^2} 
\right.\right.\nonumber\\&&\left.\left.\mbox{}
       - \frac{2}{3} \ln^2\frac{\mu^2}{m_t^2} 
     \right) 
       - \frac{7}{2}\zeta_3 T 
    \right]
  \Bigg\}
\Bigg\}
\label{eq:zetag}
\,,
\end{eqnarray}
where the contribution of the diagrams in  
Fig.~\ref{fig:Z3}(c) corresponds to the last entry
in the last line of Eq.~(\ref{eq:zetag}). 
For convenience also the pure QCD result of ${\cal O}(\alpha_s^2)$
is listed. The corresponding three-loop terms can be found 
in~\cite{CheKniSte98}.

In the remainder of this Appendix we want to provide the analytical
results for the $x_t$-enhanced corrections of order $G_Fm_t^2$ to the
coefficient functions $C_1$ and $C_2$~\cite{Ste98_higgs}.
For $C_1$ we have
\begin{eqnarray}
C_1 &=& -\frac{1}{6}\,T\,\frac{\alpha_s^{(6)}(\mu)}{\pi}\Bigg\{
  1 
  - 3 x_t
+ \frac{\alpha_s^{(6)}(\mu)}{\pi}\left[
  - C_F\,\frac{3}{4}
  + C_A\,\frac{5}{4}
  - T\,\frac{1}{3} \ln\frac{\mu^2}{m_t^2} 
\right.\nonumber\\&&\left.\mbox{}
  + x_t\left(
    C_F\,\left(
        9
      - \frac{9}{2} \ln\frac{\mu^2}{m_t^2} 
    \right)
    - C_A\,\frac{21}{4}
    +T\,\left(
      - \frac{2}{3} 
      + 2 \ln\frac{\mu^2}{m_t^2}
    \right)
  \right)
\right]
\Bigg\}
\,,
\label{eqC1}
\end{eqnarray}
where $m_t$ is the $\overline{\rm MS}$ top quark.
The ${\cal O}(\alpha_sx_t)$ terms can be found
in~\cite{DjoGam94,CheKniSte97hbb} and the
${\cal O}(\alpha_s^2)$ results were 
computed in~\cite{Inami:1983xt,Djouadi:1991tk}
The corrections of ${\cal O}(\alpha_sx_t^2)$ are taken
from~\cite{Ste98_higgs}.

For the light quarks we get for $C_2$
\begin{eqnarray}
C_{2l} &=& 1+
\left(\frac{\alpha_s^{(6)}(\mu)}{\pi}\right)^2 \left[
    \frac{5}{18} 
  - \frac{1}{3}\ln\frac{\mu^2}{m_t^2}
  + x_t \left(  \frac{7}{3} 
        - 12\zeta_3 
        + 2I_{3l}\left(
            2
          + 12\zeta_3
        \right)
  \right)
\right]
\,,
\nonumber\\
\end{eqnarray}
and in the case of the bottom quark the coefficient function reads:
\begin{eqnarray}
C_{2b} &=& C_{2d} + x_t\Bigg\{
    - 3
+
  \frac{\alpha_s^{(6)}(\mu)}{\pi} \left[
    - 7 
    - 6\ln\frac{\mu^2}{m_t^2}
  \right]
+ 
\left(\frac{\alpha_s^{(6)}(\mu)}{\pi}\right)^2 \left[
     - \frac{12169}{144} 
     + \frac{110}{3}\zeta_2 
\right.\nonumber\\&&\left.\mbox{}
     + \frac{43}{2}\zeta_3
     - \frac{89}{2}\ln\frac{\mu^2}{m_t^2} 
     - \frac{55}{4}\ln^2\frac{\mu^2}{m_t^2} 
     + n_l\left(
         \frac{241}{72}
       - \frac{4}{3}\zeta_2
       + 2\ln\frac{\mu^2}{m_t^2}
\right.\right.\nonumber\\&&\left.\left.\mbox{}
       + \frac{1}{2}\ln^2\frac{\mu^2}{m_t^2}
     \right)
  \right]
\Bigg\}
\,.
\end{eqnarray} 
In~\cite{CheKniSte97hbb} $C_{2l}$ and $C_{2b}$ are listed for general
gauge group $SU(N_c)$.


\section{\label{app:as4m4}Analytical results for $R(s)$}
\setcounter{equation}{0} 
\setcounter{figure}{0} 
\setcounter{table}{0} 

As in the literature the quadratic and quartic correction terms to
$R(s)$ are not yet available in analytic form we want to provide the
corresponding results
using the notation introduced in Eqs.~(\ref{eq:RQ}) and~(\ref{eq:RQ2}).
For completeness also the massless approximation is given. Of course,
it is the same in the $\overline{\rm MS}$ and on-shell schemes
\begin{eqnarray}
    r_0 &=& 1 + {\alpha_s\over \pi} + \left({\alpha_s\over
    \pi}\right)^2\,\bigg[ {365\over 24} - 11\,\zeta_3 + n_f\,\bigg(
  -{11\over 12} + {2\over 3}\,\zeta_3 \bigg) \bigg]
\nonumber\\&&\mbox{}
      + \left({\alpha_s\over \pi}\right)^3\,\bigg[
          {87029\over 288} 
          - {121\over 8}\,\zeta_2 
          - {1103\over 4}\,\zeta_3 
          + {275\over 6}\,\zeta_5
          + n_f\,\bigg(
              -{7847\over 216} 
              + {11\over 6}\,\zeta_2 
\nonumber\\&&\mbox{}
              + {262\over 9}\,\zeta_3 
              - {25\over 9}\,\zeta_5
              \bigg) 
          + n_f^2\,\bigg(
              {151\over 162} 
              - {1\over 18}\,\zeta_2 
              - {19\over 27}\,\zeta_3
              \bigg) 
          \bigg]
  \,,
\end{eqnarray}

\begin{eqnarray}
  r_{Q,2}^{\rm OS} &=&
    {M_Q^2\over s}\,{\alpha_s\over \pi}\,\bigg[
    12 + {\alpha_s\over \pi}\,\left(
       \frac{189}{2} + 24\lMs - \frac{13}{3} n_f
    \right)
    + \left({\alpha_s\over \pi}\right)^2\,\left(
       \frac{22351}{12} 
     - \frac{967}{2}\zeta_2
  \right.\nonumber\\&&\left.\mbox{}
     - 16\zeta_2\ln2
     + \frac{502}{3}\zeta_3
     - \frac{5225}{6}\zeta_5
     + 378\lMs 
     - 9\lMs^2 
     + n_f\left(
         -\frac{8429}{54} 
         + 42\zeta_2 
  \right.\right.\nonumber\\&&\left.\left.\mbox{}
         - \frac{466}{27}\zeta_3 
         + \frac{1045}{27}\zeta_5
         - \frac{52}{3}\lMs
         + 2\lMs^2 
          \right)
     + n_f^2\left(\frac{125}{54} - \frac{2}{3}\zeta_2 \right) 
    \right)
    \bigg]
    \,,
  \nonumber\\
r_{qQ,2}^{\rm OS} &=&
  {M_Q^2\over s}\,\left({\alpha_s\over \pi}\right)^3\,
  \bigg[ -80 +60\zeta_3 + n_f\left(\frac{32}{9}-\frac{8}{3}\zeta_3\right)
  \bigg]
  \,,
\end{eqnarray}
where $\lMs=\ln M_Q^2/s$.
\begin{eqnarray}
  r_{Q,4}^{\rm OS} &=&
    \left({M_Q^2\over s}\right)^2\,\bigg[
      -6
      + {\alpha_s\over \pi}\left(
        10 - 24\lMs
      \right) 
      + \left({\alpha_s\over \pi}\right)^2\left(
        \frac{206}{3} 
      + 218\zeta_2
      + 16\zeta_2\ln2
      + 104\zeta_3
  \right.\nonumber\\&&\left.\mbox{}
      - \frac{311}{2}\lMs
      - 15\lMs^2 
      + n_f\left(
            -\frac{35}{9} 
            - 12\zeta_2
            - \frac{8}{3}\zeta_3
            + 9\lMs 
            - 2\lMs^2 
           \right) 
      \right) 
  \nonumber\\&&\mbox{}
      + \left({\alpha_s\over \pi}\right)^3\left(
        \frac{91015}{108} 
      - \frac{76}{9}\ln^4 2 
      + \frac{2564287}{540}\zeta_2
      - \frac{4568}{9}\zeta_2\ln 2
      - \frac{128}{3}\zeta_2\ln^2 2
  \right.\nonumber\\&&\left.\mbox{}
      + \frac{56257}{18}\zeta_3
      - \frac{1439}{3}\zeta_2\zeta_3
      - \frac{1565}{6}\zeta_4
      - \frac{3770}{3}\zeta_5
      - \frac{608}{3}a_4
      + \lMs \left(-\frac{5536}{3}
            + 564\zeta_2
  \right.\right.\nonumber\\&&\left.\left.\mbox{}
            - 24\zeta_2\ln 2
            + 416\zeta_3
            \right)
      - \frac{591}{4}\lMs^2
      + \frac{15}{2}\lMs^3
      + n_f\left(
            - \frac{21011}{216} 
            + \frac{8}{27}\ln^4 2
            - \frac{3544}{9}\zeta_2
  \right.\right.\nonumber\\&&\left.\left.\mbox{}
            - \frac{176}{9}\zeta_2\ln 2
            + \frac{32}{9}\zeta_2\ln^2 2
            - \frac{2323}{9}\zeta_3
            + \frac{700}{9}\zeta_4
            + \frac{440}{9}\zeta_5
            + \frac{64}{9} a_4
            + \lMs\left(
                  \frac{2419}{12} 
  \right.\right.\right.\nonumber\\&&\left.\left.\left.\mbox{}
                + \frac{44}{3}\zeta_2
                + \frac{16}{3}\zeta_2 \ln2
                + \frac{28}{3}\zeta_3
                \right) 
            - \frac{157}{6}\lMs^2
            - \frac{2}{3}\lMs^3
            \right)
      + n_f^2\left(
             \frac{35}{18} 
           + \frac{25}{3}\zeta_2
           + \frac{112}{27}\zeta_3
  \right.\right.\nonumber\\&&\left.\left.\mbox{}
           + \lMs\left(
                 - \frac{94}{27} 
                 - \frac{8}{3}\zeta_2
                 \right)
           + \frac{13}{9}\lMs^2
           - \frac{2}{9}\lMs^3
           \right)
      \right)
  \bigg]
  \,,
  \nonumber\\
   r_{qQ,4}^{\rm OS} &=&
  \left({M_Q^2\over s}\right)^2\,
  \left({\alpha_s\over \pi}\right)^2\,\bigg[
     \frac{13}{3} - 4\zeta_3 - \lMs 
   + {\alpha_s\over \pi} \, \left(
       - \frac{4217}{48} 
       + 15\zeta_2
       + \frac{139}{3}\zeta_3
       + \frac{50}{3}\zeta_5
  \right.\nonumber\\&&\left.\mbox{}
       + \lMs\left(\frac{97}{4} - 38\zeta_3\right)
       - 2\lMs^2 
       + n_f\left(
             \frac{457}{108} 
           - \frac{2}{3}\zeta_2
           - \frac{22}{9}\zeta_3
  \right.\right.\nonumber\\&&\left.\left.\mbox{}
           + \lMs\left(
                  - \frac{13}{18} 
                  + \frac{4}{3}\zeta_3
                 \right)
            \right) 
   \right)
  \bigg]
  \,,
\end{eqnarray}
with $a_4=\mbox{Li}_4(1/2)\approx 0.517\,479$.


\end{appendix}


\vspace{1em}

\noindent
{\bf Note added:}\\
In the meantime a third independent
evaluation of the order $\alpha^2$ QED corrections to the 
muon decay became available \cite{CzaMel01_2}.


\end{document}